\documentclass[12pt,a4paper]{report}
\usepackage{a4,afterpage,amssymb,bbm,rotating,multirow,array,graphicx,psfrag}
\usepackage{amscd}

\def\Bizon{{Bizo\'{n}}}
\def\Purrer{P\"{u}rrer}
\def\Goncalves{{Gon\c{c}alves}}
\def\Gomez{{G\'{o}mez}}
\def\Universitat{Universit\"{a}t}  

\newcommand{\be}{\begin{equation}}
\newcommand{\ee}{\end{equation}}
\newcommand{\bd}{\begin{displaymath}}
\newcommand{\ed}{\end{displaymath}}
\newcommand{\ba}{\begin{array}}
\newcommand{\ea}{\end{array}}

\newcommand{\bea}{\begin{eqnarray}}
\newcommand{\eea}{\end{eqnarray}}
\newcommand{\bi}{\begin{itemize}}
\newcommand{\ei}{\end{itemize}}

\newcommand{\ccbeta}{\eta}
\newcommand{\Vr}{\frac{V}{r}}
\newcommand{\Vrn}{(\frac{V}{r})_n}

\newcommand{\lllrightarrow}{-\hspace{-1ex}-\hspace{-1ex}-\hspace{-1ex}-%
             \hspace{-1ex}-\hspace{-1ex}-\hspace{-2ex}\rightarrow}

\def\dal{\hbox{\hskip 0.5mm\hbox{\vrule width2.3mm height0.2mm
\vbox{\hrule width0.3mm height2.6mm}\hskip
-2.6mm \vbox{\hbox{\vrule width2.6mm height0.1mm}
\vskip -0.1mm\hrule width0.1mm height2.6mm}}\hskip 0.5mm}}

\newcommand{\PreserveBackslash}[1]{\let\temp=\\#1\let\\=\temp}
\begin{document}
\title{Staticity, Self-Similarity and Critical Phenomena in a
Self-Gravitating Nonlinear $\sigma$ Model}
\author{
\mbox{  }\\
\mbox{  }\\
Dissertation\\
zur Erlangung des akademischen Grades\\
Doktorin der Naturwissenschaften\\
an der Formal- und Naturwissenschaftlichen Fakult\"at\\
der Universit\"at Wien\\[2.5cm]
eingereicht von\\[0.5cm]
{\sc Christiane Lechner}\\ \mbox{ }}
\date{im Oktober 2001}   

\maketitle
{}
\newpage
\thispagestyle{empty}
\mbox{  }\\[3cm]  
\begin{flushright}  
Kunst ist sch\"on, \\
macht aber viel Arbeit.\\[0.5cm]
{\em Karl Valentin}
\end{flushright}
\newpage
{}

\begin{center}
\large{ZUSAMMENFASSUNG}
\end{center}

Ziel dieser Arbeit ist eine dataillierte Untersuchung spezieller Aspekte der
Dynamik einer Klasse nichtlinearer Materiefelder im Rahmen der allgemeinen
Relativit\"atstheorie. Von besonderem Interesse hier sind statische
L\"osungen (unter Ber\"ucksichtigung einer positiven kosmologischen
Konstante), selbst\"ahnliche L\"o\-sun\-gen sowie die Bildung schwarzer
L\"ocher.

Selbstgravitierende Materiefelder zeigen f\"ur Anfangsdaten, die an der
Schwelle zum Kollaps liegen, ein Verhalten 
(sogenannte ``kritische Ph\"anomene''), das durch
``Scaling'', Selbst\"ahnlichkeit (bzw. Statizit\"at) und
Universalit\"at charakterisiert ist .
Das hier -- in sph\"arischer Symmetrie -- untersuchte
SU(2) $\sigma$ Modell, ist in diesem Zusammenhang
von besonderem Interesse, da die dimensionslose
Kopplungskonstante in nichttrivialer Weise in die Theorie eingeht. Ziel
dieser Arbeit ist, kritische Ph\"anomene, insbesondere den Skalenexponenten
und die kritische L\"osung, in Abh\"angigkeit der Kopplung zu
untersuchen.

Die Untersuchung erfolgt in zwei Schritten. Zun\"achst
werden selbst\"ahnliche L\"osungen (unter Zuhilfenahme der Symmetrie)
numerisch konstruiert
und deren Stabilit\"at untersucht (Kapitel 4). Wir reproduzieren die
Ergebnisse von Bizon et al., die f\"ur kleine Kopplungen eine
einparametrige Familie von kontinuierlich selbst\"ahnlichen (CSS) 
L\"osungen konstruiert haben. Wir zeigen, da\ss{} die erste
Anregung dieser Familie eine instabile Mode hat. Weiters konstruieren wir
eine diskret selbst\"ahnliche (DSS) L\"osung f\"ur gro\ss e
Kopplungen, die ebenfalls eine instabile Mode hat. Wir stellen die
Hypothese auf, da\ss{} die DSS L\"osung bei einem bestimmten Wert der Kopplung
aus der ersten CSS Anregung in einer homoklinen Loopbifurkation entsteht.

Im zweiten Schritt werden einparametrige Familien von Anfangsdaten 
numerisch in der Zeit zu
entwickelt (Kapitel 5). Nahekritische Anfangsdaten erh\"alt man durch
Bisektion.
Die kritische L\"osung kann an dem
(zeitlich) intermedi\"aren Verhalten der nahekritischen 
Evolutionen abgelesen
werden. Der Skalenexponent ist durch die Masse der schwarzen L\"ocher
in Abh\"angigkeit des Parameters der Anfangsdaten bestimmt.
Die so erhaltenen Ergebnisse stimmen sehr gut mit den Eigenschaften der oben 
konstruierten
selbst\"ahnlichen L\"osungen \"uberein. F\"ur kleine Kopplungen ist
die kritische L\"osung die erste CSS Anregung. F\"ur gro\ss e
Kopplungen  ist es die DSS L\"osung.
F\"ur mittlere Kopplungen finden wir einen \"Ubergang von CSS zu
DSS in der kritischen L\"osung. Dieser \"Ubergang ist mit der Hypothese der
homoklinen Loopbifurkation konsistent. 

Zus\"atzlich wird
in dieser Arbeit \"uber statische L\"osungen des Modells (mit positiver
kosmologischer Konstante) berichtet (Kapitel 3).    
%%%%%%%%%%%%%%%%%%%%%%%%%

\begin{center}
\large{\bf Abstract}
\end{center}
%
%\addcontentsline{toc}{section}{Abstract}
%
The aim of this work is to study certain aspects of the dynamics
of a class of self-gravitating non-linear matter fields.
In particular we concentrate
on soliton solutions (in the presence of a positive cosmological constant),
self-similar solutions and the formation of black holes.

The dynamics of initial data at the threshold of black hole formation
are characterized by phenomena (so-called {\em critical phenomena})
including scaling, self-similarity (resp.
staticity) and universality. 
In this work we concentrate on SU(2) $\sigma$ models
coupled to gravity in spherical symmetry. 
These models are interesting due to a dimensionless
parameter -- the coupling -- which enters the theory non-trivially.
The aim is to investigate how critical phenomena -- in particular the
critical solution and the scaling exponent -- depend on the coupling.

We use two essentially different methods to study the 
threshold behavior: 
Making use of the symmetry both discrete (DSS) and continuous (CSS)
self-similar solutions are constructed (numerically) 
by solving boundary value problems (Chapter 4). The stability of these
solutions is studied. For small couplings, 
reproducing results of Bizon et al., we 
find a discrete one-parameter family of CSS solutions. 
Of particular interest is the first
CSS excitation. For large couplings we construct a DSS solution.
Both solutions have one unstable mode. We conjecture, that the DSS solution
bifurcates from the CSS solution in a homoclinic loop bifurcation at 
some value of the coupling constant.
 
The second method consists of evolving one parameter families of
initial data numerically (Chapter 5). By a bisection search the initial 
data are fine-tuned such that they are close to the threshold. 
The critical solution then is determined by the intermediate asymptotics
of near-critical data. The scaling exponent is determined from
the black hole mass as a function of the parameter in
the initial data. Our results 
are in very good agreement with the results on the self-similar
solutions we obtained above.
For small
couplings the critical solution is CSS, for large
couplings it is DSS and for intermediate 
couplings we find a new transition from CSS to DSS in the critical solution,
which shows ``episodic CSS''. This transition is consistent with the 
hypothesis of the homoclinic loop bifurcation.

In addition this work also contains results on 
static solutions of the model
in the presence of a positive cosmological constant $\Lambda$ (Chapter 3).

%%%%%%%%%%%%%%%%%%

\begin{center}
\large{\bf Preface}
\end{center}

This work originates in a project application to the FWF 
(Fonds zur F\"orderung Wissenschaftlicher Forschung), written by Sascha Husa
and me under the supervision of Prof.~P.~C.~Aichelburg in 1997.
The FWF kindly supported this project (at least in part), 
which started in March (resp.~May) 1998 (Project Nr. P12754-PHY). 
Michael P\"urrer 
joined our group in summer 1998 as a diploma student.
In September 1998 Sascha Husa left for Pittsburgh and 
Jonathan Thornburg took his place in the project. 

With the exception of Michael P\"urrer's work on the massless Klein-Gordon
field with a compactified radial coordinate, I try to describe 
the main results of the project as complete as posssible. 
Naturally, as the project is the
joint work of five people with woven contributions, 
not all of the work presented here was done by myself.
Whenever I used the method  of someone else in 
the group (e.g.~a code) or present results of runs, 
that were done by others, I mention this fact at the appropriate place.

I am very greatful to all five members of this group. 
Apart from everyone's scientific contributions to the project, 
I thank Prof.~P.~C.~Aichelburg for chairing our numerous
discussions, Sascha Husa for the work on the original application, 
for inviting me to Pittsburgh and for still being interested in
collaborating with me, Jonathan Thornburg for his counsel 
regarding the details
of the construction of the DSS solution and for always lending a 
helping hand in anything concerning computers, 
Michael P\"urrer for his computer assistance and the beers we had,
and myself for surviving all this.

I think we all are very greatful to Piotr Bizon, who shared 
his results in advance of publication, for various discussions and his
continuous interest in our work. I personally thank him for 
reading the manuscript of the paper on cosmological solitons and for
inviting me to Krakow and introducing me to his family.

For reading and correcting the manuscript of this thesis I thank
Prof.~P.~C.~Aichelburg and Sascha Husa, as well as Jonathan Thornburg (who
read the sections on the stability of self-similar solutions and 
the description of the DICE code), Prof.~R.~Beig for pointing out some
mistakes in the first two chapters and Walter Simon and Mark Heinzle for
suggestions concerning the introduction.

I am thankful to the gravity group at the institute with its 
broad range of interests, which gave me the chance to get into 
contact with a variety of topics through the regular seminars of the 
group. I am especially thankful to Prof.~R.~Beig for the seminar 
on ``The Cauchy problem for the Einstein Equations'', which gave me the
opportunity to study (parts of) the article of the same name in some 
detail and to present it in some length. 

Thanks to all students of the gravity group. Especially
to Natascha Riahi and Mark Heinzle, who were forced to 
listen to the weekly discussions of our project because of working in 
the same room. Thanks for housing the coffee machine, for the frequent 
conversations on all sorts of topics and for the feeling of not being the
only fool. To Walter Simon for the impetus 
to organize various gatherings and barbecues. Thanks also for the room -- it
was essential to work at some spatial distance from the center of the
typhoon! Special thanks to N.~N.~for the letters at my door.
To all of us, who helped organizing the ``Gschnas''. 
It was great fun and I think, we are rather good at this now. 
Thanks to Herbert Balasin for open ears 
and to Peter Steier for his advice 
on how to complete the PhD studies on the bureaucratic side.

[[ ... private part is cut ...]]
%%%%%%%%%%%%%%%%%%%%

\newpage
\tableofcontents
\newpage

\chapter{Introduction}
%%%%%%%%%%%%%%%%%%%%%%

The theory of general relativity describes gravity in terms of
the curvature of the four dimensional Lorentzian manifold 
representing spacetime.
The Einstein equations, $G_{\mu\nu} = \kappa
T_{\mu\nu}$, relating matter to the curvature of spacetime
involve the geometric objects $T_{\mu\nu}$, the stress energy tensor of
matter, and $G_{\mu\nu}$, the Einstein tensor. Behind this 
simple geometric formulation there 
hides a coupled system
of ten quasi-linear partial differential equations (PDEs), with the
components of the metric as dependent variables. In addition, the matter
fields themselves are subject to some field equations, 
which in turn contain the
metric and its derivatives. These field equations therefore have to 
be solved simultaneously with the Einstein equations. 
Of special physical interest is to formulate and solve the initial value
problem. 
In the ``3+1'' formulation for example, initial data consist of two symmetric 
three-tensor fields given on an
initial spacelike hypersurface, which are subject to 
four constraint equations.
The remaining six Einstein equations are used to evolve these initial data
in time. 
One question of interest is for example: given smooth initial data, what is
the long time behavior of the solution?
There are analytic results showing global existence for sufficiently
``weak'' initial 
data\footnote{In particular \cite{Christodoulou-Klainerman}. For a review see \cite{Rendall-2000-living-reviews}.}.
On the other hand, singularity theorems\footnote{See 
\cite{Penrose-singularity-theorems-1965} and \cite{Hawking-Ellis}.} 
predict that sufficiently ``strong'' 
initial data develop a singularity in finite time. 
If the cosmic censorship hypothesis holds,
these singularities should be shielded by a horizon such that they
are invisible to distant observers.
However,
according to the complexity of the equations it is clear that i)
only a very small number of exact solutions is known and ii) it is very
difficult to get an analytic handle on the equations.
It is therefore both necessary and extremely fruitful to combine 
analytic approaches with a numerical treatment of the equations.

An important step to understand the dynamics
is to study the possible ``end states'' for the given matter model.
It is reasonable to assume that regular initial data will asymptote to
a stationary solution at late times, which could be e.g.~a stationary
stable black hole, a stable soliton solution (corresponding to a star) or 
dispersion leading to Minkowski spacetime. 
Stationarity reduces the equations to an elliptic
system. If additional symmetries are imposed, the problem simplifies further.
In particular a static soliton or black hole solution, which 
is spherically symmetric, is obtained by solving a coupled system of
(nonlinear) ordinary differential equations (ODEs).
Initiated by the work of Bartnik and McKinnon \cite{Bartnik-McKinnon-1988}, 
who numerically construced static soliton solutions to the 
Einstein Yang-Mills (EYM) system 
in spherical symmetry, several self-gravitating matter models have been
studied with regard to static soliton or black hole 
solutions\footnote{Most of the work was done in spherical symmetry. 
The Einstein Yang-Mills system 
was also investigated in axis symmetry. In addition 
there are investigations 
concerning slowly rotating, that is stationary solutions to this model, 
obtained as linear rotational perturbations of the 
Bartnik McKinnon solutions and the coloured black holes.
For details and an extensive list of references see 
the review article by Volkov and Gal'tsov
\cite{Volkov-Galtsov-review-article-1998}}. 
Unfortunately most of the (non-trivial) solutions found are unstable, so
they are not relevant for the late time behavior of general initial data.
However, a few years later, it turned out that static solutions with a single
unstable mode play an important role as intermediate attractors
in type I critical collapse, as explained below.

Another fascinating field of research was started by the work of 
Choptuik \cite{Choptuik-1992-in-dInverno,Choptuik-1993-self-similarity},
who investigated the threshold of black hole formation for the 
self-gravitating massless Klein-Gordon system in spherical symmetry.
Due to the analytic work by Christodoulou\footnote{For references see
e.g.~\cite{Gundlach-1999-critical-phenomena-living-reviews}.} it was known
that ``small'' (in a well defined sense) initial data disperse, such that
in the long time evolution such data approach Minkowski spacetime,
whereas ``strong'' initial data form a black hole. 
Choptuik numerically evolved one parameter families of initial data,
that interpolated between black hole formation and dispersion.  
He fine-tuned the parameter such that a ``tiny'' change
in this parameter changed the end state from black hole formation to
dispersion or vice versa .
Speaking in the language of dynamical systems, the space of initial data
(of the model under investigation) is divided into basins of attraction, the
attractors in this case being black holes and Minkowski spacetime.
From this point of view, Choptuik studied the boundary between two such
basins of attraction.
The phenomena he found, called 
{\em critical phenomena}\footnote{For details and 
review articles see Chapter~\ref{chap::criticalCollapse}.}, can be summarized
by the keywords {\em scaling, self-similarity} and {\em universality}.
Scaling relates the black hole mass to the parameter $p$ in the initial data
via the simple power law  $m_{BH} \propto (p - p^*)^{\gamma}$, 
where $p^*$ is the critical parameter of the family.
In particular this means that 
the black hole masses can be made arbitrarily small by fine-tuning the
initial data. Furthermore, all near critical evolutions 
approach a self-similar solution at
intermediate times. Studying several families of initial data in this way
Choptuik found that 
these phenomena, especially the scaling
exponent $\gamma$ and the self-similar solution are {\em independent} of the
family. So critical phenomena are universal within a given model.
The numerical resolution of the phenomenon of
self-similarity required sophisticated methods. 
Encouraged by Choptuik's results several other models were investigated with
respect to critical phenomena. Most of the work concentrated on spherical
symmetry with the early exception of Abrahams and Evans
\cite{Abrahams-Evans-1993}, who
considered axially symmetric graviational waves.
All models exhibit the phenomena described above. The scaling exponent
$\gamma$ and the self-similar solution turned out to depend on the model.
In particular there are models were the intermediate attractor is
continuously self-similar (CSS) and other models, where it is
discretely self-similar (DSS). Up to now it is not clear what causes the
symmetry to be discrete or continuous.
The results by Bizon et al.~\cite{Bizon-Chmaj-Tabor-1999-sigma-3+1-evolution},
who studied threshold phenomena of the SU(2) $\sigma$ model in flat space,
indicate that critical phenomena are not tied to the Einstein equations, but
are rather a general feature of hyperbolic PDEs. 
On the other hand, up to now
it is not clear whether there are models in flat space which allow for a 
discretely self-similar solution, or whether
DSS is a special feature of the Einstein equations.

It is worth noting that 
Choptuik's numerical work contributed much to the general 
understanding of the Einstein equations. 
In particular it stimulated the (semi-analytic) study of self-similar
solutions and their stability properties.
It turned out that the self-similar solutions at the threshold of black hole
formation have one unstable mode. Furthermore a scaling argument
relates the eigenvalue $\lambda$ of the unstable mode to the scaling exponent 
$\gamma$ via $\gamma = 1/\lambda$.
Taking these results together, one has a good (although not rigorous)
understanding of how critical phenomena emerge.

Apart from the phenomena found by Choptuik, some models in addition 
give rise to another type of critical behavior. There
the intermediate attractor is a static (or oscillating) 
solution with one unstable mode. The black hole masses
formed by slightly super-critical data are not arbitrarily small but
are a finite fraction of the mass of the static solution. 
In analogy to statistical physics this
kind of critical behavior, where the mass as a function of the parameter $p$
is discontinuous, was called {\em type I}, whereas the phenomena found by
Choptuik, where the mass is a continuous function of $p$ are called 
{\em type II}.

We remark that the study of type II critical phenomena is entirely based
on classical general relativity, ignoring the fact, that for regions with
very strong curvature, which necessarily occur in type II critical collapse,
the classical theory should be replaced by a quantum theory. 

This work is part of a project, which originally aimed at investigating 
critical phenomena of the self-gravitating $\sigma$ model in the presence of
a cosmological constant. From the results of 
previous work \cite{sigma-on-de-Sitter} 
concerning static
solutions of the model on de Sitter background it was reasonable
to expect that these solutions would persist when gravity 
is ``turned on''. 
In particular one of the static solutions constructed in
\cite{sigma-on-de-Sitter} has one unstable mode, which would therefore
be a candidate for type I critical collapse\footnote{In the coupled
situation the ``end states''
could be de Sitter space and Schwarzschild-de-Sitter (Kottler) space.}. 
It would then
have been possible to investigate the coexistence of type I and type II
critical phenomena on one hand, and the dependece of these phenomena on a
dimensionless parameter (see below) on the other hand.
Therefore we started to investigate existence and stability 
of static solutions in the presence of a positive cosmological 
constant $\Lambda$. 
However, because unexpected new phenomena emerged, 
our attention focused on
critical phenomena in the asymptotically
flat situation. The work on static solutions with $\Lambda$
therefore is rather isolated from the rest of this thesis.
  
The main part of this thesis concentrates on type II
critical phenomena of the self-gravitating SU(2)
$\sigma$ model in spherical symmetry (without cosmological constant). 
We investigate the phenomena both by evolving one parameter families 
of initial data and doing a bisection search to fine-tune the parameter,
and by ``directly'' constructing (i.e.~by imposing the symmetry on the
equations and solving the resulting reduced problem)
the relevant self-similar solutions
and studying their stability properties.

The motivations for choosing the SU(2) $\sigma$ model as the matter model
are, that it is a very {\em simple}
model (in spherical symmetry the field equations reduce to a single
nonlinear wave equation) and that the theory contains a {\em dimensionless
parameter}, the coupling constant $\eta$. Using dimensional analysis
one can expect, that type II critical phenomena and the spectrum of
self-similar solutions depend strongly on this parameter.
This expectation is supported by previous work on the limits of strong 
\cite{Liebling-inside-global-monopoles} 
and weak coupling \cite{Bizon-Chmaj-Tabor-1999-sigma-3+1-evolution} 
and \cite{Liebling-Hirschmann-Isenberg-1999-sigma-critical}, 
where the solutions at the
threshold are DSS in the limit $\eta \to \infty$, and CSS in the limit of
vanishing coupling.
This model therefore gives the chance to find out more about the
mechanisms that are responsible e.g.~for the realization of continuous
self-similarity as opposed to discrete self-similarity.
In particular the expected transition of the critical solution from CSS to
DSS as the coupling is increased is of major interest.

This work is organized as follows: in Chapter
\ref{chap::sigma} the self-gravitating SU(2) $\sigma$
model is introduced. We give
the basic definitions and equations, that
are necessary for the work on static solutions, self-similar solutions and
critical collapse. The Einstein equations (with and without
cosmological constant) and field equations
are given with respect to coordinates, that are adapted to spherical
symmetry. In particular the time evolution code DICE (see
App.~\ref{app::dice}) is based on a characteristic formulation of the
initial value problem. The coordinates adjusted to this formulation
(Bondi coordinates) are discussed in this chapter.

Chapter \ref{chap::statSolutions} deals with static solutions of the model
in the presence of a positive cosmological constant $\Lambda$. We discuss
the static equations, investigate the possible global structures of
solutions and describe the numerical results. The limit $\eta \to
\eta_{max}$, the maximal value of the coupling constant for which solutions
exist, is carried out with care.

Chapter \ref{chap::SSSolutions} deals with self-similar solutions of the
model. We introduce the concept of self-similarity and its 
manifestation in Bondi coordinates. The main part of this chapter is
dedicated to 
the numerical construction and linear stability analysis of 
CSS solutions and a discretely self-similar solution.
The equations are given in adapted coordinates, the numerical methods for
constructing the solutions are explained, 
results are discussed and stability
of the solutions is investigated. 
In the last section we summarize which self-similar
solutions might be relevant for the dynamics, 
either as ``end states'' or as critical solutions.
Furthermore we investigate the relation between the DSS solution and 
the first CSS excitation.
We put forward the hypothesis that the DSS solution bifurcates
from the CSS solution in a {\em homoclinic loop bifurcation} at some 
critical value of the coupling constant $\eta_C \simeq 0.17$.

In Chapter \ref{chap::criticalCollapse} we describe our results concerning
type II critical phenomena. For very small couplings 
the stable CSS ground state gives rise to the formation of naked
singularities for ``intermediately strong'' initial data.
For these couplings we therefore investigate critical phenomena between
dispersion and the formation of naked singularities. For larger couplings
the end states are flat space on one hand and black holes on the other 
hand.
As expected, the critical behavior strongly depends on the coupling. 
For large couplings the critical solution
is DSS, whereas for small couplings
it is CSS. 
At intermediate couplings -- in the ``transition regime'' --
we find a new behavior, which we call ``epsiodic CSS''. 
This behavior is consistent with the hypothesis of the
homoclinc loop bifurcation, mentioned above.

Appendix \ref{app::SM} explains the shooting and matching method, which is
used for the numerical construction of both static solutions and
self-similar solutions.

In Appendix \ref{app::DFT} we explain the method of discrete Fourier
transform, which is used for the numerical construction of the DSS solution.

Appendix \ref{app::dice} describes the DICE code, that evolves the 
self-gravitating SU(2) $\sigma$ model in spherical symmetry.
This code is used to investigate critical phenomena.

Our conventions concerning curvature quantities and the signature of the
metric are those used in \cite{Wald} (for details see
App.~\ref{app::conventions}). Throughout this work the speed of light is set
to unity, $c = 1$. The index notation should be self-explaining.

%%%%%%%%%%%%%%%%%%%
%
%\documentclass[12pt,a4paper]{report}
%\usepackage{german,a4,bbm,graphicx,psfrag}
%
%
%\include{diss_macros}

%\textwidth=15cm
%\textheight=22cm
%\topmargin=-2cm
%\oddsidemargin=1cm
%\pagestyle{plain}
%\parindent=0pt                    % no indentation for paragraphs,
%\parskip=5pt plus 2pt minus 1pt   % but do a little skip

\chapter{The SU(2) $\sigma$ Model in Spherical Symmetry}\label{chap::sigma}
%%%%%%%%%%%%%%%%%%%%%%%%%%%%%%%%%%%%%%%%%%%%%%%%%%%%%%%%%%%%%%%%%%%%%%%%%%%

In this chapter we give the basic definitions and equations, that will be
used for the work on static solutions as well as on critical collapse.
We give the definition of 
the self-gravitating SU(2) $\sigma$ model
(with and without positive cosmological constant).
On spacetime we introduce two coordinate systems, that 
are adapted to spherical symmetry. For the work on static solutions
we choose coordinates such that the hypersurfaces of constant time are 
spacelike, 
whereas for the work on critical collapse
Bondi-like coordinates are used. Regularity of the metric in these
coordinates near the center of spherical symmetry is discussed.
Spherical symmetry is also imposed on the matter field via the so called
hedgehog ansatz. We write down the 
Einstein equations and the field equation in these
coordinates and give the formulae for the Misner-Sharp mass function 
as well as the Bondi mass and news function for asymptotically flat 
configurations.

\section{The SU(2) $\sigma$ Model}\label{sec::sigmaModel}
%%%%%%%%%%%%%%%%%%%%%%%%%%%%%%%%%%%%%%%%%%%%%%%%%%%%%%%%%

The SU(2) $\sigma$ model was first introduced into physics by Gell-Mann and
L\'evy \cite{Gell-Mann-Levy} in order to describe the meson fields
$\pi_+, \pi_-, \pi_0$ and $\sigma$, subject to the condition
$\pi_+^2 + \pi_-^2 + \pi_0^2 + \sigma^2 = 1$.
Geometrically it is an harmonic 
map\footnote{To be more precise, as the base manifold is spacetime with a
metric with Lorentzian signature, the map is called a {\em wave map}.} 
from spacetime
$(\mathbbm M, g)$ to the target manifold $(SU(2) \simeq \mathbbm S^3, G)$,
where $G$ is the standard metric on $\mathbbm S^3$.
Harmonic maps are well known in mathematics (see e.g. the review articles by
Eells and Lemaire \cite{Eells-Lemaire-1978, Eells-Lemaire-1988}). 
Some of their applications in physics are described by
Misner \cite{Misner-1978-harmonic-maps,Misner-1982-harmonic-maps}. 
In particular Misner points out that harmonic maps are
geometrically natural nonlinear theories
-- as are gravity and Yang-Mills theory --
and could be used to model (the much more involved) nonlinearities in 
general relativity.

Concerning this work our interest in the SU(2) $\sigma$ model
is mainly based on the fact that the theory contains a dimensionless 
parameter, while being a very simple generalization of the 
massless Klein-Gordon field. As will be argued below, both static and
critical solutions are likely to depend on this dimensionless parameter.
Therefore we are in the position to study (hopefully) a variety of 
phenomena within a comparatively simple model. 

We start by choosing coordinates $x^{\mu}$ on spacetime and 
coordinates $X^A$ on the target
manifold, and denote the map by $X$, so
\bea
(\mathbbm M, g_{\mu\nu}) & 
            \stackrel{X^A}{\lllrightarrow} &
                     (\mathbbm S^3, G_{AB})  \\
   x^{\mu}  &  \lllrightarrow &     X^A(x^{\mu}). \nonumber   
\eea
The three components $X^A(x^\mu)$ of the map 
are scalar fields on spacetime.
The harmonic map is defined to be the extremum of the action
\be\label{eq::action_HM}
S_{HM} = - \frac{f_{\pi}^2}{2} 
    \int \sqrt{-g} \ d^4  x  \ g^{\mu\nu} \partial_{\mu} X^A
    \partial_{\nu} X^B G_{AB} (X).  
\ee
Geometrically the Lagrangian is the pull back of the metric
of the target manifold, contracted with the inverse metric on spacetime. 
$f_{\pi}^2$ is the coupling constant of the fields.

Variation of the action with respect to the fields $X^A$ yields the
field equations
\be\label{eq::field_equation_HM}
  \square_g X^A + g^{\mu\nu} \ \tilde\Gamma^A_{BC} (X^D) \ \partial_{\mu} X^B
      \partial_{\nu} X^C = 0,
\ee
where $\square_g$ is the wave operator on spacetime $\square_g = g^{\mu\nu}
\nabla_{\mu} \nabla_{\nu}$, and $\tilde \Gamma^A_{BC}$ denote the Christoffel
symbols of the target manifold. For fixed base space $(\mathbbm M, g)$
Eq. (\ref{eq::field_equation_HM}) is a coupled quasi-linear system of 
wave equations for the $X^A$.

The simplest such system would be obtained, if the target manifold were
just $\mathbbm R$, 
for which the field equation for the single field $X$ would 
just reduce to the
Klein-Gordon equation. On the other hand assuming the base manifold to be
one-dimensional and the target manifold being arbitrary, the system
(\ref{eq::field_equation_HM}) would describe a geodesic on the
target manifold. In this sense harmonic maps are simple geometric 
generalizations of both the Klein-Gordon field and geodesics.

In order to incorporate gravity, the action (\ref{eq::action_HM}) has to be
supplemented by the Einstein-Hilbert action
\be\label{eq::action_total}
S_{tot} = \int \sqrt{-g} \ (\frac{1}{2\kappa} \mathcal R - 2 \Lambda) + 
          S_{HM},
\ee
where $\mathcal R$ denotes the Ricci scalar of spacetime, $\kappa$ contains
the gravitational constant $G$ by $\kappa = 8 \pi G$. In addition
we have introduced a cosmological 
constant $\Lambda$, which we will always consider positive in this work.

Variation of the total action (\ref{eq::action_total}) with respect to the
spacetime metric $g_{\mu\nu}$ yields the Einstein equations
\be\label{eq::Einstein_equation}
G_{\mu\nu} + g_{\mu\nu} \Lambda = \kappa T_{\mu\nu},
\ee

where $T_{\mu\nu}$ is the stress energy tensor of the $\sigma$ model,
obtained from varying the matter part of the action
(\ref{eq::action_HM}) with respect to the spacetime metric
\be\label{eq::stress_energy}
T_{\mu\nu} = f_{\pi}^2 \left( \nabla_{\mu}X^A \nabla_{\nu} X^B - 
    \frac{1}{2}
    g_{\mu\nu} (\nabla_{\sigma} X^A \nabla^{\sigma} X^B) \right) G_{AB}(X).
\ee
This stress energy tensor satisfies the weak, strong and 
dominant energy conditions, as defined in \cite{Hawking-Ellis}.
The weak energy condition (WEC) 
requires $T_{\mu\nu} w^{\mu} w^{\nu} \ge 0$ for
all timelike $w$ corresponding to a positive energy density
for all observers. The strong energy condition (SEC)
results from the ``timelike convergence condition'' 
$R_{\mu \nu} w^{\mu} w^{\nu} \ge 0$ for all timelike $w$, 
which for $\Lambda = 0$
translates into a condition for the stress energy tensor 
$T_{\mu\nu} w^{\mu} w^{\nu} 
\ge (1/2) w_{\mu} w^{\mu} T$. Finally the dominant energy condition
(DEC)
consists of the weak energy condition plus the requirement
that $T^{\mu\nu} w_{\nu}$ is non-spacelike for all timelike $w$.
An equivalent condition is, that the components in any orthonormal basis
satisfy $T^{\hat 0 \hat 0} \ge | T^{\hat I \hat J}|$ for all $\hat I, \hat J$.

For the WEC we choose an orthonormal basis $\{e_I\}_{I=0,1,2,3}$ with the
coordinate representation $e_{I} = e^{\mu}_I \partial_{\mu}$.
We consider the expression 
$T_{\mu\nu} e_{0}^{\mu} e_0^{\nu} = T_{\hat 0 \hat 0}$, 
which can be written as
\be
T_{\mu\nu} e_{0}^{\mu} e_0^{\nu} = \frac{1}{2}(X_* e_0)^A (X_*e_0)^B G_{AB}
    + \frac{1}{2} (X_* e_i)^A (X_*e_j)^B \delta^{ij} G_{AB} \ge 0, 
\ee
as $G_{AB}$ is Riemannian and therefore the above expression
is a sum of positive (or vanishing) terms. By $X_{*}v$ we denoted the push
forward of the vector $v$ from spacetime to the target manifold.
As the above relation is valid for any orthonormal basis, the relation
$T_{\mu\nu} w^{\mu} w^{\nu} \ge 0$ is satisfied for all timelike $w$.
 
For the SEC again we choose an orthonormal basis as above. Inserting this
into the expression $T_{\mu\nu} e_0^{\mu}e_{0}^{\nu} + (1/2)T$ one finds
that
\be
T_{\mu\nu}e_0^{\mu}e_{0}^{\nu} + \frac{1}{2}T = 0.
\ee
Again this is valid for all orthonormal bases, and therefore the matter
field satisfies the SEC borderline, as the massless Klein-Gordon
field does.

Finally to check the DEC we work with the components of $T_{\mu\nu}$ 
in an orthonormal base. We have
\bea\label{eq::stress_energy_ONB}
T^{\hat 0 \hat 0} & = & T_{\hat 0 \hat 0} = \frac{1}{2}(X_* e_0)^A (X_* e_0)^B
      G_{AB} + \frac{1}{2} (X_* e_i)^A (X_* e_j)^B \delta^{ij} G_{AB},
            \nonumber\\
T^{\hat 0 \hat i} & = & - T_{\hat 0 \hat i} = 
                - (X_{*} e_0)^A (X_{*} e_i)^B G_{AB} \nonumber\\
T^{\hat i \hat j} & = & T_{\hat i \hat j} = 
 (X_* e_i)^A (X_*e_j)^B G_{AB} - \frac{1}{2} \delta_{i j}
  \left[ -(X_*e_0)^A(X_*e_0)^B  + \delta^{kl}(X_*e_k)^A (X_*e_l)^B \right]
           G_{AB}.
         \nonumber\\
       &{}&
\eea
As $G_{AB}$ is positive definite, we have $|2 G(v,w)| \le G(v,v) + G(w,w)$.
Therefore $|T^{\hat 0 \hat i}| \le T^{\hat 0 \hat 0}$. One also easily sees,
that $| T^{\hat i \hat i}| \le T^{\hat 0 \hat 0}$ 
(without summation over $\hat i$) and that $|T^{\hat i \hat j}| \le T^{\hat
0 \hat 0}$ for $\hat i \ne \hat j$. Again the tetrad was chosen arbitrarily,
so the relations are valid in any orthonormal base.

Note that the coupling constants $f_{\pi}^2$ and $G$ only enter the 
equations (\ref{eq::Einstein_equation}) 
as the product $\eta := \frac{\kappa f_{\pi}^2}{2}$.
In units where the speed of light is set to unity $c=1$,
$G$ has dimension
of $length/mass$ and $f_{\pi}^2$ has dimension of $mass/length$, so their
product is dimensionless.
Therefore the only scale in the theory is tied to the cosmological
constant $\Lambda$, which has dimension of $1/length^2$. 
As $\Lambda$ merely sets the scale of the theory, its actual 
numerical value can be eliminated from the equations and 
only it's sign matters. 

For fixed $\Lambda$ on the other hand the presence of the 
dimensionless product of coupling constants 
$\eta := \frac{\kappa f_{\pi}^2}{2}$
provides a one parameter family of physically different theories.

For $\Lambda = 0$, the theory is scale invariant.
This implies immediately (as will be explained in Sec.
\ref{subsec::non_existence}) that for $\Lambda = 0$ this model does not admit
asymptotically flat soliton solutions \cite{Bizon-1994-gravitating-solitons}. 
On the other hand, as described in Sec. \ref{subsec::implications_matter}, 
the scale invariance for 
$\Lambda=0$ is necessary for the existence of self-similar solutions. 

\subsection{Non-existence of Asymptotically Flat Soliton Solutions and 
 Static Black Holes for $\Lambda = 0$}\label{subsec::non_existence}
%%%%%%%%%%%%%%%%%%%%%%%%%%%%%%%%%%%%%%%%%%%%%%%%%%%%%%%%%%%%%%%%%%%%%%

Using a scaling argument, it is easy to see that the system
(\ref{eq::field_equation_HM}) and (\ref{eq::Einstein_equation})
does not admit static, globally regular, asymptotically flat solutions
(so-called {\em soliton solutions}) apart from the trivial solution, which
is Minkowski spacetime with vanishing matter field.
Assume there exists a static solution $(\hat g_{\mu\nu}, \hat X^A)$ to the
system (\ref{eq::field_equation_HM}) and (\ref{eq::Einstein_equation}).
We denote the timelike Killing vector by $\xi = \partial_t$ and choose
coordinates $x^i$ in the hypersurfaces $\Sigma$ orthogonal to $\xi$.
For regular solutions $\Sigma$ is topologically $\mathbbm R^3$. Denoting the
induced metric by $\hat h_{ij}$ we can write
$ds^2 = - \hat N dt^2 + \hat h_{i j} dx^i dx^j$. Staticity in these 
coordinates
manifests itself in the independence of the metric functions 
$\hat N$ and $\hat h_{ij}$ and the fields 
$\hat X^A$ of the time coordinate $t$\footnote{For implications of a
spacetime symmetry on the fields $X^A$ see Sec.~\ref{subsec::hedgehog}}.

For static solutions, due to the existence of the Killing vector $\xi$, one
can define the energy of the field
\be\label{eq::energy_static}
E = - \int\limits_{\Sigma} \sqrt{\hat h} \ d^3 x \ n_{\mu} \ \xi^{\nu} \
            \hat T^{\mu}_{\nu}
  = - \int\limits_{\Sigma} \sqrt{\hat h} \ d^3 x \ \hat N \ \hat T^0_0.
\ee   
Asymptotic flatness guarantees the existence of this integral.
$E$ is a conserved quantity, i. e. independent of the hypersurface $\Sigma$,
which follows from $\nabla_{\mu}(\xi^{\nu} \hat T^{\mu}_{\nu}) = 0$.
Furthermore $E$ is positive definite as the lapse $\hat N$ is positive,
and the energy density 
$-\hat T^0_0$ is positive as well.
Moreover $\hat T^{0}_{0}$ vanishes iff $(X_*e_{\hat I}) = 0$ (or
equivalently $X$ is constant), i.e. if the map is trivial.

Since the model for $\Lambda = 0$ is scale invariant, there exists
a one-parameter family of solutions 
$\left( (\hat g_{\mu\nu})_\lambda (x^i) := \hat g_{\mu\nu}(\lambda x^i),
\hat X^A_{\lambda}(x^i) := \hat X^A(\lambda x^i)\right)$. The corresponding
energies $E_{\lambda}$ are obtained from $E$ by 
$E_{\lambda} = \frac{1}{|\lambda|} E$. As a static solution extremizes 
the energy (\ref{eq::energy_static}) 
we must have $dE_{\lambda}/d \lambda |_{\lambda=1} = 0$,
so $E$ has to be
zero and therefore the matter field has to vanish, $\hat X^{A} = const$.
We are left with a regular, static, asymptotically flat solution to 
the vacuum Einstein equations, which has to be Minkowski spacetime
due to the theorems by Lichnerowicz \cite{Lichnerowicz}.

It was also shown \cite{Heusler-1996-No-hair-Theorems, Heuslerbuch}, 
that this model with $\Lambda=0$ neither
admits non-trivial static asymptotically flat 
black hole solutions. 

So if one is interested in static solutions of this model, the cosmological
constant is essential, or as we shall see in Sec. \ref{subsec::eta_to_etamax},
the assumption of asymptotic flatness and $\Sigma$ being $\mathbbm R^3$
has to be dropped.

\section{Spherical Symmetry}\label{subsec::spherical_symmetry}
%%%%%%%%%%%%%%%%%%%%%%%%%%%%%%%%%%%%%%%%%%%%%%%%%%%%%%%%%%%%%%%%%

An isometry is a diffeomorphism $\Phi$ from $(\mathbbm M, g)$ to $(\mathbbm
M, g)$ ,which maps the metric to itself, i.e.
\be\label{eq::isometry}
\Phi^* g = g.
\ee
For a one parameter family of such diffeomorphisms $\Phi_\lambda$, with
$\Phi_0 = id$ one can define the generator 
$\xi = d \Phi_{\lambda}/d{\lambda} |_{\lambda = 0}$.
Eq. (\ref{eq::isometry}) then can be formulated in terms of the Lie
derivative
\be
\mathcal L_{\xi} g = 0,
\ee
or equivalently 
\be
\xi_{(\mu;\nu)} = 0.
\ee
Such a family of isometries leaves the curvature tensors
invariant (this can be seen in taking the analogous steps as in Sec.
\ref{sec::CSS-DSS}), in particular
\be
\left(\mathcal L_{\xi} G\right)_{\mu\nu} = 0.
\ee

A spacetime is said to be spherically symmetric, if it admits
the group SO(3) as a group of isometries, acting on spacelike 
two-dimensional surfaces (See e.g. \cite{Hawking-Ellis}). 
The group acts transitively but not simple
transitively: as the group is three dimensional and the spacelike surfaces
are only two-dimensional, it has SO(2) as an isotropy group.
The orbits of the group are surfaces of constant positive curvature.

It can be shown 
\cite{Hawking-Ellis} 
that the metric of a spherically symmetric spacetime can be
written as the warped product 
\be\label{eq::metric_spherical_symmetry}
d s^2 = d\tau^2 + R^2(\tau^i) \left(d\theta^2 + \sin^2\theta \ d\varphi^2 \right),
\ee
where the coordinates $\theta$ and $\varphi$ are coordinates on $\mathbbm S^2$
--
the orbits of SO(3) --
with the usual range $0 < \theta < \pi$ and $0 \le \varphi < 2\pi$.
$d \tau^2$ denotes the line element of a 
two-dimensional Lorentzian manifold (with coordinates $\tau^i$), 
and $R(\tau^i)$ is related to the area 
of the orbits of SO(3) by $A = 4 \pi R^2$.

For this work we use two different choices of coordinates
for the Lorentzian two-surfaces: for the work on static solutions
(in the presence of a positive cosmological constant), 
we choose orthogonal coordinates $(t, \rho)$ 
\be\label{eq::coordinates_static}
d s^2 = - A(t,\rho) dt^2 + B(t,\rho) d\rho^2 + R^2(t,\rho) d\Omega^2,
\ee
where $d \Omega^2 \equiv (d \theta^2 + \sin^2 \theta \ d \varphi^2)$.
Usually these coordinates are further restricted by choosing
the area of the orbits of SO(3) as a coordinate.
But this is only possible in regions, where $\nabla_{\mu} R \ne 0$.
As some of our static solutions 
will contain hypersurfaces where the gradient $\nabla_{\mu} R$
vanishes, we choose a different gauge: we set $1/B = A =: Q $, 
so
\be\label{eq::coordinates_QR}
d s^2 = - Q(t,\rho) \ dt^2 + \frac{d\rho^2}{Q(t,\rho)} + R^2(t,\rho) \
d\Omega^2.
\ee

For the investigations on critical collapse, we work with 
Bondi-like coordinates, i.e. we foliate spacetime by outgoing null cones
$u = const$, which emanate from the center of spherical symmetry
($R = 0$) and parameterize these with the area of the two-spheres 
$r:= \sqrt{A/4\pi}$, so we get
\be\label{eq::coordinates_Bondi}
d s^2 = - e^{2 \beta(u,r)} \ du \left( \Vr(u,r) \ du + 2 dr \right) + 
          r^2 d \Omega^2. 
\ee
The normal vectors to the hypersurfaces $u = const$ are null, as 
$-\nabla_{\mu} u = (-1,0,0,0)$ and $-\nabla^{\mu} u = (0,-g^{ur}, 0, 0)$, so
$\nabla_{\mu} u \nabla^{\mu} u = 0$. Furthermore, they generate  
affinely parametrized geodesics, as
\be
\nabla^{\sigma} u \nabla_{\sigma} \nabla^{\mu} u = \nabla^{\sigma} u 
\nabla^{\mu} \nabla_{\sigma} u = \frac{1}{2} \nabla^{\mu}
\left(\nabla_{\sigma} u \nabla^{\sigma} u \right) = 0. 
\ee
The areal coordinate $r$ parameterizes these null geodesics, but
is not the affine parameter.

Ingoing radial null geodesics are given as the solutions of
\be\label{eq::ingoing_null_geodesic}
\frac{d r(u)}{du} = - \frac{V(u,r(u))}{2 r(u)}.
\ee
We normalize $u$ to be proper time at the origin. We require
$\beta(u,r=0) = 0$. Regularity at the origin then enforces 
$\Vr(u,r=0)=1$ (see Sec.~\ref{subsec::regular_center}). 
The connection between these ``Bondi-like'' coordinates
and Bondi coordinates for which $\beta_B(u_B,r=\infty) = 0$ is given by
a coordinate transformation $u \to u_B(u)$, with
\be
e^{2 \beta_B (u_B,r)} du_B = e^{2 \beta(u,r)} du,
\ee
in particular at infinity we have
\be
\frac{d u_B}{d u } = e^{2 H(u)}, \qquad \textrm{with} \quad  H(u):=
                                                 \beta(u,r=\infty).
\ee
Therefore
\bea
\beta_B(u_B,r) & = & \beta(u,r) - H(u) \nonumber\\
\left(\Vr\right)_B (u_B,r)& = & \left( \Vr \right)(u,r)e^{-2 H(u)}.
\eea

For further use we give the square root of the determinant of the metric 
$\sqrt{-g} = e^{2 \beta} r^2 \sin \theta$ and the inverse metric
\be
g^{\mu\nu} = \left(
\begin{array}{cccc} 
0  & -e^{-2 \beta} & 0 & 0 \\
-e^{-2 \beta} & e^{-2 \beta} \Vr & 0 & 0 \\
0 & 0 & \frac{1}{r^2} & 0 \\
0 & 0 & 0 & \frac{1}{r^2 \sin^2 \theta}
\end{array}\right).
\ee

\subsection{A Regular Center of Spherical
Symmetry}\label{subsec::regular_center}
%%%%%%%%%%%%%%%%%%%%%%%%%%%%%%%%%%%%%%%%%%%%%%%%%%%%%%%
Clearly the coordinates (\ref{eq::coordinates_QR}) and 
(\ref{eq::coordinates_Bondi}) break down at a center of spherical
symmetry  $R=0$, $r=0$ respectively. Apart from the vanishing
volume of the two-spheres, 
which is well known from polar coordinates in flat space, 
the metric functions
$Q,R$ and $\beta, \Vr$ have to satisfy additional regularity requirements, 
if one asks for a regular center of spherical symmetry.
By definition the metric is regular ($C^k$), if its application to regular
($C^k$)
vector fields yields regular ($C^k$) functions on the manifold.
The easiest way to examine this is to switch to regular coordinates
close to the center.
 
We start with the coordinates (\ref{eq::coordinates_QR}).
First we fix the origin of the coordinate $\rho$ to be
at the center of spherical symmetry, 
$R(t, \rho = 0) = 0$.
We choose coordinates $(t,x,y,z)$, connected to $(t,\rho,\theta,\varphi)$
by $x = \rho \sin \theta \cos \varphi, y = \rho
\sin \theta \sin \varphi, z= \rho \cos \theta$ and $t=t$.
We assume $(t,x,y,z)$ to be regular coordinates in the vicinity of 
$\rho = 0$. A function then is regular, if it can be written as a regular
function of these coordinates.
By specifying the above coordinate transformation we 
also have implicitly assumed that the coordinate function $\rho$
has special regularity properties, namely $\rho$ itself is not regular, but
any even power thereof is. First note that $g(\partial_t, \partial_t) =
-Q$, which is regular if $Q$ is a regular function of $(t,x,y,z)$. 
In other words, $Q$ has to be a regular function of $t$ and $\rho^2$.
Second we consider the sum of the spatial components of the metric with
respect to the regular coordinates,
\be
g(\partial_x, \partial_x) + g(\partial_y, \partial_y) + g(\partial_z,
\partial_z) = \frac{1}{Q} + 2 \frac{R^2}{\rho^2}.
\ee 
This shows, that $R/\rho$ has to be a regular function of $t$ and $\rho^2$.
Having again a look at $g(\partial_z, \partial_z)$ ,
\be
g(\partial_z, \partial_z) = \cos^2 \theta \frac{1}{Q} + 
 \frac{\sin^2 \theta}{\rho^2} R^2,
\ee   
we see, that 
the only possibility for this expression to have a regular limit
$\rho \to 0$, is
\be
\lim\limits_{\rho \to 0} \frac{R^2 Q}{\rho^2} = 1,
\ee
so $Q(t,0) = 1/R'(t,0)$, which we can choose without loss of generality to
be $1$. Therefore near a regular center of symmetry the metric functions 
behave as follows
\bea\label{eq::metric_QR_O}
R(t, \rho) & = & \rho + O(\rho^3) \nonumber\\
Q(t, \rho) & = & 1 + O(\rho^2).
\eea

For the Bondi-like coordinates (\ref{eq::coordinates_Bondi}) we 
define in analogy the coordinates 
$t = u + r$, $x = r \sin\theta  \cos \varphi$, $y= r \sin \theta \sin \varphi$
and $z = r \cos \theta$. We have 
$g(\partial_t, \partial_t) = - e^{2 \beta} \Vr$ and
\be
g(\partial_x,\partial_x) + 
g(\partial_y,\partial_y) + g(\partial_z,\partial_z) =
2 - e^{2\beta} \Vr + 2 e^{2\beta}.
\ee
From this it follows that the metric functions $\beta$ and $\Vr$
have to be regular functions of $t$ and $r^2$. 
Looking at $g(\partial_z, \partial_z)$ 
\be
g(\partial_z, \partial_z) = e^{2\beta} (2 - \Vr) \cos^2\theta + 
  \sin^2 \theta,
\ee
we see that
$\lim\limits_{r \to 0} e^{2 \beta}(\Vr - 2) = 1$ is a necessary condition for
regularity. $\beta(u,r=0)$ has already been chosen to be unity, so
$\Vr(t, r=0) = 1$.
It remains to transform these functions of $t$ and $r^2$ back to
functions of $u = t-r$ and $r$. The first terms in a Taylor series expansion
give
\bea
\beta(u,r) & = & O(r^2) \nonumber\\
\Vr (u,r) & = & 1 + O(r^2). 
\eea
Note however, that $\beta$ and $\Vr$ if expanded in $u-u_0$ and 
$r$ don't solely contain even powers of $r$. The first non-vanishing 
term with an odd power in $r$ is e.g. the term $\dot \beta''(u_0,0) r^3$.  

\subsection{The Mass Function}\label{subsec::mass}
%%%%%%%%%%%%%%%%%%%%%%%%%%%%%%%%%%%%%%%%%%%%%%%%%
In spherical symmetry one can define the Misner-Sharp mass 
function \cite{Misner-Sharp-1964-Lagrangian-spherical-collapse} 
$m(\tau^i)$, by
\be
1 - \frac{2 m}{R} = \nabla_{\mu} R \ \nabla^{\mu} R, 
\ee
(for a recent description of the properties of this function see 
the article by Hayward \cite{Hayward-1996-Misner-Sharp-mass-function}) 
which gives 
\be\label{eq::mass_QR}
m(t,\rho) = \frac{R}{2} \left( 1 + \frac{1}{Q(t,\rho)}\ \dot R^2(t,\rho) -
   Q(t,\rho)(R'(t,\rho))^2 \right)
\ee
for the coordinates (\ref{eq::coordinates_QR}). Its interpretation 
for static solutions in the presence of a positive cosmological constant
will be described in Sec. \ref{subsec::global_structure}.
 
For the Bondi-like coordinates (\ref{eq::coordinates_Bondi})
we have
\be\label{eq::mass_ur}
m(u,r) = \frac{r}{2} \left(1 - e^{-2 \beta(u,r)} \Vr(u,r) \right).
\ee
For an asymptotically flat spacetime, the Bondi mass is obtained by
taking the limit $r \to \infty$ of $m(u,r)$ along the null hypersurfaces
$u=const$, so
\be\label{eq::mass_Bondi}
m_{Bondi}(u) = \lim\limits_{r \to \infty} m(u,r). 
\ee
In general, due to radiation (of the matter fields only in spherical
symmetry), the Bondi mass will decrease with retarded time $u$. We give an
explicit formula for the mass loss in Sec. \ref{subsec::mass_again}

The formation of an apparent horizon is signalled by the vanishing of the
expansion of outgoing null geodesics $\Theta_+ = 0$.
$\Theta_+$ can be expressed 
as the  Lie derivative of the area $A$ of 2-spheres
with respect to the tangent to outgoing null geodesics $l^{\mu}_+$,
divided by the area: $\Theta_+ = (\mathcal L_{l_+} A)/A$.

For the Bondi-like coordinates, we have already seen 
that $ - \nabla^{\mu} u$ is tangent to the outgoing radial null geodesics
$(u=const)$. The area of the 2-spheres is given by $A=4 \pi r^2$, so we have
\be
\Theta_+ = \frac{-2 g^{u r}}{r} = \frac{2 e^{-2 \beta}}{r}. 
\ee
An apparent horizon in these coordinates therefore manifests itself by
$\beta \to \infty$. The breakdown of these coordinates at an apparent
horizon is due to $r$ (in connection with $u$) ceasing to be a good 
coordinate.
As stated above the areal coordinate $r$ parameterizes the null cones
$u=const$. This is possible as long as $\nabla_{u} r \ne 0$, 
which evaluates precisely to $g^{ur} \ne 0$, which is then violated at the
apparent horizon.
 
\subsection{The SU(2) $\sigma$ Model in Spherical Symmetry}
                  \label{subsec::hedgehog}
%%%%%%%%%%%%%%%%%%%%%%%%%%%%%%%%%%%%%%%%%%%%%%%%%%%%%%%%%%%%%

Imposing a symmetry on spacetime also requires some symmetry properties
for the matter. The stress energy tensor $T_{\mu\nu}$ has to be invariant
under the isometry
\be\label{eq::symmetry_stress_energy}
\mathcal L_{\xi} T^{\mu \nu} = 0,
\ee
otherwise the symmetry of spacetime would be incompatible with the Einstein
equations.

For the SU(2) $\sigma$ model we write the stress energy tensor
(\ref{eq::stress_energy}) as 
\be
T_{\mu\nu} = (X^* G)_{\mu\nu} - \frac{1}{2} g_{\mu\nu} g^{\sigma \tau}
   (X^* G)_{\sigma \tau}.
\ee
The Lie derivative of this expression then gives
\be
\mathcal L_{\xi} T_{\mu\nu} = \mathcal L_{\xi} (X^* G)_{\mu\nu} - 
    \frac{1}{2} g_{\mu\nu} g^{\sigma \tau} \mathcal L_{\xi} (X^* G)_{\sigma
    \tau}, 
\ee
if $\xi$ is a Killing vector field.

As the Lie derivative commutes with the pull back
\be
\mathcal L_{\xi} (X^* G)_{\mu\nu} = (X^*(\mathcal L_{X_* \xi} G))_{\mu\nu},
\ee
the requirement (\ref{eq::symmetry_stress_energy}) is satisfied if
\be
\mathcal L_{X_* \xi} G_{AB} = 0.
\ee
This means that either $X_* \xi \equiv 0$ or the Killing vector field $\xi$
is mapped to a Killing vector field on the target manifold.

Applying this to spherical symmetry there are essentially two possibilities
to make the map spherically symmetric. First, if none of the fields $X^A$ 
depends on the angular variables $\theta$ and  $\varphi$ then the 
Killing vector fields $\xi_i$ of spherical symmetry 
on spacetime would be mapped to the zero vector field at the target manifold.
This way one would deal with three fields $X^A(\tau^i)$.

The second possibility, which is chosen in this work, 
uses the symmetry of the
target manifold: as the metric $G_{AB}$ is the metric of constant curvature
on $\mathbbm S^3$ it also admits SO(3) acting on 2-spheres
as a group of isometries.
We choose coordinates $(\phi, \Theta, \Phi)$ such that the line element is
given by
\be\label{eq::coordinates_S3}
ds^2_{\mathbbm S^3} = d \phi^2 + \sin^2 {\phi}(d \Theta^2 + \sin^2\Theta 
 \ d \Phi^2).
\ee 
Obviously the vector fields
\bea
\Xi_1 & = & \sin \Phi \partial_{\Theta} + \cot \Theta \cos \Phi
\partial_{\Phi}, \nonumber\\
\Xi_2 & = & - \cos \Phi \partial_{\Theta} + \cot \Theta \sin \Phi
\partial_{\Phi}, \nonumber\\
\Xi_3 & = & - \partial_{\Phi}
\eea  
are Killing vector fields on the target manifold.
We demand now, that the corresponding Killing vector fields on spacetime
$\xi_i$ are mapped to their counterparts on the target manifold, i.e.
\be
X_* \xi_i \stackrel{!}{=} \Xi_i. 
\ee
This is achieved by the so called {\em hedgehog ansatz}
\bea\label{eq::hedgehog}
\phi(x^{\mu}) = \phi(\tau^i), \qquad \Theta(x^{\mu}) = \theta,
\qquad \Phi(x^{\mu}) = \phi.
\eea
This way the field equations 
(\ref{eq::field_equation_HM}) decouple into a nonlinear wave equation
for $\phi(\tau^i)$ and two equations for $\Theta$ and $\Phi$, which are
satisfied identically.
 
In order to examine regularity of the map at the center of 
spherical symmetry, we again work with Cartesian coordinates 
$X = \phi \sin \Theta \sin \Phi = 
 \phi \sin \theta \cos \varphi = (\phi/r) x$, 
$Y= \phi \sin \Theta \cos \Phi =  \phi \sin \theta \sin \varphi = (\phi/r)
y$, $Z = \phi \cos \Theta =  \phi \cos \theta = (\phi/r) z$ (for the
coordinates (\ref{eq::coordinates_QR}) $r$ is replaced by $\rho$).
The fields $(X,Y,Z)$ are regular functions on spacetime, iff $\phi/r$
($\phi/\rho$ respectively) is a regular function.
Therefore close to the origin we get the expansions
\bea
\phi(t, \rho) & = & \rho (1 + O(\rho^2)) \label{eq::field_QR_O} \\
\phi(u, r) & = & r (1 + O(r^2))\label{eq::field_Bondi_O}. 
\eea
for fixed time $t$ or fixed retarded  time $u$.
Again for fixed $t$, $\phi$ is a regular function of $\rho^2$ whereas
for fixed $u$ odd powers of $r$ appear.

In particular this means that for all times $t$ ($u$)
the origin is mapped to a single point on $\mathbbm S^3$, which is  
the north pole as defined by (\ref{eq::coordinates_S3}).

\section{The Einstein Equations and the Field Equation}\label{sec::Einstein}
%%%%%%%%%%%%%%%%%%%%%%%%%%%%%%%%%%%%%%%%%%%%%%%%%%%%%

With the hedgehog ansatz (\ref{eq::hedgehog})
the field equations (\ref{eq::field_equation_HM}) reduce to the single
wave equation
\be\label{eq::phi}
\square_g \phi =\frac{\sin (2\phi )}{R^{2}} \qquad (\frac{\sin(2 \phi)}{r^2}
\textrm{resp.}), 
\ee
where $\square_g \phi$ reads
\bea
\square_g \phi& = & \frac{1}{R^2} \left(- \partial_t (\frac{R^2}{Q}
\partial_t ) + \partial_{\rho}(R^2 Q \partial_{\rho} ) \right) \phi\\
\square_g \phi & = & e^{-2\beta}\left(\left(\frac{2V}{r^2} +
\left(\frac{V}{r}\right)^\prime \right)\partial_r
- \frac{2}{r}\partial_u - 2\partial_u\partial_r +
\frac{V}{r}\partial_{rr}\right) \phi , 
\eea
for the coordinates (\ref{eq::coordinates_QR}), 
(\ref{eq::coordinates_Bondi}) respectively.

For the work on static solutions 
(see Chapter \ref{chap::statSolutions})
the following combinations of
the Einstein equations $G^{\mu}_{\nu} + \Lambda \delta^{\mu}_{\nu} = \kappa
T^{\mu}_{\nu}$ will turn out to be convenient:
the combinations $({}^{t}_{t})+ ({}^{\rho}_{\rho})- 2({}^{\theta}_{\theta})$, 
\bea\label{eq::ttplusrrminustwothth}
 \frac{2\,{\dot Q}(\rho ,t)^{2}}
     {Q(\rho ,t)^{3}} & - & 
   \frac{{\ddot Q}(\rho ,t)}{Q(\rho ,t)^{2}} + 
   \frac{-2 - \frac{2\,{\dot R}(\rho ,t)^{2}}
         {Q(\rho ,t)} + 
       2\,Q(\rho ,t)\,R'(\rho ,t)^{2}}{
        R(\rho ,t)^{2}} - Q''(\rho ,t) =  \nonumber\\
& = & 
2\,\eta \,\left( \frac{-2\,\sin (\phi(r,t))^{2}}
      {R(r,t)^{2}} - 
    \frac{{\dot \phi}(r,t)^{2}}{Q(r,t)} + 
    Q(r,t)\,\phi'(r,t)^{2} \right), 
\eea
$({}^{t}_{t}) - ({}^{\rho}_{\rho})$, 
\be\label{eq::ttminusrr}
 \frac{2\, \left( {\ddot R}(\rho ,t)  +   
        Q(\rho ,t)^{2}\,R''(\rho ,t) \right) }
       {Q(\rho ,t)\,R(\rho ,t)} 
 = 
  -2\,\eta \left( \frac{{\dot \phi}(\rho,t)^{2}}{Q(\rho,t)}+ 
       Q(\rho,t)\phi'(\rho,t)^{2} \right)
\ee
of the Hamiltonian constraint $({}^t_t)$
and the time evolution equations $({}^{\rho}_{\rho})$ and
$({}^{\theta}_{\theta})$, 
the time evolution equation
$({}^{\rho}_{\rho})$,
\bea\label{eq::rhorho}
\Lambda & - & R(\rho ,t)^{-2} + 
   \frac{{\dot Q}(\rho ,t)\,{\dot R}(\rho ,t)}
     {Q(\rho ,t)^{2}\,R(\rho ,t)}  -  
   \frac{{\dot R}(\rho ,t)^{2}}
     {Q(\rho ,t)\,R(\rho ,t)^{2}} - 
   \frac{2\,{\ddot R}(\rho ,t)}
     {Q(\rho ,t)\,R(\rho ,t)} +  \nonumber\\
& + & 
   \frac{Q'(\rho ,t)\,R'(\rho ,t)}
     {R(\rho ,t)} + \frac{Q(\rho ,t)\,
       R'(\rho ,t)^{2}}{R(\rho ,t)^{2}}  = \nonumber\\
& = &
   \eta \,\left( \frac{-2\,\sin (\phi(\rho ,t))^{2}}
       {R(\rho ,t)^{2}} + 
     \frac{{\dot \phi}(\rho ,t)^{2}}{Q(\rho ,t)} + 
     Q(\rho ,t)\,\phi'(\rho ,t)^{2} \right)
\eea
and the momentum constraint
$({}^{t}_{\rho})$,
\bea\label{eq::trho}
 \frac{- {\dot R}(\rho ,t)\,Q'(\rho ,t)}{Q(\rho,t)^{2} R(\rho,t)}
        + \frac{{\dot Q}(\rho ,t)\,
    R'(\rho ,t)}{Q(\rho,t)^{2} R(\rho,t)} + 
  \frac{ 2\,{\dot R}'(\rho ,t)}{Q(\rho,t) R(\rho,t)} = 
  -2\,\eta \,
  \frac{ {\dot \phi}(\rho ,t)\,\phi'(\rho ,t)}{Q(\rho,t)},
\eea
where ${}' \equiv \partial_{\rho}$ and $\dot {} \equiv \partial_t$.
Of course $G^{\phi}_{\phi} = G^{\theta}_{\theta}$ and all other components
vanish identically.
(See Chapter \ref{chap::statSolutions} on the structure of these equations 
in the static case.)

For the work on critical collapse in the coordinate frame
(\ref{eq::coordinates_Bondi})
the nontrivial Einstein equations
split up into the hypersurface equations (the
$\{rr\}$ and $\{ur\} - (V/2r)\{rr\}$ components of
$G_{\mu\nu} = \kappa T_{\mu\nu}$)
\bea
\beta'  & = & \frac{\ccbeta }{2}r(\phi' )^{2}, \label{eq::betap} \\
V' & = & e^{2\beta }(1-2\ccbeta \sin (\phi )^{2}), \label{eq::Vp}
\eea
and the subsidiary equations
$- E_{uur}
        \equiv 
                        r^2
                        \Bigl( G_{uu} - \kappa T_{uu} \Bigr)
                        - r^2 (V/r)
                          \Bigl( G_{ur} - \kappa T_{ur}
\Bigr)$  and 
$ E_{\theta\theta}
         \equiv 
                        G_{\theta\theta}
                        - 8 \pi T_{\theta\theta}
$:
\bea
E_{uur}
        & \equiv &
                2 V {\dot \beta}
                - {\dot V}
                + 2 \ccbeta \ r^2 \left[
                           {\dot \phi}^2
                           - \frac{V}{r}
                               \phi' 
                               {\dot \phi}
                    \right] = 0, 
                       \label{eq::Euur}\\
E_{\theta\theta} 
        & \equiv &
                V \bigl( r \beta'' - \beta' \bigr)
                + r \beta' V' 
                + \frac{1}{2} r V''
                - 2 r^2 \dot \beta'
                                                                \nonumber\\
        &   &
                {}
                - \ccbeta \ r^2 \phi'
                  \Bigl(
                   - \frac{V}{r} \phi' + 
                  + 2 \dot \phi
                  \Bigr) = 0,   \label{eq::Ethth}
\eea
where ${}' \equiv \partial_r$ and $\dot{} \equiv \partial_u$. 

The contracted Bianchi identity $\nabla_{\mu} G^{\mu}_{\nu} \equiv 0$ 
together with $\nabla_{\mu} T^{\mu}_{\nu} = 0$ for solutions of 
(\ref{eq::phi}) show, that the system (\ref{eq::phi}), (\ref{eq::betap})
and (\ref{eq::Vp}) is sufficient to evolve the Einstein $\sigma$ model:
The $\{\theta \theta\}$ component of the Einstein equations, 
$E_{\theta \theta} = 0$ is satisfied, if Eqs. (\ref{eq::phi}), 
(\ref{eq::betap}),
(\ref{eq::Vp}) and (\ref{eq::Euur}) are satisfied. 
This follows from the
``$r$-component'' of the Bianchi identity: $E_{\theta \theta}$ can be
expressed as an algebraic combination of the other components of the 
Einstein equations and derivatives thereof.

Furthermore the ``$u$--component'' of the Bianchi identity 
reads $(r^2 G_{uu})' \equiv
 (r V G_{ur})' - e^{2 \beta} r^2 \partial_u (e^{-2\beta} G_{ur}) + 
\frac{1}{2} r e^{2 \beta} G_{rr} \partial_u (V e^{-2 \beta})$. 
Assuming, that Eqs. (\ref{eq::phi}),
(\ref{eq::betap}) and (\ref{eq::Vp}) are satisfied, then
$(r^2 (G_{uu} - \kappa T_{uu})') = 0$. Therefore 
$r^2(G_{uu} - \kappa T_{uu}) = f(u)$. So if $G_{uu} = \kappa T_{uu}$ at some
hypersurface $r=const$ then this equation is satisfied everywhere. 
Now the regularity conditions at the origin
ensure that this equation is satisfied at the origin $r=0$,
so the $\{uu\}$ component of the
Einstein equations is satisfied everywhere if the field equation
(\ref{eq::phi}) and the hypersurface equations (\ref{eq::betap}) and
(\ref{eq::Vp}) are satisfied.
 
So the characteristic initial value problem we deal with consists of the
system (\ref{eq::phi}), (\ref{eq::betap}) and (\ref{eq::Vp}) together with
the initial conditions $\phi(u=0,r) = \phi(r)$.
The numerical treatment of this characteristic initial value problem 
is described in Appendix \ref{app::dice}.

\subsection{Mass Function, Bondi Mass and News
Function}\label{subsec::mass_again}
%%%%%%%%%%%%%%%%%%%%%%%%%%%%%%%%%%%%%%%%%%%%%%%%%%%%%%%%%

We close this section by giving explicit formulae for the mass function
$m(u,r)$ (\ref{eq::mass_ur})  
and the mass loss at infinity $\dot m_{Bondi}(u)$ in terms of the matter
field.

The mass function $m(u,r)$ can be rewritten as an integral over outgoing
null rays by the trivial identity $m(u,r) = \int\limits_0^r m'(u, \bar r) d
\bar r$. $m'(u,r)$ is given by
\be
m'(u,r) = \frac{1}{2}\left(1 - e^{-2 \beta} \Vr \right) -
      \frac{r}{2} e^{-2 \beta} \left( - 2 \beta' \Vr + (\Vr)' \right).
\ee
Using the hypersurface equations (\ref{eq::betap}) and (\ref{eq::Vp}) we obtain
\be
m'(u,r) = 
       \frac{\eta}{2} \  r^2 \ \left( e^{-2 \beta} \Vr (\phi')^2 +  
 2 \frac{\sin^2(\phi)}{r^2} \right),
\ee
and therefore 
\be\label{eq::mass_function_rho}
m(u,r) = \frac{\eta}{2} \int\limits_0^r \bar r^2 \ d\bar r
      \left( e^{-2 \beta} \frac{V}{\bar r} (\phi')^2 +  
 2 \frac{\sin^2(\phi)}{\bar r^2} \right).
\ee
This expression together with (\ref{eq::mass_ur}) will serve as an
accuracy test for the numerical code (see Appendix \ref{app::dice}).

In order to derive a formula for the mass loss at infinity
in terms of the matter
field we have to look at the behavior of the metric functions
and the field at infinity.
The Bondi mass (\ref{eq::mass_Bondi}) 
is finite if $\beta = H(u) + a_1(u)/r + O(1/r^2)$ and 
$\Vr = e^{2 H(u)} + b_1(u)/r + O(1/r^2))$. 
From (\ref{eq::mass_function_rho}) follows
further, that $\phi(u,r) = \frac{c_1(u)}{r} + O(1/r^2)$.
Using the hypersurface equation (\ref{eq::betap}) we find that $a_1(u) = 0$,
so 
\be
\beta(u,r) = H(u) + O(1/r^2).
\ee
Inserting these expansions into the
formula for the Bondi mass (\ref{eq::mass_Bondi}) we get $b_1(u) = -2
m_{Bondi} e^{2 H(u)}$, so 
\be
\Vr(u,r) = e^{2H(u)} \left(1 - \frac{2 m_{Bondi}}{r} \right) + O(1/r^2).
\ee

The derivative of the mass function $m(u,r)$ with respect to retarded time
$u$ is given by
\be
\dot m(u,r) = - \eta r^2 e^{-2 \beta} \left( \dot \phi^2 - \Vr \phi' \dot
          \phi \right),
\ee
where one of the subsidiary Einstein equations 
(\ref{eq::Euur}) has been used.
This gives for the mass loss at infinity
\be
\dot m_{Bondi}(u) = - \eta \ c_1^2(u) e^{-2 H(u)}.
\ee
Clearly $\dot m_{Bondi} \le 0$ corresponding to the energy, that is radiated
away to infinity and therefore lost.
The expression
\be
N(u):= c_1^2(u) e^{-2 H(u)}
\ee
is called the {\em news function}.

%%%%%%%%%%%%%%%

%
%\documentclass[12pt,a4paper]{report}
%\usepackage{german,a4,bbm,graphicx,psfrag}
%
%
%\include{diss_macros}

%\textwidth=15cm
%\textheight=22cm
%\topmargin=-2cm
%\oddsidemargin=1cm
%\pagestyle{plain}
%\parindent=0pt                    % no indentation for paragraphs,
%\parskip=5pt plus 2pt minus 1pt   % but do a little skip

\chapter{Static Solutions of the Self-gravitating $\sigma$ Model in the
         Presence of a Cosmological Constant}\label{chap::statSolutions}
         %%%%%%%%%%%%%%%%%%%%%%%%%%%%%%%%%%%%%%%%%%%%%%%%%%%%%%%%%%%%%%

\section{The Static Equations}\label{sec::static_equations}
%%%%%%%%%%%%%%%%%%%%%%%%%%%%%%%%%%%%%%%%%%%%%%%%%%%%%%%%%%

In this chapter we investigate the question of existence and stability of
static solutions to the self-gravitating SU(2) $\sigma$ model with positive
cosmological constant in spherical symmetry. This work is motivated by
previous work \cite{sigma-on-de-Sitter} on static solutions of the 
model on fixed de Sitter
background and by the work of Volkov 
et~al.~\cite{VSLHB-1996-cosmological-EYM-BMcK-solutions} 
and Brodbeck et~al.~\cite{BHLSV-EYM-Lambda-stability}, 
who considered the Einstein Yang-Mills system with positive 
cosmological constant.

The choice of coordinates as well as the introduction of a gauge invariant
quantity for the stability analysis follows
\cite{VSLHB-1996-cosmological-EYM-BMcK-solutions,BHLSV-EYM-Lambda-stability}. 
In addition we put some
emphasis on examining the r\^ole of the Killing horizon.
Our results closely resemble those of
\cite{VSLHB-1996-cosmological-EYM-BMcK-solutions,BHLSV-EYM-Lambda-stability} 
with the only difference,
that in the limit of maximal coupling (see Sec. \ref{subsec::eta_to_etamax}) 
the system is scale invariant. 
The results on static solutions are summarized in
\cite{sigma-coupled-with-Lambda-static}.

\subsection{Staticity}\label{subsec::staticity}
%%%%%%%%%%%%%%%%%%%%%%%%%%%%%%%%%%%%%%%%%%%%%%%

A spacetime $(\mathbbm M, g)$ is called {\em stationary}, if it admits a 
timelike Killing vector field $\xi$. 
If this Killing vector field is in addition
hypersurface orthogonal, then the spacetime is called {\em static}.
Hypersurface orthogonality is given if the Frobenius condition
\be
\xi_{[\sigma} \nabla_{\mu} \xi_{\nu]} = 0.
\ee
is satisfied.

Consider now a static, spherically symmetric spacetime (see e.g. \cite{Wald}).
If the static Killing vector field $\xi$ is unique, it has to be
invariant under the action of SO(3) (As the composition of two
isometries is again an isometry, an element of SO(3) maps $\xi$ to a
Killing vector field. Furthermore the norm is left invariant, 
so $\xi$ is mapped to a timelike Killing vector field. 
If $\xi$ is unique, then SO(3) has to map $\xi$ to itself).
From this it follows, that $\xi$ must not have components 
tangential to the orbits of SO(3), as the only vector field on $\mathbbm
S^2$, which is left invariant under the action of SO(3), is the zero vector
field. Therefore $\xi$ can be written as $\xi = \xi^t(t,\rho) \partial_t + 
\xi^{\rho}(t,\rho) \partial_{\rho}$. 
We are still free to choose the coordinates $(t,\rho)$
such that $\xi = \partial_t$. With this choice the metric
(\ref{eq::coordinates_QR}) in the presence of a hypersurface orthogonal 
Killing vector field $\xi = \partial_t$ reads
\be\label{eq::metric_QR_static}
ds^2 = - Q(\rho) dt^2 + \frac{d \rho^2}{Q(\rho)} + R^2(\rho) d\Omega^2.
\ee

Stationarity of spacetime again extends to the field $\phi$ via 
the Einstein
equations. In order to satisfy (\ref{eq::symmetry_stress_energy}) 
we set $\phi = \phi(\rho)$.

\subsection{The Static Equations}\label{subsec::static-equations}
%%%%%%%%%%%%%%%%%%%%%%%%%%%%%%%%%%%%%%%%%%%%%%%%%%%%%%%%%%%%%%%%%

Setting all time derivatives in Eqs. (\ref{eq::phi}) and
(\ref{eq::ttplusrrminustwothth})--(\ref{eq::trho}) to zero, we find that the
momentum constraint (\ref{eq::trho}) is satisfied identically.
The field equation (\ref{eq::phi}) and the combinations
(\ref{eq::ttplusrrminustwothth}) and (\ref{eq::ttminusrr}) 
yield the following autonomous system of coupled, nonlinear, second order 
ODEs
\bea\label{eq::static_equations}
(Q R^{2} \phi')' & = & \sin 2\phi, \label{eq::fpp}\\
(R^{2} Q')' & = & - 2 \Lambda R^{2},\label{eq::Qpp}\\
R'' & = & - \eta R \phi'^{2}.\label{eq::Rpp}
\eea
Furthermore Eq. (\ref{eq::rhorho}) (multiplied by $R^2$) 
is a first integral of the above system:
\be\label{eq::constofm}
2 \eta \sin^{2}\phi + R^{2}(\Lambda - \eta \ Q \phi'^{2}) + R Q' R' + Q R'^{2} 
                                     - 1 = 0.
\ee
From the $\rho$ component of the contracted Bianchi identities 
we have $(\sqrt{Q}R^2 G^{\rho}_{\rho})' = R^2 (\sqrt{Q})' G^t_t + 
\sqrt{Q} (R^2)' G^{\theta}_{\theta}$. 
Assuming Eqs. (\ref{eq::fpp})--(\ref{eq::Rpp})
to be satisfied, we have $Q R^2 (G^{\rho}_{\rho} - \kappa T^{\rho}_{\rho}) 
= const$. So if (\ref{eq::constofm}) is satisfied 
at some hypersurface $\rho = const$ it is satisfied everywhere.
Now the conditions (\ref{eq::metric_QR_O}) and 
(\ref{eq::field_QR_O}) for a regular center of
spherical symmetry yield, that
Eq. (\ref{eq::constofm})
is satisfied at $\rho = 0$ and therefore it is satisfied everywhere.

As the cosmological constant $\Lambda$ -- if non-zero -- 
sets the length scale of the theory it can be eliminated by 
switching to the dimensionless variables
$\bar \rho = \sqrt{\Lambda} \rho$ and $\bar R = \sqrt{\Lambda} R$.
Or in other words: setting $\Lambda$ to unity (say) in the above equations
means that all quantities which have dimension of length are measured in
units of $\sqrt{\Lambda}$. 

We are interested in solutions of (\ref{eq::fpp})--(\ref{eq::Rpp}) 
which have a
regular center of spherical symmetry $R=0$ (at $\rho = 0$) and a Killing
horizon $Q=0$ at some finite distance from the origin. (See Sec.
\ref{subsec::global_structure} for the global structure of such spacetimes).
The reason why we look for solutions with a horizon is the following:
turning off gravity ($\eta = 0$) Eqs. (\ref{eq::fpp}) --
(\ref{eq::Rpp}) (in combination with the regularity conditions at the origin)
describe the de Sitter spacetime (see Sec.
\ref{subsec::exact_solutions}),
which has a cosmological horizon, and we don't expect the global structure
to change when gravity is turned on slightly ($\eta$ small and positive).
Furthermore we can rule out the following possibilities: integrating 
outward from a regular center
\begin{enumerate}
\item the static region ``ends'' in a singularity,
\item the static region has a second (regular) pole $R=0$\footnote{Note that
this situation is not ruled out if $\Lambda = 0$},
\item the static region persists up to spatial infinity.
\end{enumerate}

In principle 
the first case could be produced easily by choosing an arbitrary value
$\phi'(\rho=0)$ at the regular center (See Sec. \ref{sec::num_sol}). 
Nevertheless this case is of no interest here and is 
therefore discarded. Avoiding this scenario
means, that we have to set up a boundary value problem enforcing one of the
other cases.

The second case is impossible for $\Lambda > 0$: rewriting equation
(\ref{eq::Qpp}) as an integral equation (\ref{eq::Qpp_integrated}), one
sees, that $Q'$ diverges, whenever $R$ goes to zero for a second time.

That the last case is impossible, can be seen as follows:
assume the static region extends to infinite values of $\rho$, i.e.
$Q(\rho) > 0$ for all $\rho \ge 0$. Assume further, that 
$R'(\rho) > 0$ for all $\rho \ge 0$ (this second assumption is necessary, as
$R'(\rho)$ is strictly monotonically decreasing 
(see Eq. (\ref{eq::Rpp_integrated})), 
so once it gets zero $R$ decreases until it eventually becomes zero, which
in turn would lead to the cases 1 or 2).
Then one can show that for $\rho$ large enough
$Q'$ is bounded from above by $Q'(\rho) < - const/\rho$. This means
that $Q$ has to cross zero at some finite value of $\rho$, so case 3 is
ruled out as well.

We will discuss the implications of Eqs. (\ref{eq::fpp}) --
(\ref{eq::constofm}) for $\Lambda = 0$ in
Sec. (\ref{subsec::eta_to_etamax}).  

\subsection{Exact Solutions}\label{subsec::exact_solutions}
%%%%%%%%%%%%%%%%%%%%%%%%%%%%%%%%%%%%%%%%%%%%%%%%%%%%%%%%%%%

In this section we give some simple exact solutions 
of Eqs. (\ref{eq::fpp})--(\ref{eq::Rpp}), which will be important for the
full spectrum of solutions (See Sec. \ref{sec::num_sol}), as they arise in
certain limits.

We start with the solution obtained by setting $\phi = 0$.
Then $R''(\rho) = 0$ and therefore (together with the gauge choices $R(0) =
0, R'(0) = 1$) we have $R(\rho) = \rho$.
Eq. (\ref{eq::Qpp}) then gives $Q'(\rho) = - 2 \Lambda \rho/3$ 
and $Q(\rho) = 1 - \Lambda \rho^2/3$. Together we have
\be
\phi(\rho) \equiv 0, \quad R(\rho) = \rho, \quad 
       Q(\rho) = 1 - \frac{\Lambda \rho^2}{3},
\ee
which is the de Sitter spacetime in the static frame. 

Another simple solution is obtained by setting $\phi \equiv \pi/2$.
This violates the regularity condition for the field at the origin.
Again we have $R(\rho) = \rho$, but $Q(\rho)$ cannot fulfill the regularity
requirement $Q(0) = 1/R'(0)$ at the origin any more, instead
$Q(\rho) = 1 - 2 \eta - \Lambda \rho^2/3$, so
\be
\phi(\rho) \equiv \frac{\pi}{2}, \quad R(\rho) = \rho, \quad 
             Q(\rho) = 1 - 2 \eta - \frac{\Lambda \rho^2}{3}.
\ee 
This solution behaves like the de Sitter spacetime for large values of
$\rho$, but has a conical singularity at the origin.
In the limit of vanishing coupling $\eta = 0$, where spacetime is de Sitter,
the solution $\phi \equiv \pi/2$ has diverging energy density at the origin,
but finite total energy (as measured between origin and horizon) maximizing
the energy of all regular static solutions that exist on fixed de 
Sitter background and can be viewed as a
``high excitation'' limit of this spectrum 
(see \cite{sigma-on-de-Sitter}).

For $\Lambda = 0$ we obtain a solution, which will be of interest
in the limit of maximal coupling $\eta \to \eta_{max}$ (see Sec.
\ref{subsec::eta_to_etamax}). Eq.~(\ref{eq::Qpp}) together with the
regularity conditions at the origin give $Q(\rho) \equiv 1$.
The remaining equations can be solved analytically for $\eta = 1$ to
give the static Einstein universe:
\be\label{eq::EinsteinU}
\phi(\rho) = \rho, \quad R(\rho) = \sin \rho, \quad Q(\rho) \equiv  1, \quad
\eta = 1.
\ee

\subsection{Horizons and Global Structure}\label{subsec::global_structure}
%%%%%%%%%%%%%%%%%%%%%%%%%%%%%%%%%%%%%%%%%%%%%%%%%%%%%%%%%%%%%%%%%%%%%%%%%

Clearly the coordinates (\ref{eq::metric_QR_static}) break down at 
the horizon $Q(\rho_H) = 0$. In order to examine the global structure
of spacetime we temporarily switch to outgoing 
Eddington-Finkelstein coordinates $(u,r, \theta, \phi)$, 
where the retarded time $u$ is given by
\be
u = t - \int \frac{d \rho}{Q(\rho)}.
\ee
The metric (\ref{eq::metric_QR_static}) then reads
\be\label{eq::null_metric}
ds^2 = - Q(\rho) du^2 - 2 du \ d\rho + R(\rho)^2 d\Omega^2.
\ee
Note that the coordinates (\ref{eq::null_metric}) 
cover only half of the maximally extended spacetime. 
In order to cover all of spacetime one would have to switch to
``Kruskal-like'' double null coordinates, as is done e.g. in 
\cite{VSLHB-1996-cosmological-EYM-BMcK-solutions}. In the following,
we will simplify our discussion by only talking about the Killing horizon
contained in the portion of spacetime covered here.
All statements made can be extended trivially to the complete spacetime
and in particular the second component of the horizon by time reflection.
We also remark that all solutions have the topology $S^3 \times R$.

The static Killing vector field is 
$\partial_u= \partial_t$,
The metric (\ref{eq::null_metric}) is 
regular if $Q(\rho)$ and
$R(\rho)$ are regular functions, except when $R=0$, which corresponds
either to the usual coordinate singularity of spherical symmetry, 
which has been discussed in Sec. (\ref{subsec::regular_center}) 
or to a spacetime singularity, as discussed below.

As described in Sec.~\ref{subsec::staticity} we assume the 
Killing vector field $\partial_u$, to be timelike in some neighborhood of
the center $R=0$, i.e. $g(\partial_u,\partial_u) = -Q(\rho) < 0$.
From the discussion in Sec. \ref{subsec::static-equations}
it is clear, that $Q(\rho)$ has to go to zero at some finite distance from
the origin, $Q(\rho_H) = 0$. Furthermore $Q(\rho)$ changes sign there 
(as $Q'(\rho) < 0$ the degenerate case of $Q'(\rho_H) = 0$ is impossible).
In regions, where the norm of the Killing vector is positive, 
i.e. $Q(\rho) < 0$, spacetime is dynamic. 
The slices of constant time $\rho$ are of topology
$\mathbbm R \times \mathbbm S^2$. They are generated by the  
Killing vector
fields $\partial_u$ and $\xi_i$ and are therefore 
homogeneous.
Such regions thus correspond to Kantowski-Sachs models.

The hypersurface $\rho=\rho_H$ separating the static and dynamic regions,
is characterized by $Q(\rho_H) = 0$, and is a null hypersurface
(as $\nabla_{\mu} u \nabla^{\mu} u = 0$ and $\nabla^{\sigma} u \nabla_{\sigma}
\nabla^{\mu} u = 0$). As the Killing vector field $\partial_u$
is null on and tangent to this hypersurface it is a Killing horizon.

In order to characterize this horizon further, we use the concept of 
{\em trapping horizons} introduced by Hayward
\cite{Hayward-1994-trapping-horizons}. In general for asymptotically flat 
(and asymptotically de Sitter) spacetimes the asymptotic region
can be used to classify an event horizon as a black hole horizon
(or as a cosmological horizon). Furthermore for the
(local) concept of an apparent horizon the asymptotic region is 
needed to define inward and outward directions.

The concept of a trapping horizon is based solely on the local 
behavior of null congruences. As in Sec. \ref{subsec::mass} the
expansions $\Theta_{\pm}$ of out- and ingoing null rays emanating from
the spheres $R=const$ are defined as
\be
\Theta_{\pm} = \frac{1}{R^2} \mathcal L_{\pm} R^2,
\ee
where $\mathcal L_{\pm}$ denotes the Lie derivative along the null
directions

\be
l_+ = \partial_{\rho} \qquad \textrm{and} \qquad l_-= 2 \partial_u -
      Q\partial_{\rho}
\ee
respectively, so 
\be\label{Theta_pm}
\Theta_+ = 2 \frac{R'}{R} \qquad \textrm{and} \qquad 
\Theta_- = - 2 Q \frac{R'}{R}.
\ee
The expansions can be used to define a {\em trapped surface} 
in the sense
of Penrose \cite{Penrose-structure-of-spacetime} 
as a compact spacelike 2-surface ($R=const$) 
for which
$\Theta_- \Theta_+ > 0$. If one of the expansions vanishes, 
the surface is called a {\em marginal surface}.
For a non-trapped surface $\Theta_-$ and $\Theta_+$ have opposite signs,
which is used to have a local notion of inward and outward:
the direction in which the expansion is positive is called {\em outward},
and in which the expansion is negative is called {\em inward}.

The closure of a hypersurface, which is foliated by
(non-degenerate\footnote{for $\Theta_- = 0: \mathcal L_{+} \Theta_- \ne 0$}
and non-bifurcating\footnote{for $\Theta_- = 0: \Theta_+ \ne 0$})
marginal surfaces is called a {\em trapping horizon}.
At a trapping horizon, the behavior of the non-vanishing expansion can be
used to distinguish between {\em future trapping} and {\em past trapping}
horizons, i.e. supposing $\Theta_-|_H = 0$ (as is the case in our example)
then the horizon is
\bea
\textrm{future trapping} & \quad &\textrm{if} \qquad \Theta_+|_H  <0,
                \nonumber\\
\textrm{past trapping} & \quad & \textrm{if} \qquad \Theta_+|_H  >0.
\eea
The case $\Theta_+|_H < 0$ means that in the 
trapped region both outgoing and ingoing future directed null rays converge 
whereas in the case $\Theta_+|_H > 0$, the null rays in the trapped region
converge if past directed.

Furthermore the change of the vanishing null expansion $\Theta_-$ 
in direction of $l_+$ leads to a classification of {\em outer} and 
{\em inner} trapping horizons, i.e. the horizon is
\bea
\textrm{outer trapping} & \quad &\textrm{if} \qquad 
                           \mathcal L_+\Theta_-|_H < 0,
                \nonumber\\
\textrm{inner trapping} & \quad & \textrm{if} \qquad \mathcal L_+\Theta_-|_H  >0.
\eea
The meaning of inner and outer are the following: 
For $\mathcal L_+\Theta_-|_H  <0$ starting from the non-trapped region, where
the directions inward and outward are defined as above, one has to move
{\em inward} to approach the horizon, or in other words, the horizon is an
outer boundary for the trapped region, whereas for 
$\mathcal L_+\Theta_-|_H >0$ the horizon is an inner boundary of the 
trapped region. 

A future outer trapping horizon therefore is suited to describe a black hole
horizon, a past outer trapping horizon would describe a white hole horizon.
Inner horizons on the other hand describe cosmological horizons, as they
occur for example in the de Sitter spacetime.

Our situation is as follows:
on the Killing horizon $Q=0$, $\Theta_- |_{Q=0} = 0$ while 
$\Theta_+ |_{Q=0} = 2 R_H'/R_H \ne 0$ and ${\mathcal L}_+ \Theta_- |_{Q=0} = 
- 2 Q'_H R'_H /R_H \ne 0$
(except when also $R'_H=0$, which we exclude for the moment).

As $R >0$ and $Q' >0$ for $\rho > 0$, the character of the Killing
horizon is given by the sign of $R'_H$. For $R'_H > 0$ we have $\Theta_+
|_H > 0$ and $\mathcal L_+ \Theta_- |_H < 0$. The Killing horizon therefore
is a past inner trapping horizon, and therefore a cosmological
horizon. For $R'_H < 0$ on the other hand, we have $\Theta_+|_H < 0$ and 
$\mathcal L_+\Theta_-|_H > 0$, which corresponds to a future outer trapping
horizon, and therefore to a black hole horizon.

If $R'=0$ anywhere in spacetime, then both expansions vanish. However, this
does not mark a trapping horizon, as the product $\Theta_+ \Theta_-$ does
not change sign ($\Theta_+$ as well as $\Theta_-$ both cross zero).
Instead this means, if it occurs in a non-trapped region, that the meaning
of inward and outward directions are interchanged. 

Using the expansions $\Theta_+$ and $\Theta_{-}$ (\ref{Theta_pm}) one can
rewrite the quasilocal mass (\ref{eq::mass_QR}) as
\be
m = \frac{R}{2} (1 + \frac{R^2}{4} \Theta_{+}\Theta_{-}).
\ee
so on any marginal two-sphere the quasilocal mass equals $R/2$.

Now, which kind of global structure do 
Eqs. (\ref{eq::fpp}) -- (\ref{eq::Qpp}) allow for?
We start by rewriting these equations in the integral forms
\bea
\phi'(\rho) & = & \frac{1}{Q R^{2}} \int\limits_{0}^{\rho} \sin (2 \phi) 
                   d{\bar \rho}, \label{eq::fpp_integrated}\\
Q'(\rho) & = & -\frac{2 \Lambda}{R^{2}} \int\limits_{0}^{\rho} R^{2}d {\bar
               \rho}, \label{eq::Qpp_integrated}\\
R'(\rho) & = & 1 - \eta \int\limits_{0}^{\rho} R \phi'^{2} d {\bar \rho}
                     \label{eq::Rpp_integrated}.
\eea
We see, that the first derivatives stay finite for any finite value of
$\rho$ as long as the metric functions $R$ or $Q$ don't vanish.
If we assume the horizon $Q(\rho_H) = 0$ to be regular (and non-degenerate,
$Q_H' \ne 0$), we have to enforce
$\int\limits_0^{\rho_H} \sin 2 \phi \ d\rho = 0$ and therefore 
\be\label{eq::phiHp}
\phi_H' = \frac{\sin{ 2\phi_H}}{R_H^2 Q'_H}. 
\ee
As $Q$ is monotonically decreasing with $\rho$, $\rho_H$ is the only
horizon along a $u=const$ slice.

Now the areal radius $R$ of the two-spheres can behave in two essentially
different ways: it can increase monotonically for all $\rho$ or it can have
an extremum at some $\rho_E$.

\begin{figure}
\begin{center}
\begin{psfrags}
 \begin{picture}(68,80)
  %\put(0,180){\circle{4}}
    \put(14,50){
           \begin{picture}(0,0)
           \put(19,41){$\mathcal I^+$}
           \put(19,-3){$\mathcal I^-$}
           \put(-3,38){(a)}
           \put(-6,19){Origin}
           \put(41,19){Origin}
           \put(0,0){
           \includegraphics[width=40ex]{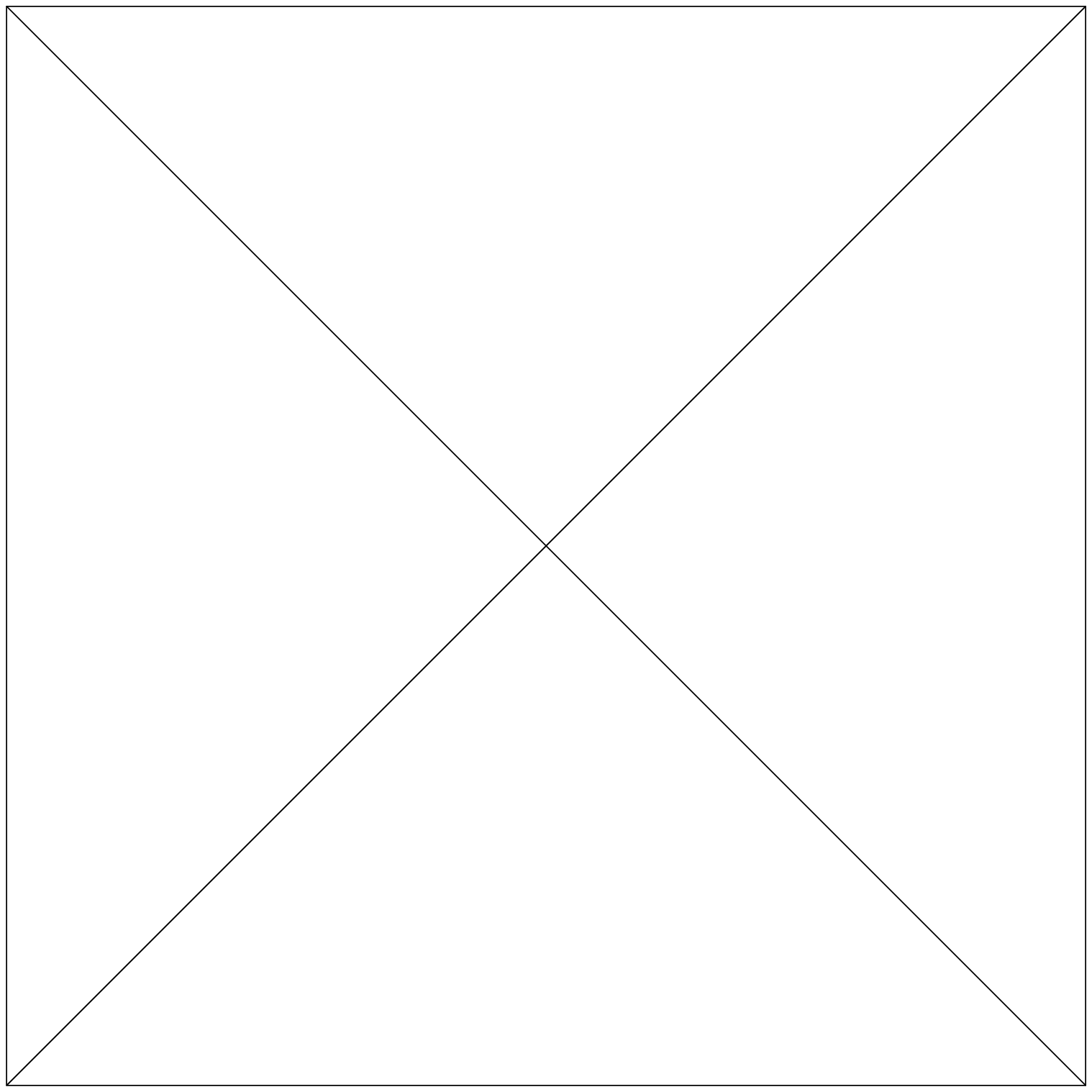}
           }
           \put(5,24){static}
           \put(30,24){static}
           \put(5,20){$\Theta_- <0$}
           \put(5,16){$\Theta_+ >0$}
           \put(16,35){dynamic}
           \put(17,31){$\Theta_- > 0$}
           \put(17,27){$\Theta_+ > 0$}
           \put(16,4){dynamic}
           \put(11,43){\vector(-1,-2){4}}
           \put(8,45){$\Theta_- = 0$}
           \end{picture}
            }
   %  \put(-120,0){\circle{2}}
     \put(-10,0){
           \begin{picture}(0,0)
           \put(0,0){
           \includegraphics[width=40ex]{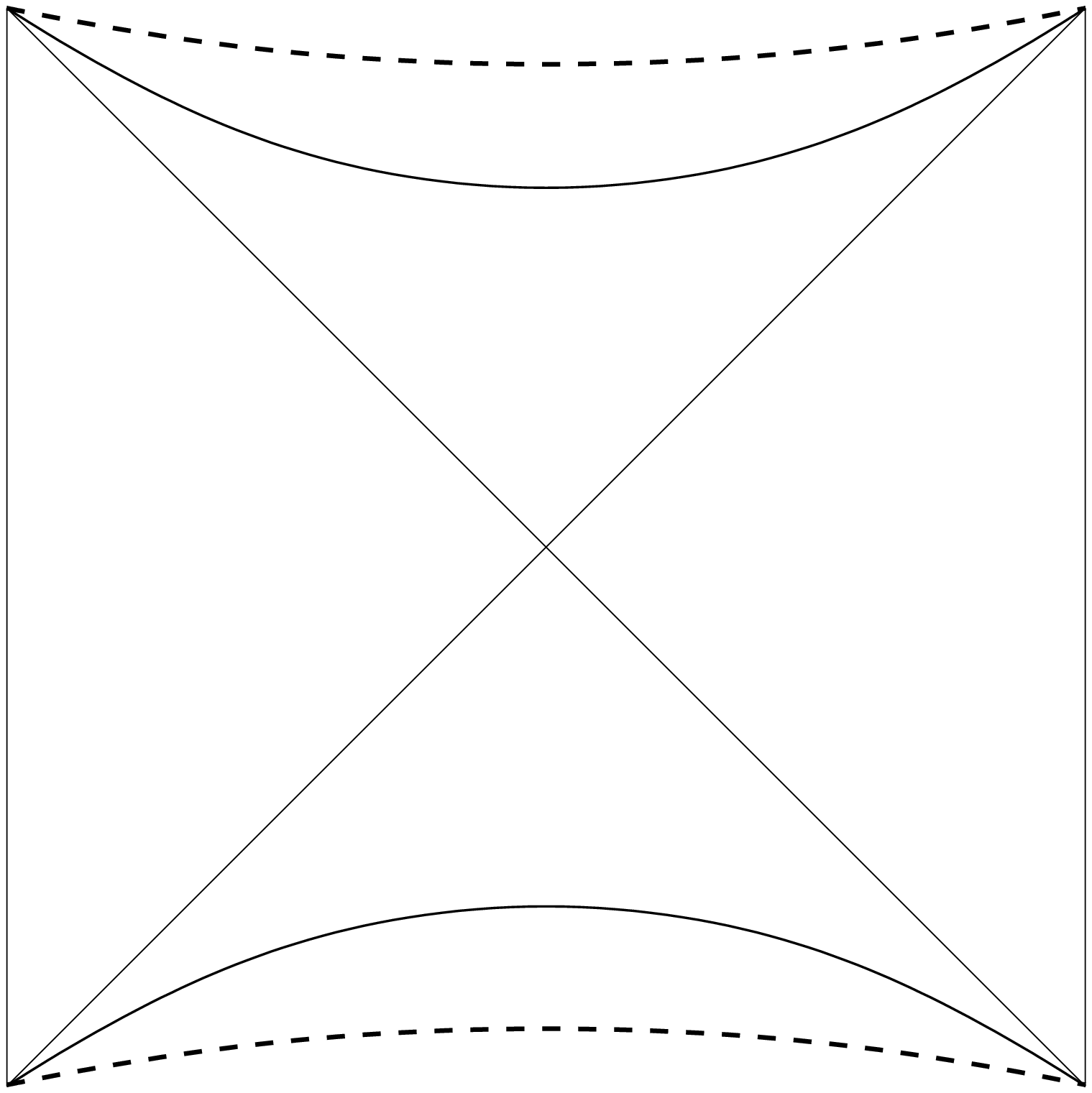}
           }
           \put(5,24){static}
           \put(30,24){static}
           \put(5,20){$\Theta_- <0$}
           \put(5,16){$\Theta_+ >0$}
           \put(18,30){$R'=0$}
           \put(15,40){Singularity}
           \put(15,-2){Singularity}
           \put(-3,38){(b)}
           \put(-3,30){\vector(2,1){8.2}}
           \put(-7,28){$\Theta_- = 0$}
           \put(6,42){\vector(1,-2){5}}
           \put(0,43){$\Theta_- >0, \Theta_+>0$}
           \put(33,42){\vector(-1,-2){3.5}}
           \put(27,43){$\Theta_-<0, \Theta_+<0$}
           \put(16,10){dynamic}
           \put(-6,19){Origin}
           \end{picture}
            }
   %   \put(120,0){\circle{10}}
     \put(36,0){
           \begin{picture}(0,0)
           \put(0,0){
           \includegraphics[width=40ex]{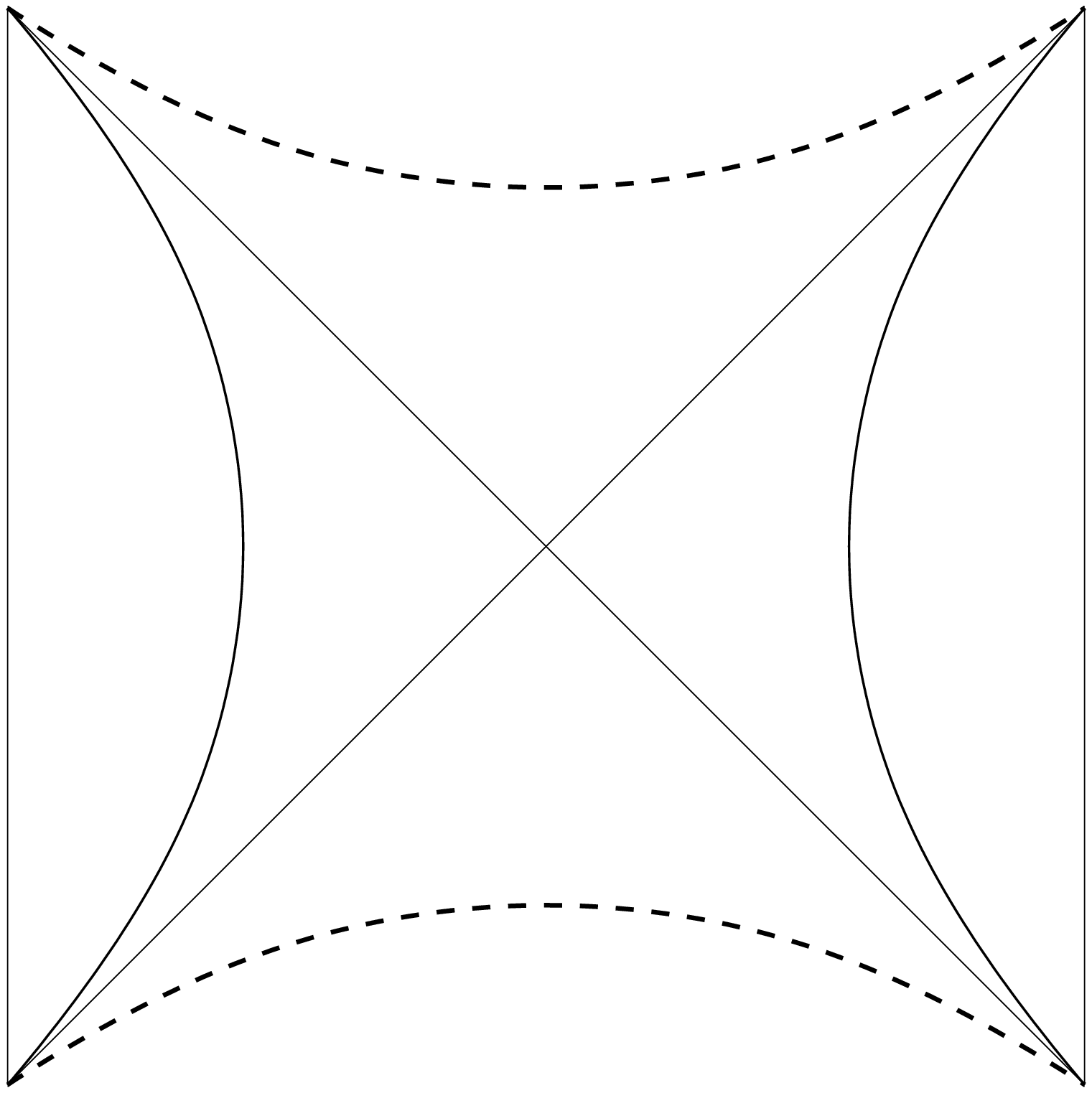}
           }
           \put(15,36){Singularity}
           \put(15,2){Singularity}
           \put(16,10){dynamic}
           \put(30,20){static}
           \put(41,19){Origin}
           \put(10.5,20){$R'=0$}
           \put(17,30){$\Theta_-<0$}
           \put(17,26){$\Theta_+<0$}
           \put(6,42){\vector(-1,-3){3}}
           \put(4,43){$\Theta_-<0, \Theta_+ >0$}
           \put(12,38){\vector(-1,-2){2.9}}
           \put(10,39){$\Theta_-=0$}
           \put(-3,38){(c)}
           \end{picture}
            }
 \end{picture}
\end{psfrags}
\end{center}
\caption{Penrose diagrams for the three different global structures of
solutions. Any point in the diagram (except where $R=0$) corresponds to a
two-sphere. The description is given in the text.
}\label{fig::penrose_diagrams}
\end{figure}

In the first case, $R'(\rho) > 0 \ \forall \rho > 0$, the solution exists for
all finite $\rho$. From Eq. (\ref{eq::Rpp}) we see, that $0 < R' \le 1
\  \forall \rho \ge 0$. Therefore $R$ diverges like $R = O(\rho)$ for large
$\rho$. From Eq. (\ref{eq::Qpp}) follows, that $Q' = O(\rho)$ and therefore
$Q = O(\rho^2)$. Eq. (\ref{eq::fpp}) shows, that $\phi$ goes to a constant
at infinity.
In this case the Killing horizon
is a past inner trapping horizon (see Fig. \ref{fig::penrose_diagrams}a)

In the second case $R'(\rho_E) = 0$, as $R''(\rho) \le 0$ for all $\rho$ and
$R'(\rho) < 0$ for $\rho > \rho_E$, $R$ goes to zero at some finite $\rho_S >
\rho_H$ (remember that we excluded the possibility of a second zero of $R$
in the static region in Sec. \ref{subsec::static-equations}). 
Again time evolution beyond the horizon exists for all $\rho < \rho_S$.
However in the limit $\rho \to \rho_S$ a spacetime singularity occurs.
This can be seen e.g. by investigating the behavior of the Kretschmann
invariant 
\bea
\mathcal R^{\mu\nu\sigma\tau} \mathcal R_{\mu\nu\sigma\tau} & = & \frac{1}{R^4} 
   \biggl( 4 + 4 R^2 Q'^2 R'^2 +
   R^4 Q''^2 + \nonumber\\ 
   & + & 8 Q R' \left( R^2 Q' R'' -
   R' \right) + 4 Q^2 \left( R'^4 + 2 R^2 R''^2
   \right) 
   \biggr).
\eea
Since by assumption $Q(\rho), Q'(\rho)$ and $R'(\rho)$ are negative in a
neighborhood of $\rho_S$, all terms are non-negative and clearly
not all denominators vanish in the limit $\rho \to
\rho_S$. Therefore the Kretschmann invariant blows up like $1/R^4$ in the
limit $\rho \to \rho_S$.

Concerning the character of the Killing horizon, two possibilities arise
depending on whether $R$ attains its maximum in the dynamic or in the
static region. If $\rho_E > \rho_H$ then the Killing horizon again is a past
inner trapping horizon (See Fig. \ref{fig::penrose_diagrams} b). 
If on the other hand $\rho_E < \rho_H$, then 
the inward and outward direction interchange in the static region and the 
horizon therefore is a future outer trapping horizon and corresponds thus
to a black hole horizon (See Fig. \ref{fig::penrose_diagrams}c).

\afterpage{\clearpage}

The three possibilities are summarized in Figs.
\ref{fig::penrose_diagrams}a - c. Fig. \ref{fig::penrose_diagrams}a 
shows the asymptotically de Sitter spacetimes. The areal radius $R$ is
increasing along a $u=const$ slice from zero at the origin to infinity at
$\mathcal I^+$. A cosmological horizon separates the static region from an
expanding dynamic region.

Fig. \ref{fig::penrose_diagrams} b shows the situation, where $R$ develops
an extremum in the dynamic region. Beyond the cosmological horizon spacetime
initially expands until it reaches its maximal spatial extension at time
$\rho_E$ and then recollapses to a singularity at finite proper time
$\tau_S = \int\limits_{\rho_H}^{\rho_S} d \rho /Q(\rho)$.

Fig. \ref{fig::penrose_diagrams} c finally shows a spacetime, where $R'=0$
in the static region. At this hypersurface the inward and outward directions
interchange. Therefore the collapsing dynamic region is enclosed by the
static region. The separating horizon can be interpreted as a black hole
horizon.

We close this section by giving an upper bound for
$R(\rho_E)$ for re-collapsing universes (Fig. \ref{fig::penrose_diagrams}b).
As already mentioned the dynamic region of our solutions 
corresponds to Kantowski-Sachs universes. Therefore we can follow the work
of Moniz \cite{Moniz-Kantowski-Sachs-universes-1993}
on the cosmic no-hair conjecture of Kantowski-Sachs models.
We re-investigate Eq.~(\ref{eq::constofm}). Remember that this 
equation is $1/R^2$ times the $({}^{\rho}_{\rho})$ component of the 
Einstein equations. As the coordinate $\rho$ plays the role 
of a time coordinate in the dynamic region, this equation is the Hamiltonian
constraint for the time evolution problem.
We rewrite Eq. (\ref{eq::constofm}) (devided by $R^2$) as
\be
R' \left( \frac{|Q'|}{R} + \frac{|Q| R'}{R^2}\right) = 
       \Lambda - \frac{1}{R^2} + \eta \left(|Q| (\phi')^2 + 2 \frac{\sin^2
\phi}{R^2}\right).
\ee
In terms of geometrical quantities on the spacelike hypersurfaces
$\rho=const$, the terms on the left hand side are
$\frac{1}{2}(K^2 - K^{ij}K_{ij})$, $K_{ij}$ being the extrinsic curvature
and $K$ the trace of $K_{ij}$.
The last term on the right hand side is the energy density
$T_{\mu\nu}n^{\mu}n^{\nu}$ and $1/R^2 = \frac{1}{2} {}^{(3)}\mathcal R$ with
${}^{(3)}\mathcal R$ being the three scalar curvature of the slices.

Note that the left hand side can only change sign if $R'$ changes sign.
So if at some instant of time $\rho_0$ the universe is expanding, 
$R'(\rho_0)> 0$, 
then the left hand side is positive.
Furthermore, as the energy densitiy is positive, we can give a lower bound
for the left hand side, namely
\be
R' \left( \frac{|Q'|}{R} + \frac{|Q| R'}{R^2}\right) \ge 
       \Lambda - \frac{1}{R^2}
\ee

Suppose now that in addition at $\rho_0$ the scalar curvature is smaller 
than
$2 \Lambda$, i.e. $R(\rho_0) > 1/\sqrt{\Lambda}$, 
then the right hand side is positive. As $R'(\rho_0)$ is positive initially
the lower bound on the right hand side increases away from zero, 
therefore making a change in sign of $R'$ impossible.
We can conclude from this, that if 
$R'(\rho_0)>0$ and $R(\rho_0) > 1/{\sqrt{\Lambda}}$ 
at some time $\rho_0$, the universe will expand
for ever and approach de Sitter space for late times.
On the other hand, a recollapsing universe must have $R(\rho) <
1/\sqrt{\Lambda}$ for all times $\rho_H < \rho < \rho_S$.

\section{Numerical Construction of Static Solutions}\label{sec::num_sol}
%%%%%%%%%%%%%%%%%%%%%%%%%%%%%%%%%%%%%%%%%%%%%%%%%%%%%%%%%%%%%%%%%%%%%%%

The problem of constructing static solutions is given by the boundary
value problem Eqs.
(\ref{eq::fpp})--(\ref{eq::Rpp}) together with the regularity conditions at
the origin  (\ref{eq::metric_QR_O}) and at the horizon 
(\ref{eq::field_QR_O}). In order to determine the global structure of such a
solution the data at the horizon (defined by the solution to the above
boundary value problem) are used as initial data for the time evolution
problem in the dynamic region.

We solve the boundary value problem numerically using a standard shooting
and matching method (see Appendix \ref{app::SM}) provided by
routine d02agf of the NAG library \cite{NAG}.
The numerical integration in the dynamic region is performed by 
routine d02cbf of the NAG library.

In all our numerical calculations we set $\Lambda = 3$.

\subsection{Phenomenology of Static Solutions for $\Lambda >
0$}\label{subsec::statSols_results}
%%%%%%%%%%%%%%%%%%%%%%%%%%%%%%%%%%%%%%%%%%%%%%%%%%%%%%%%%%%%%%%%

In order to investigate the dependence of the spectrum of solutions
on the coupling constant $\eta$, we start at $\eta = 0$ and follow a
solution to larger values of $\eta$.

\begin{figure}[h,t,b]
\begin{center}
\begin{psfrags}
 \psfrag{log(rho/rhoH)}[c][c][1][0]{$\log (\rho/\rho_H)$}
 \psfrag{pi/2}[l][c][1][0]{$\frac{\pi}{2}$}
 \psfrag{fn}[lb][rt][1][0]{$\phi_n$ }
\includegraphics[width=4in]{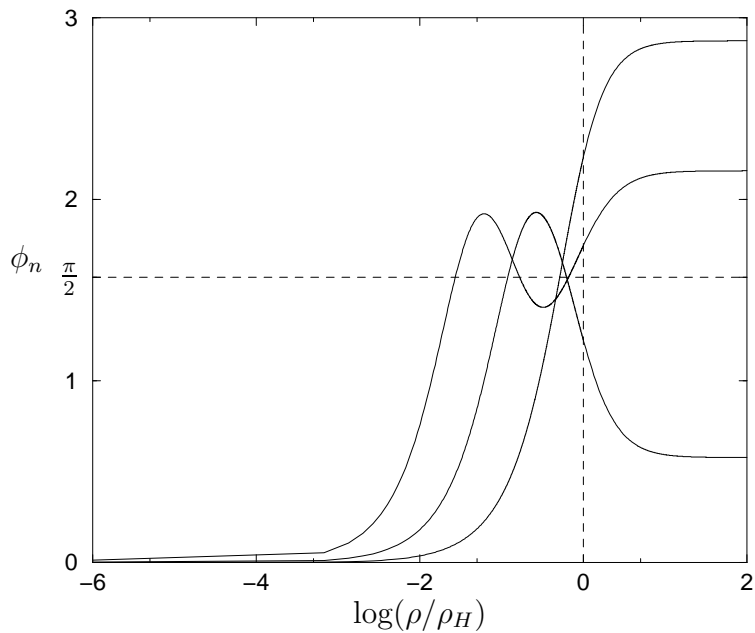}
\end{psfrags}
\end{center}
\caption{The first three excitations $\phi_n$ on fixed de Sitter background.
Within the static region the field oscillates $n$ times around $\pi/2$,
outside the horizon, it goes to a constant.
}\label{fig::static_eta0_phi}
\end{figure}

For $\eta = 0$, the vacuum Einstein equations with positive cosmological
constant yield de Sitter spacetime,
\be
R(\rho) = \rho, \qquad Q(\rho) = 1 - \frac{\Lambda \rho^2}{3}.
\ee 
It was already shown in \cite{sigma-on-de-Sitter}, 
that the remaining boundary
value problem for the field on fixed background admits a discrete
one-parameter family of regular solutions. In the static region these
solution oscillate around $\pi/2$. The number of these oscillations $n$
can be used to parameterize the family. Energy increases with the 
oscillation number and converges to the energy of the ``singular solution''
$\phi \equiv \pi/2$ (see Sec. \ref{subsec::exact_solutions}) from below.
Outside the cosmological horizon at $\rho = \sqrt{3/\Lambda}$ the field goes
to a constant at infinity (See Fig. \ref{fig::static_eta0_phi}).

%Directory:......TeX/Solutions/

\begin{figure}[h]
\begin{center}
\begin{psfrags}
 \psfrag{rho}[][][1][0]{$\rho$}
 \psfrag{0.470366<eta<0.470373}[l][c][1][0]{$0.470366 < \eta < 0.470373$}
 \psfrag{R_rho}[][][1][-90]{$R(\rho)$ }
\includegraphics[width=4in]{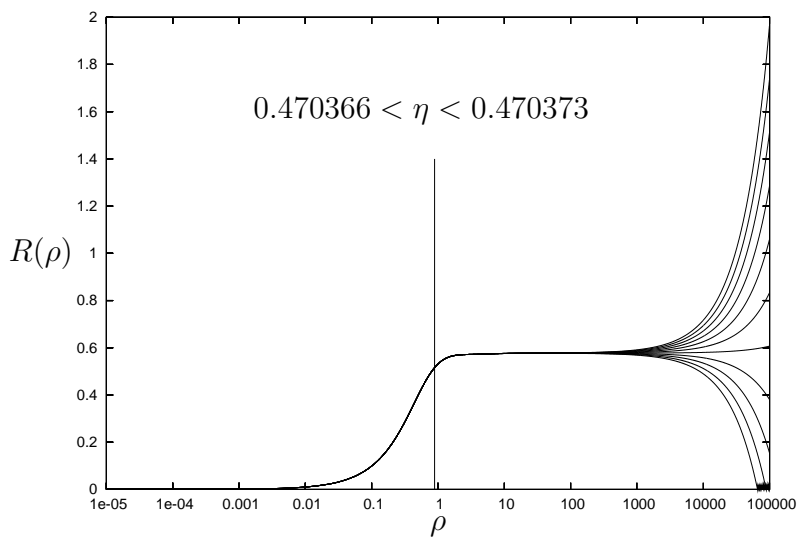}
\end{psfrags}
\end{center}
\caption{The metric function $R(\rho)$ for the first excitation at couplings
$0.470366 < \eta < 0.470373$. At $\eta_{crit}$ the global structure of
spacetime changes
from asymptotically de Sitter to a spacetime, that ends in a singularity.
The vertical line denotes the cosmological horizon at $\rho = 0.88761$.
}\label{fig::static_R_eta_crit}
\end{figure}

We concentrated on the first three excitations and investigated their
behavior as the coupling $\eta$ was turned on. We find that each member of
these excitations exists up to a maximal coupling constant $\eta_{max}(n)$.
The limit $\eta \to \eta_{max}$ needs some care and will be discussed in
Sec. \ref{subsec::eta_to_etamax}. In the following we describe
the properties of the solutions for $0 \le \eta < \eta_{max}$.
\begin{itemize}
\item For small $\eta$, $0 \le \eta < \eta_{crit}(n)$ spacetime is
asymptotically de Sitter as sketched in Fig.
\ref{fig::penrose_diagrams}a and described in the accompanying text.
The field behaves similar as in the uncoupled case.
\item For $\eta = \eta_{crit}(n)$ the areal radius $R$ does not diverge like
$\rho$ but rather goes to a constant at infinity. For even 
bigger couplings, $\eta_{crit}(n) < \eta < \eta_*(n)$ the maximum of $R$
occurs at earlier and earlier times $\rho_E$ (see Fig.
\ref{fig::static_R_eta_crit})
and time evolution ends in a
singularity, as described in Fig. \ref{fig::penrose_diagrams}b.  
\item At $\eta = \eta_*(n)$ the hypersurface, where $R$ is extremal merges
with the horizon. For bigger $\eta$, $\eta_*(n) < \eta < \eta_{max}(n)$,
$R$ attains its maximum in the static region. The situation is as in Fig.
\ref{fig::penrose_diagrams}c.
\end{itemize}  

\begin{figure}[h]
\begin{center}
\begin{psfrags}
 \psfrag{rho}[][][1][0]{$\rho$}
 \psfrag{R(rho)}[][][1][-90]{$R(\rho)$ }
 \psfrag{1oversqrtlambda}[r][c]{$\frac{1}{\sqrt{\Lambda}}$}
\includegraphics[width=4in]{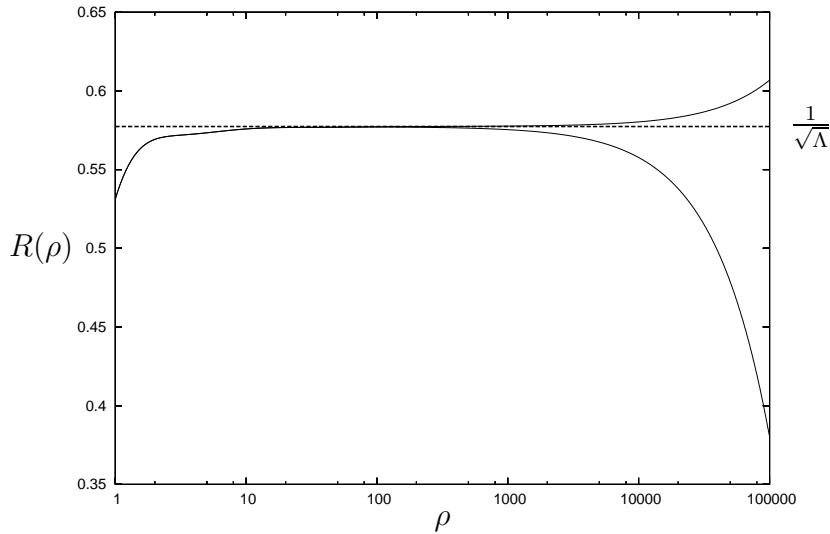}
\end{psfrags}
\end{center}
\caption{The areal radius $R(\rho)$ for the 
first excitation in the dynamic region. Plotted  are the solutions at
$\eta = 0.4703702$ and $\eta = 0.470370903$, close to $\eta_{crit}(1)$.
The dashed horizontal line denotes $1/\sqrt{\Lambda}$ ($\Lambda$ was set to
$3$ for the numerical calculation).  As can be seen, the areal radius of the
re-collapsing universe comes close to the upper bound
$1/\sqrt{\Lambda}$. We argue that therefore and due to the continuous
dependence of solutions on the coupling $\eta$, the upper bound is attained 
at infinity for $\eta = \eta_{crit}$: $R(\infty;\eta_{crit}) =
1/\sqrt{\Lambda}$.
}\label{fig::statSols_1oversqrtLambda}
\end{figure}

Figure (\ref{fig::statSols_1oversqrtLambda}) shows the areal radius
$R(\rho)$ for the first excitation for $\eta$ close to $\eta_{crit}$.
We have shown in Sec.~\ref{subsec::global_structure} 
that re-collapsing
universes must have $R(\rho) < 1/\sqrt{\Lambda}$ for all 
$\rho_H < \rho < \rho_S$. 
Fig. \ref{fig::statSols_1oversqrtLambda} shows, that $R(\rho_E)$ for the 
re-collapsing universe comes close to the upper limit. 
This suggests that the upper bound for $R$ is attained in the limit $\eta \to
\eta_{crit}$ at infinity: $R(\infty; \eta_{crit}) = 1/\sqrt{\Lambda}$.

%Directory:Working_directories/nr/Static/src/QR_gaugeinv/
%                        gauge_inv_exe/Beta_crit

%Directory:Working_directories/nr/Static/src/QR_gaugeinv/
%                        gauge_inv_exe/Beta_crit

%\afterpage{\clearpage}

The critical values of the coupling constants, $\eta_{crit}(n), \eta_*(n)$
and $\eta_{max}(n)$ depend on the excitation number $n$, as can be seen from
table \ref{tab::critical_eta}.

\begin{table}
\begin{center}
\begin{tabular}{|c|c|c|c|}
\hline
n & $\eta_{crit}$ & $\eta_*$ & $\eta_{max}$ \\
\hline
1 & 0.47037     & 0.533    &  1.0       \\
2 & 0.41981     & 0.48     &  0.74255   \\
3 & 0.41932     & 0.474    &  0.64931    \\
4 & 0.42606     & 0.4765   &  0.60260    \\
% 5 & 0.43325   & 0.4796   &  0.57530985  \\
\hline      
\end{tabular}
\end{center}
\caption{The critical values of the coupling constant for the first four
excitations. The values of $\eta_*(n)$ were read off by eye and therefore
are not as accurate as the other values. The values of $\eta_{max}(n)$ are
determined as described in Sec. \ref{subsec::eta_to_etamax}, in
particular the value of $\eta_{max}(1) = 1$ is exact.
While $\eta_{max}(n)$ decreases with the excitation number $n$ the other
critical values do not seem to share this behaviour.
}
\label{tab::critical_eta}
\end{table}

\subsection{The Limit $\eta \to \eta_{max}$}\label{subsec::eta_to_etamax}
%%%%%%%%%%%%%%%%%%%%%%%%%%%%%%%%%%%%%%%%%%%%%%%%%%%%%%%%%%%%%%%%%%%%%%%%%

\begin{figure}[p]
\begin{center}
\begin{psfrags}
 \psfrag{eta}[c][c][1][0]{$\eta$}
 \psfrag{rH/rhoH}[c][c][1][0]{$R_H/\rho_H$}
 \psfrag{E/rhoH}[c][c][1][0]{$E/\rho_H$ }
 \psfrag{fprimeO*rhoH}[c][c][1][0]{$\phi'(0) \rho_H$}
\includegraphics[width=4in]{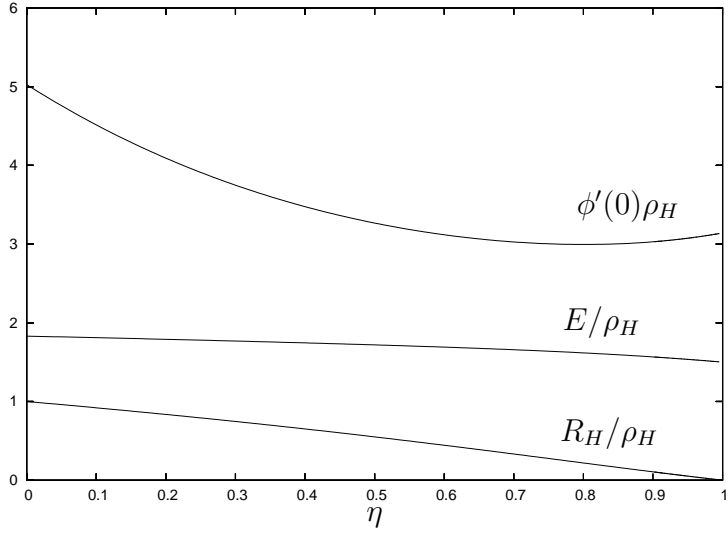}
\end{psfrags}
\end{center}
\caption{Some parameters of the first excitation -- measured in units of
$\rho_H$ -- as functions of the coupling constant. Except for $R_H$ the
parameters attain finite values in the limit $\eta \to \eta_{max}$, when
measured with respect to this unit.
}\label{fig::static_scale}
\end{figure}

Recall from Sec. \ref{subsec::static-equations} that the cosmological constant
$\Lambda$ sets the length scale in Eqs. (\ref{eq::fpp}) - (\ref{eq::Rpp}) 
and
that it can be eliminated from these equations, by introducing
the dimensionless quantities ${\bar \rho} = \sqrt{\Lambda} \rho$ and
${\bar R} = \sqrt{\Lambda} R$. This corresponds to measuring 
all quantities that have
dimension of length, as e.g. the energy $E$, the coordinate distance
of the horizon $\rho_H$ from the origin, the radial geometrical distance 
of the horizon
$d_H$ from the origin, the areal radius $R_H$ of the horizon, 
and $1/\phi'(0)$,  in units of
$1/\sqrt{\Lambda}$. 
We find that all parameters, that have dimension of length go to zero in the
limit $\eta \to \eta_{max}$ when measured with respect to this length
scale.
This indicates that $1/\sqrt{\Lambda}$ is not the appropriate length scale
for
taking this limit. We therefore switch to the alternative viewpoint with
$\rho_H$ as our length scale, and we fix $\rho_H = 1$. In this setup
$\Lambda$ depends on $\eta$ and the excitation index $n$ and goes to zero
in the limit $\eta \to \eta_{max}$. The parameters $E$, $d_H$ and
$1/\phi'(0)$ attain finite values when measured in units of $\rho_H$, whereas
$R_H/\rho_H$ goes to zero. (See Fig. \ref{fig::static_scale})
This strongly suggests, that
there exists a solution with $\eta = \eta_{max}$  
which obeys Eqs. (\ref{eq::fpp})-(\ref{eq::Rpp})
with $\Lambda = 0$ and has two centers of symmetry.
In particular this means that  
the static region of this solution has
no boundary, since any $t=const$ slice has topology ${\bf S}^3$. 

\begin{figure}[p]
\begin{center}
\begin{psfrags}
 \psfrag{beta}[c][c][1][0]{$\eta$}
 \psfrag{Rprime(rhoH)}[c][c][1][0]{$R'(\rho_H)$}
 \psfrag{f(rhoH)}[c][c][1][0]{$f(\rho_H)$ }
\includegraphics[width=4in]{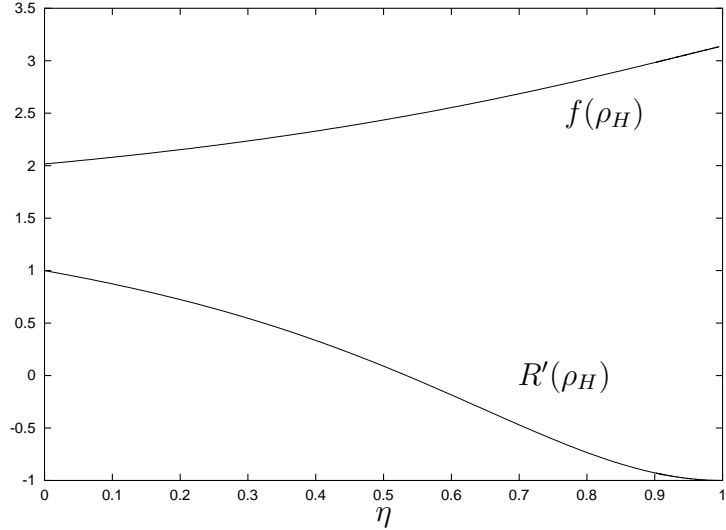}
\end{psfrags}
\end{center}
\caption{The parameters $\phi(\rho_H)$ and $R'(\rho_H)$ for
the first excitation as functions of $\eta$. In the limit $\eta \to
\eta_{max}$ $\phi(\rho_H)$ tends to $\pi$ and $R'(\rho_H)$ goes to $-1$,
which are necessary conditions for a second regular center of spherical
symmetry at $\rho_H$. 
}\label{fig::static_limit_eta_max}
\end{figure}

Furthermore, as can be seen from Fig. \ref{fig::static_limit_eta_max}, 
the dimensionless
parameter $\phi(\rho_H)$ for the first excitation tends to $\pi$, 
and $R'(\rho_H)$ tends to $-1$ in the limit $\eta \to \eta_{max}$.  
As will be shown below, 
$\Lambda = 0$ implies $Q \equiv 1$.
The limiting solution with $\Lambda = 0$ will therefore
satisfy the regularity conditions (\ref{eq::metric_QR_O}) and 
(\ref{eq::field_QR_O}) not
only at the axis $\rho = 0$ but also at the second zero of $R$, which
means that such a solution is globally regular with two (regular) centers of
spherical symmetry.
In fact, 
for the first excitation this limiting solution is just the static Einstein
universe (\ref{eq::EinsteinU}), which was already given in closed form
in Sec. \ref{subsec::exact_solutions}.

These observations allow one to determine the maximal value of the coupling
constant $\eta_{max}(n)$ not as a limiting procedure 
$\eta \to \eta_{max}$, but rather by solving the boundary value problem
Eqs. (\ref{eq::fpp}) - (\ref{eq::Rpp}) with $\Lambda =0$ and
with boundary conditions, that correspond to two regular centers 
of symmetry.

For $\Lambda = 0$ Eq. (\ref{eq::Qpp}) can be solved 
immediately to give $R^2 Q' = const$. 
According to the regularity conditions at the axis
(\ref{eq::metric_QR_O})
the constant has to vanish, which means that $Q' \equiv 0$ and 
therefore $Q \equiv 1$.
The remaining system of equations is:
\bea
(R^2 \phi')' & = &\sin (2 \phi) \label{eq::flambda0}\\
R'' & = & - \eta R \phi'^2, \label{eq::Rlambda0}
\eea
and
\be\label{constofmlabmda0}
2 \eta \sin^{2}\phi  - \eta R^{2} \phi'^{2} + R'^{2} - 1 = 0.
\ee
Note, that this system of ODEs is scale invariant, that is any solution
$R(\rho), \phi(\rho)$ leads via rescaling to the one parameter family of
solutions given by $a R(a \rho), \phi(a \rho)$. Keeping this in mind, we can
fix the scale arbitrarily, e.g. in setting the first derivative of the field
$\phi$ equal to one at the origin: $\phi'(\rho=0) = 1$. 
Thereby any solution, that
is regular at the origin, is determined {\it entirely } by the value of the
coupling constant $\eta$.
Regularity conditions at the second ``pole'' $R(\rho_P) = 0$ are the same as
at
the origin, except that $\phi$ either tends to $\pi$, if its excitation
number is odd, or to $0$, if it has even excitation number.
This can be inferred from $\pi/2 < \phi_H < \pi$ for $n$ odd and
$0 < \phi_H < \pi/2$ for $n$ even for all $\eta < \eta_{max}$, 
which is true for $\eta=0$ and according to (\ref{eq::phiHp})
there cannot be any $\eta < \eta_{max}$ where a solution with $n \ge 1$ has
$\phi_H = 0,\pi$ or $\pi/2$ (apart from the ``singular solution'' $\phi \equiv
\pi/2$)\footnote{If $\phi_H = 0, \pi$ or $\pi/2$ then
$\phi_H' = \phi_H''= 0$ and therefore all higher derivatives vanish}.
Note that this corresponds to all odd solutions
having winding number $1$, whereas even solutions are in the topologically
trivial sector.

These regularity conditions together with the invariance of the equations 
under reflection at the location of the maximal two-sphere 
$R'(\rho_E) = 0$, causes globally regular solutions $R(\rho)$ to be 
symmetric around $\rho_E$ whereas $\phi(\rho) - \pi/2$ is either antisymmetric
for $n$ odd or symmetric
for $n$ even. 
For $\phi$ symmetric the formal power series expansions of $R(\rho)$ and
$\phi(\rho)$ around $\rho = \rho_E$ gives
\bea\label{eq::fsymm}
R(\rho)& = & R(\rho_E) + O((\rho - \rho_E)^4),  \nonumber\\
\phi(\rho) & =& \arcsin \sqrt{1/2 \eta} + \frac {2 \sqrt{ 1- 1/2
\eta}}{R(\rho_E)^2 \sqrt{2\eta}} \frac{(\rho -\rho_E)^2}{2!} + O((\rho
-\rho_E)^4),
\eea
and for $\phi- \pi/2$ antisymmetric we get  
\bea\label{eq::fantisymm}
R(\rho)& = & \frac{2 \eta - 1}{\eta \phi'(\rho_E)^2} - (2\eta -1) 
     \frac{(\rho - \rho_E)^2}{2!} + O((\rho - \rho_E)^4),  \nonumber\\
\phi(\rho) & =& \frac{\pi}{2} + \phi'(\rho_E)(\rho -\rho_E) + 
        O((\rho-\rho_E)^3).
\eea
In order to solve the system (\ref{eq::flambda0}), (\ref{eq::Rlambda0}) 
we again use
the shooting and matching method on the interval [origin, $\rho_E$] using the
above Taylor series expansions to determine the boundary conditions at
$\rho = \rho_E$. Shooting parameters are now $\rho_E, \phi'(\rho_E)$ and 
$\eta$
for odd solutions and $\rho_E, R(\rho_E)$ and $\eta$ for even solutions.
The results are displayed in Table \ref{tab::Lambda0Results}.

%Directory:......TeX/Solutions/

%Directory:......TeX/Solutions/

\begin{table}[hbt]
\begin{center}

\begin{tabular}{|c|c|c|c|c|}
\hline
$n$ &$\eta_{max}$& $\rho_P = d_P$ & $E/2\pi f_{\pi}^2$ & $E/2 \pi f_{\pi}^2 d_P$
                                                                \\
\hline
1   &      1      &    $\pi$     & $3 \pi/2$          &   $ 3/2 $   \\
2   &  0.74255    &  6.74225     &  11.78039          &   1.74724   \\
3   &  0.64931    & 12.10140     &  22.43662          &   1.85405   \\
4   &  0.60260    & 19.63717     &  37.47302          &   1.90827   \\
% 5 &  0.57530985 & 29.75265     &  57.68192          &   1.93872   \\
\hline
\end{tabular}
\end{center}
\caption{Results for the first three excitations for $\Lambda = 0$. Since $Q
\equiv 1$
the coordinate distance $\rho_P$ of the two regular ``poles'' equals the
radial
geometrical distance $d_P$. The energy density $\rho_P$ and energy $E$ are
 given in units where
$\phi'(0) = 1$. The ratio $E/d_P$ can be 
compared to the results for solutions
with $\Lambda > 0$ and represents the limit $\eta \to \eta_{max}$ for
those solutions .}
\label{tab::Lambda0Results} 
\end{table}

\afterpage{\clearpage}

It is clear from (\ref{eq::fsymm}) and (\ref{eq::fantisymm}), 
that regular solutions for 
$\Lambda = 0$ can only exist if $\eta > 1/2$. Assuming now that our
numerical observations concerning the first few excitations extend to
higher excitations, we argue as follows:
since
every "branch" of the "$\Lambda > 0$ solutions" persists up to a maximal
value of $\eta$, which can be computed by solving the boundary value problem
(\ref{eq::flambda0}),(\ref{eq::Rlambda0}) 
together with regularity conditions at the 
two "poles" -- which implies $ \eta > 1/2 $ -- and since we know, 
that in the limit $\eta \to 0$ there
exists an infinite number of excitations \cite{sigma-on-de-Sitter},
we conclude that this whole family of solutions with $\Lambda > 0$
persists up to some maximal value $\eta_{max}(n)$, which is {\em greater} 
than $1/2$.
In other words, for any $\eta < 1/2 $ there exists a countably infinite
family of solutions with $\Lambda > 0$, whereas for $\eta > 1/2$ our
numerical analysis shows, that only a finite number of solutions exists.
(See Table \ref{tab::Lambda0Results}).

\subsection{Stability}\label{subsec::statSols_stability}
%%%%%%%%%%%%%%%%%%%%%%%%%%%%%%%%%%%%%%%%%%%%%%%%%%%%%%%

In order to analyze the stability properties of the above described static
solutions, small perturbations of the metric functions and the field are
considered. We set
\bea\label{eq::perturb_QR}
Q(t,\rho) & = & Q_n(\rho) + \delta Q(t,\rho), \nonumber \\
R(t,\rho) & = & R_n(\rho) + \delta R(t,\rho), \nonumber \\
\phi(t,\rho) & = & \phi_n(\rho) + \delta \phi(t,\rho).
\eea
$Q_n(\rho)$, $R_n(\rho)$ and $\phi_n(\rho)$ denote the $n$-th static
excitation. The perturbations $\delta Q(t,\rho)$, $\delta R(t,\rho)$ and
$\delta \phi(t,\rho)$ are considered to be small such that the 
equations can be linearized in these quantities. 

Inserting the ansatz (\ref{eq::perturb_QR}) into Eqs. 
(\ref{eq::ttplusrrminustwothth}) -- (\ref{eq::trho}), making use of the
static equations and linearizing in the perturbations gives a 
coupled system of linear PDEs for the perturbations. 
The momentum constraint (\ref{eq::trho}) contains only first time
derivatives, which enter each term linearly, so this equation can be
integrated with respect to time to give
\bea\label{eq::deltaQ}
\delta Q(t,\rho) & = & \frac{Q_{n}'(\rho)\delta R (t,\rho)}{R_{n}'(\rho)} - 
\frac{2 Q_{n}(t,\rho) \delta R'(t,\rho)}{R_{n}'(\rho)} - \frac{2 \eta 
Q_{n}(\rho) R_{n}(\rho) \phi_{n}'({\rho}) \delta \phi(t,\rho)}{R_{n}'(\rho)}.
\nonumber \\
 & & 
\eea
The constant of integration, a function $g(\rho)$ is determined by the
Hamiltonian constraint (making use of the other equations) 
to be
\be
g(\rho) = \frac{const}{R_n(\rho) R_n'(\rho)},
\ee
so for perturbations $\delta Q(t,\rho)$  regular at the origin,
the constant has to vanish and therefore $g(\rho) \equiv 0$. 

Eqs. (\ref{eq::ttminusrr}) and (\ref{eq::phi}) then give
\bea\label{eq::deltaRdotdot}
{\delta \ddot R} (t,\rho) & = &- Q_{n}^{2}(\rho)
     \left( \delta R ''(t,\rho) + \eta \phi_{n}'^{2}(\rho) \delta R (t,\rho)
       + 2 \beta R_{n}(\rho) \phi_{n}'(\rho) \delta \phi' (t,\rho) \right) 
\nonumber\\
& & 
\eea
and
\bea\label{eq::deltaphidotdot}
& & - \frac{R_{n}^{2}(\rho) \delta {\ddot \phi}(t,\rho)}{Q_{n}(\rho)} +  
  \left(R_{n}^2(\rho) Q_{n}^2(\rho) \delta \phi'(t,\rho) \right)' + 
\nonumber\\ 
& + & \left( \frac{R_n^2(\rho)\phi_{n}'(\rho)}{R_{n}'(\rho)} \left(  
 Q_{n}'(\rho)\delta R (t,\rho) - 
  2 Q_{n}(\rho) \delta R'(t,\rho) - 2 \eta 
Q_{n}(\rho) R_{n}(\rho) \phi_{n}'(\rho) \delta \phi(t,\rho) \right) \right)' 
+ \nonumber\\
& + & \left(2 R_{n}(\rho)Q_{n}(\rho)\phi_{n}'(\rho) \delta R (t,\rho) \right)'
=
 2 \cos (2 \phi_{n}(\rho)) \delta \phi (t,\rho).
\eea

Now it is important to note, that the form (\ref{eq::coordinates_QR})
still contains a certain gauge freedom. Therefore the perturbations
(\ref{eq::perturb_QR}) may -- in addition to physical modes -- 
also represent pure gauge modes.
We investigate this, by considering small coordinate
transformations 
\bea
t & \to & \bar t = t + \epsilon \chi^t(t,\rho)\\
\rho & \to & \bar \rho = \rho + \epsilon \chi^\rho(t,\rho).
\eea    

As described in Sec.~\ref{subsec::gauge_modes} 
the perturbations (\ref{eq::perturb_QR}) then
transform
up to order $\epsilon$ according to the Lie derivative along $\chi$ of the
``background solution''. In detail we have
\bea\label{eq::gaugemodes_QR}
\delta \bar Q & = &  \delta Q - \epsilon 
       (Q_{n}' \  \chi^{\rho} + 2 Q_{n} \ \chi^{t}_{,t})
\nonumber\\
\delta \bar R & = &  \delta R - 
       \epsilon \  R_{n}' \chi^{\rho}                       \nonumber\\ 
\delta {\bar \phi}  & = &  \delta \phi - \epsilon \phi_{n}' \chi^{\rho},
\eea 
where $\chi$ is subject to the following conditions:
\bea
\frac{1}{Q_{n}} \chi^{\rho}_{,t} & = & - Q_{n} \chi^{t}_{,\rho} \label{eq::chi1}
 \\
\chi^{t}_{,t} & = & - \chi^{\rho}_{,\rho} \label{eq::chi2}. 
\eea
Condition (\ref{eq::chi1}) comes from the fact, that the shift is zero in
this gauge and therefore ${\mathcal L}_{\chi} (g_{n})_{\rho t} = 0$, 
and condition (\ref{eq::chi2}) arises from the fact, that
$\delta g_{tt} = Q_{n}^{2} \delta g_{\rho \rho}$ and therefore
 ${\mathcal L}_{\chi} (g_{n})_{t t} =
Q_{n}^2 {\mathcal L}_{\chi} (g_{n})_{\rho \rho}$.
Combining Eqs. (\ref{eq::chi1}) and
(\ref{eq::chi2}) yields a wave equation for $\chi^{\rho}$:
\be\label{eq::chirho}
Q_{n}^2 \chi^{\rho}_{,\rho \rho}  - \chi^{\rho}_{,tt} = 0. 
\ee
Gauge transformations that respect the choice (\ref{eq::perturb_QR}) 
therefore
are determined by the single function $\chi^{\rho}$, which is 
subject to Eq. (\ref{eq::chirho}).

Pure gauge modes are perturbations, that can be removed by a coordinate
transformation, i.e. they have to be of the form
\bea\label{eq::pure_gaugemodes_QR}
\delta Q & - & \epsilon 
       (Q_{n}' \  \chi^{\rho} + 2 Q_{n} \ \chi^{t}_{,t})
\nonumber\\
\delta R & = &
       \epsilon \  R_{n}' \chi^{\rho}                       \nonumber\\ 
\delta \phi & = & \epsilon \phi_{n}' \chi^{\rho},
\eea 
with $\chi^{\rho}$ and $\chi^t$ subject to (\ref{eq::chi1},\ref{eq::chi2}).
Clearly pure gauge modes satisfy the perturbation equations
(\ref{eq::deltaRdotdot})--(\ref{eq::deltaphidotdot}).

On the other hand, one can try to combine the perturbations in such a way,
that the combination is invariant under a coordinate transformation.
One can show, that such a quantity has to be of the form
\be\label{eq::gaugeinvxi0}
\xi (\rho, t) = a(\rho) (R_{0}'(\rho) \ \delta f (\rho,t) - 
      f_{0}'(\rho) \  \delta R (\rho,t) ), 
\ee 
where $a(\rho)$ is an arbitrary function of $\rho$, which we choose to be
$1/R_n(\rho)$. Using the definition for $\xi$ and Eqs.
(\ref{eq::deltaRdotdot}), (\ref{eq::deltaphidotdot}) we obtain a single 
pulsation equation for the gauge-invariant quantity $\xi$:
\bea\label{eq::pulsationeq}
& - & \frac{R_n(\rho)^2\,R_n'(\rho)}
     {Q_n(\rho)}{\ddot \xi} (t,\rho)
+ 
   Q_n(\rho)\,R_n(\rho)^2\,R_n'(\rho)\,\xi''(t,\rho)
 + \biggl( 2\,\eta \,Q_n(\rho)\,R_n(\rho)^3\,
       \phi_n'(\rho)^2 +                    \nonumber\\
    & + & R_n(\rho)^2\,Q_n'(\rho)\,R_n'(\rho) + 
      4\,Q_n(\rho)\,R_n(\rho)\,R'_n(\rho)^2  \biggr) \,
    \xi'(t,\rho) +  \nonumber\\
& + & 
       \biggl( \eta \,R_n(\rho)^3 \,\phi_n'(\rho)^2\,
       Q_n'(\rho) - 2\,\cos (2\,\phi_n(\rho))\,R_n'(\rho) + \nonumber\\
& + &
      2\,\eta \,Q_n(\rho)\, R_n(\rho)^2\,\phi_n'(\rho)^2\,
       R_n'(\rho) + 
      2\,Q_n(\rho)\,R_{n}'(\rho)^3 - \nonumber \\  
&-&   R_n(\rho)\,\left( 2\,\eta \,\sin (2\,\phi_n(\rho))\,
          \phi_n'(\rho) - Q_n'(\rho)\,R_n'(\rho)^2 \right)  \biggr)
        \xi(t,\rho)
        = 0.
\eea
As the coefficients do not depend on time, we can work with Fourier modes
\be
\xi(t,\rho) = e^{i \sigma t} \ y(\rho),
\ee
which turns Eq. (\ref{eq::pulsationeq}) into a linear second order ODE
\bea\label{eq::thisistheeq}
& & Q_n(\rho)^2\,R_n(\rho)^2\,R_n'(\rho)\,y''(\rho) 
  + 
   \biggl( 2\,\eta \,Q_n(\rho)^2\,R_n(\rho)^3\,
       \phi_n'(\rho)^2 + \nonumber \\
& + & Q_n(\rho)\,R_n(\rho)^2\,Q_n'(\rho)\,
       R_n'(\rho) +
       4\,Q_n(\rho)^2\,R_n(\rho)\,R_n'(\rho)^2
       \biggr) \,y'(\rho)   
 +  \nonumber\\
& + & 
  \biggl( -2\,\eta \,Q_n(\rho)\,R_n(\rho)\,\sin (2\,\phi_n(\rho))\,
       \phi_n'(\rho) + 
       \eta \,Q_n(\rho)\,R_n(\rho)^3\,
       \phi_n'(\rho)^2\,Q_n'(\rho) - \nonumber\\
& - &  2\,\cos (2\,\phi_n(\rho))\,Q_n(\rho)\,R_n'(\rho) +  
      2\,\eta \,Q_n(\rho)^2\,R_n(\rho)^2\,
       \phi_n'(\rho)^2\,R_n'(\rho) + \nonumber\\
& + &
      Q_n(\rho)\,R_n(\rho)\,Q_n'(\rho)\,R_n'(\rho)^2 + 
      2\,Q_n(\rho)^2\,R_n'(\rho)^3 \biggr) \, y(\rho)
  = - \sigma^2\,R_n(\rho)^2\,R_n'(\rho) y(\rho). \nonumber\\
& & 
\eea
Again this equation has regular singular points at the origin where $R(0) =
0$ and at the horizon $Q(\rho_H) = 0$. The corresponding regularity
conditions for $y(\rho)$ are 
\be
y'(0) = 0, \qquad y(\rho) \sim (\rho_H - \rho)^{\alpha}, \quad 
\textrm{with} \quad \alpha = - \frac{\sqrt{-\sigma^2}}{Q_H'},
\ee
for negative $\sigma^2$.

If the background solution has $R_n' > 0$ for all $\rho < \rho_H$, then 
Eq. \ref{eq::thisistheeq} together with the boundary conditions constitutes
an eigenvalue problem with eigenvalue $\sigma^2$ and eigen vector
$y(\rho)$. As we are interested in unstable modes, we look for negative
eigenvalues $\sigma^2$.

If on the other hand the background geometry contains a maximal two-sphere
$R'(\rho_E) = 0$ within the static region $\rho_E < \rho_H$, then Eq.
(\ref{eq::thisistheeq}) has an additional singular point within the range
of integration. The behavior of the two independent solutions in the 
vicinity of this singular point is 
\bea
y_1(\rho) &=& (\rho - \rho_E)^3 \sum\limits_{n=0}^{\infty} 
                    a_n (\rho - \rho_E)^n \nonumber\\
y_2(\rho) &=& \sum\limits_{n=0}^{\infty} b_n x^n,
\eea 
so the general solution stays regular near $\rho = \rho_E$.
Nevertheless, the coefficients of the first and zeroth derivative of
$y(\rho)$ in Eq. (\ref{eq::thisistheeq}) are singular, which renders the
standard numerical shooting and matching methods impossible. 

Nevertheless we tried to solve this problem, using a standard relaxation 
method (routine d02raf of the NAG library \cite{NAG}) 
on one hand. On the other hand, we
discretized Eq. (\ref{eq::thisistheeq}) by hand, thereby turning the Sturm
Liouville eigen value problem into an algebraic eigen value problem.
We 
only present preliminary results
here: for $\eta = 0$ stability was already analyzed in 
\cite{sigma-on-de-Sitter}. 
It turned out, that the $n$-th static excitation has
$n$ unstable modes. Our preliminary investigation of the first three
excitations in the coupled case gives,
that these solutions keep their unstable modes when gravity is turned on 
until in the limit of maximal coupling one negative eigenvalue crosses
zero. The occurrence of one ``indifferent mode'' in the limit $\eta =
\eta_{max}$ is due to the fact, that there the equations are 
scale invariant, i.e. for each coupling $\eta_{max}(n)$ we have a one
parameter family of solutions $(R_{\lambda} (\rho) = \lambda R_n(\lambda
\rho), \phi_{\lambda}(\rho) = \phi_n(\lambda)(\rho))$. 
For $\lambda = 1 + \epsilon$ we can write
\bea
R_{\lambda}(\rho) & = &R_n(\rho) + \epsilon (R_n(\rho) + R'_0(\rho) \rho)\\
\phi_{\lambda}(\rho) & = & \phi_n(\rho) + \epsilon \phi'_n(\rho) \rho,
\eea
which just corresponds to an ``indifferent'' mode with $\sigma = 0$.

%%%%%%%%%%%%%%%%%%%%%%%
%
%\documentclass[12pt,a4paper]{report}
%\usepackage{german,a4,bbm,graphicx,psfrag}
%
%
%\include{diss_macros}

%\textwidth=15cm
%\textheight=22cm
%\topmargin=-2cm
%\oddsidemargin=1cm
%\pagestyle{plain}
%\parindent=0pt                    % no indentation for paragraphs,
%\parskip=5pt plus 2pt minus 1pt   % but do a little skip

\chapter{Self-similar Solutions of the Self-gravitating $\sigma$
Model}\label{chap::SSSolutions}
%%%%%%%%%%%%%%%%%%%%%%%%%%%%%%%%%%%%%%%%%%%%%%%%%%%%%%%%%%%%%%%%%%%%%%%

As self-similar solutions usually govern type II critical phenomena
(see Chap.~\ref{chap::criticalCollapse}), it is of great advantage to know
about their properties like existence and stability from a ``direct
construction''. The ``direct construction'' -- in contrast to a construction
that uses time evolutions of near critical data (see Chapter
\ref{chap::criticalCollapse}) -- 
profits from the symmetry, which is imposed on the equations.
This
way the problem of constructing self-similar solutions reduces to a
boundary value problem for ODEs in the case of continuously self-similar
solutions (see Sec. \ref{sec::CSSsolutions}) and to a time-periodic boundary
value problem for PDEs in case of discrete self-similar solutions 
(see Sec.~\ref{sec::DSSsolutions}). 
CSS solutions of the $\sigma$ model have already been constructed by Bizon
\cite{Bizon-1999-existence-of-self-similar-sigma-CSS-solutions}
and Bizon and Wasserman
\cite{Bizon-Wasserman-2000-CSS-exists-for-nonzero-beta}. 
We reproduce their results here.
Stability is analyzed via the usual
method for CSS solutions (Sec.~\ref{subsec::CSSstability_BVP}). In addition
J.~Thornburg \cite{JT-matrix-analysis} proposed and carried out a
method, that is based on discretization of the field 
and uses the full (nonlinear)
field equations (Sec.~\ref{subsec::Matrix_Analysis}). 

We summarize our results in Table \ref{table:CSS-DSS-relevant}. In
Sec.~\ref{subsec::CSS-DSS} finally we compare the CSS (first excitation)
and DSS solution in a certain range of couplings, where both solutions
exist. Our observations lead us to conjecture, that the DSS solution
bifurcates from the first CSS excitation in a homoclinic loop bifurcation at
$\eta \simeq 0.17$. To our knowledge up to now such a bifurcation has not
been observed in the context of self-similar solutions to self-gravitating
matter fields.

\section{Continuous and Discrete Self-Similarity}\label{sec::CSS-DSS}
%%%%%%%%%%%%%%%%%%%%%%%%%%%%%%%%%%%%%%%%%%%%%%%%%%%%%%%%%%%%%%%%%%%%%

A spacetime $(M,g)$ is said to be {\it discretely self-similar (DSS)} 
if it admits a
diffeomorphism $\Phi_\Delta: M \to M$, which leaves the metric
invariant up to a constant scale factor, that is 
\be\label{def::DSS}
\left. (\Phi_{\Delta}^* g  )\right|_p  = 
       e^{2  \Delta} \left. g \right|_p \qquad \forall p \in M,
\ee 
where $\Delta$ is a real constant.

A spacetime, that admits a one-parameter family of such diffeomorphisms,
parametrized by $\Delta$ and with $\Phi_0 = id_M$, is called
{\it continuously self similar (CSS)}.
The generating vector field  
$\xi = \frac{d}{d \Delta} \left. \Phi_{\Delta}\right|_{\Delta=0}$ is a 
special 
case of a conformal Killing vector field. It obeys the conformal Killing
equation  
\bea\label{def::homothetic}
{\mathcal L}_{\xi} \left. g \right|_p & = & 
      \lim\limits_{\Delta \to 0} \frac{1}{\Delta}
      \left( \left. g\right|_p  - 
             \left. ((\Phi_{\Delta}^{-1})^* g) \right|_p \right) = \nonumber\\ 
      & = & \lim\limits_{\Delta \to 0} \frac{1}{\Delta}
      \left( \left. g\right|_p  - 
             \left. e^{-2 \Delta }g \right|_p \right) 
      = 2 \left. g \right|_p.
\eea
with a constant factor in front of the metric at the right hand side.
A vector field satisfying (\ref{def::homothetic}) is called 
{\em homothetic}.

At any point $p \in {\mathbbm M}$ we can compare the original metric
$ \left. g\right|_p$ and the metric $\bar g$ that results from pulling back
$\left. g \right|_{\Phi_{\Delta}(p)}$ to p, 
$\left. {\bar g} \right|_p = \left. (\Phi_{\Delta}^* g) \right|_p
= e^{2 \Delta} \left. g \right|_p$. 
As $g$ and $\bar g$ only differ by a constant rescaling both metrics
yield the same covariant derivative, $\nabla_{g} =
\nabla_{{\bar g}}$. Therefore the Riemann as well as the Ricci tensors
are identical, $\left. \mathcal R^{\sigma}_{\mu \tau \nu}(\bar g) \right|_p = 
\left. \mathcal R^{\sigma}_{\mu \tau \nu} (g) \right|_p$ and
$\left. \mathcal R_{\mu \nu}(\bar g) \right|_p = 
\left. \mathcal R_{\mu \nu} (g) \right|_p. $
The Ricci scalar scales as the inverse metric,
\be
\left. \mathcal R(\bar{g})\right|_p = 
\left. {\bar g}^{\mu\nu} \mathcal R_{\mu\nu}(\bar{g}) \right|_p = 
\left. e^{-2 \Delta} g^{\mu\nu} \mathcal R_{\mu\nu}(g) \right|_p = 
\left. e^{-2 \Delta} \mathcal R(g)\right|_p,  
\ee    
and therefore the Einstein tensors for both metrics agree:
\be
\left. G_{\mu\nu}(\bar{g}) \right|_p = \left. G_{\mu\nu}(g)\right|_p.
\ee
Now the Riemann, Ricci and Einstein tensor 
and the scalar curvature (summarized as $T$) behave under the pull-back
of a general diffeomorphism $f:M \to M$ as
\be\label{pullbackofmetrictensors}
\left. (f^* T (g))\right|_p = 
   \left. T(f^*g)\right|_p.
\ee
For a DSS spacetime we therefore have
\bea\label{geomDSS}
\left. (\Phi_{\Delta}^* \mathcal R)^{\sigma}_{\mu \tau \nu}\right|_p & = & 
                \left.
\mathcal R^{\sigma}_{\mu\tau\nu}\right|_p,  \nonumber\\
\left. (\Phi_{\Delta}^* \mathcal R)_{\mu\nu}\right|_p & = & \left.
\mathcal R_{\mu\nu}\right|_p,  \nonumber\\
\left. (\Phi_{\Delta}^* \mathcal R)\right|_p & = & e^{-2 \Delta} \left.
\mathcal R\right|_p , \nonumber\\
\left. (\Phi_{\Delta}^* G)_{\mu\nu}\right|_p & = & \left.
G_{\mu\nu}\right|_p.
\eea
For a CSS spacetime the above considerations directly yield 
\bea
\mathcal L_{\xi} \mathcal R^{\sigma}_{\mu\tau\nu} &=& \mathcal L_{\xi}
\mathcal R_{\mu\nu} = 0, \nonumber\\
\mathcal L_{\xi} \mathcal R &=& (\mathcal L_{\xi} g^{-1})^{\mu\nu} 
       \mathcal R_{\mu\nu} = -2 \mathcal R,
\nonumber\\
\mathcal L_{\xi} \ G_{\mu\nu} &=& 0.
\eea

\subsection{Adapted Coordinates}\label{subsec::adapted}
In order to simplify the discussion,
we introduce coordinates, that are adapted to the symmetry (see
e.g. \cite{Gundlach-1996-scaling-in-critical-collapse}).

In order to formally construct such a coordinate system for a DSS spacetime,
we choose a hypersurface $\Sigma$, with $\Sigma \cap \Phi_{\Delta}(\Sigma) = 
\emptyset$, provide it with coordinates $(z^i)$, $i=1,2,3$ and label it 
with $\tau_0$.
(Up to now this hypersurface can be spacelike, null or timelike).
We apply the diffeomorphism to $\Sigma$, label $\Phi_{\Delta}(\Sigma)$ with
$\tau_0 - \Delta$ and choose the coordinates in this hypersurface such that
$\Phi_{\Delta}(\tau_0,z^i) = (\tau_0-\Delta, z^i)$.
Next we choose coordinates $(\tau, z^i)$ in between these two hypersurfaces,
with $\tau_0 - \Delta \le \tau \le \tau_0 $ and their restriction to $\Sigma$ and
$\Phi_{\Delta}(\Sigma)$ being $(\tau_0,z^i)$, respectively
$(\tau_0-\Delta,z^i)$. Then we use the diffeomorphism to copy 
this coordinate patch to the other regions of spacetime.   
Of course this construction is very far from being unique. 

Per construction the diffeomorphism maps a point $p=(\tau, z^i)$ to the
point $\Phi_{\Delta}(p) = (\tau - \Delta, z^i)$. So the Jacobian
is the identity, 
$\frac{\partial \Phi_{\Delta}^{\mu}}{\partial x^{\nu}} =
\delta^{\mu}_{\nu}$.

Using the definition of the pull-back as well as Eq.~(\ref{def::DSS})
in these coordinates we obtain for $p=(\tau,z^i)$
\bea\label{DSSmetric1}
\left. (\Phi^*_{\Delta} g)_{\mu\nu} \right|_p & = & 
    \frac{\partial \Phi_{\Delta}^{\alpha}}{\partial x^{\mu}}
    \frac{\partial \Phi_{\Delta}^{\beta}}{\partial x^{\nu}} 
            \left. g_{\alpha \beta} \right|_{\Phi_{\Delta}(p)} =
      g_{\mu\nu} (\tau-\Delta,z^i) =  \nonumber\\
    & \stackrel{(\ref{def::DSS})}{ = } &
     e^{2 \Delta} g_{\mu\nu} (\tau,z^i).  
   \eea
As this is valid for any point $p \in M$ we can conclude that the metric 
is conformal to a metric $\tilde g$, which is periodic in the coordinate
$\tau$, with the conformal factor being an exponential in $\tau$,
\be\label{DSSmetric2}
g_{\mu\nu}(\tau,z^i) = e^{-2 \tau} \ {\tilde g}_{\mu\nu}(\tau,z^i) 
    \qquad \textrm{with} \quad
   {\tilde g}_{\mu\nu}(\tau + \Delta, z^i) = {\tilde g}_{\mu\nu}(\tau, z^i).
\ee

For a CSS spacetime again we 
choose a hypersurface $\Sigma$, 
with $\Sigma \cap \Phi_{\Delta}(\Sigma) = \emptyset$ $\forall \Delta$, so
the homothetic Killing vector field is transversal to $\Sigma$.
We choose coordinates
($z^{i}$) on this hypersurface and transport them across spacetime
via the one-parameter family of diffeomorphisms. 
We parameterize the orbits
$\gamma^{\mu}$ of
the homothetic KVF $\xi$ with $-\tau$, so $\xi = - \partial_{\tau}$. 
The freedom in the construction in this case consists of choosing the
hypersurface and applying diffeomorphisms within $\Sigma$.

In analogy to (\ref{DSSmetric1}) we have
\bea
({\mathcal L}_{\xi} g)_{\mu\nu} = - \partial_{\tau} g_{\mu\nu}
\stackrel{(\ref{def::homothetic})}{ = } 2 g_{\mu\nu},
\eea
and therefore  
\be\label{CSSmetric2}
g_{\mu\nu}(\tau,z^i) = e^{-2 \tau} {\tilde g}_{\mu\nu}(z^i).
\ee

\subsection{Self-Similarity in Spherical Symmetry}
\label{subsec::SSinSpherSymm}
%%%%%%%%%%%%%%%%%%%%%%%%%%%%%%%%%%%%%%%%%%%%%%%%%%

The fact that the above introduced adapted coordinates are not unique is
no draw back for our purposes. What we are interested in 
is to find out whether our numerically evolved spacetimes 
contain self-similar regions. 
In order to do so, we need to know, what self-similarity looks like in
the coordinates our code uses, namely the Bondi-like coordinates in spherical
symmetry, defined in Sec.~(\ref{subsec::spherical_symmetry})).
\begin{equation}\label{eq::Bondi_again}
ds^2 = - e^{2 \beta(u,r)} du ( \frac{V}{r}(u,r) du + 2 dr) + r^2 d\Omega^2.
\end{equation}
Assuming that (\ref{eq::Bondi_again}) describes a self-similar spacetime,
we seek for a coordinate transformation 
$(u,r) \to (\tau(u,r), z(u,r))$, such that the resulting metric is of the
form $e^{-2\tau}{\tilde g}_{\mu\nu}$, where 
${\tilde g}_{\mu\nu}$ is periodic in $\tau$ for a DSS spacetime and
independent of $\tau$ in the case of a CSS spacetime.
In the following a ``$\ {\tilde {}} \ $'' means, that the function
is periodic in $\tau$ with period $\Delta$.

The first observation is, that the coordinate transformation does not
involve the angles $\theta$ and $\varphi$. So the ``${\mathbbm S}^2$''-part
of the metric is unchanged. We immediately get 
\bea
r(\tau,z) &=& e^{-\tau} {\tilde R}(\tau,z) \qquad \textrm{for DSS},
      \label{rtauz_DSS} \\
r(\tau,z) & = & e^{-\tau} R(z)    \qquad \textrm{for CSS}. \label{rtauzCSS}
\eea
This means that the diffeomorphism maps an $r=const$ hypersurface to the
hypersurface $e^{\Delta} r = const$.

In the following we exploit the relations (\ref{def::DSS})
and (\ref{def::homothetic}) with respect to the Bondi coordinates, 
in order to find the behavior of $u=const$ hypersurfaces under the
diffeomorphism as well as determining the $\tau$ dependence of the
metric functions $\beta$ and $\frac{V}{r}$.

For the DSS spacetime 
we write $\Phi_{\Delta}(u,r) = (\Phi_{\Delta}^u(u,r),
\Phi_{\Delta}^r(u,r)) = (\Phi_{\Delta}^u(u,r), e^{\Delta} r)$,
the last equality following from (\ref{rtauz_DSS}). 
Therefore $\partial \Phi_{\Delta}^r/\partial u = 0$, 
$\partial \Phi_{\Delta}^r/\partial r = e^{\Delta}$.
We first examine the $rr$ component of (\ref{def::DSS}):
\be
(\Phi^*_{\Delta} g)_{rr} = e^{2 \Delta} g_{rr} = 0 
\ee
and so 
\be
\frac{\partial \Phi_{\Delta}^u}{\partial r} 
\left( \frac{\partial \Phi_{\Delta}^u}{\partial r} g_{uu} + 2 e^{\Delta}
g_{ur}\right) = 0.
\ee
This formula states, that the null vector $\nabla^{\mu}u$, which 
is tangent to the outgoing null geodesics generating the $u=const$
hypersurfaces, is mapped again to a null vector.
If the first factor vanishes, 
$\frac{\partial \Phi^u_{\Delta}}{\partial r} = 0$, the push forward of 
$\nabla^{\mu} u |_p$ is parallel to $\nabla^{\mu}u|_{\Phi_{\Delta}(p)}$.
If the expression in the parentheses vanishes the push forward would be parallel
to the ingoing null geodesic vector $\nabla^{\mu} v|_{\Phi_{\Delta}(p)}$.
Here we are interested in those diffeomorphisms 
that are connected to the identity map. Therefore we want the first 
factor to vanish, so
$\frac{\partial \Phi^u_{\Delta}}{\partial r} = 0$ and 
$\Phi^u_{\Delta} = \Phi^u_{\Delta}(u)$.
This shows, that $u=const$ hypersurfaces are mapped to $u=const$
hypersurfaces. 

We invoke now e.g. the $(u,r)$ component of (\ref{def::DSS}) in order to 
get more information on $\Phi_{\Delta}^u(u)$.
\bea\label{DSSeq2}
\left. (\Phi^*{\Delta} g )_{ur} \right|_p & = & 
        \left. e^{2 \Delta} g_{ur}\right|_p \nonumber\\
\left(\frac{\partial \Phi^u_{\Delta}}{\partial u} \right) e^{\Delta}
g_{ur}(\Phi_{\Delta} (p)) & = & e^{2 \Delta} g_{ur}(p).
\eea
Consider now a point $p$ at the origin, i.e. $p = (u, r=0)$. As the origin
is mapped to itself (which follows from (\ref{rtauz_DSS})) the image of the
point $p$ is $\Phi_{\Delta}(p) = (\Phi_{\Delta}^u(u),0)$. As described
in Sec.~(\ref{subsec::regular_center}) 
the components of the metric with respect to Bondi-like
coordinates are fixed at the origin, due to regularity at the center as well
as the choice of retarded time being proper time at the origin.
In particular we have $\beta(u,r=0) = 0$. Inserting this into
Eq.~(\ref{DSSeq2}) we obtain
\be
\left(\frac{\partial \Phi^u_{\Delta}}{\partial u} \right)
 = e^{\Delta}.
\ee 
Integrating gives
\be
\Phi^u_{\Delta}(u) = e^{\Delta} u + const.
\ee
Reinserting this into (\ref{DSSeq2}) we get
\be\label{DSSeq2b}
e^{2 \beta(\tau - \Delta,z)} = e^{2 \beta(\tau,z)},
\ee
and for the $(u,u)$ component of (\ref{def::DSS})
we have
\bea\label{DSSeq1}
\left. (\Phi^*_{\Delta} g )_{uu} \right|_p & = & 
        \left. e^{2 \Delta} g_{uu}\right|_p \nonumber\\
\left(\frac{\partial \Phi^u_{\Delta}}{\partial u} \right)^2
g_{uu}(\Phi_{\Delta} (p)) & = & e^{2 \Delta} g_{uu}(p) \nonumber\\
e^{2 \beta(\tau - \Delta, z)} \frac{V}{r}(\tau - \Delta,z) & = &
e^{2 \beta(\tau , z)} \frac{V}{r}(\tau,z).
\eea

From (\ref{DSSeq2b}) and (\ref{DSSeq1}) we now see, that 
both metric functions are periodic functions in $\tau$
\bea\label{metricPeriodic}
\beta(u,r) &=& {\tilde \beta}(\tau(u,r),z(u,r)), \nonumber\\
\frac{V}{r}(u,r) & = & {\tilde \frac{V}{r}} (\tau(u,r), z(u,r)).
\eea

Note, that this relation is valid for any set of adapted coordinates
constructed as described in Sec.~(\ref{subsec::adapted}).

For a CSS spacetime we use analogous arguments. We start by writing
the homothetic Killing vector as  
\begin{equation}
- \left. \partial_{\tau} \right|_{z}\equiv \xi = 
       \xi^u(u,r) \partial_u + \xi^r (u,r) \partial_r. 
\end{equation}  
Of course we have $\xi^u = - \partial_{\tau} u$ and $\xi^r = - \partial_{\tau} r$
(both partial derivatives taken at constant $z^{i}$).
(\ref{rtauzCSS}) immediately gives the $r$ component of $\xi$,
\be
\xi^r = - \left. \partial_{\tau} r(\tau,z) \right|_z = r(\tau,z).
\ee

As above we examine the $rr$ component of (\ref{def::homothetic}),
\begin{equation}
{\mathcal L}_{\xi} g_{rr} = \xi^{\mu} g_{rr,\mu} + 
             2 \ \xi^{\mu}_{,r} g_{\mu r}
           = 2 \ \xi^u_{,r} g_{ur} = 0.
\end{equation} 
So we have $\xi^u_{,r} = 0$ and  $\xi^u = \xi^u (u)$. 
Now the $u u$ component of the 
homothetic Killing equation (\ref{def::homothetic}) reveals
\bea\label{CSSEq1}
{\mathcal L}_{\xi} g_{u u} & = & \xi^{\mu} g_{u u,\mu} + 
             2 \  \xi^{\mu}_{,u} g_{\mu u}  
                 = 2 \ g_{u u} \nonumber\\
   & & \xi^{\mu} g_{u u,\mu} +  
             2 \  \xi^{u}_{,u} g_{u u}  
                 = 2 \ g_{u u}.
\eea              
% .
Again we use the fact that the origin is an orbit of the homothetic
Killing vector, so $\xi |_{r=0} = \partial_u |_{r=0}$, and $g_{uu}|_{r=0} =
1$. So (\ref{CSSEq1}) gives 
\be
\xi^u_{,u} = 1 \quad \Rightarrow \quad \xi^u(u) = u + const.
\ee
Therefore Eq.~(\ref{CSSEq1}) gives
\be
\partial_{\tau} g_{u u} = 0,
\ee
and the $(u r)$ component of (\ref{def::homothetic}) yields
\bea
{\mathcal L}_{\xi} g_{u r} & = & \xi^{\mu} g_{u r,\mu} + 
               \xi^{\mu}_{,u} g_{\mu r} + 
               \xi^{\mu}_{,r} g_{u \mu }  
                 = 2 \ g_{u r} \nonumber\\
      & &   \xi^{\mu} g_{u r, \mu} + 2 g_{u r} = 2 \ g_{u,r} \nonumber\\
    &  & \partial_{\tau} g_{u r} = 0.
\eea
So in analogy to (\ref{metricPeriodic}) we have for a CSS spacetime 
\bea\label{metricConstant}
\beta(x^{\mu}) &=& \beta (z), \nonumber\\
\frac{V}{r}(x^{\mu}) & = & \frac{V}{r} (z).
\eea

We now explicitly write down a set of adapted
coordinates for DSS and CSS spacetimes, which will be used later.

As we have seen, the diffeomorphisms map a $u=const$ hypersurface to another
$u=const$, different form the first. Therefore the $u=const$ hypersurfaces
are valid candidates for $\tau=const$ hypersurfaces, i.e. we may set 
$\tau = \tau(u)$.

For a DSS spacetime the function $u(\tau)$ has to obey
\be\label{utauDSS}
e^{\Delta} u(\tau) + const = u(\tau - \Delta).
\ee
The simplest function $u(\tau)$ satisfying (\ref{utauDSS})
is given by
\be
u^* - u = e^{-\tau}, 
\ee
where we have set $u^* = const \ e^{\Delta}$.

For the CSS spacetime we integrate
\be
 - \frac{d u}{d \tau} = \xi^u = u + const, 
\ee
which gives
\be\label{utauCSS}
\ln (|u + const|) = - \tau + c_1,
\ee
$c_1$ -- being a simple shift in $\tau$ -- can be chosen arbitrarily, 
and so we
set $c_1 = 0$. If we again introduce $u^* = - const$, we have
\be
u^* - u = e^{-\tau},
\ee
the left hand side coming from resolving the absolute value in (\ref{utauCSS})
for $u < u^*$, which is the region we will be interested in in critical
collapse situations.

Finally we parameterize the $\tau = const$ hypersurfaces as follows:
for a DSS spacetime we set
\be\label{eq::r_tau_z_DSS}
r(\tau,z) = e^{-\tau} {\tilde R}(\tau,z) = e^{-\tau} \ z \ \zeta(\tau)
 \quad \textrm{with} \quad \zeta(\tau + \Delta) = \zeta(\tau),
\ee
where we require $\zeta(\tau) > \dot \zeta(\tau)$ for all $\tau$, such that
$\partial_{\tau} r(\tau,z) < 0$ for all $\tau$.
There is still a coordinate freedom contained in $\zeta(\tau)$. 
We keep this in order to describe the null hypersurface, which is mapped to
itself via the diffeomorphism (the so called self-similarity horizon) 
with $z=const$. The resulting condition
on $\zeta(\tau)$ will be given below.

For a CSS spacetime we set
\be
r = e^{-\tau} R(z) = e^{-\tau} z.
\ee

Summarizing, we chose the following adapted coordinates for a DSS spacetime
\bea\label{coordDSS}
\tau(u) = - \ln(u^*-u) &  & u(\tau) = u^* - e^{-\tau} \nonumber\\
                       & \Longleftrightarrow&   \nonumber\\  
z(u,r) = \frac{r}{(u^* - u) \zeta(\tau)} & & r(\tau,z) = e^{-\tau} z
                    \zeta(\tau),
\eea
and for a CSS spacetime
\bea\label{coordCSS}
\tau(u) = - \ln(u^*-u) &  & u(\tau) = u^* - e^{-\tau} \nonumber\\
                       & \Longleftrightarrow&   \nonumber\\  
z(u,r) = \frac{r}{(u^* - u)} & & r(\tau,z) = e^{-\tau} z.
\eea

By construction, the hypersurfaces $\tau = const$ are null.
For $u \to -\infty$ we have $\tau \to -\infty$, 
whereas for $u \to u^*$, $\tau \to +\infty$.

The hypersurfaces $z=const$, along which the diffeomorphism acts
all meet in the point $(u=u^*,r=0)$. Clearly the adapted coordinates 
get singular there. Due to the symmetry, which squeezes the geometry
into smaller and smaller spacetime regions, this point also
has diverging curvature, as can easily be seen as follows:
We examine the scalar curvature $\mathcal R$ in adapted coordinates.
From (\ref{geomDSS}) we have
\bea
\left. (\Phi^*_{\Delta} \mathcal R)\right|_p & = & 
           \left. e^{-2 \Delta} \mathcal R \right|_p
           \nonumber\\
\mathcal R(\Phi_{\Delta}(p)) & = & e^{-2 \Delta} \mathcal R(p) \nonumber\\
\mathcal R(\tau - \Delta, z) & = & e^{-2 \Delta} \mathcal R(\tau, z)
\eea
and therefore
\be
\mathcal R(\tau,z) = e^{2 \tau} {\tilde \mathcal R}(\tau,z) \quad \textrm{with} \quad
         {\tilde \mathcal R}(\tau - \Delta,z) = {\tilde \mathcal R}(\tau,z). 
\ee
Note that ${\tilde \mathcal R}$ does in general not agree with the 
scalar curvature built from the metric ${\tilde g}$, introduced in Sec.
(\ref{subsec::adapted}), although this would have a periodic 
$\tau$ dependence as well.

In particular, as ${\tilde \mathcal R}(\tau,z)$ has a periodic 
$\tau$ dependence, it
is bounded for all $\tau$ if it is bounded in one ``segment'' between
$\Sigma$ and $\Phi_{\Delta}(\Sigma)$.
So moving along $z=const$ we find the 
scalar curvature blowing up
like $e^{2 \tau}$ for $\tau \to \infty$, or $u \to u^*$ 
(unless ${\tilde \mathcal R}$ vanishes identically of course).
As the origin is a line of constant $z$, 
this blowup occurs at $(u=u^*,r=0)$.
We call the point $(u^*,0)$ the {\em culmination point}, and $u^*$ the
culmination time.

Another point is worth to note. 
Consider the square of the vector field $\partial_{\tau}$.
From (\ref{coordDSS}) we have
\be
\partial_{\tau}  = (u^* - u) \partial_u  + 
              r (- 1 + \frac{\dot \zeta}{\zeta}) \partial_r    
\ee
and so
\bea\label{gtautau}
g(\partial_{\tau}, \partial_{\tau}) & = &  (u^* - u)^2 g_{uu} + 
    2 r (u^* - u) (- 1 + \frac{\dot \zeta}{\zeta}) g_{u r} = \nonumber\\
& = & - e^{-2\tau} e^{2 \beta(\tau,z)} (\frac{V}{r}(\tau,z)
           - 2 z\zeta(\tau) + 2 z {\dot \zeta}(\tau)) ).
\eea
At the origin (z = 0), $\partial_{\tau}$ is timelike,  
which is clear, as the diffeomorphism maps the origin onto itself.
Furthermore, as mentioned after (\ref{eq::r_tau_z_DSS})
we assume $\zeta(\tau) > \dot \zeta(\tau)$ for all $\tau$.

Moving away from the origin $z$ grows and goes to infinity 
for $u \to u^*$. As we assume $\frac{V}{r}$ to be bounded, the expression
in the parentheses in (\ref{gtautau}) will vanish at some
value of $z$, which depends on $\tau$. We can now use the 
additional gauge freedom, which is contained in $\zeta(\tau)$ in order
to have $\partial_\tau$ getting null on a hypersurface of constant $z$, e.g.
at $z=1$. The resulting condition on $\zeta(\tau)$
reads
\be
\left(\frac{V}{r}(\tau,1)
           - 2 \zeta(\tau) + 2 {\dot \zeta}(\tau)\right) = 0. 
\ee 
So this hypersurface, $z=1$,  is a null hypersurface, which is mapped
onto itself via the diffeomorphism. It is called the past 
{\em self-similarity horizon} (SSH). In fact this self-similarity horizon
is just the backwards light cone of the culmination point.

There exists another null hypersurface which is mapped to itself by the
diffeomorphism. This is the future light cone of the culmination point, and
is called the future SSH. As the culmination point 
corresponds to a spacetime singularity, the region beyond this 
future SSH is not determined
by the solution given for $u < u^*$.

\begin{figure}[h,t,b]
\begin{center}
\begin{psfrags}
 \psfrag{uconst}[]{$u = const$}
 \psfrag{tauconst}[]{$\tau = const$}
 \psfrag{r0}[]{$r=0$}
 \psfrag{t}[]{$time$}
 \psfrag{r}[]{$space$}
 \psfrag{z0}[]{$z=0$}
 \psfrag{uustar}[]{$u = u^*$}
 \psfrag{tauinfty}[]{$\tau = \infty$}
 \psfrag{zinfty}[]{$z = \infty$}
 \psfrag{zconst}[]{$z = const$}
 \psfrag{Origin}[]{Origin}
 \psfrag{Culmination}[]{Culmination}
 \psfrag{point}[]{point}
 \psfrag{future SSH}[]{future SSH}
 \psfrag{past SSH}[]{past SSH}
\includegraphics[width=5in]{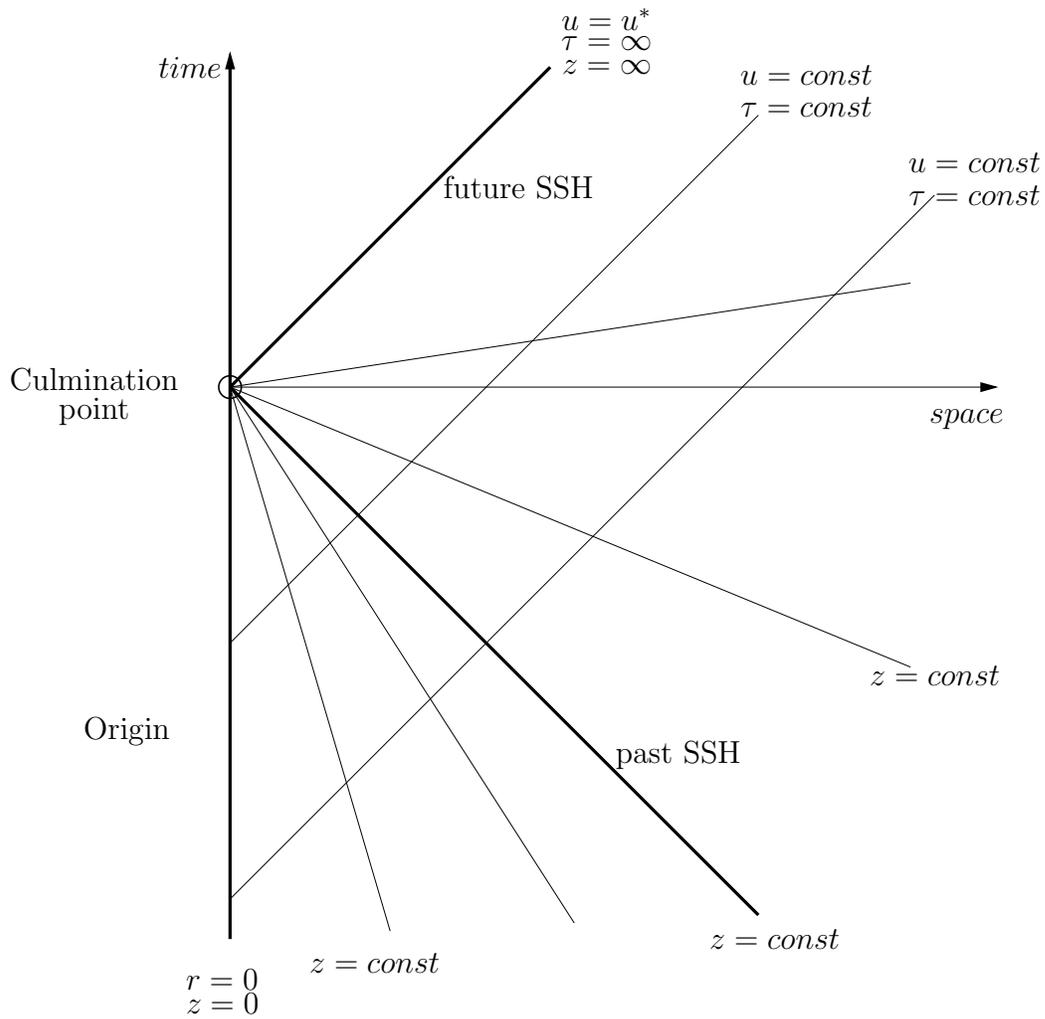}
\end{psfrags}
\end{center}
\caption{A schematic diagram of a self-similar spacetime. 
The adapted coordinates
$(\tau, z)$ defined in (\ref{coordCSS}) cover only the region $u < u^*$.
In this region the metric functions $\beta$ and $\frac{V}{r}$ are constant
along $z=const$, thereby shrinking their profile to zero
at the culmination point $(u=u^*,r=0)$, where (unless in flat space) 
a spacetime singularity occurs. The region within the backwards light cone
of the culmination point is the one of interest for critical collapse
situations.}\label{fig::CSSspacetime}
\end{figure}

For a CSS spacetime the statements ``periodic in $\tau$'' have to be replaced
by ``independent of $\tau$''.
Furthermore with our choice of coordinates the location of the past SSH
is given by
\be
\frac{V}{r}(z_H) = 2 z_H.
\ee
Fig. \ref{fig::CSSspacetime} shows a schematic diagram of a 
self-similar spacetime.

\afterpage{\clearpage}

Note, that a self-similar spacetime is not asymptotically flat
(unless spacetime as a whole is flat). This can be seen by going to infinity
on a spacelike hypersurface  $z = const$ outside the past SSH. 
As the metric functions are periodic in $\tau$ (resp. constant)  
along these hypersurfaces, they do not fulfill the fall-off conditions 
required by asymptotically flatness. 

We close this section by writing down the line element for a DSS (CSS) 
spacetime with respect to the adapted coordinates 
(\ref{coordDSS}) (respectively (\ref{coordCSS})).
\bea\label{metricDSS}
ds^2_{DSS} & = & e^{-2 \tau} \left[ - e^{2 \beta(\tau,z)}
    \left\{ 
\frac{V}{r}(\tau,z) - 2 z \zeta(\tau) + 2 z {\dot \zeta}(\tau)
\right\} d \tau^2 - e^{2 \beta(\tau,z)} 2 \zeta(\tau) d\tau d z  + \right.
 \nonumber\\
  & & + \left.  z^2 \zeta(\tau)^2 d\Omega^2 \right],
\eea
and
\be\label{metricCSS}
ds^2_{CSS} = e^{-2 \tau} \left[ - e^{2 \beta(z)}
    \left\{ 
\frac{V}{r}(z) - 2 z 
\right\} d \tau^2 - e^{2 \beta(z)} 2 d\tau d z  + 
 z^2 d\Omega^2 \right]. 
\ee

\subsection{Implications for the matter
field}\label{subsec::implications_matter}
%%%%%%%%%%%%%%%%%%%%%%%%%%%%%%%%%%%%%%%%%%%%%%%%%

We turn now to the conditions that the symmetries (\ref{def::DSS})
and (\ref{def::homothetic}) imply for the matter field.
It is clear that the matter will have to share (at least part of) the symmetry
imposed on the  stress energy tensor via the Einstein equations, i.e.
we have for a DSS spacetime
\be
(\Phi_{\Delta}^*T)_{\mu\nu} = T_{\mu\nu},
\ee
and for a CSS spacetime
\begin{equation}
{\mathcal L}_{\xi} T_{\mu\nu} = 0.
\end{equation}

As defined in Sec.~\ref{sec::sigmaModel} 
the stress energy tensor of a harmonic map
is
\be
T_{\mu\nu}(x) = f_{\pi}^2 
            \left( \nabla_{\mu} X^A(x) \nabla_{\nu} X^B(x) - \frac{1}{2}
             g_{\mu\nu}(x) (g^{\sigma \tau}(x) \nabla_{\sigma} X^A(x)
             \nabla_{\tau} X^B(x)) \right) G_{AB}(X(x)).
\ee
The pull-back of this tensor under $\Phi_{\Delta}$ is given by
\bea\label{stressenergy1}
\left. (\Phi_{\Delta}^*T)_{\mu\nu} \right|_p & = & 
           \frac{\partial \Phi_{\Delta}^{\alpha}}{\partial x^{\mu}}
           \frac{\partial \Phi_{\Delta}^{\beta}}{\partial x^{\nu}}
           \left[ \nabla_{\alpha} X^A(\Phi_{\Delta}(p)) 
                  \nabla_{\beta} X^B(\Phi_{\Delta}(p)) \right. - \nonumber\\
              &  - & \left. \frac{1}{2} g_{\alpha\beta}(\Phi_{\Delta}(p)) 
                   (g^{\gamma \delta}(\Phi_{\Delta}(p)) 
                   \nabla_{\gamma} X^A(\Phi_{\Delta}(p))
                   \nabla_{\delta} X^B(\Phi_{\Delta}(p))) 
           \right] \nonumber\\
        & & G_{AB}(X(\Phi_{\Delta}(p))). 
\eea
As the inverse metric transforms like
\be
\left. ((\Phi_{\Delta})_* g^{-1})^{\gamma \delta} \right|_{\Phi_{\Delta}(p)} =
    \left. e^{2 \Delta} g^{\gamma \delta} \right|_{\Phi_{\Delta}(p)} ,       
\ee

we have
\bea
& & g^{\gamma \delta}(\Phi_{\Delta}(p)) 
                   \nabla_{\gamma} X^A(\Phi_{\Delta}(p))
                   \nabla_{\delta} X^B(\Phi_{\Delta}(p))
= \nonumber\\
& = & e^{-2 \Delta} g^{\sigma \tau}(p) 
    \frac{\partial \Phi_{\Delta}^{\gamma}}{\partial x^{\sigma}}
           \frac{\partial \Phi_{\Delta}^{\delta}}{\partial x^{\tau}}
                   \nabla_{\gamma} X^A(\Phi_{\Delta}(p))
                   \nabla_{\delta} X^B(\Phi_{\Delta}(p)).
\eea
The factor $e^{-2 \Delta}$ cancels with $e^{2 \Delta}$ coming from
$\frac{\partial \Phi_{\Delta}^{\alpha}}{\partial x^{\mu}}
  \frac{\partial \Phi_{\Delta}^{\beta}}{\partial x^{\nu}}
g_{\alpha\beta}(\Phi_{\Delta}(p))$ in the 
second term of (\ref{stressenergy1}). In order to simplify things, we
contract (\ref{stressenergy1}) with $g^{\mu\nu}(p)$
\be
g^{\mu\nu}(p) \left. (\Phi_{\Delta}^*T)_{\mu\nu} \right|_p  =  
    - 2 g^{\mu\nu} \frac{\partial \Phi_{\Delta}^{\alpha}}{\partial x^{\mu}}
           \frac{\partial \Phi_{\Delta}^{\beta}}{\partial x^{\nu}}
           \nabla_{\alpha} X^A(\Phi_{\Delta}(p)) 
                  \nabla_{\beta} X^B(\Phi_{\Delta}(p))
          G_{AB}(X(\Phi_{\Delta}(p))),
\ee
and switch to adapted coordinates to get
\bea
 & &    g^{\mu\nu}(\tau,z,\omega)
           \nabla_{\mu} X^A(\tau-\Delta,z,\omega) 
                  \nabla_{\nu} X^B(\tau-\Delta,z,\omega)
          G_{AB}(X(\tau-\Delta,z,\omega)) = \nonumber\\
 &= &   g^{\mu\nu}(\tau,z,\omega)
           \nabla_{\mu} X^A(\tau,z,\omega) 
                  \nabla_{\nu} X^B(\tau,z,\omega)
          G_{AB}(X(\tau,z,\omega)), 
\eea
where we abbreviated the angular coordinates $\theta$ and $\varphi$
with $\omega$.
So the self-gravitating harmonic map, leading to a discretely 
self-similar spacetime has to be either periodic or ``anti-periodic''
in the adapted time coordinate $\tau$,
\be\label{XDSS}
X^A(x^\mu) = {\tilde X}^A(\tau, z,\omega) \qquad \textrm{with} \quad 
                         {\tilde X}^A(\tau - \Delta,z,\omega) = 
                        \pm {\tilde X}^A(\tau,z,\omega), 
\ee
and independent of $\tau$ for a CSS spacetime,
\be\label{XCSS}
X^A(x^{\mu}) = X^A(z,\omega).
\ee 

Concerning the sign in (\ref{XDSS}) we adopt the following convention:
if the metric functions are periodic with period $\tilde \Delta$ and the
field is ``anti-periodic'' with respect to this period, then of course it is
periodic with respect to twice the period, i.e. $2 \tilde \Delta$.
In this case we say that the solution is DSS with period $\Delta = 2 \tilde
\Delta$, and has the additional symmetry 
$\beta(\tau + \Delta/2,z)=\beta(\tau,z)$ 
etc. and $X^A(\tau + \Delta/2,z) = -X^A(\tau,z)$.

Of course the conditions (\ref{XDSS}) and (\ref{XCSS}) are only 
necessary conditions for the existence of a regular 
self-similar spacetime\footnote{By regular in this context, we mean regular
within the backwards light cone of the culmination point}. 
 
Remember, that the self-gravitating nonlinear $\sigma$ model
is a {\em scale invariant} theory. We stress that this scale invariance
is a necessary condition for the existence of self-similar solutions.
Consider for example vacuum with a cosmological constant $\Lambda$.
\be\label{vacuumLambda}
G_{\mu\nu} + \Lambda g_{\mu\nu} = 0.
\ee
Here the cosmological constant $\Lambda$ introduces a length scale.
Applying the pull-back $\Phi_{\Delta}^*$ to Eq.~(\ref{vacuumLambda})
we get using Eqs.~(\ref{geomDSS}) and (\ref{def::DSS})  
\be
\left. (\Phi^*_{\Delta} (G + \Lambda g)_{\mu\nu} \right|_p = 
      G_{\mu\nu} + \Lambda e^{2 \Delta} g_{\mu\nu}|_p, 
\ee 
which does not satisfy (\ref{vacuumLambda}) anymore.

Another example of a model with a length scale would be the self-gravitating
nonlinear $\sigma$ model with an additional potential in the Lagrangian, 
e.g. $V(X) = X^A X^B G_{AB}(X)$. For dimensional reasons this potential has
to be multiplied by a constant of dimension $(1/length)^2$, which
breaks the scale invariance. As this potential term appears 
in the stress energy tensor multiplied by the metric $g_{\mu\nu}$,
it would get a factor $e^{2 \Delta}$ under the action of
the pull-back $\Phi_{\Delta}^*$,
which does not cancel. 
As this is the only term which transforms that way, again the Einstein
equations cannot be satisfied.

Nevertheless a model with a length scale can admit asymptotic
self-similarity, i.e. it might display self-similarity at scales which are
small compared to the length scale of the theory.
This can be seen by writing the equations in adapted coordinates
and neglecting the terms which contain a factor $e^{-\tau}$ or any power
thereof. (As $\tau \to \infty$ denotes the region close to the culmination
point, spatial extensions are already very small). 
This concerns the terms that are tied to the length scale, therefore
the remaining theory again is scale invariant. In this context such terms
are called {\em asymptotically irrelevant}.
An example is the Einstein-Yang-Mills system, which admits 
an asymptotically DSS solution at the (type II) threshold of black hole
formation \cite{Choptuik-Chmaj-Bizon-1996-EYM-critical-behavior}. 
This solution also has been constructed
directly (using the asymptotic symmetry) by Gundlach
\cite{Gundlach-1996-scaling-in-EYM-critical-collapse}.

\section{Numerical Construction of CSS Solutions}\label{sec::CSSsolutions}
%%%%%%%%%%%%%%%%%%%%%%%%%%%%%%%%%%%%%%%%%%%%%%%%%%%%%%%%%%%%%%%%%%%%%%%%%

This section deals with the (numerical) construction of CSS solutions
of the self-gravitating SU(2) $\sigma$-model in spherical symmetry
using the hedgehog ansatz introduced in Sec.~\ref{subsec::hedgehog}.
This problem has already be studied by Bizon
\cite{Bizon-1999-existence-of-self-similar-sigma-CSS-solutions}
for the simpler case of fixed Minkowski background ($\eta = 0$), and by
Bizon and Wasserman for the coupled case
\cite{Bizon-Wasserman-2000-CSS-exists-for-nonzero-beta}.

In \cite{Bizon-1999-existence-of-self-similar-sigma-CSS-solutions} a 
discrete one-parameter family of CSS
solutions was constructed numerically, the existence was proven 
analytically and the stability properties of the solutions were given.
In \cite{Bizon-Wasserman-2000-CSS-exists-for-nonzero-beta} this family of 
solutions was shown to persist up to a coupling constant 
$\eta_{max} = 0.5$ by
means of the numerical construction. Furthermore, the
analytic continuation beyond the past SSH was studied numerically, showing
that for each member of the family there exists a critical value of the
coupling $\eta_n^*$ beyond which the analytic continuation contains
marginally trapped surfaces.

In order to be able to compare the directly constructed 
CSS solution to critical solutions obtained by a bisection search
of time evolved data, 
we re-did the (numerical) calculations of 
\cite{Bizon-1999-existence-of-self-similar-sigma-CSS-solutions}
and \cite{Bizon-Wasserman-2000-CSS-exists-for-nonzero-beta}, 
reproducing their results.
Furthermore we studied the stability of the solutions for nonzero couplings,
which has not been considered 
in \cite{Bizon-Wasserman-2000-CSS-exists-for-nonzero-beta}. 

We will report here on both the numerical construction of the 
solutions and their analytic continuation beyond the past SSH
(in order to make everything self-contained) 
as well as on their stability properties.

\subsection{The ``CSS equations''}\label{subsec::CSSequations}
%%%%%%%%%%%%%%%%%%%%%%%%%%%%%%%%%%%%%%%%%%%%%%%%%%%%%%%%%%%%%%

We start by combining  the symmetry requirement (\ref{XCSS})
with the hedgehog ansatz. For a CSS solution of the self-gravitating
nonlinear $\sigma$ model we have with $X^A = (\phi, \Theta, \Phi)$
\be
\phi(x^{\mu}) = \phi(z), \qquad 
\Theta(x^{\mu}) = \theta, \qquad 
\Phi(x^{\mu}) = \varphi.
\ee
Transforming now Eqs.~(\ref{eq::phi}), (\ref{eq::betap}) and 
(\ref{eq::Vp}) to the adapted coordinates
defined in (\ref{coordCSS}) and dropping all derivatives with respect to
$\tau$, we get
\begin{eqnarray}
{\beta}' & = & \frac{\eta}{2} \, z  \,  (\phi')^2 \label{CSSbetap}\\
(\frac{V}{r}){}' & = & -\frac{1}{z}\left( - {e^{2 \beta}} + 
         2 \, \eta \, {e^{2 \beta}}
\sin^2(\phi) + \frac{V}{r} \right)\label{CSSVoverrp} 
\end{eqnarray}
and
\be\label{CSSphippSimple}
 \left(z^2 (\Vr - 2z) \phi' \right)' + 2 z^2 \phi' = \sin(2 \phi) e^{2 \beta}
\ee
or 
\begin{eqnarray}\label{CSSphipp}
\phi'' & = &  
 \frac{1}{z^2\,
       \left( -\frac{V}{r} + 2\,z \right) }
   \left(  -{e^{2 \beta}}\,
          \left\{ \sin (2\,\phi ) + 
          z \,
             \left( -1 + 2\,\eta \,{\sin (\phi)}^2 \right) \,\phi' \right\} 
          + \right. \nonumber \\
        & + & \left. z\, \left\{ \left( \frac{V}{r}   - 4 \,  
              \,z \right) \phi'  \right\} \right),
\end{eqnarray}
where ${}' \equiv \partial_z$.

The subsidiary Einstein equation ``Euur'' (\ref{eq::Euur}) gives in addition
\be\label{CSSEuur}
(\frac{V}{r}){}' = \eta z \left(2 z - \frac{V}{r} \right) (\phi')^2.
\ee
Eqs.~(\ref{CSSVoverrp}) and (\ref{CSSEuur}) can be combined 
to give an algebraic relation for $\frac{V}{r}$:
\be\label{CSSalgebraicV}
\frac{V}{r} = \frac{- e^{2 \beta} (1 -2  \eta \sin ^2 \phi) + \eta 2 z^3
             (\phi')^2}{-1 + \eta z^2 (\phi')^2}.
\ee
As discussed already in Sec.~\ref{subsec::spherical_symmetry}, 
regularity at the origin ($z= 0$)
as well as the gauge choice for $u$ requires 
\be\label{CSSregOrigin}
\phi(0) = 0, \qquad \beta(0) = 0, \qquad  \frac{V}{r}(0) = 1.
\ee
The only free parameter, which determines the solution is
\be
\phi'(0) = b.
\ee

According to (\ref{CSSEuur}) $\frac{V}{r}$ is decreasing for $z>0$
and eventually  equals $2z$ at some $z_H$. This marks a singular point
of the equations and corresponds physically to the past self-similarity
horizon discussed in Sec.~\ref{subsec::SSinSpherSymm}. 

Eq.~(\ref{CSSEuur}) shows, that at the horizon in addition to 
$\Vr |_{H} = 2 z_H$ , we have $\left( \Vr \right)' |_{H} = 0$.
Eq.~(\ref{CSSVoverrp}) then yields the following relation for $\beta_H$
\be\label{CSSbetaH}
e^{2 \beta_H} = \frac{2 z_H}{1 - 2 \eta \sin^2 (\phi_H)},
\ee
where $\beta_H, \phi_H$ denote the fields $\beta$ and $\phi$ evaluated at the
horizon $z_H$.

Furthermore following from Eq.~(\ref{CSSphipp}) regularity at the horizon
requires $\sin 2 \phi_H = 0$, which can be resolved to 
$\phi_H = 0$ (mod $\pi$) or $\phi_H = \pi/2$ (mod $\pi$).
The first case is impossible for the following reason:
assume $\phi_H = 0$, so $e^{2\beta_H} = 2 z_H$ which in turn equals
$\Vr|_H$. So we would have $e^{2\beta_H} = \Vr|_H$. From Eqs.~(\ref{CSSbetap})
and (\ref{CSSEuur}) we know that $\beta' \ge 0$, whereas $\Vr' \le 0$
between origin and horizon. As $e^{2 \beta}$ equals $\Vr$ at the origin,
the two metric functions cannot attain the same value at the horizon
unless they are constant functions, which is the case for vanishing coupling
$\eta = 0$.

Summarizing we get
\be\label{CSSregHorizon}
\phi_H = \frac{\pi}{2},  \qquad e^{2\beta_H} = \frac{2 z_H}{1 - 2\eta}.
\ee 

One sees immediately, that $\beta_H$ can only be finite if
$\eta < 0.5$. CSS solutions, regular at both origin and past SSH, can
therefore only exist for small couplings $\eta \in [0, 0.5)$.

Note that solutions to (\ref{CSSbetap})--(\ref{CSSphippSimple}) satisfying
(\ref{CSSregOrigin}) and (\ref{CSSregHorizon}) 
(if they exist) are analytic in $z$ and the shooting
parameters. (In the vicinity of the singular points $z=0$ and $z_H$ one can
use Prop.~1 of \cite{Breitenlohner-Forgacs-Maison-EYM-existence} 
to show analyticity, for other $z$ it follows from
the analyticity of the right hand sides of the equations.)  

In order to construct regular solutions numerically, we proceed as follows.
We consider the boundary value problem, consisting of 
the coupled system of four first order ODEs Eqs.~(\ref{CSSbetap}), 
(\ref{CSSphipp}) and (\ref{CSSEuur})
subject to the boundary conditions (\ref{CSSregOrigin}) and
$\phi_H = \frac{\pi}{2}$. The relation for $\beta_H$ can be dropped here,
as it results from a subsidiary equation, which is not used for the
calculation.
(An alternative would be to substitute Eq.~(\ref{CSSEuur}) by
the algebraic relation (\ref{CSSalgebraicV}), thereby reducing 
the system to three first order ODEs.)
The four free shooting parameters at the boundaries
are
\be
\phi'(0) = b, \qquad z_H, \qquad \phi'_H, \qquad \beta_H. 
\ee
For fixed $\eta$ the problem is solved numerically using the shooting
and matching routine d02agf from the NAG Library \cite{NAG}.
We start at $\eta = 0$ by pinning down one member of the discrete one
parameter family of solutions reported by Bizon 
\cite{Bizon-1999-existence-of-self-similar-sigma-CSS-solutions} and 
follow this
solution to higher values of the coupling constant.
As we are mainly interested in the ground state and first excited state,
we only constructed the first few solutions. 

\subsection{Phenomenology of Numerically Constructed Solutions}
%%%%%%%%%%%%%%%%%%%%%%%%%%%%%%%%%%%%%%%%%%%%%%%%%%%%%%%%%%%%%%%

As in \cite{Bizon-Wasserman-2000-CSS-exists-for-nonzero-beta} we 
find that the solutions 
present on Minkowski background stay regular between origin and past SSH
up to a maximal value of the coupling constant $\eta_{max} = 0.5$.
At this maximal value the metric function $\beta$ diverges at the horizon,
which can be inferred from (\ref{CSSregHorizon}). A more complete picture of
this phenomenon can be obtained if one considers the analytic continuation
beyond the past SSH, which will be discussed in the next section.

\begin{figure}[p]
\begin{center}
\begin{psfrags}
 \psfrag{eta}[]{$\eta$}
 \psfrag{phipO}[][][1][-90]{$\phi'(0)$}
\includegraphics[width=4in]{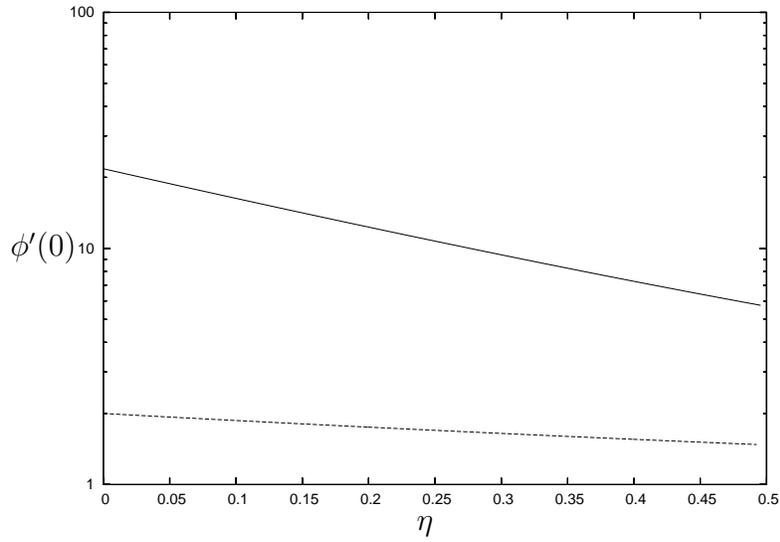}
\end{psfrags}
\end{center}
\caption{The shooting parameter $\phi'(0)$ for the ground state (dashed line)
and the first excitation (solid line) as a function
of the coupling constant $\eta$.}\label{fig::CSS_eta_phipO}
\end{figure}

\begin{figure}[p]
\begin{center}
\begin{psfrags}
 \psfrag{eta}[]{$\eta$}
 \psfrag{betaH}[][][1][-90]{$\beta_H$}
\includegraphics[width=4in]{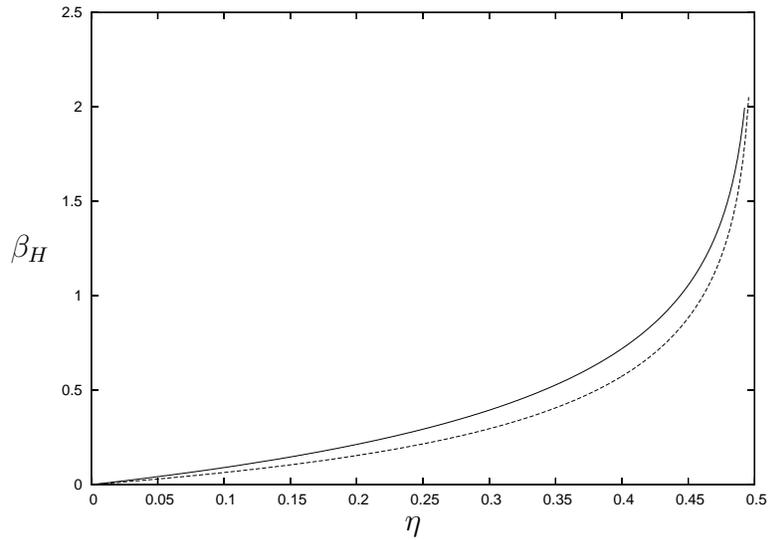}
\end{psfrags}
\end{center}
\caption{The shooting parameter $\beta_H$ for the ground state (solid line)
and the first excitation (dashed line) as a function of the 
coupling constant $\eta$. 
$\beta_H$ rises for both solutions as one approaches
the maximal coupling $\eta_{max} = 0.5$ and diverges in the limit, as can be
inferred from Eq. \ref{CSSregHorizon}.}\label{fig::CSS_eta_betaH}
\end{figure}

Figure \ref{fig::CSS_eta_phipO} shows the shooting parameter
$\phi'(0)$ for the ground state and the first excited state as functions
of the coupling. This parameter alone determines the solution. 
Figure \ref{fig::CSS_eta_betaH} shows the value of $\beta_H$ for the two 
solutions again as functions of the coupling.

\subsection{Analytic Continuation beyond the past
SSH}\label{subsec::analytic_cont_CSS}
%%%%%%%%%%%%%%%%%%%%%%%%%%%%%%%%%%%%%%%%%%%%%%%%%%%%%%

In order to get a more global picture of the solutions described
above, we study the analytic continuation of the solutions beyond the
past SSH. 
We do this by integrating  Eqs.~(\ref{CSSbetap}), 
(\ref{CSSphipp}) and (\ref{CSSEuur}) towards larger values of $z$,
with initial conditions imposed at the horizon $z_H$, 
that are taken from the solutions to the boundary value problem discussed
above. 
Note, that -- as the
past SSH's domain of dependence is zero -- in general 
it is not enough to give data there. 
This procedure only works by the means
of analyticity.

As already mentioned above, we have $(\frac{V}{r})' = 0$ at the horizon.
From this it follows, that $2 z - \frac{V}{r} > 0$ for $z$ close to but 
larger than $z_H$. Therefore $(\frac{V}{r})' > 0$ for $z > z_H$ and
$\frac{V}{r}$ is monotonically increasing in the region beyond the horizon. 
So it might happen, that $\Vr$ eventually equals $2 z$ at some
value $z_S$, which means that the equations have a second (or counting the
origin a third) singular point there.

It was already shown in
\cite{Bizon-Wasserman-2000-CSS-exists-for-nonzero-beta} 
that this second singular point does not
correspond to a second self-similarity horizon (i.e.~a null surface that is
mapped to itself by the diffeomorphisms), but rather a spacelike
hypersurface where $2m/r \to 1$.
We will repeat the argument here.

Consider the expression
\be
h(z) = \frac{e^{-2\beta}}{2z - \frac{V}{r}}.
\ee
This function is positive for $z_H < z < z_S$ and diverges, when approaching
the past SSH from above:
\be
\lim\limits_{z \to z_H^+} h(z) = + \infty.
\ee
Using Eqs.~(\ref{CSSbetap}) and (\ref{CSSEuur}) we find for the derivative
\bea
h'(z) & = &  \frac{e^{-2 \beta}}{(2z - \frac{V}{r})^2} \left(
       - 2 (2z - \frac{V}{r}) \beta ' - (2 - (\frac{V}{r})')) 
       \right) = \nonumber\\
       & = & - \frac{2 e^{- 2\beta}}{(2 z - \frac{V}{r})^2},
\eea
which is less than zero in the region we are interested in.
From this it follows, that 
\be
\lim\limits_{z \to z_S^-} h(z)  \quad \textrm{exists},
\ee
and therefore as the denominator vanishes, we have
\be
\lim\limits_{z \to z_S^-} e^{-2 \beta(z)} = 0.
\ee
From Eq.~(\ref{CSSalgebraicV}) one finds, that the combination
$(\Vr - 2 z) \ e^{-2 \beta} (\phi')^2$ stays finite in the limit
$z \to z_S$. Making use of this one can show
that the  scalar curvature, the Kretschmann invariant, the square of the
Ricci tensor $R_{\mu\nu} R^{\mu\nu}$ and the Weyl invariant
are bounded when $z \to z_S$. Up to now it is not clear whether $\phi$
itself has a limit. If so, the above invariants have a limit as well.

As in \cite{Bizon-Wasserman-2000-CSS-exists-for-nonzero-beta}
our numerical integration gives the following results:
for each member of the one parameter family, there exists
a critical value of the coupling constant $\eta_n^*$ such that for smaller
values of the coupling, the solution extends smoothly up to the 
future SSH, whereas for stronger couplings the geometry contains
marginally trapped surfaces at $z = z_S > z_H$. The coordinate location of
these marginally trapped surfaces is a decreasing function of the coupling
constant and eventually merges with the location of the past SSH in the
limit $\eta \to \eta_{max}$. Furthermore the critical value
of the coupling increases with the excitation number $n$.

\begin{figure}[p]
\begin{center}
\begin{psfrags}
 \psfrag{z}[]{$z$}
 \psfrag{phi}[][][1][-90]{$\phi$}
\includegraphics[width=4in]{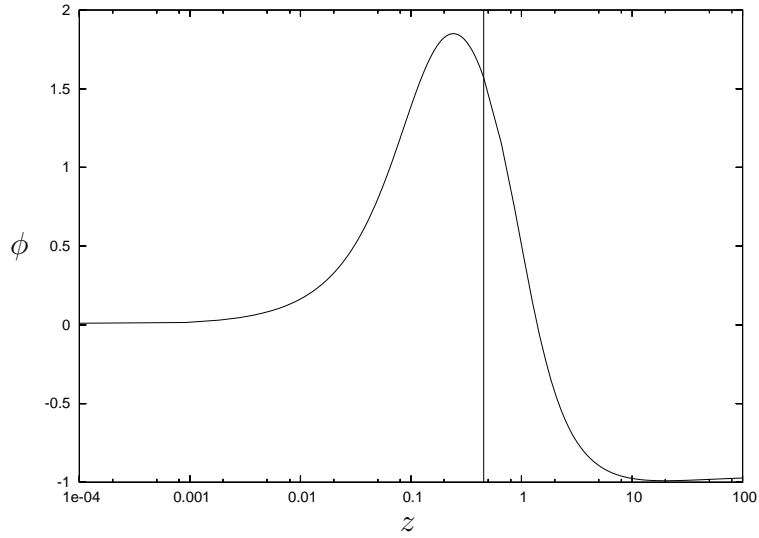}
\end{psfrags}
\end{center}
\caption{The field $\phi$ of the first excitation for $\eta = 0.1$.
The vertical line marks the horizon at $z_H = 0.45457$.}\label{fig::CSS_1_phi}
\end{figure}

\begin{figure}[p]
\begin{center}
\begin{psfrags}
 \psfrag{z}[]{$z$}
 \psfrag{2z}[]{$2 \ z$}
 \psfrag{Vr}[]{$\Vr$}
 \psfrag{2mr}[]{$\frac{2m}{r}$}
\includegraphics[width=4in]{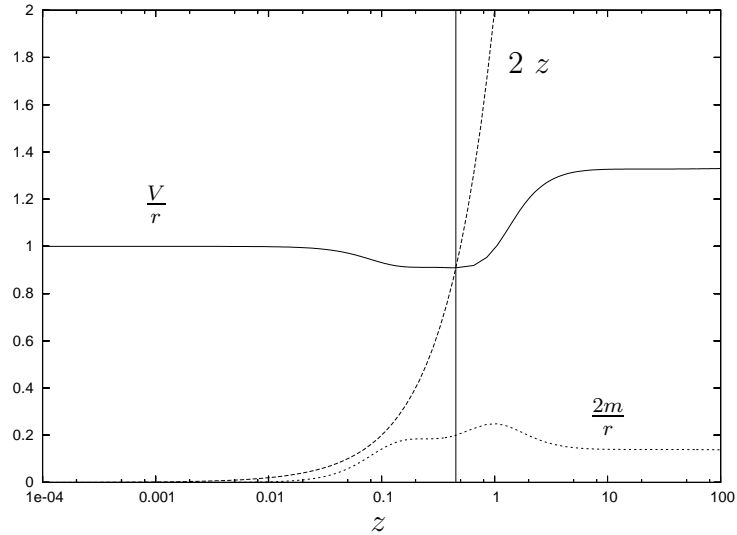}
\end{psfrags}
\end{center}
\caption{The same situation as above. Plotted are 
the metric function $\Vr$ as well as $\frac{2m}{r}$ for the first 
excitation at $\eta = 0.1$. $\Vr$ crosses the line $2 z$
at the location of the horizon. As $\eta$ is well below the critical value
$\eta_1^* \simeq 0.152$, $\Vr$ stays well below $2 z$ outside the horizon
and $\frac{2m}{r}$ is far from being unity anywhere.}\label{fig::CSS_1_Vr_2mr}
\end{figure}

\begin{figure}[p]
\begin{center}
\begin{psfrags}
 \psfrag{z}[]{$z$}
 \psfrag{phi}[][][1][-90]{$\phi$}
\includegraphics[width=4in]{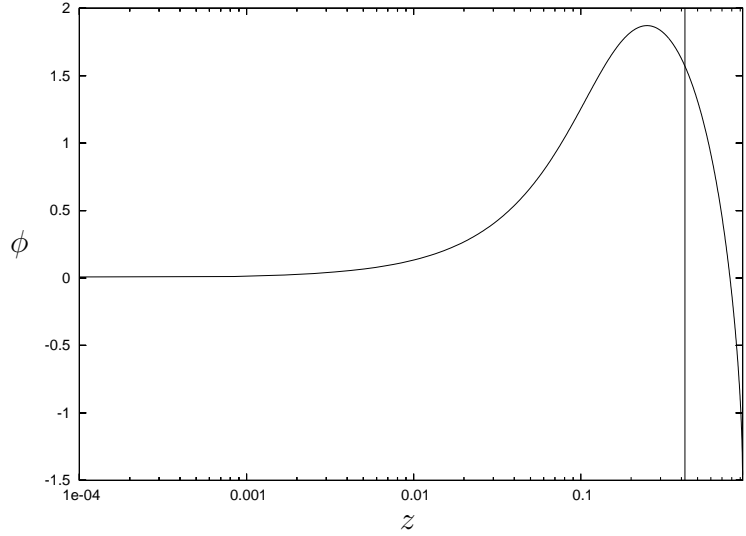}
\end{psfrags}
\end{center}
\caption{The field $\phi$ of the first excitation for $\eta = 0.175$.
The vertical line marks the horizon at $z_H =0.4198$. 
The right boundary of the plot is at the second singular point 
$z_S = 0.9292$.}\label{fig::CSS_1_0.175_phi}
\end{figure}

\begin{figure}[p]
\begin{center}
\begin{psfrags}
 \psfrag{z}[]{$z$}
 \psfrag{2z}[]{$2 z$}
 \psfrag{Vr}[]{$\Vr$}
 \psfrag{2mr}[]{$\frac{2m}{r}$}
\includegraphics[width=4in]{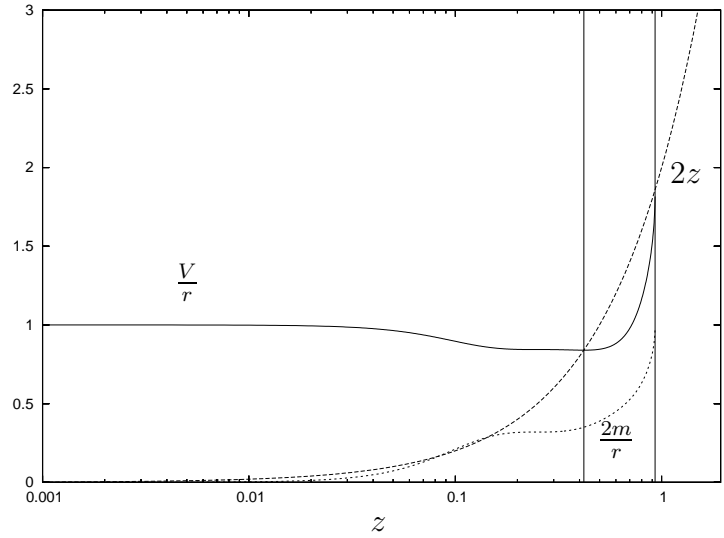}
\end{psfrags}
\end{center}
\caption{The same situation as above. Plotted are 
the metric function $\Vr$ as well as $\frac{2m}{r}$ for the first 
excitation at $\eta = 0.175$. $\Vr$ crosses the line $2 z$
at the location of the horizon. As $\eta$ is above the critical value
$\eta_1^* \simeq 0.152$, $\Vr$ equals $2 z$ for a second time
at $z_S = 0.9292$.
At the same time $\frac{2m}{r}$ tends to 1.
}\label{fig::CSS_1_0.175_Vr_2mr}
\end{figure}

\begin{figure}[h]
\begin{center}
\begin{psfrags}
 \psfrag{eta}[]{$\eta$}
 \psfrag{z_S}[]{$z_S$}
 \psfrag{z_H}[]{$z_H$}
 \psfrag{zS}[]{}
 \psfrag{zH}[]{}
\includegraphics[width=4in]{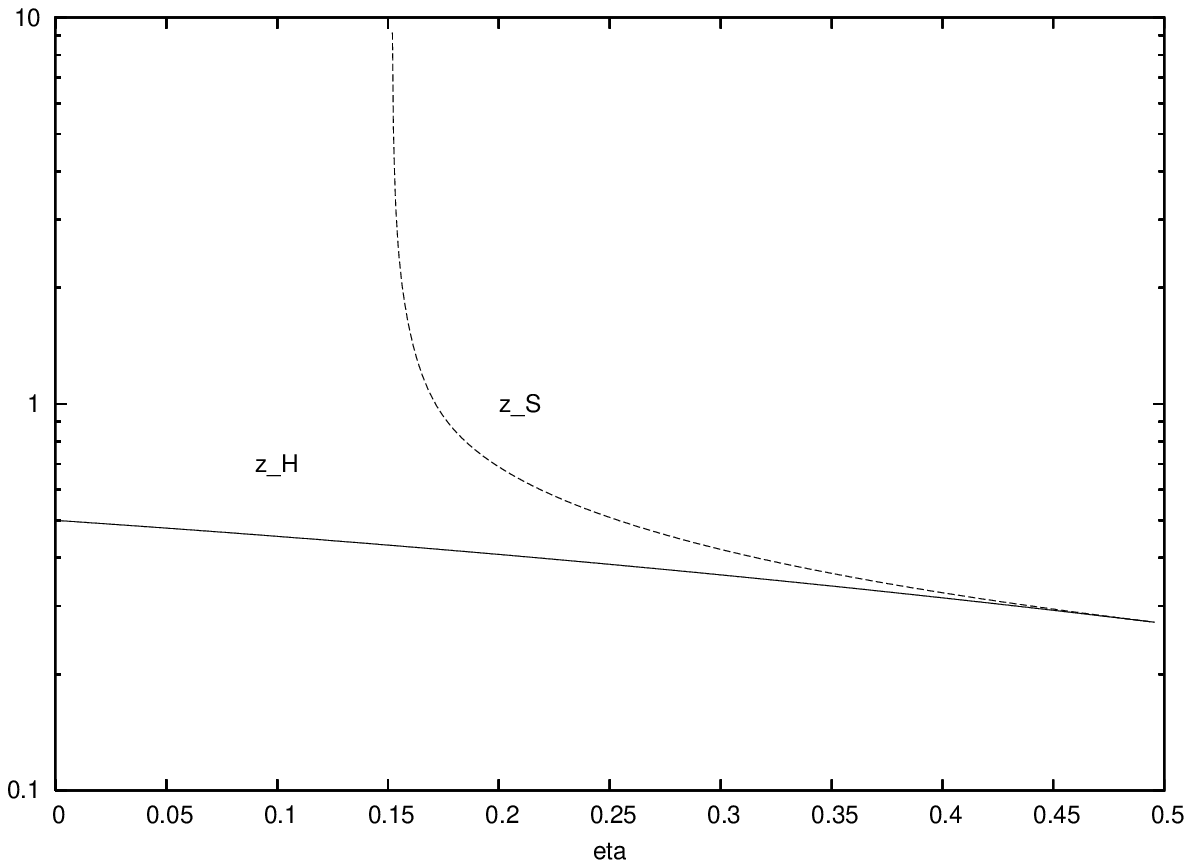}
\end{psfrags}
\end{center}
\caption{The coordinate locations of the past self-similarity horizon
$z_{H}$ and the second singular point $z_{S}$ 
for the first CSS excitation as functions of the coupling constant $\eta$.
For $\eta < \eta_1^* \ {\simeq} \ 0.152$ the analytic extension 
of the CSS solution is regular up to the future SSH ($z \to \infty$).
For bigger $\eta$ the solution develops an apparent horizon 
at the spacelike hypersurface $z=z_S = const$. This 
hypersurface finally merges with the past SSH in the limit $\eta \to 0.5$.
The other excitations behave in a similar way, where 
$\eta_n^*$ is an increasing function of the excitation number $n$. 
}\label{fig::eta_zS_zH}
\end{figure}

\section{Stability of CSS Solutions}
%%%%%%%%%%%%%%%%%%%%%%%%%%%%%%%%%%%%
In order to answer the question, whether any of the above constructed
solutions may play a role as a critical solution in gravitational collapse,
it is essential to study the stability properties.

In \cite{Bizon-1999-existence-of-self-similar-sigma-CSS-solutions} 
the stability of the one-parameter family of
CSS solutions on Minkowski background was studied. The results reported
are, that the $n$-th excitation has $n$ unstable modes, in particular the
ground state is stable and the first excitation has one unstable mode.
This suggests, that the ground state plays the role of a global attractor
in the time evolution of strong enough initial data, and that the first 
excitation might be a critical solution at the border of two 
different end states.
These predictions were verified in 
\cite{Bizon-Chmaj-Tabor-1999-sigma-3+1-evolution}. We will talk about 
these critical phenomena in more detail in Sec.~\ref{sec::smallcoupling}.

As the limit $\eta \to 0$ is a regular limit concerning the existence of CSS
solutions, one expects, that the stability properties of the solutions 
do not change, when gravity is switched on, at least as long as the coupling
constant is small.
 
We report here on the stability analysis for the coupled case, 
which we performed in two essentially different ways, 
and show, that indeed the stability properties of the CSS solutions 
do not change for small couplings.

The first method is the standard way to determine the stability properties 
of CSS solutions.
The CSS solution is perturbed in a small spherically symmetric, time
dependent way. The perturbations then are decomposed into modes
with an exponential (and oscillatory) time dependence.
Inserting this ansatz 
into the linearized field equations   
gives a coupled system of ODEs, with the same
singular points as for the background solution. Regularity then requires the
perturbations to be solutions to a boundary value problem, or to be more
precise to a linear eigenvalue problem, the real part of the 
eigenvalue, if bigger than zero, being responsible for the 
exponential growth of the unstable mode.
The advantage of this method is that one only deals with ODEs, which can be
integrated rather accurately. One main disadvantage is, that if using a
shooting and matching method, one needs good initial guesses for the
shooting parameters.
In other words, one can never be sure, that one obtains {\em all} of
the relevant, i.~e.~unstable, modes, unless one has further
theoretical arguments. 
Although in all similar situations, where the stability of an
``expected-to-be'' critical CSS solution was analyzed, the unstable modes
were all real, there is no theory guaranteeing this. 
For the system on fixed Minkowski background it was shown in 
\cite{Bizon-1999-existence-of-self-similar-sigma-CSS-solutions}, 
that the perturbation operator can be brought 
into a self-adjoint form (using orthogonal coordinates), therefore
the eigenvalues have to be real. Furthermore a theorem for Sturm-Liouville
operators could be applied, which determined the number of
eigenvalues giving rise to unstable modes.\footnote{We
mention that the continuous spectrum of the operator in 
\cite{Bizon-1999-existence-of-self-similar-sigma-CSS-solutions}
would seem to be unstable in adapted coordinates $(\tau, z)$. In fact these
modes are
growing as fast as the gauge mode, which shows, that the growth is due to
the shrinking of adapted coordinates and therefore these modes
are not considered as unstable. We also mention that members of the
continuous spectrum oscillate infinitley many times in the vicinity of the
horizon. They therefore cannot be detected by the methods described below.}
Unfortunately a similar analysis does not 
exist for the coupled system. 

We therefore present an alternative method to compute the
unstable modes. This method was proposed and carried out by 
J.~Thornburg \cite{JT-matrix-analysis}. The method uses the full
nonlinear field equations. It is based on the observation, that 
a numerical time evolution maps the discretized field, i.e. 
the N-dimensional vector, on the initial slice
to a discretized field on a later slice. If the initial configuration is
close to a CSS solution and the time step is small 
enough\footnote{The time step under consideration has to be small such that
unstable modes don't drive the solution out of the linear regime. The time
step may consist of several numerical time steps of the evolution code.}, 
then the 
relation between the deviations from the CSS solution at the initial
and final slices is linear, i.e. determined by an $N \times N$ matrix.
The unstable modes can then be extracted from the eigenvalues and
eigenvectors of this matrix. 
The main advantage of this method is, that it should give {\em all}
the unstable modes of the CSS solution that are ``seen'' by the time
evolution code, and therefore, if the number of grid points $N$ 
is big enough, all the unstable modes of the continuum problem.
A further advantage of this method is, that it uses an already existing
evolution code (the DICE code, see App.~\ref{app::dice}).
One minor disadvantage of the method is, that the numerical answers are
expected to be less accurate than the answers obtained from the ODE
boundary value problem. 
Unfortunately up to now this method suffers from a more serious drawback,
namely it fails to converge with respect to resolution. More precisely,
increasing the number of grid points, the numerical results move further and
further away from the predicted ones (an example is the gauge mode,
described below). 
What is even worse, is that we do not know, why this method
does not converge. One possible reason could be that for higher resolutions
the method gets increasingly ill-conditioned, that is increasingly sensitive
to small numerical errors in the numerical time evolution.
Nevertheless, as we get answers for a certain number of
grid points ($N = 500$) which are in good agreement with the results of the
other method, we are inclined to believe these results (for this number of
grid points) and deduce the stability properties from them.

\subsection{Unstable Modes from a Boundary Value Problem}
     \label{subsec::CSSstability_BVP}
%%%%%%%%%%%%%%%%%%%%%%%%%%%%%%%%%%%%%%%%%%%%%%%%%%%%%%%%%%%%%%
In order to analyze the stability of the CSS solutions, we proceed
in the usual way. Consider small time-dependent
radial perturbations of the CSS solution 
\bea\label{CSSperturbations}
\phi(\tau,z) & = & \phi_n(z) + \delta \phi(\tau,z), \\
\beta(\tau,z) & = & \beta_n(z) + \delta \beta(\tau,z), \\
\Vr(\tau,z) & = & \Vrn(z) + \delta V(\tau,z),
\eea
where $\phi_n, \beta_n, \Vrn$ denote the $n$-th CSS excitation.
(Note that by $\delta V$ we denote the perturbation of the metric function
$\Vr$ not of $V$ alone).
As the perturbations are supposed to be small, we can linearize 
Eqs.~(\ref{eq::phi}), (\ref{eq::betap}), (\ref{eq::Vp}) 
in these perturbations. Together with the requirement
that the background solutions $\phi_n, \beta_n, \Vrn$ solve the CSS
equations (\ref{CSSbetap}), (\ref{CSSphipp}) and
(\ref{CSSVoverrp})  (or alternatively (\ref{CSSEuur})), one obtains the
following linear system of PDEs:
\bea\label{CSSperturbationEqs1}
 \delta \beta'(\tau ,z) & = & 
  z\,\eta \,\phi_{n}'(z)\,\delta \phi'(\tau ,z),  \\
\delta V'(\tau ,z)  & = &   
  - \frac{1}{z}\biggl( 2\,e^{2\,\beta_n(z)}\,
        \left( -1 + 2\,\eta \,{\sin (\phi_n(z))}^2
          \right) \,\delta \beta(\tau ,z) + \delta V(\tau ,z) +
    \nonumber\\
&+ &  
       2\,e^{2\,\beta_n(z)}\,\eta \,
        \sin (2\,\phi_n(z))\,\delta \phi (\tau ,z) \biggr), 
\eea
\bea\label{CSSperturbationEqs2}
0 & = &  2\,e^{2\,\beta_n(z)}\,
       \left( 2\,z - \Vrn(z) \right) \,\delta \beta(\tau,z)\,
  \biggl[ \sin (2\,\phi_n(z)) + 
     z \,\left( -1 + 2\,\eta \,{\sin (\phi_n(z))}^2
\right) \, \phi_n'(z) \biggr]  + \nonumber\\
& + &  
\delta V(\tau ,z)\,\biggl[ 2\,z^2\,\phi_n'(z) + e^{2\,\beta_n(z)}\,
   \left( \sin (2\,\phi_n(z)) +  
   z \,\left( -1 + 2\,\eta \,{\sin (\phi_n(z))}^2\right) \,
 \phi_n'(z) \right)  \biggr]  + \nonumber\\
   & + & \left( 2\,z - \Vrn(z) \right) \,
  \biggl( 2\,e^{2\,\beta_n(z)}\,\delta \phi (\tau ,z)\,
 [ \cos (2\,\phi_n(z)) +
            z \,\eta \,\sin (2\,\phi_n(z))\,\phi_n'(z)]  + \nonumber\\
& + &
         z\,\biggl\{ \left[ 4\,z + e^{2\,\beta_n(z)}\,
                \left( -1 + 2\,\eta \,{\sin (\phi_n(z))}^2 \right)  - 
               \Vrn(z) \right] \,\delta \phi'(\tau ,z) + \nonumber\\
& + & 
     z\,\left( 2\,z - \Vrn(z) \right) \,
             \delta \phi''(\tau ,z) + 2\,\delta {\dot \phi}(\tau,z) \biggr\} 
         \biggr)  + \nonumber\\
& + & 
    \left[ 2\,z^2\,\left( 2\,z - \Vrn(z)\right) \right] 
   \delta {\dot \phi}'(\tau ,z). 
\eea

We now decompose the general perturbations (\ref{CSSperturbations})
into modes of the form
\bea\label{CSSperturbmodes}
\delta\phi(\tau,z) & = & e^{\lambda \tau} y(z), \nonumber\\
\delta \beta (\tau,z) & = & e^{\lambda \tau} g(z), \nonumber\\
\delta V (\tau,z) & = & e^{\lambda \tau} h(z).
\eea
Inserting this ansatz into Eqs.~(\ref{CSSperturbationEqs1}),
(\ref{CSSperturbationEqs2}) yields a coupled
system of ODEs for the perturbations $y(z), g(z)$ and $h(z)$.
This system again suffers from two singular points, the origin $z=0$
and the past SSH $z = z_H$.

Regularity at the origin restricts the perturbations to  
\be\label{CSSperturborigin}
y(0) = 0, \qquad g(0) = 0, \qquad
h(0)= 1,
\ee
whereas the first spatial derivative of the field $y'(0)$ is unconstrained.

At the horizon the requirement of regularity relates $y'_H$ to the 
boundary values of the other fields
\bea\label{CSSperturbhorizon}
y'_H & = & \frac{1}{\lambda 2 z_H (-1 + 2 \eta)} \biggl(
(-4 + (2 - 4 \eta)\lambda)y_H + 4 z_H (\phi_n')_H(-1 + 2 \eta)  g_H -
\nonumber\\
& - &  
(\phi_n')_H (1 + 2\eta + z_H^2 \eta (1-2\eta) (\phi_n')_H^2) h_H 
\biggr).
\eea

Eqs.~(\ref{CSSperturbationEqs1}), 
(\ref{CSSperturbationEqs2}) together with the 
boundary conditions (\ref{CSSperturborigin}), (\ref{CSSperturbhorizon}) 
constitute an eigenvalue problem, with eigenvalue $\lambda$ and
eigenfunctions $(y,y',g,h)$. 
As already mentioned the
eigenvalues and eigenfunctions may be complex. 
Nevertheless, as the coefficients in Eqs.
(\ref{CSSperturbationEqs1}),(\ref{CSSperturbationEqs2}) 
are all real the eigenvalues and eigenvectors, if complex, 
come in complex conjugate pairs.

The resulting problem again is a boundary value problem, consisting of eight
linear first order ODEs (Eqs.~(\ref{CSSperturbationEqs1}, 
(\ref{CSSperturbationEqs2})) separated into real
and imaginary part) and the eight boundary conditions
(real and imaginary parts of Eqs.~(\ref{CSSperturborigin}) and 
(\ref{CSSperturbhorizon})). The parameters that have to be matched
are 
\be
\lambda, \qquad y_H, \qquad g_H, \qquad h_H,
\ee
(again real and imaginary parts thereof).
As the problem is linear and homogeneous the solutions are only fixed up to
an overall scale. We fix that, by setting $y'(0) = 1$.

This boundary value problem again was solved numerically, 
using the shooting and matching routine d02agf of the 
NAG-library (\cite{NAG}).
The background solution was computed first and interpolated when necessary 
in order to provide the coefficients in
Eqs.~(\ref{CSSperturbationEqs1}), (\ref{CSSperturbationEqs2}).

\subsection{Gauge Modes}\label{subsec::gauge_modes}
%%%%%%%%%%%%%%%%%%%%%%%%%%%%%%%%%%%%%%%%%%%%%%%%%%%

Before reporting on the numerical results, we determine the gauge modes,
that result from a certain arbitrariness in relating the adapted coordinates
$(\tau, z)$ to Bondi coordinates $(u,r)$.
Recalling from Sec.~\ref{subsec::SSinSpherSymm} 
(Eq.~(\ref{coordCSS})) we have
\bea
\tau(u) = - \ln(u^*-u) &  & u(\tau) = u^* - e^{-\tau} \nonumber\\
                       & \Longleftrightarrow&   \nonumber\\  
z(u,r) = \frac{r}{(u^* - u)} & & r(\tau,z) = e^{-\tau} z.
\eea
A shift in the culmination time $u^* \to {\bar u}^* = u^* + \epsilon$ 
(corresponding to a shift of the origin of $u$) 
yields another pair of adapted coordinates $({\bar \tau}, {\bar z})$,
defined as above with respect to ${\bar u}^*$.
The relation between these two sets of adapted coordinates is given by
\bea\label{taubartau}
{\bar \tau} & = & \tau - \ln (1 + \epsilon e^{\tau}), \nonumber\\ 
{\bar z} & = & \frac{z}{1 + \epsilon e^{\tau}}.
\eea
For small $\epsilon$ we linearize to get
\bea\label{taubartaulin}
{\bar \tau} & = & \tau - \epsilon e^{\tau} = \tau + \epsilon
        \chi^{\tau}(\tau,z)\nonumber\\ 
{\bar z} & = & z - \epsilon z e^{\tau} = z + \epsilon \chi^z(\tau,z),
\eea
where we have introduced the generating vector field $\chi^{\mu}(\tau,z)$.
In the following we will treat the vector field $\chi^{\mu}$ as general
and show in the end, that the form (\ref{taubartaulin}) is indeed the only
possible one.\footnote{In particular the following arguments rule out 
a coordinate transformation, that resets the origin
of $\tau$, $\tau \to \tau + c$ : 
such a coordinate transformation violates (\ref{eq::chi_tau}) and 
therefore does not correspond to a regular perturbation of the CSS solution
within our choice of coordinates (especially the choice of $u$ being proper
time at the origin).}

Such a small coordinate transformation introduces a (small) change in the
perturbations according to
\bea
\delta {\bar g}_{\mu\nu} & = & \delta g_{\mu\nu} - \epsilon {\mathcal L}_{\chi}
   (g_0)_{\mu\nu}, \nonumber\\
\delta {\bar \phi} & = & \delta \phi - \epsilon {\mathcal L}_{\chi} \phi_0,
\eea
where in our case the objects with the subscript $0$ denote a CSS solution.
Pure gauge modes are characterized by
\bea\label{CSSgaugemodesgen}
 \delta g_{\mu\nu} & = & \epsilon {\mathcal L}_{\chi}
   (g_0)_{\mu\nu}, \nonumber\\
 \delta \phi & = & \epsilon {\mathcal L}_{\chi} \phi_0,
\eea
i.e. modes of this form can be removed by a small change in the coordinates
according to (\ref{taubartaulin}).

We first note, that the coordinate transformations are not completely
arbitrary, but have to ensure that a hypersurface ${\bar \tau} = const$ 
still is a null cone, which is reflected by
\be
{\mathcal L}_{\chi} (g_0)_{z z} = 0,
\ee 
or
\be
\chi^{\tau}_{,z} = 0 \qquad \Rightarrow \chi^{\tau} = \chi^{\tau}(\tau).
\ee
The second observation is, that via \ref{CSSperturbations} we
fixed the ``${\mathbbm S}^2$ - part'' of the metric to be $e^{-2\tau} z^2$,
which gives
\be
{\mathcal L}_{\chi} (g_0)_{\theta \theta} = 0,
\ee  
or
\bea
\chi^{\tau} (g_0)_{\theta \theta , \tau} & + & 
\chi^z (g_0)_{\theta \theta, z}  = 0,
\nonumber\\
\chi^z & = & z \chi^{\tau}.
\eea

For the remaining components of (\ref{CSSgaugemodesgen})
we get
\bea\label{CSSgaugemodesgen2}
{\mathcal L}_{\chi} (g_0)_{\tau \tau} & = &  - e^{-2 \tau} \biggl[
   2 (- \chi^{\tau} + \chi^{\tau}_{,\tau} ) e^{2 \beta_0} 
       ((\Vr)_0 - 2 z) + \chi^z (e^{2 \beta_0} ((\Vr)_0 - 2 z))'
   + \nonumber\\
& & + 2 \chi^z_{,\tau} e^{2 \beta_0})\biggr] =   \nonumber\\
& = &  
   - e^{-2 \tau} \biggl[
   2 (- \chi^{\tau} + \chi^{\tau}_{,\tau} ) e^{2 \beta_0} 
       ((\Vr)_0 - 2 z) + z \chi^{\tau} (e^{2 \beta_0} ((\Vr)_0 - 2 z))'
 + \nonumber\\
& & + 2 z \chi^\tau_{,\tau} e^{2 \beta_0})\biggr]              \nonumber\\
{\mathcal L}_{\chi} (g_0)_{\tau z} & = & 
   - e^{-2 \tau} \left[  (- 2 \chi^{\tau} + 
      \chi^{\tau}_{,\tau} + \chi^{z}_{,z}) e^{2 \beta_0} + 2 \chi^{z}
      \beta_0' e^{2\beta_0} \right] = \nonumber\\
& = &  - e^{- 2 \tau} \left[  (-  \chi^{\tau} + 
      \chi^{\tau}_{,\tau}) e^{2 \beta_0} + 2 z \chi^{\tau}
      \beta_0' e^{2\beta_0} \right]                   \nonumber\\  
{\mathcal L}_{\chi} \phi_0 & = &  \chi^{z} \phi_0'(z)  = 
                  z \chi^{\tau} \phi'_0(z).                 
\eea
On the other hand the perturbations of the metric functions $\beta$ and
$\Vr$ are related to the perturbations of the components 
of the metric with respect to $(\tau,z)$ coordinates via
\bea\label{thedeltas}
\delta g_{\tau \tau} & = & - e^{-2 \tau} e^{2 \beta_0} 
       \left[ 2 ((\Vr)_0 - 2 z) \delta \beta + \delta V \right], \nonumber\\ 
\delta g_{\tau z} & = & - 2 e^{-2 \tau} e^{2 \beta_0} \delta \beta. 
\eea
The regularity requirements at the origin discussed in the last
section, namely $\delta \beta(\tau,0) = \delta V (\tau,0) = 0$, 
give $\delta g_{\tau \tau} (\tau, 0) = \delta g_{\tau z}(\tau, 0) = 0$.
Comparing this to (\ref{CSSgaugemodesgen2}) gives the further restriction on
the generating vector field $\chi^{\mu}$
\be\label{eq::chi_tau}
\chi^{\tau}_{,\tau} = \chi^{\tau},
\ee
and therefore
\be\label{chigauge}
\chi^{\tau} = e^{\tau}, \qquad \chi^{z} = z e^{\tau},
\ee
which is exactly the coordinate transformation introduced
above (\ref{taubartaulin}).

Finally we combine (\ref{CSSgaugemodesgen2}),(\ref{thedeltas}) with
(\ref{chigauge}). 
In a stability analysis, as described above, we therefore expect to find
the gauge mode
\bea
\delta\phi & = &  \epsilon e^{\tau} z \phi_n' \nonumber\\
\delta\beta & = & \epsilon e^{\tau} z \beta_n' \nonumber\\
\delta V & = &    \epsilon e^{\tau} z \Vrn'.
\eea

\subsection{Unstable Modes from a Matrix
               Analysis}\label{subsec::Matrix_Analysis}
%%%%%%%%%%%%%%%%%%%%%%%%%%%%%%%%%%%%%%%%%%%%%%%%%%%%%%%%
The following method was proposed and carried out by J.~Thornburg
\cite{JT-matrix-analysis}.
It uses the full (nonlinear) time evolution equations. As already
mentioned, the method
has the advantage, to give {\em all} the unstable modes, in contrast
to the shooting and matching method described above, where only
those unstable modes can be found, which lie close to the 
``initial guess''.

Consider Eqs.~(\ref{eq::phi}), (\ref{eq::betap}) and (\ref{eq::Vp}). 
As described in Sec.~\ref{sec::Einstein}
it is sufficient to prescribe the matter field $\phi$ at the initial null
cone $u=u_0$ (or $\tau = \tau_0$). The metric functions on the initial slice
then are determined by $\phi$ via the hypersurface equations (\ref{eq::betap})
and (\ref{eq::Vp}). Eq.~(\ref{eq::phi}) is then used to evolve the 
configuration (together with Eqs.~(\ref{eq::betap}) and (\ref{eq::Vp}) 
in order to update the metric functions). 
Altogether the time evolution maps the field $\phi(\tau_0,z)$
at the initial slice to a field configuration 
$\phi(\tau_0 + \Delta\tau,z)$ at a later slice:
\be
\phi(\tau_0 + \Delta\tau,z) = F(\phi(\tau_0,z)),
\ee
where $F$ is the nonlinear operator representing time evolution.
Of course the operator $F$ depends on the time step $\Delta \tau$.
Clearly, if the initial data correspond to the CSS solution,
$F$ acts as the identity operator,
\be
F(\phi_{CSS}(\tau_0,z)) = \phi_{CSS}(\tau,z).
\ee

In an actual numerical time evolution in (1+1) dimensions 
any smooth function of $z$ 
is represented by it's values at the grid points, i.e. 
the function $\phi(\tau_0,z)$ is replaced by the N-dimensional
vector $(\phi_i^0)_{i=1,N}$, where $\phi_i^0 = \phi(\tau_0, z^i)$, and
$N$ is the number of grid points.
The numerical time evolution then maps this vector
to the corresponding vector at the next time step:
\be
\phi_i^1 = F_i(\phi_j^0).
\ee

Consider now a small generic perturbation of the CSS solution,
$\phi_{CSS}(\tau_0,z) + \delta \phi(\tau_0,z)$. 
If the time step $\Delta \tau$ is small enough, 
this configuration is mapped to another small perturbation of the CSS
solution at time $\tau_0 + \Delta \tau$. 
Translating this to the language of a finite grid we have
\be
\phi_i^1 = (\phi_{CSS})_i^1 + (\delta \phi)_i^1 = 
F_i((\phi_{CSS})_j^0 + (\delta \phi)_j^0).
\ee
For a small perturbation this can be linearized to give
\be
(\phi_{CSS})_i^1 + (\delta \phi)_i^1 = 
F_i((\phi_{CSS})_j^0) + 
\left. \frac{\partial F_i}
{\partial \phi_j^0} \right|_{(\phi_{CSS}^0)} (\delta \phi)_j^0 = 
(\phi_{CSS})_i^0 + \left. \frac{\partial F_i}
{\partial \phi_j^0} \right|_{(\phi_{CSS}^0)} (\delta \phi)_j^0.
\ee
So for the perturbation the following linear relation holds
\be
(\delta \phi)_i^1 = \left. \frac{\partial F_i}
{\partial \phi_j^0} \right|_{(\phi_{CSS}^0)} (\delta \phi)_j^0.
\ee
The Jacobian on the right hand side is a $N \times N$ matrix,
which depends on the size of the time step $\Delta \tau$.
Therefore it has $N$ eigenvalues $\{ {\tilde \lambda}_i \}_{i=1,N}$,
which are functions of the time step $\Delta \tau$.
If the Jacobian is diagonalizable we can switch to a basis such that
\be
(\delta \phi)_i^1 = {\tilde \lambda}_i (\delta \phi)_i^0, \quad
\textrm{with} \quad {\tilde \lambda}_i = {\tilde \lambda}_i(\Delta \tau).
\ee
For the quotient of differences we get
\be
\frac{\delta \phi_i(\tau_0 + \Delta \tau) - \delta \phi_i(\tau_0)}
{\Delta \tau} = \frac{{\tilde \lambda}_i - 1}{\Delta \tau} \delta
    \phi_i(\tau_0).
\ee
In order that the limit $\Delta \tau \to 0$ exists, the denominator
on the right hand side has to be proportional to $\Delta \tau$. We set
\be\label{lambdatilde}
{\tilde \lambda}_i -1 = \lambda_i \ \Delta \tau,
\ee
$\lambda_i$ being constants.

In the limit $\Delta \tau \to 0$ we have
\be
\delta {\dot \phi}_i (\tau) = \lambda_i \delta \phi_i(\tau)
\ee
and therefore
\be
\delta \phi_i(\tau) = e^{\lambda_i (\tau - \tau_0)} \delta \phi_i(\tau_0).
\ee
This is valid with respect to the eigenbasis of the Jacobian.
Transforming back to the original basis we have
\be\label{CSSperturbmodesdiscrete}
(\delta \phi_k)_i(\tau) = e^{\lambda_k (\tau - \tau_0)} (\delta \phi_k)_i
(\tau_0),
\ee
where $\delta \phi_k$ denotes the $k$-th eigenvector of the Jacobian.
The perturbation (\ref{CSSperturbmodesdiscrete}) now is of the same 
form as the modes in Eq.~(\ref{CSSperturbmodes}), with the 
eigenvalues $\lambda$ being related to the eigenvalues of the time evolution
Jacobian via Eq.~(\ref{lambdatilde}).

The Jacobian was computed using the DICE code 
(described in Appendix \ref{app::dice}). 
As this code uses grid points, that are
freely falling along ingoing null geodesics, the evolved field 
$\phi$ had to be interpolated to give the appropriate values at constant
$z$. Technically the Jacobian is computed, by first evolving the 
CSS solution for one time step, and then perturbing each grid point 
separately, according
to the N perturbations $(\delta \phi_k)_i^0 = \epsilon
\delta_{ik}$, $k = 1,N$, with $\epsilon$ small.
Each of these perturbations is evolved again for one time step, so
in the end the $N \times N$ numbers $F_i(\phi_k)$ are known.
The Jacobian results from this by a forward differencing. 
In order to compute the eigenvalues and eigenvectors of 
the matrix a linear algebra package (EISPACK \cite{EISPACK-book}) is used.

Finally we want to point out, that although this method
only yields $N$ modes of the perturbation operator, according to the $N$
grid points involved, we expect to find {\em all} the relevant 
unstable modes. The reason for this is, that we expect the 
unstable modes to vary on scales that are large compared to the grid 
spacing.

\subsection{Numerical Results of the Stability
           Analysis}\label{CSSstability_results}
%%%%%%%%%%%%%%%%%%%%%%%%%%%%%%%%%%%%%%%%%%%%%%%%%%%%%%%%%%%%%%%%

Before we present our results here, we want to stress two points:
First we are interested in the stability of the CSS ground state and the
first CSS excitation, because we want to get information on the possible
roles they might play in the context of critical phenomena.
With respect to this, only the stability for small couplings 
($\eta \lesssim \eta_*(0) \sim 0.69$ for the ground state and
$\eta \lesssim 0.2$ for the first excitation)
is of interest.

Second we trust the results of the boundary value problem 
Sec.~\ref{subsec::CSSstability_BVP}. On the other hand, as already
mentioned, the numerical scheme of the matrix 
analysis Sec.~\ref{subsec::Matrix_Analysis}
does not show convergence with resolution, and results should be taken with
some care. Nevertheless, as we will show below, for $N=500$
grid points (and $\eta$ not too large), 
the results for the gauge modes and the unstable mode for the
first excitation are in very good agreement with the 
theoretical predictions and the results of the boundary value problem. 
Therefore -- for these couplings -- we are inclined to trust the results of
the matrix analysis, in particular the number of unstable modes.

\begin{figure}[p]
\begin{center}
\begin{psfrags}
 \psfrag{eta}[]{$\eta$}
 \psfrag{Relambdai}[r][c][1][-90]{$Re(\lambda_i)$}
\includegraphics[width=3.5in]{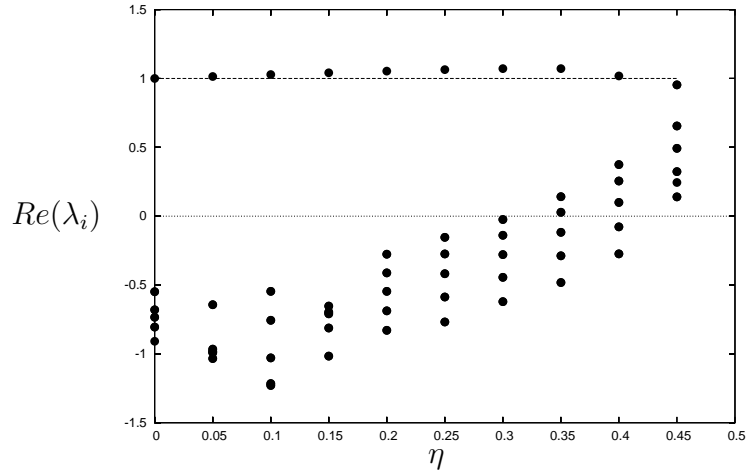}
\end{psfrags}
\end{center}
\caption{Dots represent the real parts of the first few eigenvalues
of perturbations of the groundstate as obtained by a matrix analysis
with $N=500$. 
The line $\lambda = 1$ represents the predicted eigenvalue of the
gauge mode. For small $\eta$ the matrix analysis gives no unstable mode.
For bigger $\eta$ (e.g. at $\eta = 0.4$)
the additional positive eigenvalues could not be confirmed by the
shooting and matching method.  
}\label{fig::CSS_500_groundst_eigenvalues}
\end{figure}

\begin{figure}[p]
\begin{center}
\begin{psfrags}
 \psfrag{eta}[]{$\eta$}
 \psfrag{Relambdai}[r][c][1][-90]{$Re(\lambda_i)$}
\includegraphics[width=3.5in]{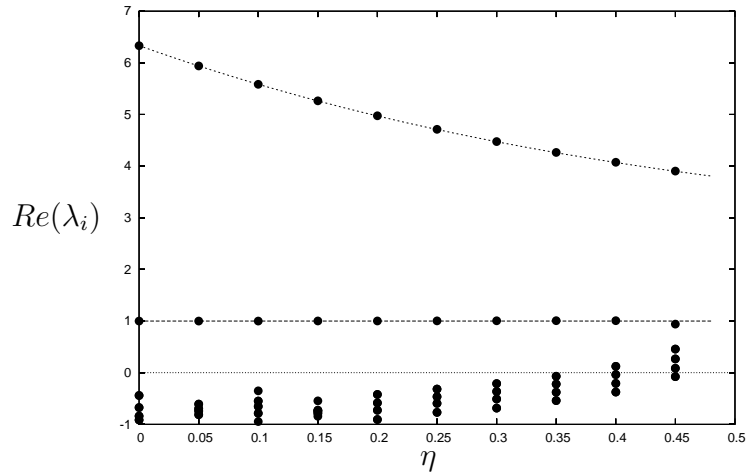}
\end{psfrags}
\end{center}
\caption{The same situation as in 
Fig.~\ref{fig::CSS_500_groundst_eigenvalues} for the first CSS excitation
The dashed line starting at
$\eta = 0$ with $6.33$ represents the eigenvalue of the unstable mode
obtained from the boundary value problem. 
The agreement of results of the two methods is very good.
The matrix analysis gives one unstable mode for $\eta \le 0.35$. The
additional positive eigenvalues for larger $\eta$ (e.g.~$\eta = 0.4$)
could not be confirmed by the
shooting and matching method.
}\label{fig::CSS_500_1stExcit_eigenvalues}
\end{figure}

Given these caveats, we numerically find that the ground state is stable,
whereas the first excitation has one unstable mode 
(both results for $\eta$ not too large 
(See Figs.~\ref{fig::CSS_500_groundst_eigenvalues} 
and \ref{fig::CSS_500_1stExcit_eigenvalues})).
 
\begin{figure}[p]
\begin{center}
\begin{psfrags}
 \psfrag{z}[]{$z$}
 \psfrag{deltaphi}[][][1][-90]{$y_{gauge}$}
 \psfrag{groundstate}[]{ground state}
 \psfrag{1stExcitation}[]{1st excitation}
\includegraphics[width=4in]{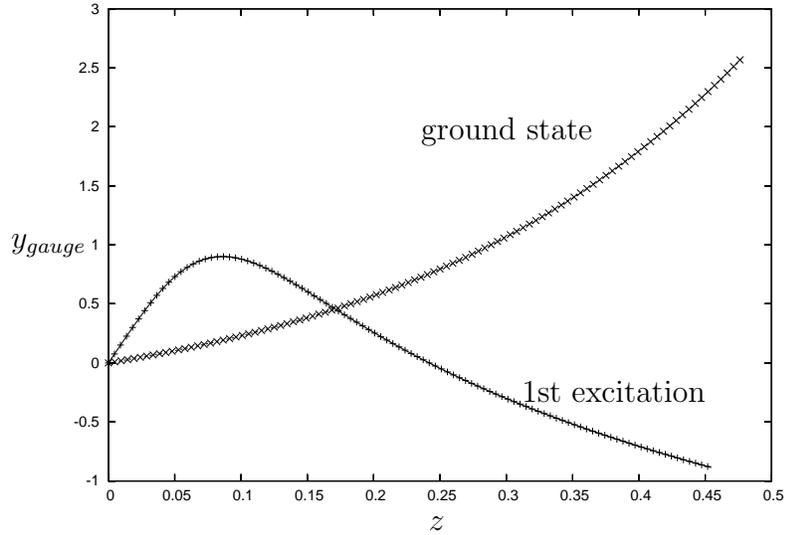}
\end{psfrags}
\end{center}
\caption{The gauge modes for the ground state and the first excitation
at a coupling $\eta = 0.1$. Plotted are the predicted functions, 
$y_{gauge} = z \phi_n'$, (solid lines), and the corresponding results of the
shooting and matching method (``x'', ``+'').
}\label{fig::CSS_gaugemodes_0.1}
\end{figure}

Fig. \ref{fig::CSS_gaugemodes_0.1} compares the gauge modes -- obtained
via the boundary value problem of Sec.~\ref{subsec::CSSstability_BVP} --
of the ground state and first excitation 
at a coupling $\eta = 0.1$ to the theoretical 
predictions $y_{gauge} = z \phi_n'$. 
Note that in this figure and in the
following ones the overall scale is chosen arbitrarily.

\begin{figure}[p]
\begin{center}
\begin{psfrags}
 \psfrag{z}[]{$z$}
 \psfrag{deltaphi}[][][1][-90]{$y$}
 \psfrag{unstable mode}[]{unstable mode}
 \psfrag{gauge mode}[]{gauge mode}
\includegraphics[width=4in]{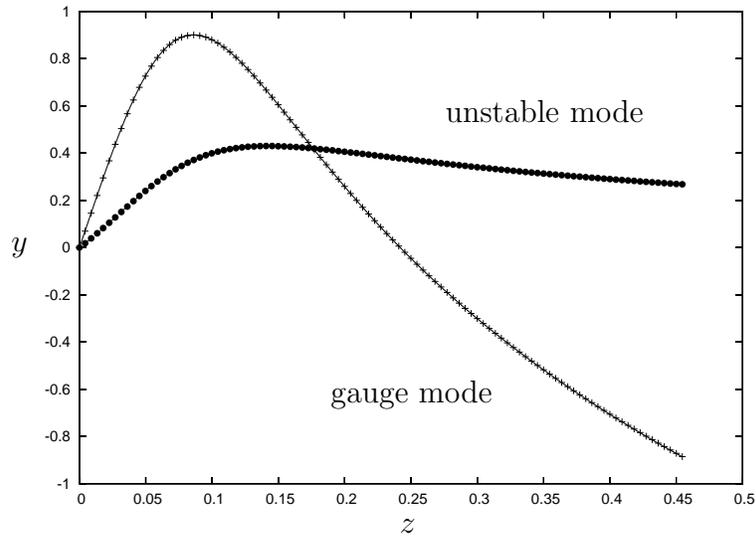}
\end{psfrags}
\end{center}
\caption{The eigenfunctions of the unstable mode 
($\lambda = \lambda_1 =5.58463$)
and the gauge mode ($\lambda = 1$) 
for the first CSS excitation at a coupling $\eta = 0.1$. 
Plotted are the result of a 500 points matrix analysis
(dots, only every 5th point is plotted). 
These are compared to the results from the boundary value problem (lines):
the gauge mode is compared to the predicted eigenfunction
$y_{gauge}(z) = z \phi_1'(z)$ and the unstable mode is compared to the
result $y(z)$ of the boundary value problem described in
Sec.~\ref{subsec::CSSstability_BVP}.  
}\label{fig::CSS_500_1stExcit_0.1_eigenf}
\end{figure}

\begin{figure}[p]
\begin{center}
\begin{psfrags}
 \psfrag{eta}[]{$\eta$}
 \psfrag{1overlambda}[r][c][1][-90]{$1/\lambda_1(\eta), f(\eta)$}
\includegraphics[width=4in]{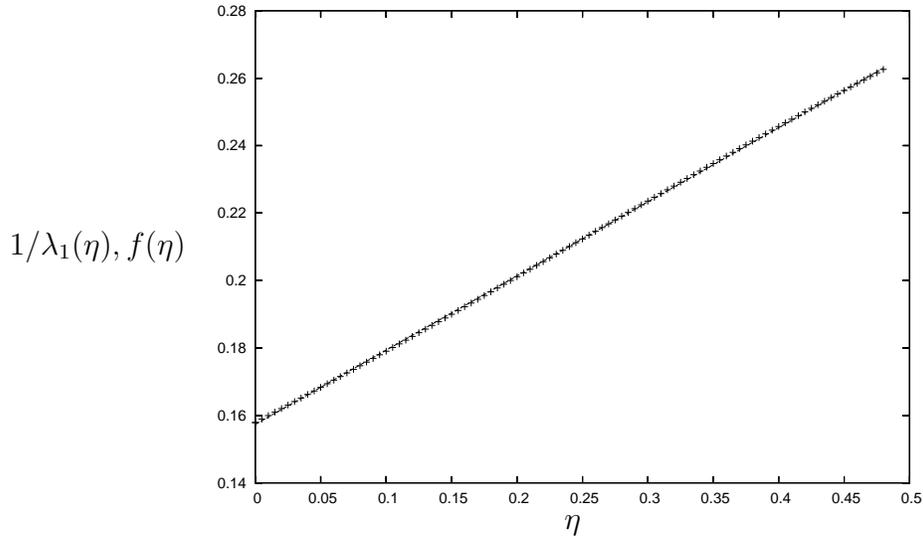}
\end{psfrags}
\end{center}
\caption{Plotted is $1/\lambda_1$ (dots), where $\lambda_1$ 
is the eigenvalue of the unstable mode of the first CSS excitation,
obtained by the shooting and matching method of 
Sec.~\ref{subsec::CSSstability_BVP}, as a function of $\eta$.
This is very well fitted by the straight line
$f(\eta) = 0.21997 \eta + 0.15736$ (solid line). See the next figure 
for the error of this fit. 
}\label{fig::CSS_one_over_lambda}
\end{figure}

\begin{figure}[p]
\begin{center}
\begin{psfrags}
 \psfrag{eta}[]{$\eta$}
 \psfrag{fminusfit}[r][l][1][-90]{$1/\lambda_1(\eta) - f(\eta)$}
\includegraphics[width=4in]{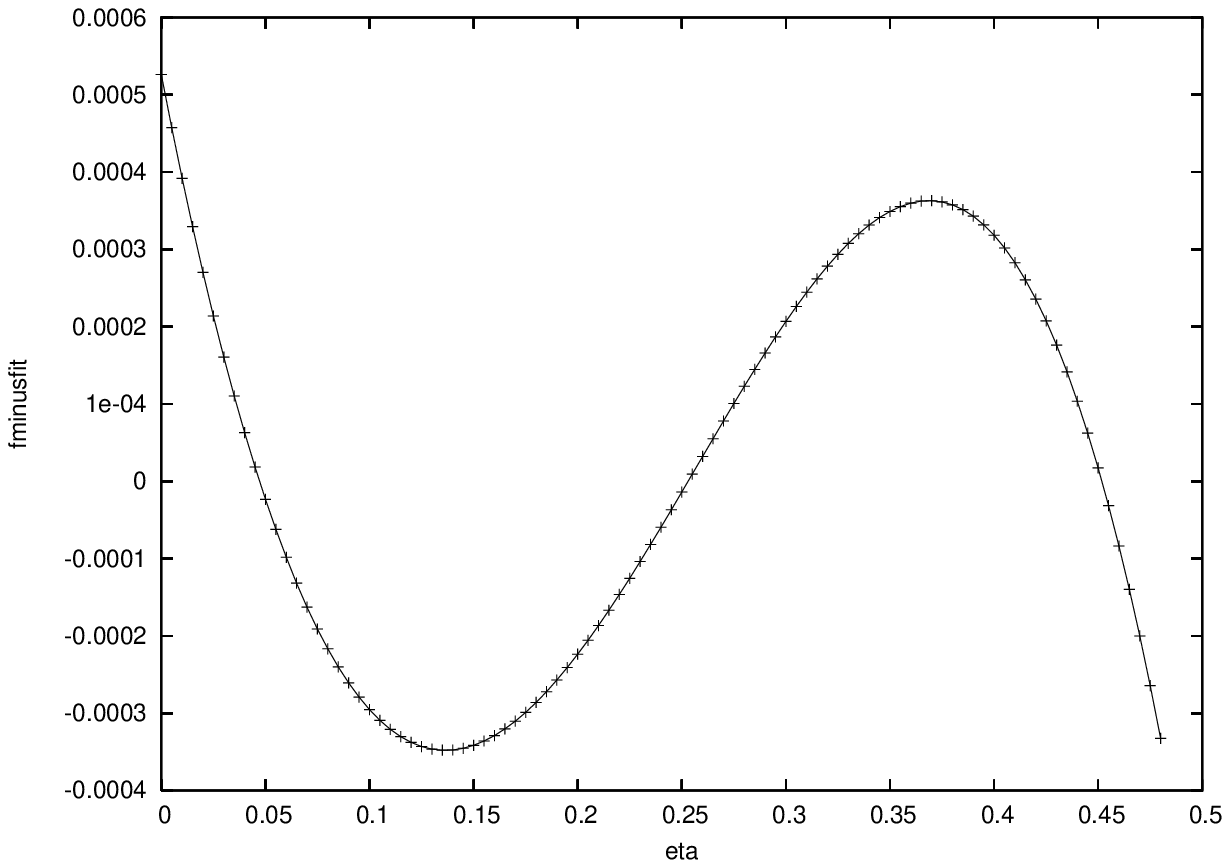}
\end{psfrags}
\end{center}
\caption{The error $1/\lambda_1(\eta) - f(\eta)$, where $f(\eta)$ is the
straight line defined in Fig.~\ref{fig::CSS_one_over_lambda}.
}\label{fig::CSS_fitting_error}
\end{figure}
\afterpage{\clearpage}

Fig.~\ref{fig::CSS_500_1stExcit_0.1_eigenf}
compares the eigenfunctions of the gauge mode and the unstable mode
-- obtained via the matrix
method of Sec.~\ref{subsec::Matrix_Analysis} --
of the first CSS excitation for $\eta = 0.1$ to the results
of the boundary value problem. The agreement is very good.

Figs. \ref{fig::CSS_500_groundst_eigenvalues} and
\ref{fig::CSS_500_1stExcit_eigenvalues} show the real parts of the 
first few eigenvalues of the groundstate and first excitation.
The agreement of the results of the matrix method and the boundary value
problem is good for the groundstate at low couplings and very good for the
first excitation. 

Finally,
note that the inverse eigenvalue of the unstable mode of 
the first CSS excitation $1/\lambda_1$ as a function of coupling $\eta$
is very well approximated by a straight line 
(see Figs.~\ref{fig::CSS_one_over_lambda} and \ref{fig::CSS_fitting_error}).

\section{Numerical Construction of DSS Solutions}\label{sec::DSSsolutions}
%%%%%%%%%%%%%%%%%%%%%%%%%%%%%%%%%%%%%%%%%%%%%%%%%%%%%%%%%%%%%%%%%%%%%%%%%

This section deals with the numerical construction of DSS solutions to the
self gravitating nonlinear $\sigma$ model.
Due to the 
(periodic) time dependence of the 
metric functions and the field,
the numerical construction of DSS solutions is considerably 
more involved than the corresponding construction of CSS solutions.
Concerning self-gravitating matter fields
there are essentially two papers (not counting subsequent ones using the
same methods) that deal with the problem of constructing time-periodic
solutions to a boundary value problem (in space).
Seidel and Suen \cite{Seidel-Suen-1991-breathers} construct
solutions to the Einstein-Klein-Gordon system with mass, which are 
oscillating (periodic in time). 
Gundlach
\cite{Gundlach-1996-scaling-in-critical-collapse}
constructs a DSS solution to the massless Einstein-Klein-Gordon system.
Both methods use Fourier series, 
as suggested by the periodic time dependence, but their ``implementations''
are different, as will be explained below.

Following Gundlach closely, we present a method, that involves 
discrete Fourier transform, 
pseudo spectral methods and the reduction to a boundary value problem
for ODEs. 

The procedure of construction can be roughly outlined
as follows: we are looking for solutions to Eqs.~(\ref{eq::phi}),
(\ref{eq::betap}) and (\ref{eq::Vp}), rewritten in
adapted coordinates (\ref{coordDSS}), 
that are regular between origin and past SSH and that
are periodic in the adapted time coordinate $\tau$.
We expand the metric functions as well
as the field into Fourier series in time, where the coefficients are
functions of the spatial variable $z$. Inserting these Fourier series into 
the equations yields a coupled system of ODEs for the Fourier coefficients.
This system has singular points at the origin and the past SSH, so together
with the boundary conditions required by regularity we have to solve 
an ODE boundary value problem for the Fourier coefficients.
As the Fourier series consist of infinite many terms and as the problem is
nonlinear, one has to truncate the Fourier series at some appropriate
maximal frequency. This then yields a boundary value problem for a finite
system of ODEs, which again can be solved by the means of a shooting and
matching routine.

Seidel,Suen and Gundlach use different methods for explicitly setting up the
ODEs.
Seidel and Suen truncate the series and plug the truncated series into the
equations (see Appendix \ref{app::DFT}). 
Comparing coefficients yields the desired coupled system of ODEs
for the coefficients. The disadvantage of this method is, that these direct
expressions quickly get horribly complicated as one allows for more and more
coefficients. 

Gundlach on the other hand uses so called {\em pseudo-spectral methods},
that is, he does part of the computations in ``real'' space and part 
in Fourier space (see Appendix \ref{app::DFT}). 
Basically the variables are the Fourier coefficients, but
in order to set up the ODEs, the coefficients are transformed
back to real space, there the algebraic manipulations are carried out in
order to define the derivatives of the functions, which then in turn are
transformed back to Fourier space. This requires a pair of (backward and
forward) Fourier transformations at each integration step.
Employing a {\em fast Fourier transform (FFT)} 
instead of the ordinary  discrete Fourier transform (DFT) 
is essential for reducing computational cost.

\subsection{The DSS Equations}\label{subsec::DSSequations}
%%%%%%%%%%%%%%%%%%%%%%%%%%%%%%%%%%%%%%%%%%%%%%%%%%%%%%%%%%

We start by transforming Eqs.~(\ref{eq::phi}),(\ref{eq::betap}) and
(\ref{eq::Vp}) to the adapted coordinates
(\ref{coordDSS}) defined in Sec.~\ref{subsec::SSinSpherSymm}. This gives
\begin{eqnarray}
{\beta}' & = & \frac{\eta}{2} \, z  \,  (\phi')^2 \label{DSSbetap}\\
(\frac{V}{r}){}' & = & -\frac{1}{z}\left( - {e^{2 \beta}} + 
         2 \, \eta \, {e^{2 \beta}}
\sin^2(\phi) + \frac{V}{r} \right)\label{DSSVoverrp} 
\end{eqnarray}
and
\begin{eqnarray}\label{DSSphipp}
\phi'' & = &  
 \frac{1}{z^2\,
       \left( -\frac{V}{r} + 2\,z\,\zeta - 2\,z\, {\dot \zeta} \right) }
   \left(  -{e^{2 \beta}}\,
          \left\{ \sin (2\,\phi )  
       +   z \,
             \left( -1 + 2\,\beta \,{\sin (\phi)}^2 \right) \,\phi' \right\} 
          + \right. \nonumber \\
        & + & \left. z\, \left\{ \left( \frac{V}{r} + 4
          z\, {\dot\zeta}  - 4 \,  
             \zeta \,z \right) \phi' -2 \zeta \left( {\dot \phi} + 
             z\,{\dot \phi}' \right)  \right\} \right),
\end{eqnarray}
where ${}' \equiv \partial_z$ and $\dot {} \equiv \partial_{\tau}$.
Looking for DSS solutions means that we require
$\beta = \beta(\tau,z), \Vr = \Vr(\tau,z), \phi = \phi(\tau,z), \zeta =
\zeta(\tau)$, where all the time dependencies being periodic, with some
period $\Delta$, the determination of which is part of the problem.

Furthermore we fix the coordinate freedom contained in the function $\zeta$,
by 
\be\label{zetadot}
{\dot \zeta}(\tau) - \zeta(\tau) + 
              \frac{1}{2}\frac{V}{r}(\tau,1)= 0,
\ee 
which makes the hypersurface $z=1$ null, as explained in Sec.
\ref{subsec::SSinSpherSymm}. Assuming that $\Vr(\tau,1)$ is given, Eq.
(\ref{zetadot}) is a first order ODE for $\zeta(\tau)$, with periodic
boundary conditions.

As we will have to deal with an equation similar to Eq.~(\ref{zetadot}) 
again, we have a short look at the more general equation
\be\label{fdot}
\dot f(\tau) + g(\tau) f(\tau) + h(\tau) = 0,
\ee
where $g(\tau)$ and $h(\tau)$ are given functions, periodic in $\tau$ with
period $\Delta$, and $f(\tau)$ is the unknown, which is required to be
periodic too with the same period.

The general solution to Eq.~(\ref{fdot}) is given by
\be\label{fgeneral}
f(\tau) = e^{-G(\tau)} \left(- \int\limits_{\tau_0}^{\tau} e^{G(\bar \tau)}
  h(\bar \tau) d \bar \tau  + c \right), \qquad G(\tau) =
\int\limits_{\tau_0}^{\tau} g(\bar \tau) d \bar \tau, 
\ee
where the function $G(\tau)$ as defined above was introduced for
abbreviation. The constant $c$ entering the formula now has to be determined
from the required periodicity of $f(\tau)$.
The behavior of $G(\tau)$ under a shift of $\Delta$ is
\be
G(\tau + \Delta) = G(\tau_0 + \Delta) + G(\tau),
\ee
so 
\be\label{fperiodic}
f(\tau + \Delta) = f(\tau) - c \ e^{-G(\tau)}\left(1 - e^{-G(\tau_0 +
\Delta)}\right) - e^{-G(\tau)} e^{-G(\tau_0 + \Delta)}
\int\limits_{\tau_0}^{\tau_0 + \Delta} e^{G(\bar \tau)} h(\bar \tau) d
\bar\tau.
\ee
And so, if $G(\tau_0 + \Delta) \ne 0$, the constant is determined to be
\be
c = \frac{1}{1-e^{G(\tau_0+ \Delta)}} \int\limits_{\tau_0}^{\tau_0 + \Delta}
 e^{G(\bar \tau)} h(\bar \tau) d \bar \tau,
\ee
and therefore the unique periodic solution to Eq.~(\ref{fdot}) is given by
\be\label{fsolution}
f(\tau) = e^{-G(\tau)} \left(-\int\limits_{\tau_0}^{\tau} e^{G(\bar \tau)}
h(\bar \tau) d \bar \tau + \frac{1}{1 - e^{G(\tau_0 + \Delta)}} 
\int\limits_{\tau_0}^{\tau_0 + \Delta} e^{G(\bar \tau)} h(\bar \tau) d \bar
\tau  \right).
\ee

In case $g(\tau)$ has no zero frequency, i.e. $G(\tau_0 + \Delta) = 0$,
there is no periodic solution to Eq.~(\ref{fdot}) unless the last integral
in (\ref{fperiodic}) vanishes as well, in which case there is a
one-parameter family of periodic solutions to Eq.~(\ref{fdot})
of the form (\ref{fgeneral}).

According to this the solution to Eq.~(\ref{zetadot}) can be given in closed
form:
\be\label{zetasol1}
\zeta(\tau) =  \frac{e^{\tau}}{2} 
     \left(- \int\limits_{\tau_0}^{\tau} e^{-\bar \tau} 
\frac{V}{r}(\bar \tau,1)
d \bar \tau + 
\frac{1}{1-e^{-\Delta}} \int\limits_{\tau_0}^{\tau_0 + \Delta} 
e^{-\bar \tau} \frac{V}{r}(\bar \tau,1) d \bar \tau  \right).
\ee
There is an alternative way to compute $\zeta$, namely as Eq.~(\ref{zetadot})
is linear in all periodic functions, the Fourier coefficients of $\zeta$ 
can be computed directly from the Fourier coefficients of $\Vr(\tau,1)$. We
will give details on this in Sec.~\ref{subsec::DSScode}.

Finally we rewrite 
Eqs.~(\ref{DSSbetap}) -- (\ref{DSSphipp}), using the
logarithmic coordinate $y = \ln z$ instead of $z$. This is useful for
numerical purposes, as the main variations in the solutions occur close to
the origin. We have 
\begin{eqnarray}
{\beta}' & = & \frac{\eta}{2} \,  (\phi')^2 \label{DSSbetap_log}\\
(\frac{V}{r}){}' & = & -\left( - {e^{2 \beta}} + 
         2 \, \eta \, {e^{2 \beta}}
\sin^2(\phi) + \frac{V}{r} \right)\label{DSS_Voverrp_log} 
\end{eqnarray}
and
\begin{eqnarray}\label{DSSphipp_log}
\phi'' & = &  
 \frac{1}{
       \left( -\frac{V}{r} + 2\, e^{y}\,\zeta - 2\, e^{y}\, {\dot \zeta}
\right)
 }
   \left(  -{e^{2 \beta}}\,
          \left\{ \sin (2\,\phi ) +    
             \left( -1 + 2\,\eta \,{\sin (\phi)}^2 \right) \,\phi' \right\} 
          + \right. \nonumber \\
        & + & \left.  \left\{ \left(2
          e^y \, {\dot\zeta}  - 2 \,  
             \zeta \,e^y \right) \phi' -2 \zeta e^y \left( {\dot \phi} + 
             \,{\dot \phi}' \right)  \right\} \right),
\end{eqnarray}
where ${}'$ now denotes $\partial_y$ and $y = \ln(z)$.

Note that $\tau$ does not enter Eqs.~(\ref{DSSbetap}) -- (\ref{DSSphipp}), 
((\ref{DSSbetap_log})--(\ref{DSSphipp_log}) respectively) explicitly. 
Therefore, given a solution,
abbreviated by $Z(\tau,z)$, all expressions resulting from this  
by a constant shift in $\tau$, $Z(\tau + const,z)$ are solutions to the
system as well.

\subsection{Regularity at Origin and Past SSH}\label{subsec::DSSregularity}
%%%%%%%%%%%%%%%%%%%%%%%%%%%%%%%%%%%%%%%%%%%%%%%%%%%%%%%%%%%%%%%%%%%%%%%%%%%

Regularity at the origin $z=0$, $y = -\infty$ is as usual given by
\be
{\beta}(\tau,z=0) = 0, \qquad \frac{V}{r}(\tau,z=0) = 1, 
\qquad \phi(\tau,z=0) = 0,
\ee
and $\phi'(\tau, z=0)$ is a free (periodic) function of $\tau$.
Using the logarithmic coordinate $y = ln(z)$, we start our numerical
integration at some finite $y_0$. Near the origin $z=0$ 
(i.e. $y \to - \infty$)
the functions 
behave as
\bea
\beta(\tau,y) & = & O(e^{2 y}) \nonumber\\
\Vr(\tau,y) &   = & O(e^{2 y}) \nonumber\\
\phi(\tau,y) & = & \phi_1(\tau) e^{y} + O(e^{2 y}) \nonumber\\
\phi'(\tau, y) & = & \phi_1(\tau) e^{y} + O(e^{2 y}), 
\eea
where $\phi_1(\tau) = \partial_z \phi(\tau, z=0)$.

At the past SSH $z=1$ (i.e.~$y = 0$) the denominator in
(\ref{DSSphipp_log}) vanishes. Regularity therefore enforces the nominator to
vanish as well, that is
\bea\label{DSSboundaryODE}
  0& = & -{e^{2 \beta_H}}\,
          \left\{ \sin (2\,\phi_H ) 
       +   \left( -1 + 2\,\eta \,{\sin (\phi_H)}^2 \right) \,
          \phi'_H \right\} 
          + \nonumber \\
        & + &  \left\{ \left( (\frac{V}{r})_H + 4
          {\dot\zeta}  - 4 \,  
             \zeta \right) \phi'_H -2 \zeta \left( {\dot \phi}_H + 
             {\dot \phi}'_H \right)  \right\} ,
\eea
where the expressions with subscript $H$ are periodic functions of $\tau$.
Suppose now, we are given the functions $\beta_H, (\Vr)_H$ and $\phi_H$,
then Eq.~(\ref{DSSboundaryODE}) is an ODE in time $\tau$ for $\phi'_H$.
In fact, if we abbreviate
\begin{eqnarray}
g & = & \frac{1}{2\zeta} \left( (\frac{V}{r})_H + e^{2 \beta_H} (-1 + 2 \eta
          \sin^2 \phi_H)\right) \\
h & = & \frac{1}{2 \zeta} e^{2 \beta_H} \sin(2 \phi_H) + {\dot \phi}_H,
\end{eqnarray}
and use Eq.~(\ref{zetadot}) to replace $\dot \zeta$,
then Eq.~(\ref{DSSboundaryODE}) is of the same form as Eq.~(\ref{fdot}),
described in the last section, and has the unique periodic solution
(\ref{fsolution}).

\subsection{Additional Symmetry}\label{subsec::Delta_over_2_symmetry}
%%%%%%%%%%%%%%%%%%%%%%%%%%%%%%%%%%%%%%%%%%%%%%%%%%%%%%%%%%%%%%%%%%%%

From critical searches (see Chap.~\ref{chap::criticalCollapse} for details), 
we know, that the critical solution for large couplings, 
which is DSS, not only is
periodic with period $\Delta$, but has an additional
symmetry, namely
\bea\label{Delta_2}
{\beta}(\tau + \Delta/2, z) & = &   {\beta}(\tau,z), \nonumber \\
\Vr (\tau + \Delta/2, z) & = &   \Vr (\tau,z), \nonumber \\
\zeta   (\tau + \Delta/2, z) & = &   \zeta   (\tau,z), \nonumber \\
\phi    (\tau + \Delta/2, z) & = & - \phi    (\tau,z),
\eea
which means that the metric functions as well as $\zeta$ consist only of
even frequencies\footnote{even multiples of $1/\Delta$.}, whereas the 
field and it's derivatives contain only odd
frequencies.
As we are interested in the direct construction of the DSS solution, 
which is the critical solution in a certain range of coupling constants, 
we impose this additional symmetry on the solution by the means of 
its Fourier coefficients. 

\subsection{Numerical Construction of DSS solutions via an ODE boundary
            value problem}\label{subsec::DSScode}
%%%%%%%%%%%%%%%%%%%%%%%%%%%%%%%%%%%%%%%%%%%%%%%%%%%%%%%%%%%%%%%%%%%%%%%%

According to the required periodicity in $\tau$ of the solution we 
expand the metric functions, $\zeta$, and the field into Fourier
series. The truncation of theses series is performed according to 
the discrete Fourier transform described in detail in Appendix
\ref{app::DFT}. We denote the number of ``collocation points'' 
in ``$\tau$ space'' by $N$, assuming, that it is an integer multiple of
4. So the equally spaced points in $\tau$ space are given by  
$\tau_k = k \Delta /N$. 

In principle these $N$ points give rise to $N$ Fourier coefficients, but
the additional symmetry introduced in the last section, 
Sec.~\ref{subsec::Delta_over_2_symmetry} reduces this number 
by a factor of 2.
So each function consists of $M = N/2$ Fourier coefficients, where we
assume, that $M$ is even.
The expansions then are given by
\bea\label{expansions_even}
\beta_N(\tau_k,z) & = &\beta_0(z) + \sum\limits_{l=1}^{M/2 -1} (\beta
cos)_{2l}(z)
\cos(\frac{4\pi l k}{N}) + \sum\limits_{l=1}^{M/2 -1} (\beta sin)_{2l}(z)
\sin(\frac{4\pi l k}{N}) + \nonumber\\
      & & + (\beta cos)_{M}(z) \cos(\pi k)
\nonumber\\
(\Vr)_N(\tau_k,z) & = &V_0(z) + \sum\limits_{l=1}^{M/2 -1} (Vcos)_{2l}(z)
\cos(\frac{4\pi l k}{N}) + \sum\limits_{l=1}^{M/2 -1} (Vsin)_{2l}(z)
\sin(\frac{4\pi l k}{N}) + \nonumber\\ 
      & & + (Vcos)_{M}(z) \cos(\pi k)
\nonumber\\
\zeta_N(\tau_k,z) & = &\zeta_0(z) + \sum\limits_{l=1}^{M/2 -1} (\zeta
cos)_{2l}(z)
\cos(\frac{4\pi l k}{N}) + \sum\limits_{l=1}^{M/2 -1} (\zeta sin)_{2l}(z)
\sin(\frac{4\pi l k}{N}) + \nonumber\\
      & & + (\zeta cos)_{M}(z) \cos(\pi k) \nonumber\\
& & {}
\eea
and
\be
\phi_N(\tau_k,z) = \sum\limits_{l=1}^{M/2} (\phi cos)_{2l-1}(z)
\cos(\frac{2\pi (2 l -1) k}{N}) + \sum\limits_{l=1}^{M/2} 
(\phi sin)_{2l-1}(z)
\sin(\frac{2\pi (2 l -1)k}{N}) ,
\ee
and the expansion for $\phi'$ follows directly from the one of $\phi$.
The coefficients for a function with only even frequencies are given by
e.g.
\begin{eqnarray}
\beta_0 &=& \frac{1}{N} \sum\limits_{k=0}^{N-1} \beta_N(\tau_k) \\
(\beta cos)_{2l} &=& 
       \frac{2}{N} \sum\limits_{k=0}^{N-1} \beta_N(\tau_k) 
       \cos(\frac{4 \pi l k}{N})
       \quad l= 1,\dots M/2 -1 \\
(\beta sin)_{2l} &=& \frac{2}{N} \sum\limits_{k=0}^{N-1} \beta_N(\tau_k) 
      \sin(\frac{4 \pi l k}{N})
      \quad l= 1,\dots M/2 -1 \\
(\beta cos)_{M} &=& \frac{1}{N} \sum\limits_{k=0}^{N-1} \beta_N(\tau_k) 
      \cos(\pi k),
\end{eqnarray}
and for a function with odd frequencies by
\begin{eqnarray}
(\phi cos)_{2l-1} &=& 
       \frac{2}{N} \sum\limits_{k=0}^{N-1} \phi_N(\tau_k) 
        \cos(\frac{2 \pi (2 l-1) k}{N})
         \quad l= 1,\dots M/2  \\
(\phi sin)_{2l-1} &=& \frac{2}{N} \sum\limits_{k=0}^{N-1} \phi_N(\tau_k)
\sin(\frac{2 \pi (2l-1) k}{N}) \quad l= 1,\dots M/2. 
\end{eqnarray}

Our variables that have to solve the coupled system of ODEs
now are the $4 M$ Fourier coefficients of $\beta, \Vr, \phi$ and $\phi'$. 
Note that $\zeta$ is not part of the system, as it can be computed
whenever $\Vr$ is given at the horizon.
The boundary conditions consist of boundary conditions for the coefficients
of $\beta, \Vr$ and $\phi$ at the origin, and on the other hand
of the conditions on the coefficients of $\phi'$ at the horizon, so in total
we have $4 M$ boundary conditions imposed on our system of $4 M$ 
first order ODEs.

The free (shooting) parameters consist of the coefficients of
$\beta, \Vr$ and $\phi$ at the horizon, which make a total of $3 M$.
Furthermore the $M$ coefficients of 
$\phi_1(\tau) = \partial_z \phi(\tau,z=0)$ are free parameters.
Nevertheless, as noted in Sec.~\ref{subsec::DSSequations} the equations
(\ref{DSSbetap})--(\ref{DSSphipp}) are translation invariant in $\tau$.
A constant shift in $\tau$ therefore transforms a given solution 
again to a solution. 
On the other hand, a constant shift in $\tau$ just changes the Fourier
coefficients in a well defined way. We can therefore chose
one coefficient of $\phi_1(\tau)$ arbitrarily, 
so we are left with only $M-1$ shooting parameters at the
origin. 
Finally there is the period $\Delta$, which is the last free parameter.
So in total we have again $4 M$ shooting parameters.  

In order to solve the boundary value problem for the ODEs again a shooting
and matching method is used (see Appendix \ref{app::SM}). We describe now in
detail, how the system of ODEs is ``set up'' numerically.
The first step consists of providing good initial guesses for the shooting
parameters. Given these guesses, Eq.~(\ref{zetadot}) is solved for $\zeta$.
As this equation is linear in the periodic functions, it can be solved
directly in Fourier space. Applying the rules for differentiation
in Fourier space, as given in Appendix \ref{app::DFT} to $\zeta$ and comparing
coefficients, Eq.~(\ref{zetadot}) has the following solution in Fourier
space
\bea
\zeta_0 & = & \frac{1}{2} \left. V_0 \right|_{y=0}\\
(\zeta cos)_{2 l} & = & \frac{1}{2 \left(1 + (4 \pi l/\Delta)^2 \right)}
                        \left( \left. (Vcos)_{2l}\right|_{y=0} + 
                        \frac{4 \pi l}{\Delta} 
                        \left. (Vsin)_{2l}\right|_{y=0} \right)  \\
(\zeta sin)_{2 l} & = & \frac{1}{2 \left(1 + (4 \pi l/\Delta)^2 \right)}
                         \left( - \frac{4 \pi l}{\Delta} 
                          \left. (Vcos)_{2l}\right|_{y=0} +  
                          \left. (Vsin)_{2l}\right|_{y=0} \right) \\
(\zeta cos)_{N/2} & = & \frac{1}{2} \left. (Vcos)_{N/2}\right|_{y=0}.
\eea
In the next step, the coefficients of $\phi'_H$ have to be computed from 
the coefficients of $\zeta$ and the other variables at the horizon.
To do this, the variables are transformed back to real space. There formula
(\ref{fsolution}) is used to obtain $\phi'_H$ as a function of $\tau$.
Transformation to Fourier space then yields the desired coefficients.
Note, that each time such a ``backward-forward transformation pair'' is used,
the number of coefficients is first increased by some factor and after
the operations in real space have been performed and 
the forward transformation has been applied, the number of coefficients is
reduced to the original size again. This is a way to reduce aliasing
errors, as explained in more detail in App.~\ref{app::DFT}.

After these first steps are completed, all the variables, i.e. the Fourier
coefficients, are known at the boundaries, the coefficients of $\zeta$ are
determined and the period $\Delta$ has some definite value.
In order to integrate the ODEs, the spatial derivatives of the Fourier
coefficients have to be determined. This is achieved again by a
transformation to real space. There the right hand sides of Eqs.
(\ref{DSSbetap_log}) -- (\ref{DSSphipp_log}) are evaluated, and a
transformation back to Fourier space yields the desired derivatives of the
coefficients. Again for de-aliasing the number of coefficients is first
increased and after all the operations reduced again.
These operations have to be carried out at each integration step,
which is determined internally by the NAG routine's integrator.
This huge amount of Fourier transformations for a single integration to the
matching point, necessitates the use of the fast Fourier transform.

The shooting and matching routine d02agf again was taken from the NAG library
(\cite{NAG}). We mention one technical detail, concerning the magnitude
of the variables. As the solution is expected to be smooth, the ``higher''
Fourier coefficients should decrease exponentially in magnitude.
As the NAG routine uses some mixture of absolute and relative error
to determine the local error of the solution, it is necessary to rescale the
variables to approximate unity.

As the shooting and matching method needs a good initial guess, we use the
results of a ``dice-critical-search''. (For details on the setup of a
critical search and on the following see Chap.~\ref{chap::criticalCollapse}). 
As our initial guess we take the critical solution at $\eta = 100$,
which is clearly a DSS solution. Since these data are given in 
terms of Bondi coordinates we have to switch to adapted coordinates.
We compute $u^*$ and the initial guess for the echoing period $\Delta$
from $\max (2m/r)$. The past self-similarity horizon theoretically is
determined as the backwards light cone of the culmination point.
Another way to determine the SSH is to look for the ingoing null geodesic,
along which the metric functions $\beta$ and $\Vr$ as well as the field
$\phi$ are periodic functions of $\tau$. This gives us the initial guesses
for the shooting parameters at the horizon.
From $\Vr$ at the horizon, we compute $\zeta$ and finally 
$\partial_r \phi(\tau, r=0)$ can be converted into $\phi_1(\tau)$, as the
relation between Bondi coordinates and adapted coordinates is already fixed.

We start at $\eta = 100$ with $N= 16$, and follow the solution
down to smaller values of $\eta$.
As will be reported in the next section, the echoing period $\Delta$
increases ``sharply'' for $\eta$ below $0.5$. This increase of the period
will necessitate a larger and larger number of coefficients.

Given a solution obtained with a certain number of Fourier coefficients
$M$,
we can double the number of coefficients with the second half set to zero.
This way we obtain a reasonable initial guess for the problem with $2M$
coefficients.

\subsection{Numerical Results}\label{subsec::DSScode_results}
%%%%%%%%%%%%%%%%%%%%%%%%%%%%%%%%%%%%%%%%%%%%%%%%%%%%%%%%%%%%%

Applying the method described in the last section, Sec.~\ref{subsec::DSScode},
we find good numerical evidence, that the system 
Eqs.~(\ref{DSSbetap_log}) -- (\ref{DSSphipp_log})
admits a discretely self similar solution for 
$100 > \eta \gtrsim 0.17$. The smallest value of $\eta$, where we
constructed a DSS solution, was $\eta = 0.17262$. At this coupling constant
already a large number of coefficients is necessary. 
Lowering $\eta$ further would require at least
$N = 256$, i.e. $128$ Fourier
coefficients per dependent variable. 
With this number of coefficients -- in addition for de-aliasing the
number of coefficients was increased and decreased by a factor of 8 -- 
a single Newton iteration on an Alpha (ev6 processor) already
takes several days.

%Directory: /DSS/src/New_Period
\begin{figure}[p]
\begin{center}
\begin{psfrags}
 \psfrag{eta}[]{$\eta$}
 \psfrag{Delta}[r][l][1][-90]{$\Delta$}
\includegraphics[width=3.5in]{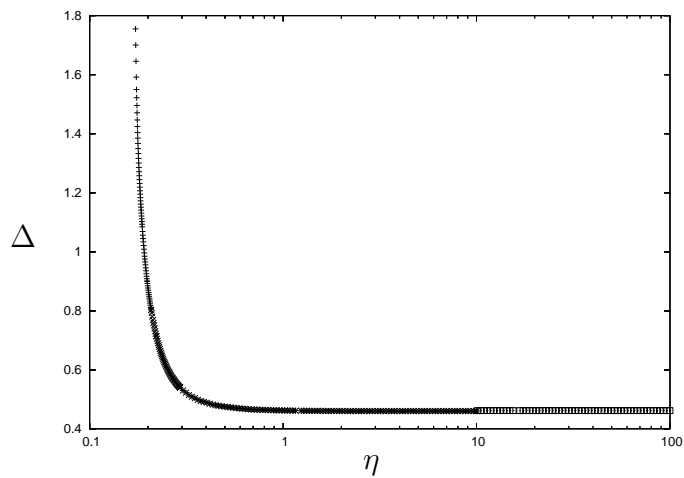}
\end{psfrags}
\end{center}
\caption{The period $\Delta$ of the DSS solution as a function of the
coupling constant $\eta$. The number of coefficients used to produce the
results of this figure were: $M = 8 \ (N = 16)$ for $10 < \eta < 100$,
$M = 16 \ (N = 64)$ for $0.2933 < \eta < 10$, $M = 32 \ (N = 64)$ for
$0.2079 < \eta < 0.2933$ and $M = 64 \ (N = 128)$ for $0.1726 < \eta < 0.2079$.
}\label{fig::DSS_Period}
\end{figure}

%Directory: /DSS/src/New_Period
\begin{figure}[p]
\begin{center}
\begin{psfrags}
 \psfrag{eta}[]{$\eta$}
 \psfrag{Delta}[r][l][1][-90]{$\Delta$}
 \psfrag{error}[l][c][1][-90]{$|\Delta - f(\eta)|$}
\includegraphics[width=3.5in]{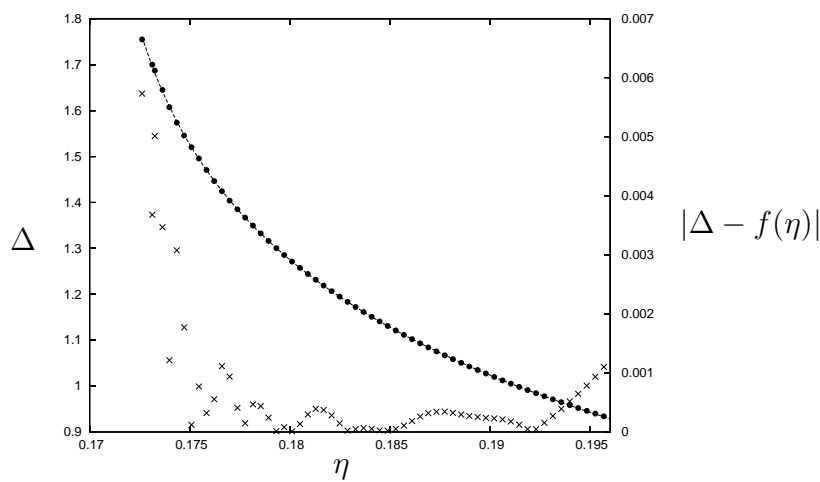}
\end{psfrags}
\end{center}
\caption{``+'' represent the period $\Delta$ of the DSS solution 
in the region of smallest $\eta$, where we constructed the solutions.
All the solutions are obtained with 64 coefficients ($N = 128$).
In the region $0.1726 \le \eta \le 0.195$ these data were fitted against the 
function $f(\eta) = - a \ln(\eta - \eta_C) + b$. 
The fit determines the critical coupling to be $\eta_C \simeq 0.17$, and the
constant 
$a=0.36278$. 
The fitted function $f(\eta)$ is plotted as dashed line. On the right axis
the error is plotted. Presumably due to an insufficient number of coefficients
the error increases towards the lower end of the interval. 
}\label{fig::DSS_Period_fit}
\end{figure}

We find the following behavior: Fig.~\ref{fig::DSS_Period} shows that below
$\eta \sim 0.3$ the period $\Delta(\eta)$ rises sharply with decreasing
$\eta$. In Sec.~\ref{subsec::CSS-DSS} we will give an argument that we
expect $\Delta(\eta)$ to behave like $-a \ln(\eta-\eta_C) + b$ for $\eta$
close to a critical coupling constant $\eta_C$. 
Fig.~\ref{fig::DSS_Period_fit} shows a fit of
$\Delta(\eta)$ against this function for $0.1726 \le \eta \le 0.195$.
According to this fit the period $\Delta(\eta)$ would blow up at $\eta
\simeq 0.17$.

\begin{figure}[p]
\begin{center}
\begin{psfrags}
 \psfrag{k}[]{$k$}
 \psfrag{eta.eq.0.1854}[]{$\eta = 0.1854$}
 \psfrag{eta.eq.0.2933}[]{$\eta = 0.2933$}
 \psfrag{maxphi}[r][l][1][-90]{$\max\limits_{y} |\phi_{coeff}|$}
\includegraphics[width=3.5in]{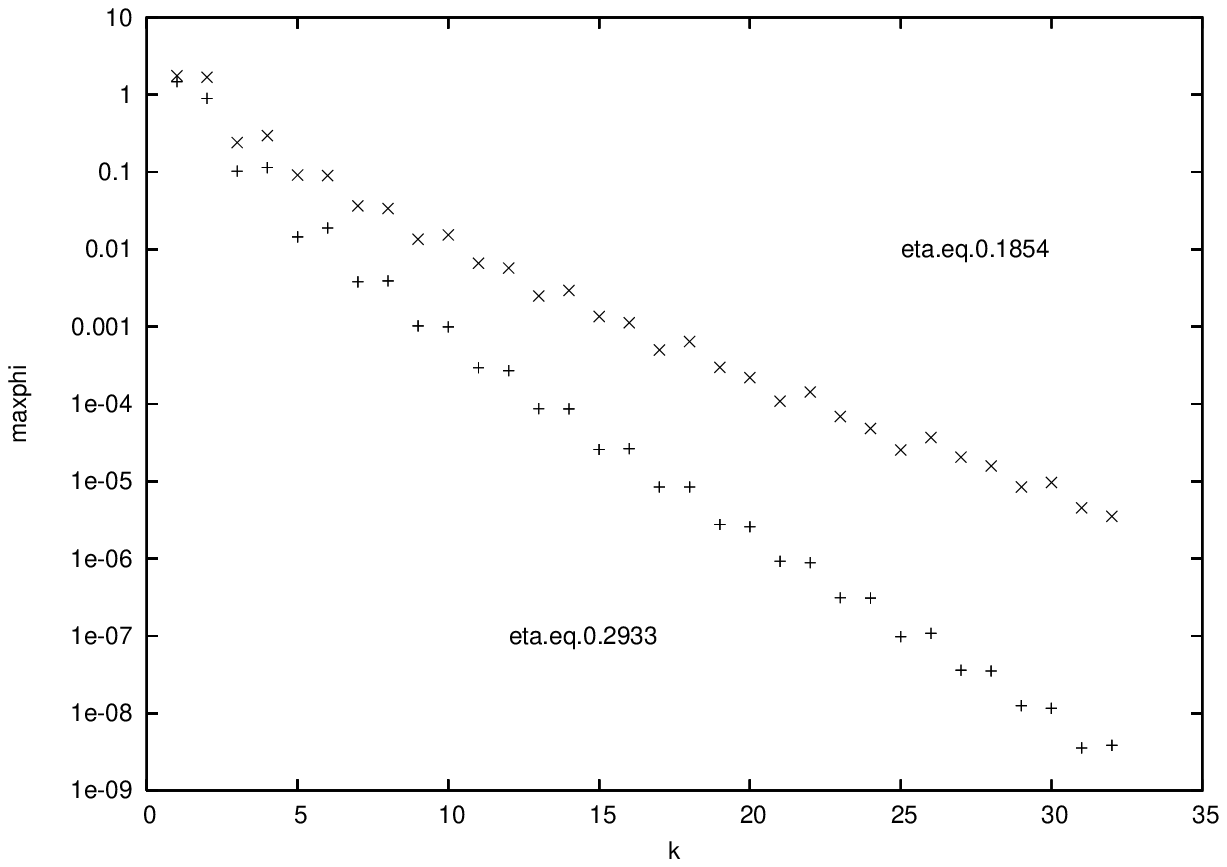}
\end{psfrags}
\end{center}
\caption{This figure illustrates the exponential fall-off of the Fourier
coefficients. Plotted is the maximum over $y=\ln(z)$ of the Fourier
coefficients of $\phi$ versus the coefficient number $k$ 
for the two coupling constants $\eta = 0.2933$
and $\eta = 0.1854$. The solutions were computed using $M = 32$ Fourier
coefficients. For both couplings the magnitude of the coefficients
decreases exponentially. However, the slope of this decrease
is steeper for larger couplings.
}\label{fig::DSS_max_phi_falloff}
\end{figure}

\begin{figure}[p]
\begin{center}
\begin{psfrags}
 \psfrag{eta}[]{$\eta$}
 \psfrag{maxphi}[r][l][1][-90]{$\max\limits_{y} |\phi_{coeff}|$}
\includegraphics[width=3.5in]{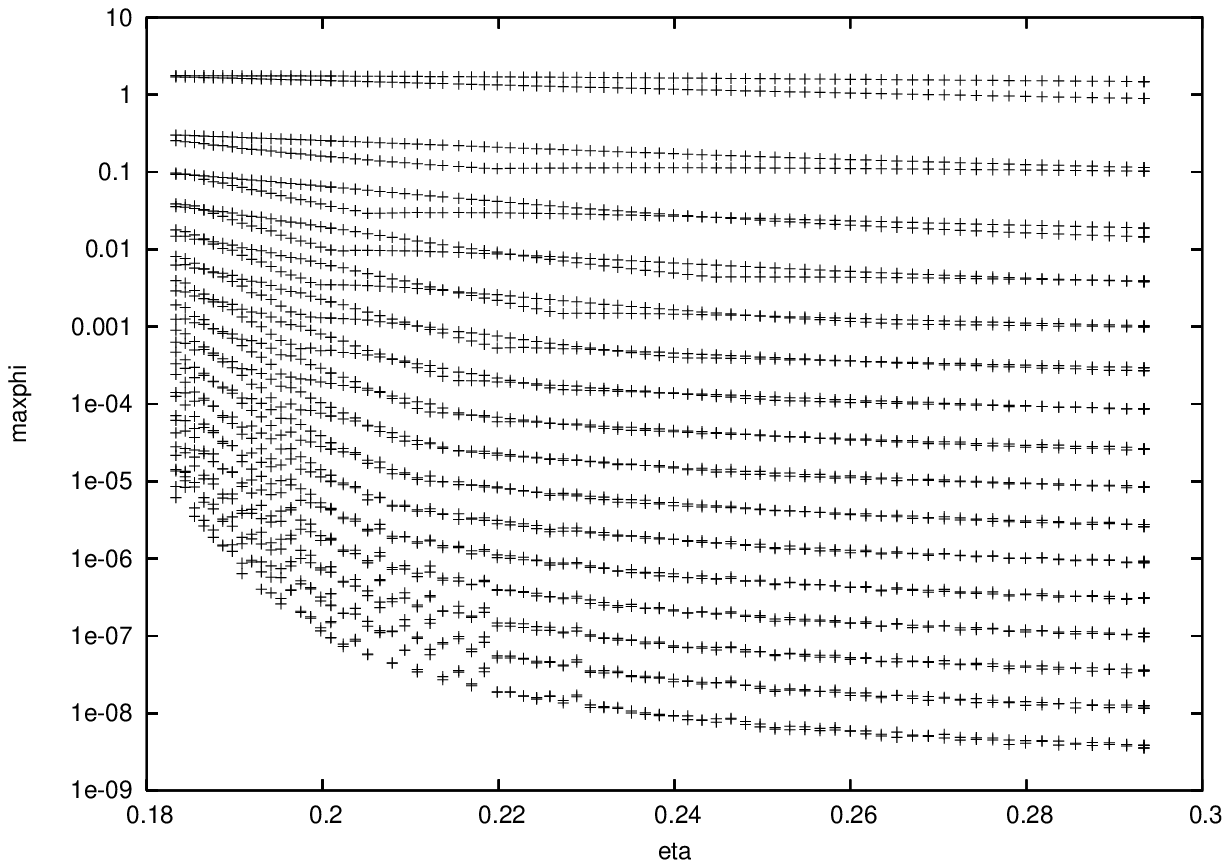}
\end{psfrags}
\end{center}
\caption{The magnitude of the Fourier coefficients as a function of the
coupling constant $\eta$. All solutions were computed using $M=32$ Fourier
coefficients. Plotted is again the maximum over the spatial variable
$y=\ln(z)$ of the coefficients of $\phi$.
See also Fig. \ref{fig::DSS_Euur_coeffs_eta.eq.0.2933_0.1833} for a
comparison of errors at the lower and upper end of the $\eta$ interval
shown in this figure.
}\label{fig::DSS_eta_max_phi}
\end{figure}

The rise in $\Delta(\eta)$ also means, that more and more coefficients
are needed in order to represent the solution to a given accuracy (see
Sec.~\ref{subsec::DSSconvergence}). 
Figs.~\ref{fig::DSS_max_phi_falloff} and 
\ref{fig::DSS_eta_max_phi} illustrate this fact. 
Fig.~\ref{fig::DSS_max_phi_falloff} shows that at a fixed coupling the
coefficients decay exponentially with the coefficient number.
Nevertheless the slope of the decay decreases with decreasing coupling.
Fig.~\ref{fig::DSS_eta_max_phi} shows the coefficients of the field $\phi$
as functions of the coupling $\eta$. With decreasing $\eta$ the coefficients
grow.

%Directory: /DSS/src/Results_64/Phi_max_coeffs

%Directory: /DSS/src/Results_64/Phi_max_coeffs

\subsection{Convergence with the Number of Fourier
Coefficients}\label{subsec::DSSconvergence}
%%%%%%%%%%%%%%%%%%%%%%%%%%%%%%%%%%%%%%%%%%%%%%%%%%%

As a test for accuracy, we can use the ``supplementary'' combination of
the Einstein equations,  
Eq.~(\ref{eq::Euur}), which for an exact solution should evaluate to zero.
Transformation to self-similar coordinates and multiplication by
$e^{-\tau} \zeta(\tau)$ yields   
\bea\label{DSSEuur}
0 & = &
  2\,\dot \beta\,\Vr\,\zeta - 
  \dot \Vr \,\zeta
  - ( \zeta - \dot \zeta) \biggl[ e^{2\,\beta}\, - \Vr - 2 \eta e^{2 \beta}
      \sin^2 (\phi) + \eta \Vr (\phi')^2 - \nonumber\\
  & - &    4 \eta e^y \zeta \dot \phi \phi' -
 2 \eta e^y (\zeta - \dot \zeta) (\phi')^2 \biggr] + 
  2\,e^y\,\eta\,{\dot \phi}^2\,{\zeta}^2    
 - 2\eta  \Vr \,\zeta  \dot \phi\,\phi', \nonumber\\
{}
\eea
where ${}' = \partial_y$.

To compute the right hand side of (\ref{DSSEuur}), the Fourier coefficients
of the numerically computed solution and their derivatives with respect to
$\tau$ are taken, and (\ref{DSSEuur}) is evaluated in $\tau$ space.
A Fourier transformation yields the coefficients of the expression.  
Again, in order to diminish aliasing errors, the coefficients are increased
before the first transformation and
decreased again after the back transformation.

\begin{figure}[p]
\begin{center}
\begin{psfrags}
 \psfrag{yeqlnz}[]{$y = \ln(z)$}
 \psfrag{Euurzi}[r][l][1][-90]{$| Euur |$}
\includegraphics[width=3.5in]{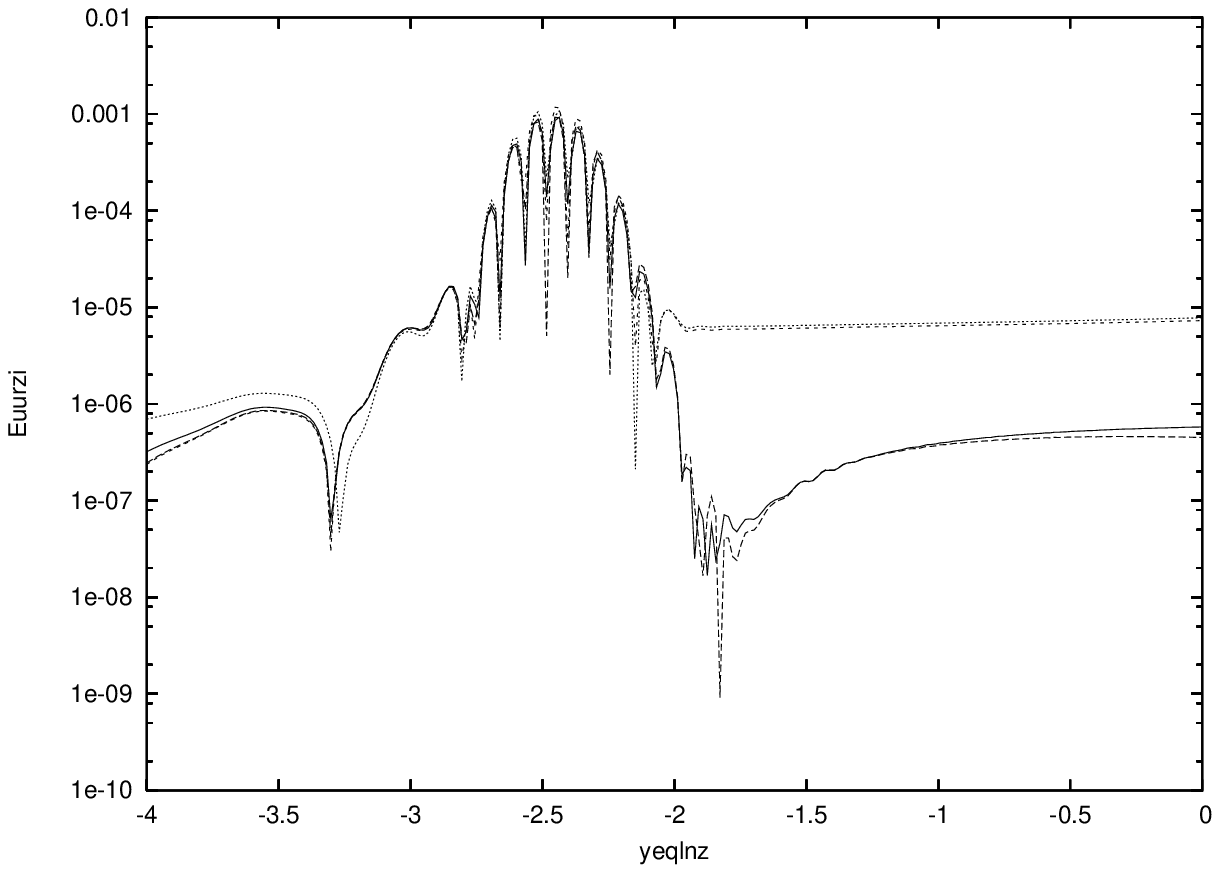}
\end{psfrags}
\end{center}
\caption{The error of the subsidiary Einstein equation 
(\ref{DSSEuur}) for $\eta =
0.29336$. The solution was obtained with $N=32$. Plotted is the absolute
value of the expression 
(\ref{DSSEuur}) as a function of the spatial variable $y$
at the time steps $\tau_i = i \Delta/N$ for $i=0,4,8,12$, thereby
spanning almost half the period. The echoing period was computed to be
$\Delta = 0.5403$. 
}\label{fig::DSS_eta.eq.0.2933_Euur_tau_32}
\end{figure}

\begin{figure}[p]
\begin{center}
\begin{psfrags}
 \psfrag{yeqlnz}[]{$y = \ln(z)$}
 \psfrag{Euurzi}[r][l][1][-90]{$| Euur |$}
\includegraphics[width=3.5in]{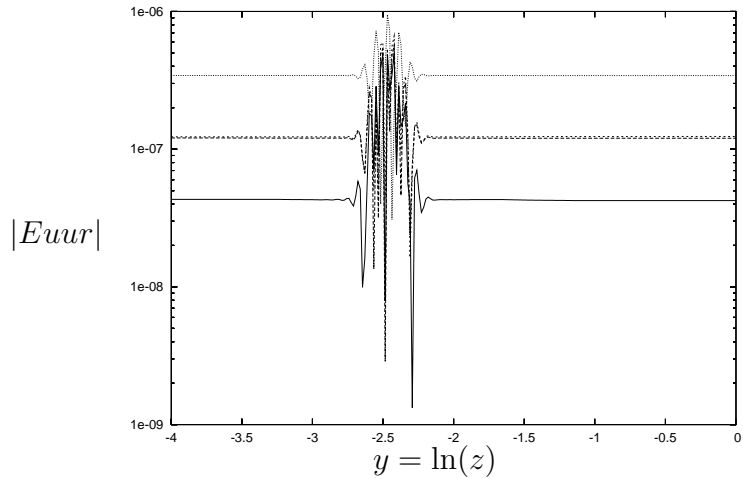}
\end{psfrags}
\end{center}
\caption{
The same situation as in Fig. \ref{fig::DSS_eta.eq.0.2933_Euur_tau_32}, where
this time the solution was computed using $N=64$.
Plotted is the absolute value of the expression 
(\ref{DSSEuur}) as a function of the spatial variable $y$
at the time steps $\tau_i = i \Delta/N$ for $i=0,8,16,24$, i.e. at 
(approximately) the same
time steps as in Fig. \ref{fig::DSS_eta.eq.0.2933_Euur_tau_32}.
The period computed with $N = 64$ is $\Delta = 0.5399$.
From these figures it is clear that the error is reduced by several orders of
magnitude, when increasing the number of Fourier coefficients.
}\label{fig::DSS_eta.eq.0.2933_Euur_tau_64}
\end{figure}

%Directory:on merlin: DSS/src/21_128_exe/
\begin{figure}[p]
\begin{center}
\begin{psfrags}
 \psfrag{y}[]{$y = \ln(z)$}
 \psfrag{Euur}[r][l][1][-90]{$| Euur |$}
\includegraphics[width=3.5in]{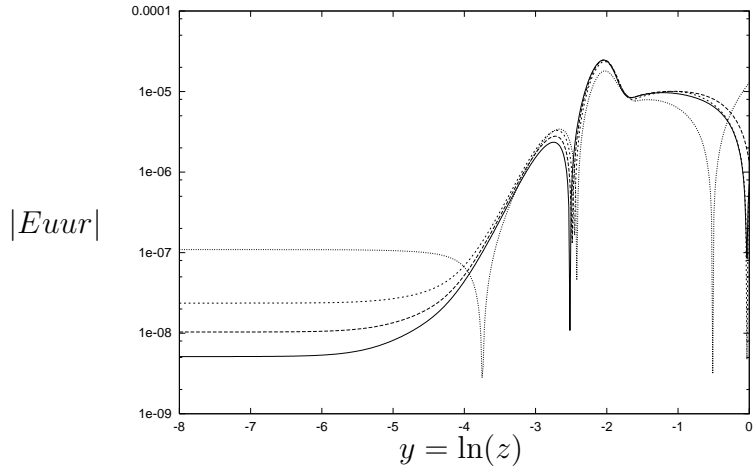}
\end{psfrags}
\end{center}
\caption{The error of the subsidiary Einstein equation (\ref{DSSEuur}) 
for $\eta = 0.1726$. The solution was obtained with $N=128$. 
Plotted is the absolute
value of expression 
(\ref{DSSEuur}) as a function of the spatial variable $y$
at the time steps $\tau_i = i \Delta/N$ for $i=0,16,32,48$.
The computed value of the period is $\Delta = 1.7551$. 
}\label{fig::DSS_eta.eq.0.1726_Euur_tau_128}
\end{figure}

\begin{figure}[p]
\begin{center}
\begin{psfrags}
 \psfrag{yeqlnz}[]{$y = \ln(z)$}
 \psfrag{Euuryi}[r][l][1][-90]{$|Euur|$}
\includegraphics[width=3.5in]{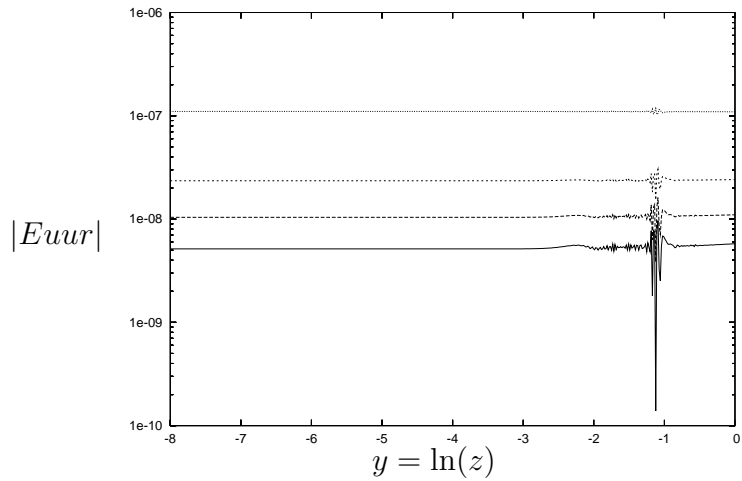}
\end{psfrags}
\end{center}
\caption{
The same situation as in Fig. \ref{fig::DSS_eta.eq.0.1726_Euur_tau_128}, 
where this time the solution was computed using $N=256$.
Plotted is the absolute value of expression 
(\ref{DSSEuur}) as a function of the spatial variable $y$
at the time steps $\tau_i = i \Delta/N$ for $i=0,32,64,96$.
With this number of coefficients the period was computed to be
$\Delta = 1.7521$
Again the error is reduced by several magnitudes, by doubling the number of
Fourier coefficients.
Note that at this small coupling constant ($\eta = 0.1726$) more Fourier
coefficients are needed to obtain a similar accuracy than in Figs. 
\ref{fig::DSS_eta.eq.0.2933_Euur_tau_32} and 
\ref{fig::DSS_eta.eq.0.2933_Euur_tau_64}
for $\eta = 0.2933$.}\label{fig::DSS_eta.eq.0.1726_Euur_tau_256}
\end{figure}

%Directory:DSS/src/Convergence_tests/Figures
\begin{figure}[p]
\begin{center}
\begin{psfrags}
 \psfrag{k}[]{$k$}
 \psfrag{maxEuurk}[r][l][1][-90]{$\max\limits_{y} | Euur_{coeff} |$ }
 \psfrag{N32}[]{$N = 32$}
 \psfrag{N64}[]{$N=64$}
\includegraphics[width=3.5in]{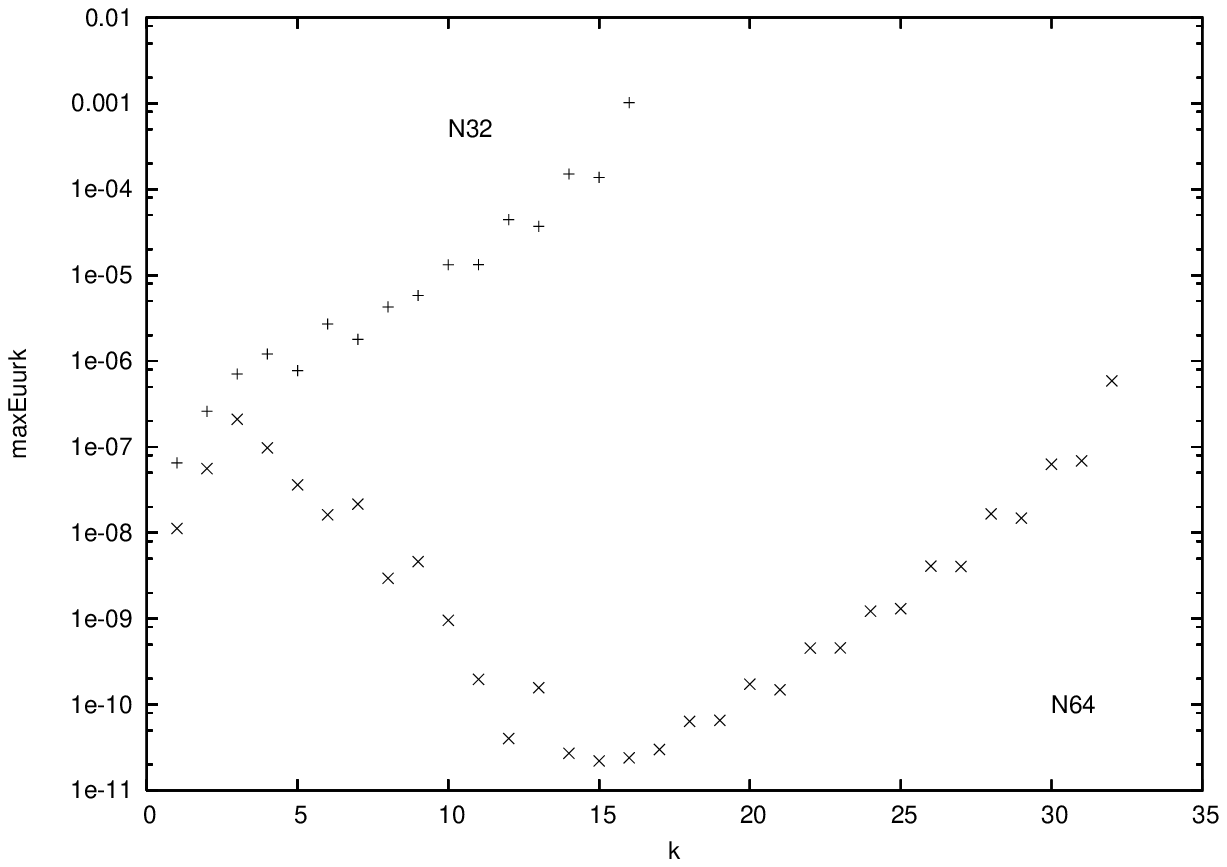}
\end{psfrags}
\end{center}
\caption{For $\eta = 0.29336$ the Fourier coefficients of the expression
(\ref{DSSEuur}) are compared for solutions obtained with $N=32$ (denoted by
``+'')and $N=64$ (denoted by ``x'').
Shown is the maximum over the spatial coordinate $y = \ln(z)$ of the 
absolute value of the Fourier
coefficients plotted against the coefficient number.
}\label{fig::DSS_eta.eq.0.2933_Euur_coeffs}
\end{figure}

%Directory: DSS/src/Convergence_tests/eta.eq.0.1726_128/
\begin{figure}[p]
\begin{center}
\begin{psfrags}
 \psfrag{k}[]{$k$}
 \psfrag{maxEuurcoeffs}[r][l][1][-90]{$\max\limits_{y} | Euur_{coeff} |$ }
 \psfrag{N128}[]{$N = 128$}
 \psfrag{N256}[]{$N = 256$}
\includegraphics[width=4in]{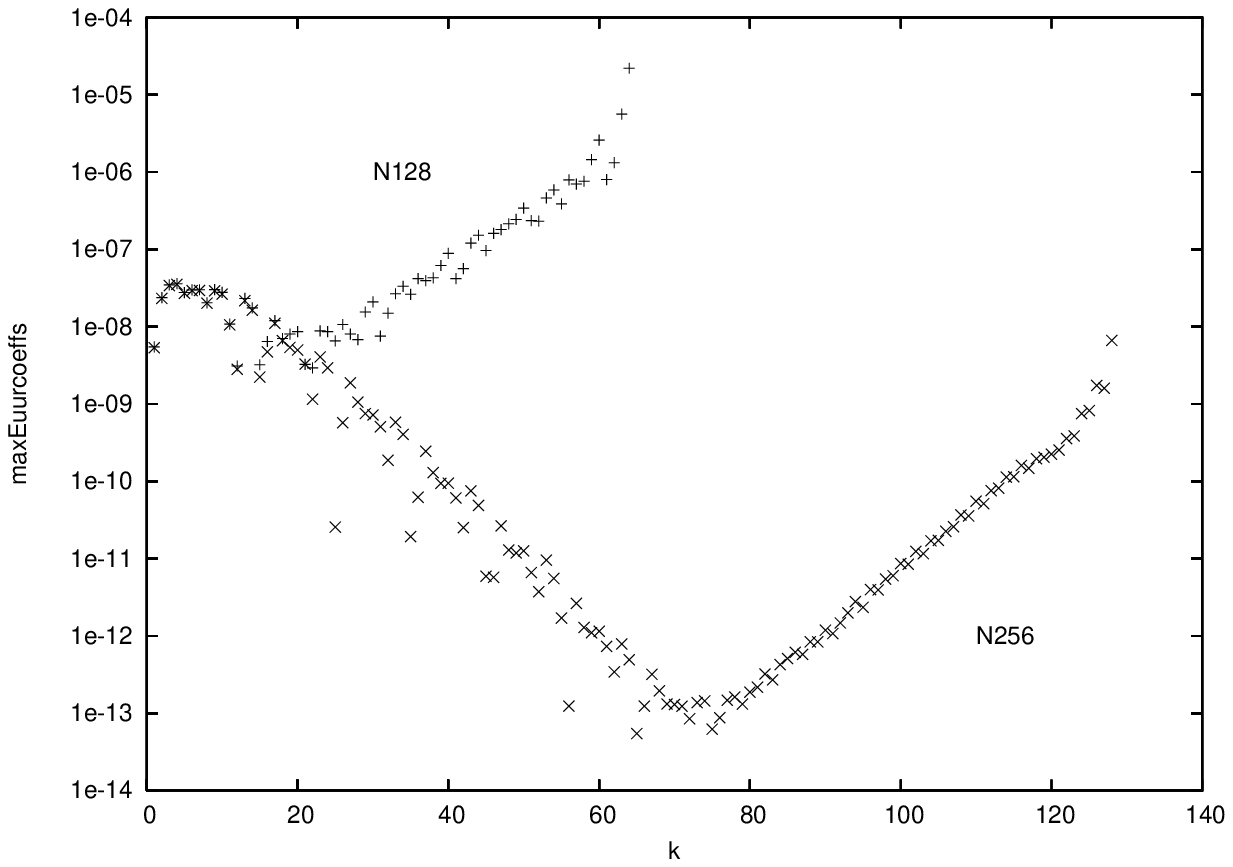}
\end{psfrags}
\end{center}
\caption{For $\eta = 0.17262$ the Fourier coefficients of the expression
(\ref{DSSEuur}) are compared for solutions obtained with $N=128$ 
(denoted by ``+'') and $N=256$ (denoted by ``x'').
Shown is the maximum over the spatial coordinate $y = \ln(z)$ of the 
absolute value of the Fourier
coefficients plotted against the coefficient number.
}\label{fig::DSS_eta.eq.0.1726_Euur_coeffs}
\end{figure}

%Directory:src/Results_64/eta.eq.0.1833
\begin{figure}[h]
\begin{center}
\begin{psfrags}
 \psfrag{k}[]{$k$}
 \psfrag{N.eq.64}[]{$N = 64$}
 \psfrag{eta.eq.0.2933}[]{$\eta = 0.2933$}
 \psfrag{eta.eq.0.1833}[]{$\eta = 0.1833$}
 \psfrag{Euurcoeff}[r][l][1][-90]{$\max\limits_{y} | Euur_{coeff} |$ }
\includegraphics[width=3.5in]{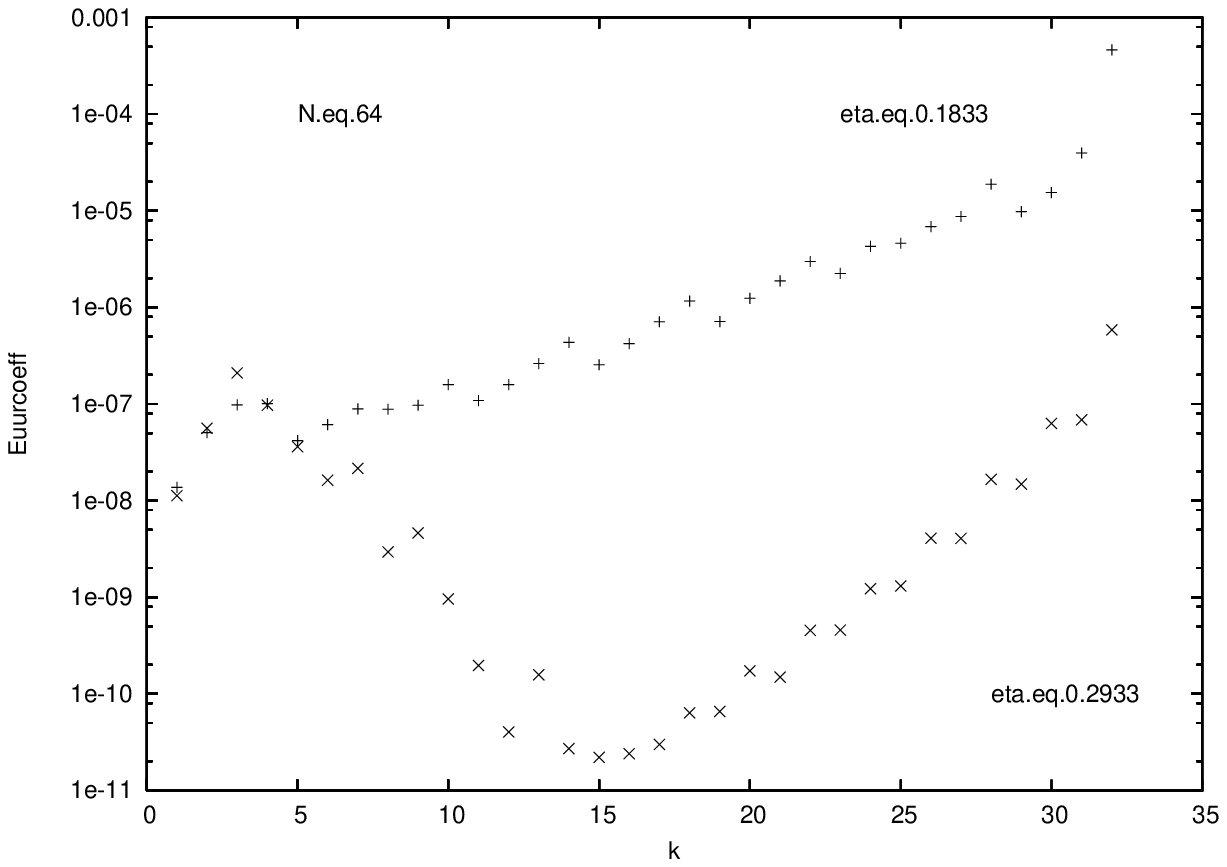}
\end{psfrags}
\end{center}
\caption{A comparison of the error for a fixed number of coefficients
($M=32, N = 64$) for two different values of the coupling constant $\eta$
Plotted is the maximum over $y$ of the Fourier coefficients of the expression
(\ref{DSSEuur})  against the coefficient number.
Compare these errors to Fig. \ref{fig::DSS_eta_max_phi}, where the magnitude
of the coefficients of the variable $\phi$ are shown.
As the latter increase relative to the first coefficients, 
the solution obtained with a fixed number of
coefficients gets less accurate.
}\label{fig::DSS_Euur_coeffs_eta.eq.0.2933_0.1833}
\end{figure}

%\afterpage{\clearpage}

Figs.~\ref{fig::DSS_eta.eq.0.2933_Euur_tau_32} -- 
\ref{fig::DSS_eta.eq.0.1726_Euur_coeffs} show this error for the two
couplings $\eta = 0.2933$ and $\eta = 0.1726$. At each coupling the
solutions computed with a given number of coefficients
are compared to those, obtained by using twice this number. One can see,
that doubling the number of coefficients shrinks the error by several orders
of magnitude. Furthermore at $\eta = 0.1726$ a larger number of 
coefficients is needed in order to keep the error at the same magnitude
as for higher couplings.
Fig.~\ref{fig::DSS_Euur_coeffs_eta.eq.0.2933_0.1833} shows that keeping the
number of Fourier coefficients fixed, the error increases with decreasing
$\eta$.

\subsection{Stability of the DSS Solution}\label{subsec::StabilityDSS}
%%%%%%%%%%%%%%%%%%%%%%%%%%%%%%%%%%%%%%%%%%%%%%%%%%%%%%%%%%%%%%%%%%%%%

The stability of DSS solutions might be analyzed in a similar way
as the stability properties of CSS solutions.
Denoting the metric functions and the field by $Z(\tau,z)$  
we write the perturbed DSS solution as
\be
Z(\tau,z) = \tilde Z_{DSS}(\tau,z) + \delta Z(\tau,z),
\ee 
where $\tilde Z_{DSS}(\tau,z)$ is the DSS solution which is periodic
in $\tau$. The equations are then linearized in the perturbations
$\delta Z$. The main difference to the corresponding problem for CSS
solutions is, that the coefficients are not independent of time, but
depend on time $\tau$ in a periodic way. Therefore one sets
\be\label{eq::perturbationsDSS}
\delta Z(\tau,z) = e^{\lambda \tau} \delta \tilde Z(\tau,z),
\ee   
where $\delta \tilde Z(\tau,z)$ is periodic in $\tau$ with the 
period $\Delta$ of the DSS solution.
Inserting this ansatz into the linearized equations again yields a
time-periodic boundary value problem 
(the boundary conditions originating from regularity at the origin 
and the past SSH). This problem can be solved in the same way as above.
Due to lack of time we had to postpone these computations. 

Instead J.~Thornburg adapted the ``matrix analysis''
(Sec.~\ref{subsec::Matrix_Analysis}) to perturbations of DSS solutions.
The method works as described in Sec.~\ref{subsec::Matrix_Analysis}
with the only difference, that one has to evolve the (slightly perturbed)
DSS solution for a whole period (or alternatively due to the additional
symmetry for half a period) instead of integrating only for one time step.
The reason for this is, that the perturbations depend not only exponentially
on time $\tau$ but also periodically, as can be seen from 
(\ref{eq::perturbationsDSS}). 
In order to extract the eigenvalues $\lambda$, one has to make sure, that
the unknown function $\delta \tilde Z$ drops out of the problem, 
which is the case, 
when slices are compared, that are half a period apart.

We mention, that again we expect gauge modes to be detected by this method.
As for CSS perturbations, there is one gauge mode with $\lambda = 1$. But
now due to the translation invariance in $\tau$ of the DSS equations, an
additional mode\footnote{We note that this mode is not a gauge mode, i.e.~it
cannot be removed by a coordinate transformation (see the footnote
in Sec.~\ref{subsec::gauge_modes}).} 
with $\lambda = 0$ should show up, with the
eigenfunction being of the form
\be
\delta \tilde Z (\tau,z) = \partial_{\tau} \tilde Z_{DSS}(\tau,z).
\ee
Again the method suffers from non-convergence with grid resolution. But 
for a certain number of grid points the gauge modes are reproduced rather
well.

Using this method Jonathan Thornburg investigated stability of the DSS
solution for some values of the coupling constant.
He reports \cite{JT-private-communication-2001} that the DSS solution has 
one unstable mode for $\eta \ge 0.18$. For $\eta = 0.1726$ 
the numerical results are not reliable.

\section{The Spectrum of Self-Similar Solutions Relevant for Type II
         Critical Collapse}
%%%%%%%%%%%%%%%%%%%%%%%%%%%%%%%%%%%%%%%%%%%%%%%%%%%%%%%%%%%%%%%%%%%%

We summarize the results on existence, properties and stability of
self-similar solutions in Table \ref{table:CSS-DSS-relevant}.
As will be explained in Chap.~\ref{chap::criticalCollapse} the stability
properties of the solutions studied here 
will be essential for the dynamics
of the model. The stable CSS ground state probably is
a (singular) end state of time evolution for a certain class of initial
data. The first CSS excitation and the DSS solution having one unstable mode
are candidates for intermediate attractors for near critical initial data in
type II critical collapse. As is clear from Table
\ref{table:CSS-DSS-relevant} the spectrum of these solutions strongly
depends on the coupling. For large $\eta$ we have constructed
the DSS solution, which probably ceases to exist at $\eta \simeq 0.17$.
The CSS solutions on the other hand only exist for $0 \le \eta < 0.5$.
Furthermore the existence of marginally trapped surfaces in the analytic
continuations of the CSS solutions will be relevant (see
Chap.~\ref{chap::criticalCollapse} and the remark in
\cite{Bizon-Wasserman-2000-CSS-exists-for-nonzero-beta}).

\renewcommand{\arraystretch}{2}
\setlength{\tabcolsep}{1mm}

\begin{sidewaystable}
\begin{center}
\begin{tabular}{|>{\centering}p{2.0cm}||
      >{\centering}m{2.5cm}|>{\centering}m{2.5cm}|
      >{\centering}m{2.5cm}|>{\centering}m{3cm}|
      >{\centering}m{2.5cm}|>{\PreserveBackslash\centering}m{7.0cm}|}
\hline
& $0 \le \eta \lesssim 0.069$ & $0.069 \lesssim \eta \lesssim 0.15$ & 
         $0.15 < \eta < 0.17$ & $ 0.17 \lesssim \eta \lesssim 0.4 $ &
         $0.4 \lesssim \eta < 0.5 $ & $0.5 < \eta < \infty$ \\ 
\hline\hline
\multirow{3}{2.0cm}{CSS groundstate} &
    \multicolumn{5}{>{\centering}m{13cm}|}{CSS ground state exists} &  \\ 
\cline{2-7}
   & regular up to future SSH 
   & \multicolumn{4}{>{\centering}m{10.5cm}|}{analytic extension beyond
     past SSH contains marginally trapped surfaces} & \\
\cline{2-7}
   & \multicolumn{4}{>{\centering}m{10.5cm}|}{stable} 
   & ?? &                                  \\ 
\hline\hline
\multirow{3}{2.0cm}{CSS 1st excitation}  
    & \multicolumn{5}{>{\centering}m{13cm}|}{CSS 1st excitation exists} &  \\ 
\cline{2-7}   
   & \multicolumn{2}{>{\centering}m{5cm}|}{regular up to future SSH} 
   & \multicolumn{3}{>{\centering}m{8cm}|}{analytic extension beyond
     past SSH contains marginally trapped surfaces} & \\  
\cline{2-7}
   & \multicolumn{4}{>{\centering}m{10.5cm}|}{one unstable mode}  
   & ?? &  \\
\hline\hline
\multirow{2}{2.0cm}{DSS solution} 
   & \multicolumn{3}{>{\centering}m{7.5cm}|}{} 
   & \multicolumn{3}{>{\centering}m{12.5cm}|}{DSS solution exists} \\
\cline{2-7}
   & \multicolumn{3}{>{\centering}m{7.5cm}|}{} 
   & \multicolumn{3}{>{\centering}m{12.5cm}|}{??\hfill 
          one unstable mode\hfill.} \\
\hline\hline   
\end{tabular}
\caption{The spectrum of self-similar solutions relevant for 
type II critical collapse}\label{table:CSS-DSS-relevant}
\end{center}
\end{sidewaystable}

\renewcommand{\arraystretch}{1}
\setlength{\tabcolsep}{2mm}

\subsection{Comparison of CSS and DSS solutions}\label{subsec::CSS-DSS}
%%%%%%%%%%%%%%%%%%%%%%%%%%%%%%%%%%%%%%%%%%%%%%%%%%%%%%%%%%%%%%%%%%%%%%

\begin{figure}[p]
\begin{center}
\begin{psfrags}
 \psfrag{lnz}[]{$ z$}
 \psfrag{eta.eq.0.1726}[]{$\eta = 0.1726$}
 \psfrag{CSS-SSH}[]{$\textrm{\scriptsize SSH}_{\textrm{\tiny CSS}}$}
 \psfrag{DSS-SSH}[c][c]{$\textrm{\scriptsize SSH}_{\textrm{\tiny DSS}}$}
 \psfrag{phiCSS, phiDSS}[r][l][1][-90]{$\phi_{CSS}, \phi_{DSS}$ }
\includegraphics[width=4in]{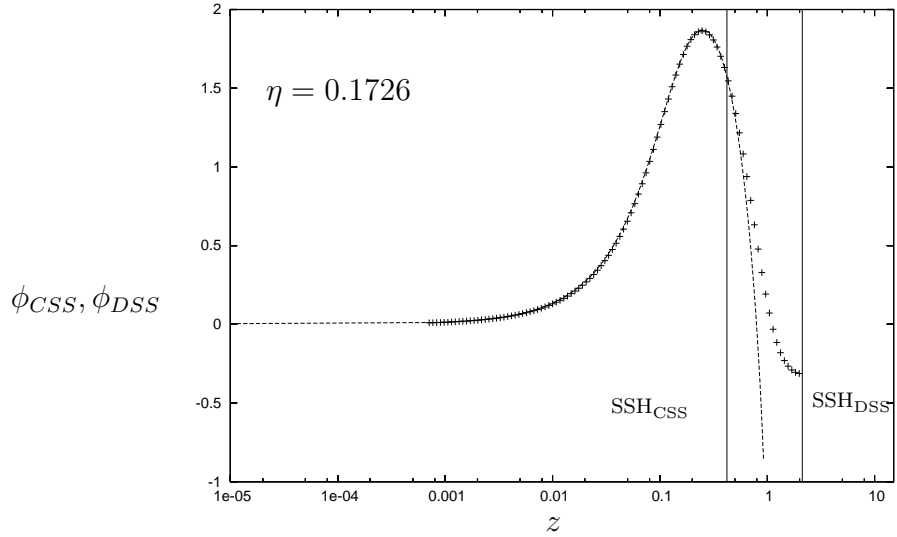}
\end{psfrags}
\end{center}
\caption{Comparison of CSS and DSS solution at the coupling $\eta = 0.1726$,
the lowest at which we explicitly constructed the DSS solution.
Plotted are the CSS solution (dashed line) and the DSS solution (dots)
at a special instant of time, where the amplitude
of the field $\phi_{DSS}$ is maximal. The vertical lines denote the intersections
of the slice $\tau = const$ with the past SSH of the CSS, respectively 
DSS solution. At this coupling both solutions agree
rather well -- though not exactly -- 
up to the past SSH of the CSS solution. 
}\label{fig::CSS-DSS_eta.eq.0.1726}
\end{figure}

The results of our numerical simulations of type II critical collapse for
intermediate couplings (described in Sec.~\ref{sec::transitionregion})
led us to compare the DSS solution with the first CSS excitation
in the range $0.1726 \le \eta \lesssim 0.18$. In this range of couplings
both solutions exist. Comparing the profiles of the two solutions one sees,
that at $\eta = 0.1726$ one can find a DSS phase such that the DSS and CSS
solution resemble each other strongly in some fraction of the DSS
``backwards light cone''\footnote{By this sloppy formulation we mean the
region on a null slice $\tau = const$ ($u=const$) bounded by the
intersection of the past SSH with this slice.}. This resemblance is rather
good up to the past SSH of the CSS solution (which does not agree with that
of the DSS solution). 
Fig.~\ref{fig::CSS-DSS_eta.eq.0.1726} illustrates the situation.

%Directory:Dice/New_runs/eta.eq.0.1726/1_Width/Critical_Solution_1000

%Directory:MarkusH/Try/0.1805_exe
\begin{figure}[p]
\begin{center}
\begin{psfrags}
 \psfrag{z}[]{$z$}
 \psfrag{eta.eq.0.1805}[]{$\eta = 0.1805$}
 \psfrag{CSS_SSH}[]{$\textrm{\scriptsize SSH}_{\textrm{\tiny CSS}}$}
 \psfrag{DSS_SSH}[c][c]{$\textrm{\scriptsize SSH}_{\textrm{\tiny DSS}}$}
 \psfrag{phi_CSSphi_DSS}[r][l][1][-90]{$\phi_{CSS}, \phi_{DSS}$ }
\includegraphics[width=4in]{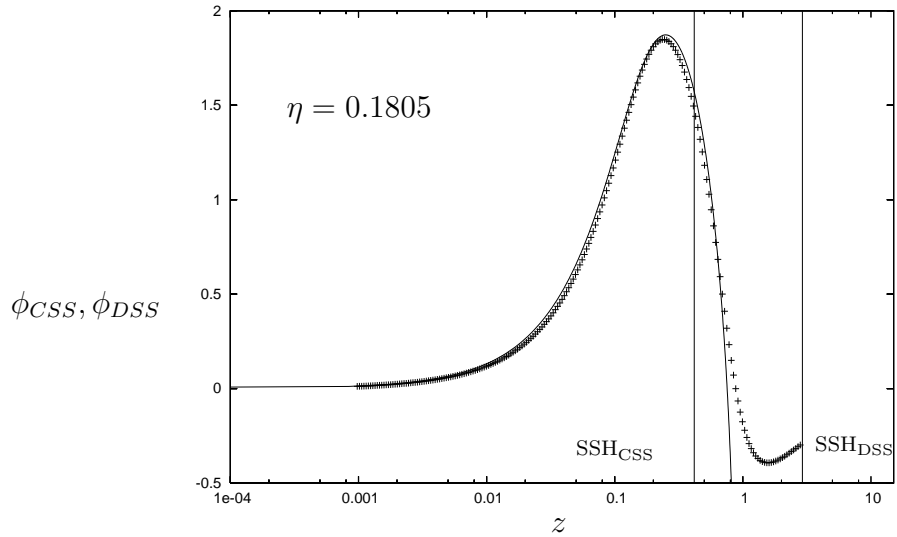}
\end{psfrags}
\end{center}
\caption{The same situation as in Fig. \ref{fig::CSS-DSS_eta.eq.0.1726}, for
a coupling of $\eta = 0.1805$. At this coupling one cannot find 
an instant of time in the DSS solution, for which the shape of the field
resembles that of the CSS solution as closely as for $\eta = 0.1726$.  
}\label{fig::CSS-DSS_eta.eq.0.1805}
\end{figure}

For $\eta = 0.1805$ on the other hand the agreement (for the ``best fitting''
DSS phase) is not as good as can be seen from 
Fig.~\ref{fig::CSS-DSS_eta.eq.0.1805}. 

For the following considerations we have to introduce concepts from
the theory of dynamical systems\footnote{Textbooks for dynamical
systems are e.g.~\cite{Arnold-1983} and \cite{Arrowsmith-Place}. 
\cite{Temam-infinite-dynamical-systems} deals with infinite dimensional
dynamical systems}, 
which will also be useful for understanding
critical phenomena (see Sec.~\ref{sec::critcoll_intro}). Consider 
the (characteristic) initial value problem for the $\sigma$ model in
spherical symmetry. As described in Sec.~\ref{sec::Einstein}, a complete set
of initial data is given by the field $\phi$ at the initial null surface,
$\phi_0(r) = \phi(u_0,r)$. These data then are evolved by the means of
Eqs.~(\ref{eq::phi}), (\ref{eq::betap}) and (\ref{eq::Vp}).
This system can be viewed as an infinite dynamical system in the following
way: phase space is the set of all (asymptotically flat) initial data. An
initial configuration $\phi_0(r)$ thus corresponds to one point in the
(infinite dimensional) phase space. Time evolution (Eqs.~(\ref{eq::phi}),
(\ref{eq::betap}) and (\ref{eq::Vp})) of the initial data $\phi_0(r)$
corresponds to a trajectory
(an orbit) in phase space. 

In adapted coordinates the CSS solution is independent of time.
Time evolution maps these data onto themselves, so this solution is a {\em
fixed point} of the system\footnote{neglecting the fact, that the CSS
solution is not asymptotically flat}. An initial configuration that
corresponds to the DSS solution, is mapped onto itself after one 
period $\Delta$. The DSS solution therefore can be viewed as a 
{\em limit cycle} 
of the system.

Our dynamical system depends on a parameter, the coupling constant $\eta$.
Existence and stability properties of fixed points and limit cycles in a
dynamical system may depend on such an ``external'' parameter.
In particular the number of fixed points and limit cycles might change at
some critical value of the parameter $\eta_C$. This ``process'' is called
{\em bifurcation}. There are so-called local bifurcations, where the
appearance of a new fixed point (or limit cycle) 
is connected to a change in stability of the already existing fixed point.
And there are {\em global} bifurcations, where the fixed point
keeps its stability properties (for the possibilities of global 
bifurcations in two dimensions 
see e.g.~Chap.~8.4 in \cite{Strogatz-nonlinear-dynamics-chaos}). 

\begin{figure}[t]
\begin{center}
\begin{psfrags}
 \psfrag{etaltetaC}[]{$\eta < \eta_C$}
 \psfrag{eta.eq.etaC}[]{$\eta = \eta_C$}
 \psfrag{eta.gt.etaC}[]{$\eta > \eta_C$}
\includegraphics[width=5.5in]{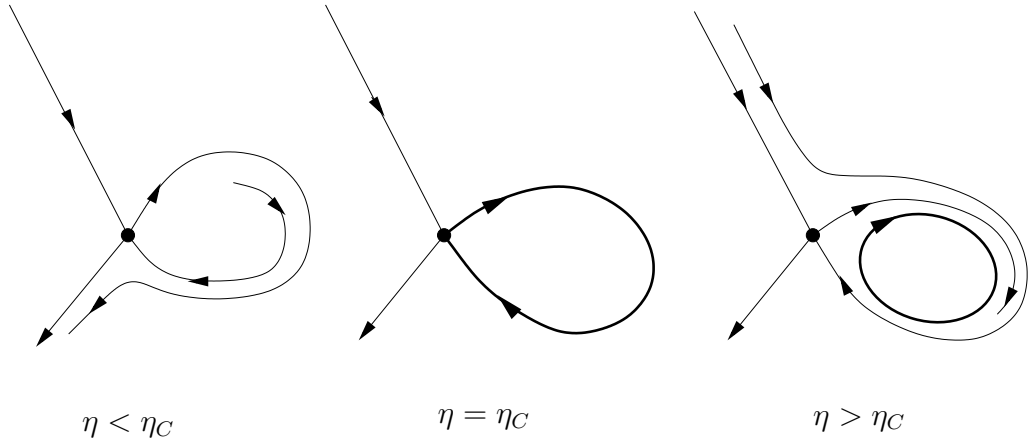}
\end{psfrags}
\end{center}
\caption{Example of an homoclinic loop bifurcation: at $\eta < \eta_C$ a
fixed point with one unstable direction exists. At $\eta = \eta_C$ the 
unstable manifold and the stable manifold merge to form a homoclinic loop.
For $\eta > \eta_C$ the homoclinic loop separates from the fixed point as a
limit cycle. The fixed point does not change stability throughout.
}\label{fig::homoclinic}
\end{figure}

Here we are interested in {\em homoclinic loop bifurcations}, which are
global bifurcations.
Fig.~\ref{fig::homoclinic} shows a schematic picture of a homoclinic 
loop bifurcation
(for the simple case of phase space being two dimensional): 
for $\eta < \eta_C$ a fixed point with one unstable direction exists. 
Increasing the parameter towards $\eta_C$ the stable and unstable
manifold bend more and more towards each other until at $\eta=\eta_C$ they
merge and a homoclinic loop develops: one can ``leave'' the fixed point
along the unstable manifold and return to it via the stable manifold. Of
course such a ``motion'' would take infinite time. For $\eta > \eta_C$ the
homoclinic loop separates from the fixed point as a limit cycle. Stable and
unstable manifold of the fixed point break apart. During this ``process''
the fixed point does not change stability. In principle the emerging limit
cycle can be either stable or unstable.
Approaching the critical value of the parameter from above, it is clear that
the period of the limit cycle diverges in the limit $\eta \to \eta_C^+$.
For a homoclinic loop bifurcation 
\cite{Strogatz-nonlinear-dynamics-chaos} gives the scaling of the amplitude
as $O(1)$ and of the period of the limit cycle as $O(\ln(\eta - \eta_C))$.

%Directory:MarkusH/Try/0.1805_exe

Returning to our situation, we concentrate on the ``vanishing'' of the DSS
solution at $\eta \sim 0.17$.
We summarize some features of this process:
\begin{enumerate}
\item CSS and DSS solution ``come close'' in phase space as one approaches
$\eta \sim 0.17$ from above. They lie ``farther apart'' for bigger $\eta$.
\item The first CSS excitation does not change stability around $\eta \sim
0.17.$
\item The DSS period $\Delta$ rises sharply and seems to diverge at $\eta
\simeq 0.17$.
\item The amplitude of the DSS oscillations is $O(1)$.
\end{enumerate}
This suggests, that the DSS solution ``emerges'' from the CSS 
solution at $\eta \sim 0.17$ in a bifurcation.
From 2.~one might conclude, that the bifurcation is not a local
bifurcation (as would be e.g. a Hopf bifurcation) but rather a global 
one . 
3.~and 4.~suggest that the bifurcation is a {\em homoclinic loop
bifurcation}\footnote{3.~and 4.~would also fit to an ``infinite loop
bifurcation'', 
but we consider this as unlikely.}${}^,$\footnote{
The DSS solution has the additional symmetry 
$\phi_{DSS}(\tau + \Delta/2,z) = - \phi_{DSS}(\tau,z)$. Therefore, if 
there is a phase at which the DSS solution resembles the CSS solution
$\phi_{CSS}$, within the same cycle there is another phase (separated by
$\Delta/2$) at which it resembles $-\phi_{CSS}$. Strictly 
speaking a {\em heteroclinic loop} connecting $\phi_{CSS}$ to 
$-\phi_{CSS}$ forms at the bifurcation point. The bifurcation therefore
should be called {\em heteroclinic} loop bifurcation. Nevertheless we prefer
to stick to the term homoclinic here, because we think this expresses the
essential features.}.

Assuming, that we really deal with a
homoclinic loop bifurcation at $\eta_C \simeq 0.17$,
we can give the following arguments for the
behavior of the period $\Delta$ of the DSS solution:
For $\eta$ slightly bigger than $\eta_C$, where CSS and DSS are already
``close'', we separate the period into the time $T$, the DSS solution spends
in the vicinity of the CSS solution and the remainder $T_{rem}$.
As the DSS solution has the additional symmetry $\phi(\tau + \Delta/2,z) =
- \phi(\tau,z)$, the DSS solution comes close to the CSS solution twice
(to $\phi_{CSS}$ and $-\phi_{CSS}$) during one period. Therefore we can
write
\be
\Delta = 2 T + 2 T_{rem}.
\ee
If DSS is close to CSS we can expand the DSS solution in terms of the CSS
solution and its perturbations:
\be\label{eq::DSS_linearized}
\phi_{DSS}(\tau,z) = \phi_{CSS}(z) + \delta \phi_{unstable}(\tau,z) + \delta
\phi_{stable}(\tau,z).
\ee 
Note that in this equation the coordinates $\tau$ and $z$ are adapted to the
symmetry of the CSS solution, in particular the DSS solution is not periodic
in the coordinate $\tau$. This fact does not matter here, as we are only
interested in the local behavior in the vicinity of the CSS solution.

We define $\tau_1$ to be the moment of time, where the stable modes have
shrinked to order $\epsilon$ ($||\delta \phi_{stable}|| = \epsilon$ 
in some suitable norm). According to the definition of a homoclinic 
loop bifurcation, the admixture
of the unstable mode in (\ref{eq::DSS_linearized}) depends on $\eta$ and
goes to zero as $\eta$ tends to $\eta_C$.  Therefore we 
can always find a value $\eta_0$ such 
that the norm of the unstable mode at this moment of time $\tau_1$
is less than $\epsilon$ for all $\eta_C < \eta < \eta_0$. 
We define
$\tau_2 > \tau_1$ to be the moment of time, where the unstable mode
has grown to order $\epsilon$ ($||\delta \phi_{unstable}|| = \epsilon$).
From the stability analysis we know, that the CSS solution has one unstable
mode with eigenvalue $\lambda_1$, which does not depend strongly on $\eta$. 
Writing $\delta \phi_{unstable} = A_0 e^{\lambda_1 \tau} y(z)$,
the time $T$ elapsing between $\tau_1$ and $\tau_2$ is given by 
\be\label{eq::T}
T = - \frac{1}{\lambda_1} \ln \tilde A_0 + \frac{1}{\lambda_1} \ln
        \frac{\epsilon}{||y||},
\ee
where $\tilde A_0$ denotes the amplitude of the unstable mode at the time 
$\tau_1$.

Now the only expression in (\ref{eq::T}) 
that depends on the parameter $\eta$ is the amplitude $\tilde A_0$. 
(We neglect the $\eta$-dependence of $\lambda_1$ as $\lambda_1$ is only
slowly varying with $\eta$). By
definition it should go to zero for $\eta \to \eta_C$. If we assume further
that $\tilde A_0$ is a regular function of $\eta - \eta_C$, namely
$\tilde A_0(\eta) = a (\eta - \eta_C) + O((\eta - \eta_C)^2)$, we obtain the
following formula
\be
T = - \frac{1}{\lambda_1} \ln(\eta - \eta_C) + const.
\ee

We may assume further that for $\eta$ close to $\eta_C$,
the remaining part of the period can be approximated by a constant,
$T_{rem} \simeq  const$.
Therefore we have
\be\label{eq::Delta_of_eta}
\Delta(\eta) = - \frac{2}{\lambda_1} \ln(\eta - \eta_C) +
const.\footnote{This argument was pointed out by C. Gundlach
\cite{Gundlach-March-2001} to us, however
on the basis that a second unstable mode of the CSS solution appears at the
bifurcation.} 
\ee

Fig.~\ref{fig::DSS_Period_fit} shows the period $\Delta$ fitted against
the function $f(\eta) = - a \ln(\eta - \eta_C) + b$. As stated there, the
fit gives $\eta_C \simeq 0.17$ and $a = 0.36278$. According to  
Eq.~(\ref{eq::Delta_of_eta}) this would correspond to an unstable eigenvalue
$\lambda_1 = 5.51298$. The ``true'' 
value of $\lambda_1$ at $\eta = 0.17$, computed with the shooting and 
matching method as in Sec.~\ref{subsec::CSSstability_BVP}, 
is $\lambda_1 = 5.14282$. The relative difference of these quantities
is $\sim 7 \%$. This correspondence of numbers gives a strong
support to the hypothesis of the homoclinic loop bifurcation.

%%%%%%%%%%%%%%%%%%%%%

%\documentclass[12pt,a4paper]{report}
%\usepackage{german,a4,bbm,graphicx,psfrag}
%
%
%\include{diss_macros}

%\textwidth=15cm
%\textheight=22cm
%\topmargin=-2cm
%\oddsidemargin=1cm
%\pagestyle{plain}
%\parindent=0pt                    % no indentation for paragraphs,
%\parskip=5pt plus 2pt minus 1pt   % but do a little skip

\chapter{Type II Critical Behavior of the Self-gravitating $\sigma$ Model
%%%%%%%%%%%%%%%%%%%%%%%%%%%%%%%%%%%%%%%%%%%%%%%%%%%%%%%%%%%%%%%%%%%%%%%%%%%
}\label{chap::criticalCollapse}
%%%%%%%%%%%%%%%%%%%%%%%%%%%%%%

This chapter finally deals with type II critical phenomena of the
self-gravitating SU(2) $\sigma$ model in spherical symmetry.
This model has already been studied in its limits of strong coupling ($\eta
\to \infty$) by Liebling \cite{Liebling-inside-global-monopoles} and weak
coupling ($\eta = 0$) independently by Bizon 
et al.~\cite{Bizon-Chmaj-Tabor-1999-sigma-3+1-evolution}
and Liebling 
et al.~\cite{Liebling-Hirschmann-Isenberg-1999-sigma-critical}.
They find type II critical behavior governed by self-similar solutions
in conformity with Table \ref{table:CSS-DSS-relevant}. 
From these results and our knowledge of
self-similar solutions (Chap.~\ref{chap::SSSolutions}) we expect critical
phenomena to depend strongly on the coupling. In particular we expect the
critical solution to change from CSS to DSS in some intermediate regime of
couplings around $\eta \sim 0.17$ (see Sec.~\ref{sec::transitionregion}).

As is clear from Chap.~\ref{chap::SSSolutions} (especially Table
\ref{table:CSS-DSS-relevant}), for small couplings we have to consider the
possibility of the formation of naked singularities -- according to the
stable CSS ground state -- for a certain class of
initial data. This is investigated in Sec.~\ref{sec::endstates}.

In agreement with Table \ref{table:CSS-DSS-relevant} we essentially find
three different types of critical behavior: for small couplings the critical
solution is CSS (see Sec.~\ref{sec::smallcoupling}), while for large couplings
we have DSS critical behavior (see Sec.~\ref{sec::largecoupling}). 
And for some intermediate range of couplings 
$0.15 \lesssim \eta \lesssim 0.18$
we find that the intermediate asymptotics of near critical evolutions 
show a behavior which we call ``episodic CSS'':
at intermediate times 
we see a repeated approach to the first CSS excitation.
These ``episodes'' are part of an approach to the DSS solution at couplings
where the latter exists, and still have some resemblance with
discrete self-similarity at couplings where we think the DSS solution does
not exist anymore (see Sec.~\ref{sec::transitionregion}).
With increasing coupling the ``CSS episodes'' get less pronounced, while at
the same time the number of episodes (or cycles) increases.
To our knowledge this sort of transition from CSS to DSS as the critical
solution, which is in very good agreement with our results obtained by a
direct construction of the self-similar solutions and the hypothesis of a
homoclinic loop bifurcation of Chapter \ref{chap::SSSolutions}, 
has not been observed in the context of critical phenomena of 
self-gravitating systems up to now.

The results for large couplings are summarized in 
\cite{Husa-Lechner-Puerrer-Thornburg-Aichelburg-2000-DSS}. Part of the
phenomena in the transition region is described in
\cite{Thornburg-Lechner-Puerrer-Aichelburg-Husa-MG9-Proceedings}. 
A more complete description is given in
\cite{Lechner-Thornburg-Husa-Aichelburg-prl-2001}.

In order to avoid confusion originating from the inconsistent use of the 
notion ``the critical solution'' in the literature, 
we fix our convention as follows: by the {\em critical solution} 
we denote the intermediate attractor, which by the means of its 
``stable manifold'' separates two different
end states in phase space. Solutions that approach the critical solution in
some intermediate regime of time are called {\em near critical solutions}
(evolutions, data etc.). The member of a family of initial data 
with $p = p^*$ is called {\em critical data}.   
With this nomenclature we have constructed the respective ``critical
solutions'' in Chapter \ref{chap::SSSolutions} and deal with the evolution
of ``near critical data'' in this chapter. 
It is the aim of the bisection procedure 
(see Sec.~\ref{sec::Crit_Search_Setup}) to approximate 
``critical data'' as close as possible, but of course they are not 
realized numerically.

\section{Introduction to Critical Phenomena}\label{sec::critcoll_intro}
%%%%%%%%%%%%%%%%%%%%%%%%%%%%%%%%%%%%%%%%%%%%%%%%

There is a couple of excellent review articles on critical phenomena.
An elementary introduction to critical phenomena is given by Choptuik,
who pioneered the work on this field, in
\cite{Choptuik-1998-self-similarity-review-in-GR15}.
Gundlach's reviews
\cite{Gundlach-1998-critical-phenomena-review,
Gundlach-1999-critical-phenomena-living-reviews} 
give more details as well as an overview of the models
studied and the phenomena found. We also mention the review by 
Brady and Cai
\cite{Brady-Cai-1999-critical-phenomena-review}.
At the moment the most recent reference lists can be found in the 
article by Wang
\cite{Wang-2001-review-ofcritical-collapse} and on Choptuik's home-page
\cite{Matt_Choptuik_homepage}. 

The field of study can be explained as follows: in their long time evolution
isolated (asymptotically flat) self-gravitating systems are supposed to
evolve to some stationary end state, e.g. to a black hole, a stable star or
flat space. According to this small number of distinct kinds of 
end states, the space
of initial data is divided into basins of attraction. The ``boundaries'' of
these basins and their ``vicinities'' are the scope of studies of critical
phenomena.

In the simplest models (e.g.~the massless Klein-Gordon field 
studied by Choptuik in spherical symmetry), where 
only two different end states are possible, namely black holes and Minkowski
space,  
``small'' initial data, i.e. initial data, that do not deviate too
much from Minkowski data, will finally disperse to infinity, leaving flat
space behind, whereas for ``strong'' initial data, part of the 
mass present in the initial slice will be trapped and a black hole
will form. 

%Directory:Figures/
\begin{figure}[h]
\begin{center}
\begin{psfrags}
 \psfrag{r}[]{$r$}
 \psfrag{psub}[l][c][1][0]{$p_{sub}$}
 \psfrag{psuper}[l][c][1][0]{$p_{super}$}
 \psfrag{pcrit}[l][c][1][0]{$p^*$}
 \psfrag{phi0}[r][l][1][-90]{$\frac{\phi_0(r;p)}{r^2}$}
\includegraphics[width=3.5in]{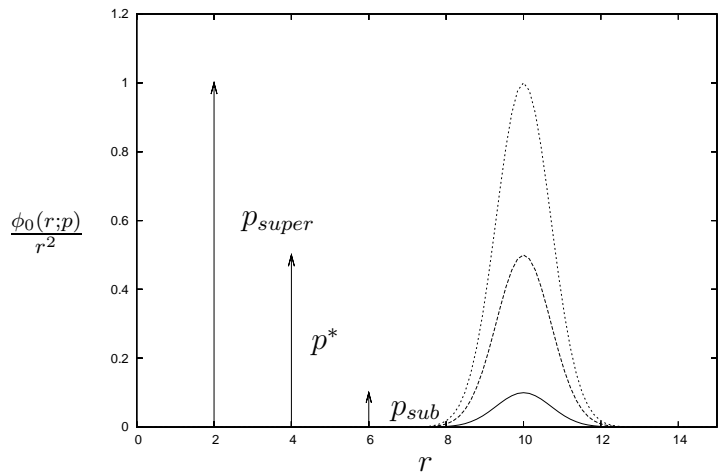}
\end{psfrags}
\end{center}
\caption{A family of Gaussian initial data $\phi_0(r;p) =
p r^2 exp(-(r-r_0)^2/\sigma^2)$, where the width $\sigma$ and 
the center $r_0$ are fixed and the amplitude $p$ parameterizes the family. 
Depending on the value of  $p$, the data will finally evolve to a 
black hole or disperse.
}\label{fig::gaussian}
\end{figure}

Technically one constructs a one parameter family of initial data, 
parametrized by $p$, such that for large values of $p$ the data evolve 
to a black hole, whereas
for small values of $p$ the data disperse.
E.g. for the SU(2) $\sigma$ model 
in spherical symmetry, and working on null slices,
Eqs.~(\ref{eq::phi}), (\ref{eq::betap}) and (\ref{eq::Vp}) 
show, that a complete set of initial data is given by the shape of the 
field $\phi_0(u=0,r)$ at the initial null slice. 
A one parameter family of initial data then can be
modeled e.g. by a Gaussian with fixed width and center and
the amplitude serving as the parameter.  
For a family constructed this way, there will be a value of the parameter
$p$, denoted by $p^*$, which separates initial data
that disperse ({\em sub-critical data} with $p_{sub} < p^*$) 
from those, that form a black hole 
({\em super-critical data} with $p_{super} > p^*$). 
Phenomena that occur for initial data with $p \simeq p^*$ are called
{\em critical phenomena} (See Fig.~\ref{fig::gaussian}). 
One of the original key questions was, whether the
black hole mass for slightly super-critical data 
can be made arbitrarily small (such that the
black hole mass as a function of the parameter $p$ would be continuous)
or has a finite
value (such that there would be a {\em mass gap}).
The answer is, that depending on the model under investigation, both
behaviors can occur. In analogy to statistical physics one distinguishes
between two types of critical phenomena, {\em type I} where the black hole
mass shows a mass gap, and {\em type II}, where the black hole mass is
a continuous function of $p - p^*$.

In the following we will concentrate on type II critical phenomena and only 
refer to the other possibility at the end of this section.
The first model, for which critical phenomena have been investigated, was the
self-gravitating massless Klein-Gordon field. This was done numerically by 
M. Choptuik 
\cite{Choptuik-1992-in-dInverno,Choptuik-1993-self-similarity}.
In order to resolve all the features, including self-similarity, 
he had to develop a sophisticated numerical algorithm, 
which refines the numerical grid, when variations occur 
on too small scales to be resolved.
Other models have been studied, including e.g. gravitational waves in 
axial
symmetry, perfect fluids, the Einstein-Yang-Mills system etc.
For the most recent lists of references see
\cite{Wang-2001-review-ofcritical-collapse} and the bibliography on 
Matt Choptuik's home-page \cite{Matt_Choptuik_homepage}.

These investigations showed, 
that the behavior of near critical evolutions 
can be characterized by three main features

\begin{itemize}
\item Scaling, 
\item Self-similarity,
\item Universality.
\end{itemize}

Scaling relates the black hole mass of super-critical data, as well as other
quantities, that have dimension of length or any power thereof to the
parameter $p$ in the initial data.
One finds, that the black hole mass scales as
\be\label{eq::scaling_gen}
m_{BH} \sim (p - p^*)^{\gamma},
\ee
where the exponent $\gamma$, called the {\em critical exponent},
is {\em independent} of the family of initial data, although it depends in
general on the model.

The second general feature of type II critical phenomena is that near
critical evolutions spend their intermediate asymptotics in the vicinity of
a self-similar solution (which can be either continuous or discrete,
depending on the model), before they actually decide whether 
to disperse or to form a black hole. 

The third important point is universality, which means {\em independence} 
of the above features of the family of initial data.

\begin{figure}[p]
\begin{center}
\begin{psfrags}
 \psfrag{BH}[]{black hole}
 \psfrag{flatspace}[]{flat space}
 \psfrag{unstabledirection}[c][r][1][0]{unstable direction}
 \psfrag{stablemanifold}[l][r][1][0]{``stable manifold''}
 \psfrag{initialdata}[r][c][1][0]{initial data}
 \psfrag{fof}[r][c][1][0]{family of} 
 \psfrag{psuper}[l][c][1][0]{$p_{super}$}
 \psfrag{psub}[]{$p_{sub}$}
 \psfrag{pstar}[]{$p^*$}
 \psfrag{CSS}[l][c][1][0]{CSS}
\includegraphics[width=5.0in]{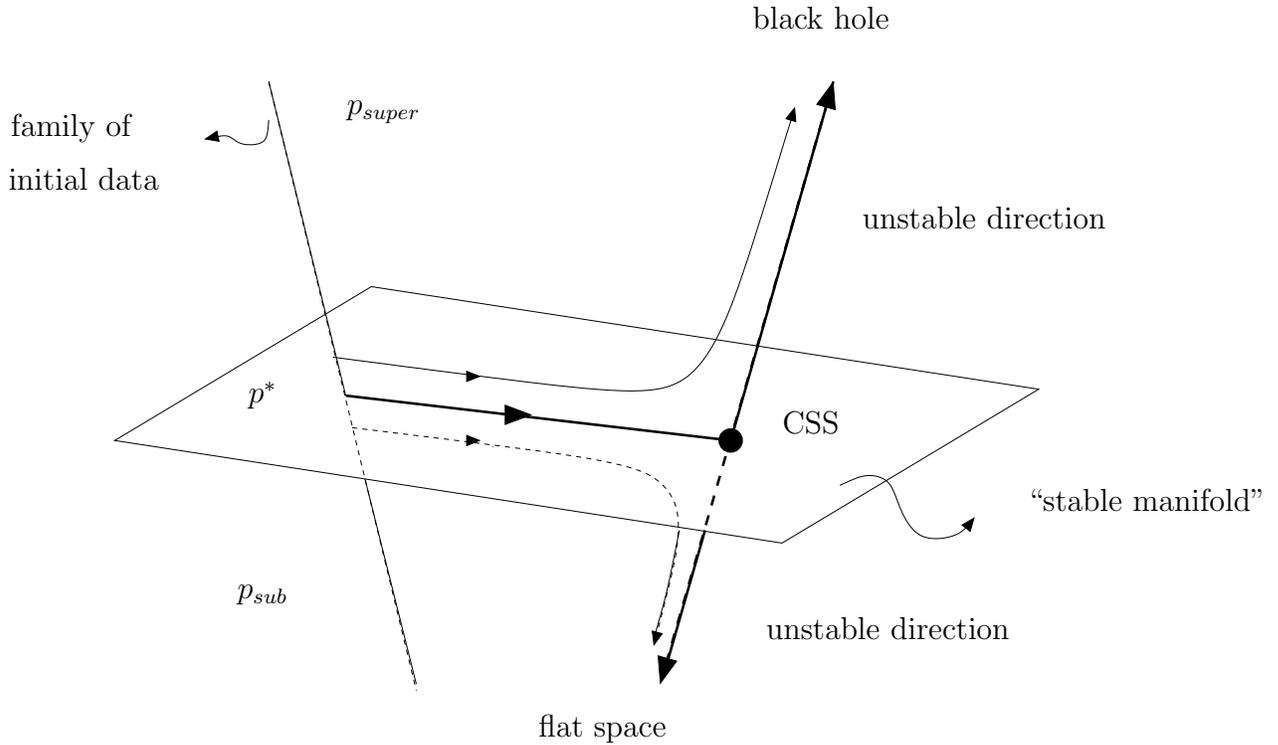}
\end{psfrags}
\end{center}
\caption{A schematic picture of phase space. Every point in this figure
corresponds to one configuration $\phi(r,u=const)$. In adapted coordinates
the CSS solution is a fixed point of the system, it therefore is
drawn as a point (large
solid circle). The CSS solution has one unstable mode, 
the ``stable manifold'' therefore is of co-dimension one. 
The straight line at the left of the figure
represents a one parameter family of initial data, with parameter $p$.
The value $p^*$ corresponds to those initial data, that ``start out'' 
on the ``stable manifold'' and are completely attracted 
to the CSS solution (where they arrive only in the limit $\tau \to \infty$).
For $p > p^*$ the configuration is initially attracted by the CSS solution
via the stable modes until the (initially very small) admixture of the
unstable mode takes over and pushes the solution towards black hole
formation. For $p< p^*$ the final state is flat space. 
Any one parameter family of initial data, cutting the
``stable manifold'', will show the same near critical phenomena. 
Two remarks are in order here: a DSS solution, being periodic in the adapted
time $\tau$, corresponds to a limit cycle, and should therefore be drawn as
a cycle, with near critical data spiraling in and out. 
Second, this sketch does not claim, that
the ``stable manifold'' is indeed a manifold. It is just a 
very helpful abstract picture, modeling the essential facts, that have been
observed.
}\label{fig::phase_space.fig}
\end{figure}

An explanation of these phenomena can be given in the language of 
dynamical systems. Suppose the matter model admits a 
self-similar solution. For simplicity we concentrate on 
continuously self-similar solutions $\phi_{CSS}(z)$.
Suppose further, that the CSS solution has exactly one unstable mode
with eigenvalue $\lambda$ and that an initial configuration, which
corresponds to the CSS solution plus a small admixture of the 
unstable mode leads to either black hole formation or 
dispersion, depending on the overall sign of the perturbation.
In the simplest case the ``stable manifold'' of this solution 
divides the phase space into sub and super-critical data.
Fig.~\ref{fig::phase_space.fig} shows a sketch of this scenario
using a ``phase space picture''.

%Directory:Figures

\afterpage{\clearpage}

Then general near critical initial data are attracted by the 
CSS solution via the stable modes, until they are close to the CSS 
solution. In this vicinity the solution can 
be written as a small perturbation of the CSS solution
\be\label{eq::close_to_CSS}
\phi(u,r) = \phi_{CSS}(\frac{r}{u^*-u}) + 
     C(p) (u^* - u)^{-\lambda} y(\frac{r}{u^*-u}) 
+ \delta \phi_{stable} (u,r),
\ee
where the eigenvalue $\lambda$ is real for all known examples and positive.
The amplitude of the unstable mode contains information on the initial data,
in particular it depends on the parameter $p$.
For $p = p^*$ the unstable mode is tuned out completely, and the
configuration evolves towards the CSS solution, therefore $C(p*) = 0$.
For near critical data, we have 
$C(p) = \frac{d C}{d p}(p^*) (p - p^*) + (O((p-p^*)^2))$.

From (\ref{eq::close_to_CSS}) we can estimate the time for which the solution
stays in the vicinity of the CSS solution. Fix $\tau_1$ to be some instant
of time, where near critical solutions are already in the vicinity 
of the CSS solution, and therefore (\ref{eq::close_to_CSS})
is valid.
Let furthermore $\tau_2 = \tau_1 + T$ be the instant of time when the 
amplitude of the 
unstable mode has grown to be $\epsilon$, then the time $T$ spent in the
vicinity of the CSS solution is given by
\be
\epsilon = {\tilde C}(p) e^{\lambda T},
\ee
with ${\tilde C}(p) = C(p) e^{\lambda \tau_1}$, or
\be\label{eq::life-time}
T = \frac{1}{\lambda} \ln (\frac{const}{p-p^*}).
\ee
So a near critical solution spends longer and longer time
in the vicinity of the CSS solution, when $p$ comes closer and closer to 
$p^*$, until for $p=p^*$, the 
logarithmic time goes to infinity.

In order to explain the scaling of the black hole mass,
we ``redefine'' our family of initial data in the following way:
we fix the time $u_p$, where the stable modes are already negligible compared 
to the unstable mode, and the amplitude of the unstable mode has grown to be
$\epsilon$
\be\label{eq::epsilon}
\epsilon = C(p) (u^* - u_p)^{-\lambda}.
\ee
So
\be
\epsilon = \frac{d C}{d p}(p^*) (p - p^*) (u^* - u_p)^{-\lambda}
\ee
or resolved for $u^* - u_p$
\be
u^* - u_p = const (p-p^*)^{1/\lambda}
\ee
where the constant contains $\epsilon$ and some information on the 
original family of initial data, but is independent of $p$.
This way we have constructed a family of initial data
$\phi_0(r;p) = \phi_{CSS}(r/(u^* - u_p)) + \epsilon y(r/(u^* - u_p))$ , which
depends on $r$ only via the ratio $r/(u^* - u_p)$.
As the field equations are scale invariant, a solution $\phi(u,r)$ with
initial conditions $\phi_0(r)$ implies the existence of a one parameter
family of solutions $\phi_{\sigma}(u,r) = \phi(\sigma u, \sigma r)$ with
initial conditions $(\phi_{\sigma})_0(r) = \phi_0(\sigma r)$.
From this it follows, that the evolution of our one-parameter family of
initial data gives a one-parameter family of solutions
of the form 
\be\label{selfsimilarity}
\phi_p(u,r) = \phi(\frac{u}{u^* - u_p}, \frac{r}{u^* - u_p}).
\ee
This is valid for the whole future evolution of the data, even when the
linearity assumptions break down.  

Assume now, that for super-critical data the solution 
$\phi(u,r)$ has an apparent horizon at $r_H(u)$.
If we fix $z=r/(u^*-u)$, then the apparent horizon is located 
at $(u_H, r_H = z \ (u^* - u_H))$ and so the metric
function $\beta(r/(u^*-u), u)$ diverges when $u \to u_H$. 
Therefore the rescaled solutions $\beta_p(r/(u^*-u),u) = 
\beta(r/(u^*-u),u/(u^*-u_p))$
diverge if $u \to (u^* - u_p) u_H$, or $r \to (u^* - u_p) r_H$.
The mass of the apparent horizon $m(z;p) = r_H(z;p)/2$, 
measured at constant $z$, therefore
scales as $(u^* - u_p)$. 

Remark: the above analysis, of writing near critical data 
as the CSS solution plus a perturbation is valid only up to some finite 
radius. Outside this region the near critical solutions -- being
asymptotically flat -- will differ considerably from a small perturbation of
the CSS solution. 
Nevertheless, if the region outside does not influence
the black hole mass 
the latter scales as
\be
m_{BH} \sim (u^* - u_p) = const (p-p^*)^{1/\lambda}.
\ee
So we have derived the scaling law (\ref{eq::scaling_gen}) 
with the additional information,
that the critical exponent $\gamma$ is related to the eigenvalue of the
unstable mode via
\be
\gamma = \frac{1}{\lambda}.
\ee
This relation was derived independently by 
Koike et al.~\cite{Koike-Hara-Adachi-1995-scaling-in-critical-collapse} and
Maison \cite{Maison-1996-scaling-law}. 
(Evans and Coleman
\cite{Evans-Coleman-1994-suggest-gamma-from-stability-analysis} 
first suggested to
look at the linear stability of the CSS solution in order to get an estimate
on the critical exponent $\gamma$.)
Another well defined quantity as pointed out by Garfinkle and Duncan
\cite{Garfinkle-Duncan-1998-critical-curvature-scaling}, 
which exhibits scaling, is the maximum of the
Ricci scalar at the axis for sub-critical data, 
the maximum taken over a whole evolution, $\max\limits_{u} {\mathcal R}(u,0)$.
As this quantity has dimension of $1/length^2$, it should scale as
\be\label{eq::Ricci_scaling}
\max\limits_{u} \mathcal R(u,0) \sim (p^* - p)^{-2/\lambda}.
\ee

If a DSS solution is the critical solution, the scaling law
undergoes some modification, in that a small wiggle is overlaid.
The derivation, as first given independently by 
Gundlach \cite{Gundlach-1996-scaling-in-critical-collapse} and 
Hod and Piran \cite{Hod-Piran-1997-fine-structure-of-mass-scaling-law},
is analogous to the CSS case, a first difference arising
in (\ref{eq::close_to_CSS}), where now the DSS solution and its unstable
mode have an additional periodic dependence on $\tau$.
The family of initial data constructed as above,
therefore depends periodically on the parameter $\tau_p$,
\be
\phi_0(r;p) = \phi_{DSS}\left(\frac{r e^{\tau_p}}{\zeta(\tau_p)}, \tau_p \right)
                 + \epsilon
 y_1\left(\frac{r e^{\tau_p}}{\zeta(\tau_p)},\tau_p\right).
\ee
Again the equations are scale invariant, therefore the solutions 
to initial conditions $\phi_0(r;p)$ behave as
\be
\phi_p(r,u) = \phi\left(\frac{r e^{\tau_p}}{\zeta(\tau_p)}, \frac{u
     e^{\tau_p}}{\zeta(\tau_p)};\tau_p \right).
\ee
Any quantity of dimension length should therefore scale as 
$e^{-\tau_p}\zeta(\tau_p)f(\tau_p) = e^{-\tau_p} \tilde f(\tau_p)$, where
$f, \zeta$ and $\tilde f$ are periodic functions of their argument $\tau_p$.
So we have
\be
m_{BH} = c_1 (p - p^*)^{1/\lambda} \tilde f (- \ln c_1 -
\frac{1}{\lambda} \ln(p - p^*) )
\ee
or
\be
\ln m_{BH} = \ln c_1 + \frac{1}{\lambda} \ln(p-p^*) + 
\hat f(-\ln c_1 - \frac{1}{\lambda} \ln(p-p^*)),
\ee
where $\hat f = \ln(\tilde f)$.
Now $\hat f$ is periodic in $\ln(p - p^*)$ with period 
$\Delta/\gamma$, or since the metric functions have the additional symmetry
of consisting only of even frequencies, and therefore have a period of
$\Delta/2$, the period of $\hat f$ is rather half this value, i.e.
$\frac{\Delta}{2 \gamma}$.
Note that there is only one constant $c_1$, which depends on the family.
The scaling exponent $\gamma = 1/\lambda$ and the periodic function $\hat
f$ are universal.

We close this section with some words on type I critical phenomena, which
have been observed in several models, e.g.~in the Einstein-Yang-Mills 
system \cite{Choptuik-Chmaj-Bizon-1996-EYM-critical-behavior} or the 
massive minimally coupled scalar field
\cite{Brady-Chambers-Goncalves-1997-typeI-oscillating}.
In this type of transition the intermediate asymptotics is governed by an
unstable (the stable manifold being of co-dimension one) static solution
or a solution that is oscillating in time.
Again the ``life-time of the critical solution'', that is the time a near
critical solution spends in the vicinity of the intermediate attractor,
scales according to (\ref{eq::life-time}).
The major difference  to a type II collapse is that the black hole mass
for slightly super-critical data is finite, i.e. the black hole mass as a
function of the parameter $p - p^*$ is discontinuous. 
The magnitude of the mass gap or the fraction of the mass of the
intermediate attractor that is radiated away by slightly super-critical
data depends on the model.

\section{Limits of Weak and Strong Coupling}\label{sec::limits}
%%%%%%%%%%%%%%%%%%%%%%%%%%%%%%%%%%%%%%%%%%%%%%%%%%%%%%%%%%%%%%

The self-gravitating SU(2) $\sigma$ model has already been investigated
with respect to critical phenomena for the two extremes of coupling,
$\eta = 0$ and $\eta \to \infty$.

The case $\eta = 0$ corresponds to the SU(2) $\sigma$ model on fixed
Minkowski background and was investigated independently by 
Bizon et al.~\cite{Bizon-Chmaj-Tabor-1999-sigma-3+1-evolution} and 
Liebling et al.~\cite{Liebling-Hirschmann-Isenberg-1999-sigma-critical}.
They looked at the threshold between dispersion and blow up (of the first
derivative of $\phi$ with respect to $r$) at the 
origin, which is governed by the CSS ground state. 
They found that the solution at the threshold is the first CSS excitation.
A quantity that can be used to examine the scaling, is the maximum (over
time) of the energy density at the origin. It was found, that this quantity
shows scaling with an exponent $\gamma = 1/\lambda_{CSS}$, where
$\lambda_{CSS}$ is the eigenvalue of the unstable mode of the CSS solution
for $\eta = 0$ \cite{Bizon-2000-scaling-priv-comm}.

The case $\eta \to \infty$ on the other hand corresponds to the
self-gravitating $\sigma$ model with three dimensional flat 
target manifold $({\mathbbm R}^3, \delta_{AB}$). This is natural, as
the coupling $\eta$ corresponds to the inverse of the curvature of the target
manifold, the curvature goes to zero, as $\eta$ tends to infinity.
It is also easy to see, by e.g. looking at 
Eqs.~(\ref{eq::phi}), (\ref{eq::betap}) and (\ref{eq::Vp}). 
Defining $\tilde \phi = \sqrt{\eta} \ 
\phi$, rewriting the equations with respect to $\tilde \phi$ and taking the
limit $\eta \to \infty$ gives the following system of equations
\bea
\beta' & = & \frac{1}{2} r^2 (\tilde \phi')^2, \nonumber\\
\Vr' & = & e^{2 \beta} (1 - 2 \tilde \phi^2), \nonumber\\
\dal \tilde \phi & = & \frac{2 \tilde \phi}{r^2},
\eea
which are precisely the equations for the $\sigma$ model with flat
target manifold in the hedgehog ansatz.

This model was investigated by Liebling
\cite{Liebling-inside-global-monopoles}, where in
addition he considered a potential. As was explained in 
Sec.~\ref{subsec::implications_matter}, such a
potential is asymptotically irrelevant, the critical behavior of the models
with and without this potential should therefore be the same.
Liebling found that the critical solution at the threshold of black hole
formation is DSS with an echoing period $\Delta = 0.46$ 
and a scaling exponent
$\gamma = 0.119$. Note, that the value of the echoing period nicely fits to
the period of the DSS solution at $\eta = 100$, described in Sec.
\ref{subsec::DSScode_results}.

From these rather different critical phenomena at the limits of very small
and very large couplings, one can infer that there will be a transition
of the critical solution from CSS to DSS as the coupling is increased.

\section{Possible End States}\label{sec::endstates}
%%%%%%%%%%%%%%%%%%%%%%%%%%%%%%%%%%%%%%%%%%%%%%%%%%%

As is clear from the last section, criticality is only defined
with respect to the end states. 
Usually these two different end states 
would be black hole formation and dispersion to infinity.
If the model allows also for other stable stationary configurations, 
e.g. for stable static solutions, then these 
can be considered as possible end states as well. 

As was already mentioned in Sec.~\ref{subsec::non_existence}, 
the SU(2) $\sigma$ model does not allow
for static asymptotically flat solutions. Therefore it is natural to
investigate the transition between black hole formation and 
dispersion.

On the other hand, as we have seen in Chap.~\ref{chap::SSSolutions},
the spectrum of CSS solutions contains a {\em stable} ground state.
Furthermore on fixed Minkowski 
background this stable ground state
governs the ``long time behavior'' of strong initial data, as
was described in \cite{Bizon-Chmaj-Tabor-1999-sigma-3+1-evolution} 
and the intermediate asymptotics of 
near critical data between this singularity formation and dispersion 
is ruled by the first excitation of this CSS family.

As the ground state and its stability properties persist, when gravity is
turned on, it is reasonable to expect, that a naked singularity
is a possible ``end state'' for strong (but not too strong) initial data,
at least for couplings less than $\eta \simeq 0.069$. 

\begin{figure}[p]
\begin{center}
\begin{psfrags}
 \psfrag{r}[]{$r$}
 \psfrag{eta.eq.0.01}[]{$\eta = 0.01$}
 \psfrag{phi}[r][l][1][-90]{$\phi$}
\includegraphics[width=3.3in]{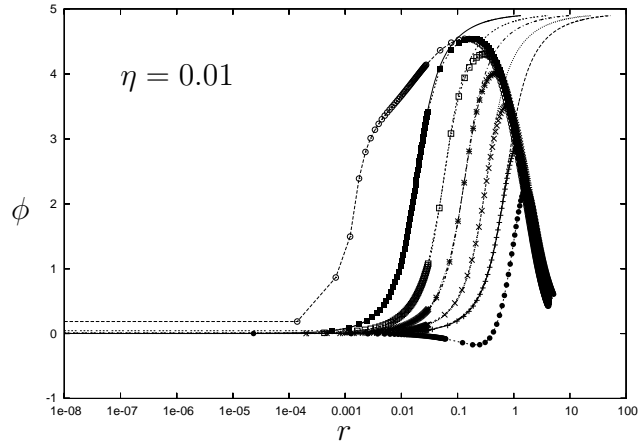}
\end{psfrags}
\end{center}
\caption{Late time behavior of initial data as described in the text
for $\eta = 0.1$. This figure shows the evolved field $\phi$ (lines-points)
-- moving from right to left as time proceeds --
for several time steps.
The solution clearly comes close to the CSS ground state
(lines), and stays there until the grid resolution at the origin becomes
too sparse and the evolved data break away from a solution of
Einstein equations (this last time step shown in the plot is already 
``after'' the culmination time $u^*$). 
The culmination time $u^*$ of the CSS solution was determined by
the fit of a single time step to be $u^* = 11.346$.
For the other time steps the CSS solution was shifted according
to $r = z (u^* - u)$. 
The past SSH of the CSS ground state is located 
where the field equals $\pi/2 \simeq 1.58$. 
This shows that the region, where the evolved data and the CSS
ground state agree, extends some way outside the past SSH.
}\label{fig::CSS_endstate_phi}
\end{figure}

\begin{figure}[p]
\begin{center}
\begin{psfrags}
 \psfrag{r}[]{$r$}
 \psfrag{eta.eq.0.01}[]{$\eta = 0.01$}
 \psfrag{2mr}[r][l][1][-90]{$\frac{2m}{r}$}
\includegraphics[width=3.3in]{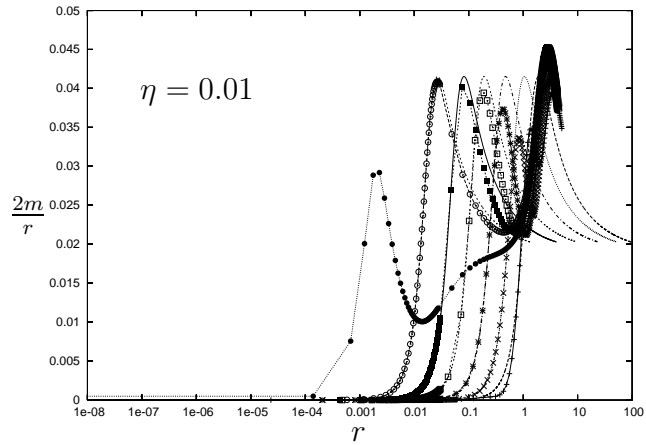}
\end{psfrags}
\end{center}
\caption{The same scenario as in Fig. \ref{fig::CSS_endstate_phi}, where
this time $\frac{2m}{r}$ is plotted.
Note that $\frac{2m}{r}$ is far from being unity everywhere in the evolved
grid. This shows, that the singularity, which is approached via the CSS
ground state, is in general not shielded by an apparent horizon,   
}\label{fig::CSS_endstate_2mr}
\end{figure}

\begin{figure}[p]
\begin{center}
\begin{psfrags}
 \psfrag{r}[]{$r$}
 \psfrag{eta.eq.0.05}[]{$\eta = 0.05$}
 \psfrag{phi}[r][l][1][-90]{$\phi$}
\includegraphics[width=3.5in]{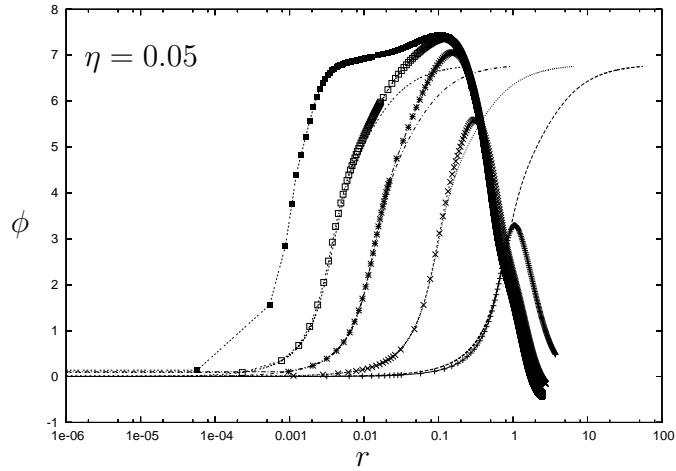}
\end{psfrags}
\end{center}
\caption{For $\eta = 0.05$ the same initial data as in Figs.
\ref{fig::CSS_endstate_phi} and \ref{fig::CSS_endstate_2mr} are evolved. 
This figure shows the evolved field $\phi$ (lines-points)
for several time steps between $u = 10.1448$ and 
$u = 11.2416$. 
Again the solution clearly comes close to the CSS ground state
(lines) and breaks away due to insufficient resolution near the origin.
(clearly the ``latest'' time step plotted 
suffers from insufficient resolution)
The culmination time $u^*$ is determined via the fit to be 
$u^* =11.24085$. 
}\label{fig::CSS_endstate_phi_0.05}
\end{figure}

\begin{figure}[p]
\begin{center}
\begin{psfrags}
 \psfrag{r}[]{$r$}
 \psfrag{eta.eq.0.05}[]{$\eta = 0.05$}
 \psfrag{2mr}[r][l][1][-90]{$\frac{2m}{r}$}
 \includegraphics[width=3.5in]{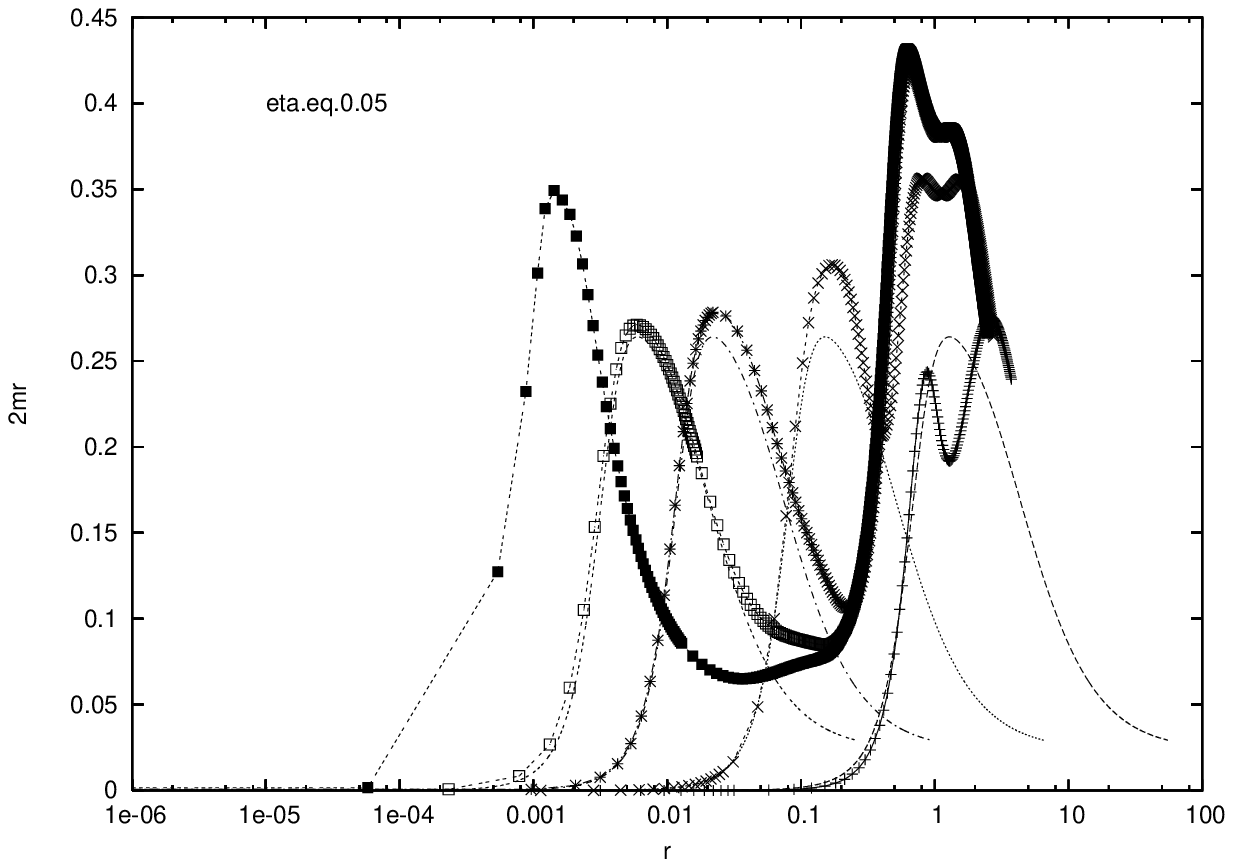}
\end{psfrags}
\end{center}
\caption{Same data as in Fig. \ref{fig::CSS_endstate_phi_0.05}.
This time $\frac{2m}{r}$ is plotted. Note that although growing in time,
$\frac{2m}{r}$ does not come close to one anywhere in the slice
before the culmination time is reached. This means that these initial data
lead to the formation of a naked
singularity.
}\label{fig::CSS_endstate_2mr_0.05}
\end{figure}

As a first step we tested, whether ``semi-strong'' (but otherwise arbitrary)
data really develop towards the ground state. 
For $\eta = 0.01$ and $\eta = 0.05$ we evolved a Gaussian
(\ref{eq::initial_data_gaussian})
with width $\sigma
= 1.0$ and center $r_0 = 5.0$ and chose an amplitude $p = 0.03$, which is  
neither too big (such that there is no black hole formation) 
nor too small (such that the data don't 
disperse), but otherwise arbitrary. What we find is, that
for such data (chosen rather arbitrarily) the solution evolves towards
the CSS ground state (see Figs.~\ref{fig::CSS_endstate_phi} and
\ref{fig::CSS_endstate_phi_0.05}) 
and stays there, until the lack of resolution
near the origin causes the numerical data to break away.
(The numerical evolution then does not represent a solution to the Einstein
equations anymore). Furthermore  
we find, that $\frac{2m}{r}$ stays away from 1 during the evolution.
(see Figs \ref{fig::CSS_endstate_2mr} and \ref{fig::CSS_endstate_2mr_0.05})
We can conclude from this, that indeed a naked CSS-singularity
is the generic end state of ``intermediately strong'' initial data.

Due to numerical difficulties, we were not able to determine the 
value of the coupling constant, from which on black hole formation 
is the only possible end state for ``strong''
initial data. Presumably it lies close to $\eta_0^* \simeq 0.069$, where the
CSS ground state has marginally trapped surfaces.
We can only state, that for $\eta \ge 0.09$ we detect black holes as
super-critical end states, which show the expected scaling, and that
for $\eta \ge 0.1$ the black hole masses show second order convergence.
(For details see Sec.~\ref{sec::smallcoupling}).

%Directory:/Dice/New_runs/eta.eq.0.01/CSS_Singularity_Big_router

%Directory:/Dice/New_runs/eta.eq.0.01/CSS_Singularity_Big_router

%Directory:/Dice/New_runs/eta.eq.0.05/CSS_Singularity

%Directory:/Dice/New_runs/eta.eq.0.05/CSS_Singularity

\afterpage{\clearpage}

\section{The first CSS excitation for $0.15 \lesssim \eta < 0.5$}
         \label{sec::CSS_large_eta}
%%%%%%%%%%%%%%%%%%%%%%%%%%%%%%%%%%%%%%%%%%%%%%%%%%%%%%%%%%%%%%%%%
In Section \ref{subsec::analytic_cont_CSS} we demonstrated, that the first
CSS excitation, if continued analytically beyond the past SSH, contains
marginally trapped surfaces for $\eta \gtrsim 0.152$. This fact might
prevent this solution to play the role of a critical solution between
dispersion and black hole formation for couplings $0.15 \lesssim \eta <
0.5$. 

In order to investigate this, we matched a certain class of asymptotically
flat data\footnote{These data are given in the following way: at the past
SSH the CSS solution is matched to a cubic polynomial such that the
resulting data are $C^2$. At some distance away from the past SSH the cubic
polynomial is matched to a Gaussian, the matching again being 
$C^2$. The two free parameters for these data are the location of the second
matching point and the width of the Gaussian, all the other parameters of
these data are used to achieve the required smoothness at the matching
points.} to the first CSS solution,
the matching point being the past SSH and the matching condition being such
that the resulting data were $C^2$. J. Thornburg evolved these data
(with several values for the parameters) numerically for $\eta = 0.2$ and 
found, that they developed an apparent horizon outside the past SSH. 
It is reasonable to assume, that these data will show the same behavior 
for couplings $0.2 < \eta < 0.5$. For couplings $0.15 < \eta < 0.2$ the
numerical evolution does not yield definite results, because at these
couplings it is easy to construct data, which do not form an apparent
horizon before the evolved data break away from the CSS solution 
due to numerical errors, which correspond to an excitation of the 
unstable mode. If one could eliminate (or diminish) these numerical errors,
it is likely, that alos for these couplings all data of the above
described class would form an apparent horizon. At this point
we cannot decide this.
 
If the behavior observed for $\eta = 0.2$ is generic, it is
clear that the CSS solution cannot be found by the means of a critical
search between dispersion and black hole formation, because the solution
itself (matched to asymptotically flat data) evolves to a black hole -- and
so would small perturbations independently of the sign of the admixture of
the unstable mode. In other words, the CSS solution 
does not lie at the boundary between dispersion and black hole formation.

Indeed, as stated in Sec.~\ref{sec::largecoupling} the first CSS solution
does not show up as a critical solution for $\eta \ge 0.2$. In the
transition region $0.15 \lesssim \eta \lesssim 0.18$ as described in
Sec.~\ref{sec::transitionregion}, the CSS solution appears in the ``CSS
episodes'' of near critical evolutions, but clearly it is not ``the critical
solution'', i.e.~the intermediate attractor, whose unstable mode is tuned
out by bisection.

\section{Critical Searches - Setup and Extraction of Results}
                 \label{sec::Crit_Search_Setup}
%%%%%%%%%%%%%%%%%%%%%%%%%%%%%%%%%%%%%%%%%%%%%%%%%%%%%%%%%%%%%%%%%%%%%

In order to investigate critical behavior we used several 
families of initial data, namely a ``Gaussian''
\be\label{eq::initial_data_gaussian}
\phi_0(r) = p \ r^2 e^{-(r - r_0)^2/\sigma^2},
\ee
with the center $r_0$ usually set to be $5.0$, the width $\sigma$
fixed to be either $1.0$ or $2.0$ and the amplitude $p$ serving as the
parameter. For large couplings also the family 
\be\label{eq::initial_data_4thpower}
\phi_0(r) = - 4 A r^2 \left(\frac{r - r_0}{p}\right)^3 
                e^{-(r-r_0)^4/p^4},
\ee
generated by keeping the center $r_0 = 5.0$ and the amplitude $A$ fixed
and
varying the width $p$ as the parameter 
was used (see \cite{Husa-Lechner-Puerrer-Thornburg-Aichelburg-2000-DSS}).

Furthermore we tried a ``double Gaussian'',
\be\label{eq::double_gaussian}
\phi_0(r) = p \ r^2 e^{-(r - r_0)^2/\sigma^2} + 
            A \ r^2 e^{-(r-r_2)^2/\sigma_2^2},
\ee
with the second Gaussian fixed ($A = 0.001$, $r_2 = 7.0$, $\sigma_2 = 0.5$),
width and center of the first Gaussian fixed, $ \sigma = 1.0$, $r_0 = 5.0$
and $p$ being the parameter.

For a fixed value of the parameter the initial data $\phi_0$ were 
evolved using the DICE code (see App.~\ref{app::dice}) 
until for large couplings
either a black hole formed (for the numerical criterion for a 
black hole see App.~\ref{app::dice}) or the field dispersed 
(most of the times measured via $\max\limits_{r} 2m/r $ less than 
some small value, e.g. $10^{-4}$, which might of course depend on the
coupling; see also remark below).
For very small couplings, where we expected naked singularities as 
super-critical end states, an evolution was defined to be super-critical, 
whenever the errors grew above some limit\footnote{Note: it is always a
great pleasure to declare the limitations in accuracy of a numerical code
a ``physical state''.}.

Starting with some value $p \in [p_{min},p_{max}]$, the interval chosen such
that $p_{min}$ leads to dispersion, while $p_{max}$ leads to a super-critical 
end
state, the parameter $p$ was driven towards $p^*$ by bisection:
for $p_1 > p^*$, the interval $[p_{min},p_1]$ was halved to give the new value
of the parameter
$p_2 = (p_{min} + p_1)/2$, at the same time the interval was reset
to $[p_{min}, p_{max} = p_1]$ and so on.
(This description applies if increasing the parameter makes the initial data
stronger, as is the case e.g. for $p$ being the amplitude of a Gaussian. For
$p$ being e.g.~the width of a Gaussian the parameter has to be decreased
to make the initial data stronger.)
Such a bisection search is limited by floating point errors, so
a critical search finished, when $(p_{N} - p_{N-1})/p_{N-1} < 10^{-14}$.

Given the result of a critical search, the critical value of the parameter
$p^*$ was approximated by $p^* = (p_{sub} + p_{super})/2$, where
$p_{sub}$ was the biggest sub-critical and $p_{super}$ the smallest
super-critical value obtained.

For large couplings, where the critical solution is DSS, the echoing period
$\Delta$ and the culmination time $u^*$ were determined simultaneously
from $(\max\limits_{r} \frac{2m }{r})(u)$. This function of time
reflects the periodicity of the DSS solution 
in logarithmic time $\tau = - \ln (u^* -u)$ 
(see Fig. \ref{fig::Max_2mr_eta.eq.100}). A perl script, written by Jonathan
Thornburg, extracted $\Delta/2$ and $u^*$ using the minima 
of $(\max\limits_{r} \frac{2m }{r})(u)$, at times $u_i$.
The times $\delta u_n$ elapsing between the two adjacent
minima at $u_n$ and $u_{n+1}$ are given by
$\delta u_n = e^{- (n-1) \Delta/2} \delta u_1$, a least squares fit
of $\ln(\delta u_n)$ to the straight line  
$-(n-1) \Delta + const$ gives the echoing period $\Delta/2$.
Furthermore for an exact DSS solution these times $\delta u_n$
sum up in a geometric series 
to give $u^* = u_n + \delta u_n/(1 - e^{-\Delta})$ (for any $n$), from which
$u^*$ can be calculated.

In order to examine the scaling of the black hole mass for super-critical
data and of the Ricci scalar at the axis for sub-critical data, 
a whole series of time evolutions was done, starting close to $p^*$
and increasing (decreasing) the parameter to $\log(|p-p^*|) \sim -10$,
with steps equally spaced in $\ln(|p-p^*|)$.
For the details of measuring the black hole mass see App.~\ref{app::dice}.
To extract the scaling exponent $\gamma$ from the black hole masses, a perl
script (written by Jonathan Thornburg) least squares fitted
$\ln m_{BH}(x)$ to the straight line $\gamma x + k$ with $x = \ln(p-p^*)$.
For an extraction of $\gamma$ from the scaling of the Ricci scalar
usually $\ln m_{BH}(x)$ was fitted to the straight line by naked eye.

\section{Critical Phenomena for Large Couplings}\label{sec::largecoupling}
%%%%%%%%%%%%%%%%%%%%%%%%%%%%%%%%%%%%%%%%%%%%%%%%%%%%%%%%%%%%%%%%%%%%%%%%%%

\begin{figure}[p]
\begin{center}
\begin{psfrags}
 \psfrag{-ln(u*-u)}[]{$-\ln (u^* - u)$}
 \psfrag{eta.eq.100}[l][r][1][0]{$\eta = 100$}
 \psfrag{max2mr}[r][l][1][-90]{$\max\limits_{r} \frac{2m}{r}$}
\includegraphics[width=3.5in]{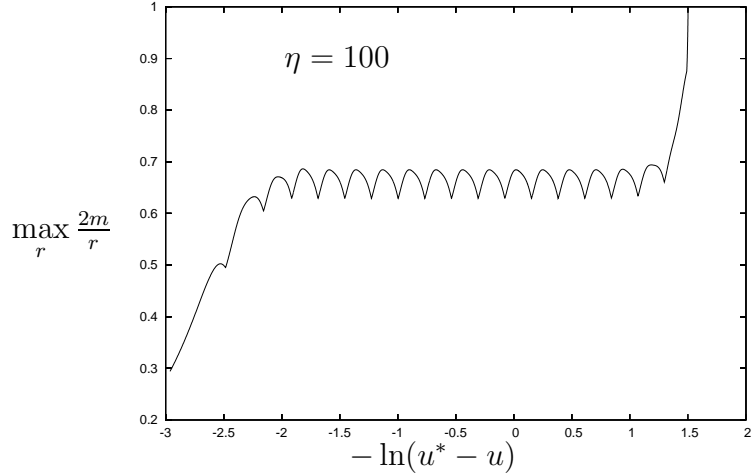}
\end{psfrags}
\end{center}
\caption{$\max\limits_{r} \frac{2m}{r}$ as a function of $\tau=\ln(u^*-u)$.
The echoing period $\Delta$ was computed to be 
$0.4599$. This is the easiest way to extract the echoing period from a near
critical solution.  
}\label{fig::Max_2mr_eta.eq.100}
\end{figure}

\begin{figure}[p]
\begin{center}
\begin{psfrags}
 \psfrag{-ln(u*-u)}[]{$\tau = -\ln (u^* - u)$}
 \psfrag{eta.eq.100}[lb][rc][1][0]{$\eta = 100$}
 \psfrag{R*exp(-2*tau)}[r][l][1][-90]{$\frac{2}{\eta} \mathcal R(\tau, 0) \ e^{-2
\tau}$}
\includegraphics[width=3.5in]{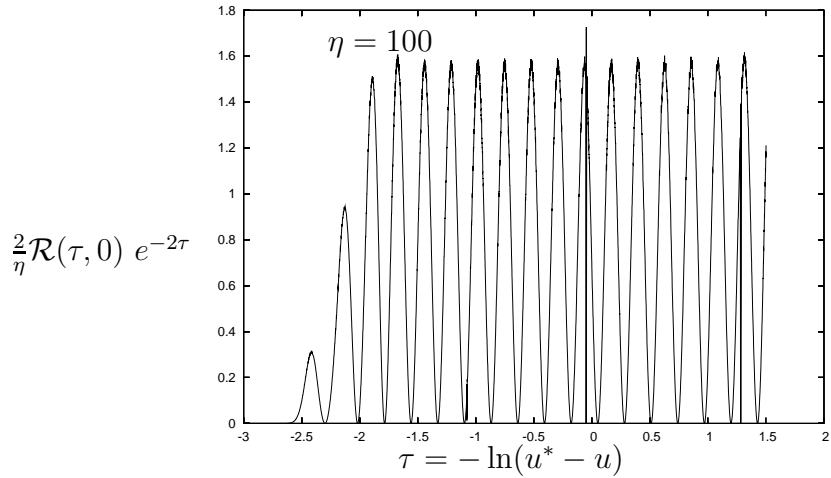}
\end{psfrags}
\end{center}
\caption{The Ricci scalar at the center of spherical symmetry as a function
of $\tau = - \ln (u^* - u)$.
As was discussed in Sec.~\ref{subsec::SSinSpherSymm} 
the Ricci scalar behaves like
$\mathcal R(\tau,z) = e^{-2 \tau} {\tilde R}(\tau,z)$, with 
$\tilde R$ being periodic
in $\tau$. (Remark: the vertical lines in the middle of the figure and near
the right end are errors, that occur at each grid refinement, but which do
not seem to influence the time evolution.)
}\label{fig::R_axis.eta.eq.100}
\end{figure}

\begin{figure}[p]
\begin{center}
\begin{psfrags}
 \psfrag{ln(p-pstar)}[]{$\ln (p - p^*)$}
 \psfrag{eta.eq.100}[lb][rc][1][0]{$\eta = 100$}
 \psfrag{ln(m)}[r][l][1][-90]{$\ln(m_{BH})$}
 \psfrag{ln(m) - f}[l][r][1][-90]{$\ln(m_{BH}) - f$}
\includegraphics[width=4.5in]{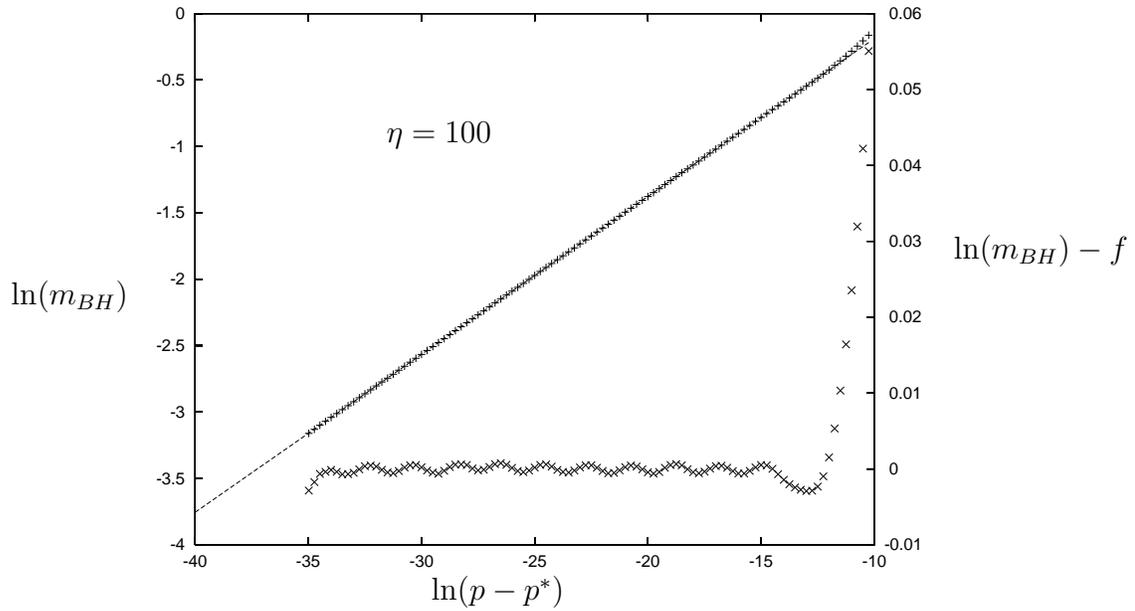}
\end{psfrags}
\end{center}
\caption{The scaling of the black hole mass $m_{BH}$. Plotted are the
masses obtained from a series of time evolutions (dots) together with the
straight line $f(\ln(p - p^*)) = \gamma \ln(p - p^*) + k$, where $\gamma =
0.1189$, versus the left axis. At the right axis the difference of these
functions is plotted. This reveals the ``fine structure'' of the mass
scaling, oscillations with period $\Delta/ 2 \gamma$. 
}\label{fig::Mass_scaling_eta.eq.100}
\end{figure}

\begin{figure}[p]
\begin{center}
\begin{psfrags}
 \psfrag{ln(p-pstar)}[]{$\ln (p^* - p)$}
 \psfrag{eta.eq.100}[lb][rc][1][0]{$\eta = 100$}
 \psfrag{lnR}[r][l][1][-90]{$\ln(\max\limits_{u} \mathcal R(u,0))$}
\includegraphics[width=3.5in]{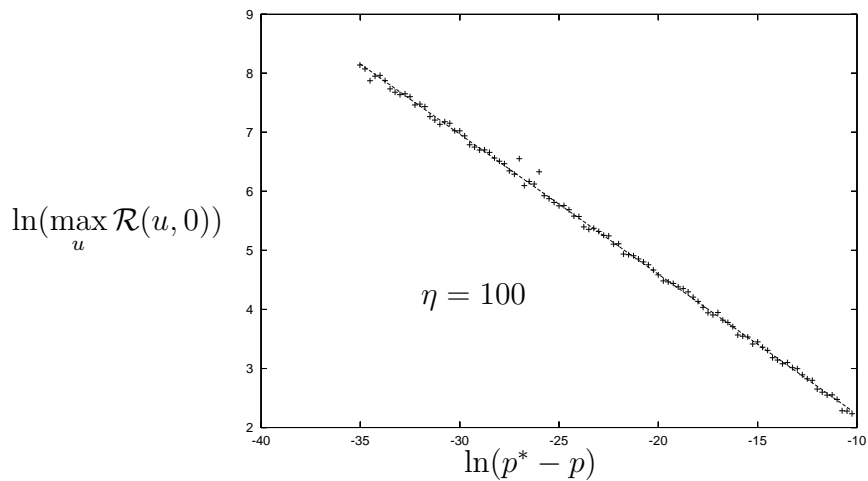}
\end{psfrags}
\end{center}
\caption{The scaling of $\max\limits_{u} \mathcal R(u,0)$ for sub 
critical evolutions
as a function of $\ln(p^* - p)$. The overlaid straight line has slope $-2
\gamma$. (Remark: the ``escaping'' 
points in the middle of the graph presumably stem
from a grid refinement.)
}\label{fig::Ricci_scaling_eta.eq.100}
\end{figure}

\begin{sidewaysfigure}[p]
\begin{center}
\begin{psfrags}
 \psfrag{lnr}[]{$\ln (r)$}
 \psfrag{eta.eq.100}[]{$\eta = 100$}
 \psfrag{phi}[r][l][1][-90]{$\phi_{crit}, \phi_{DSS}$}
\includegraphics[width=8in]{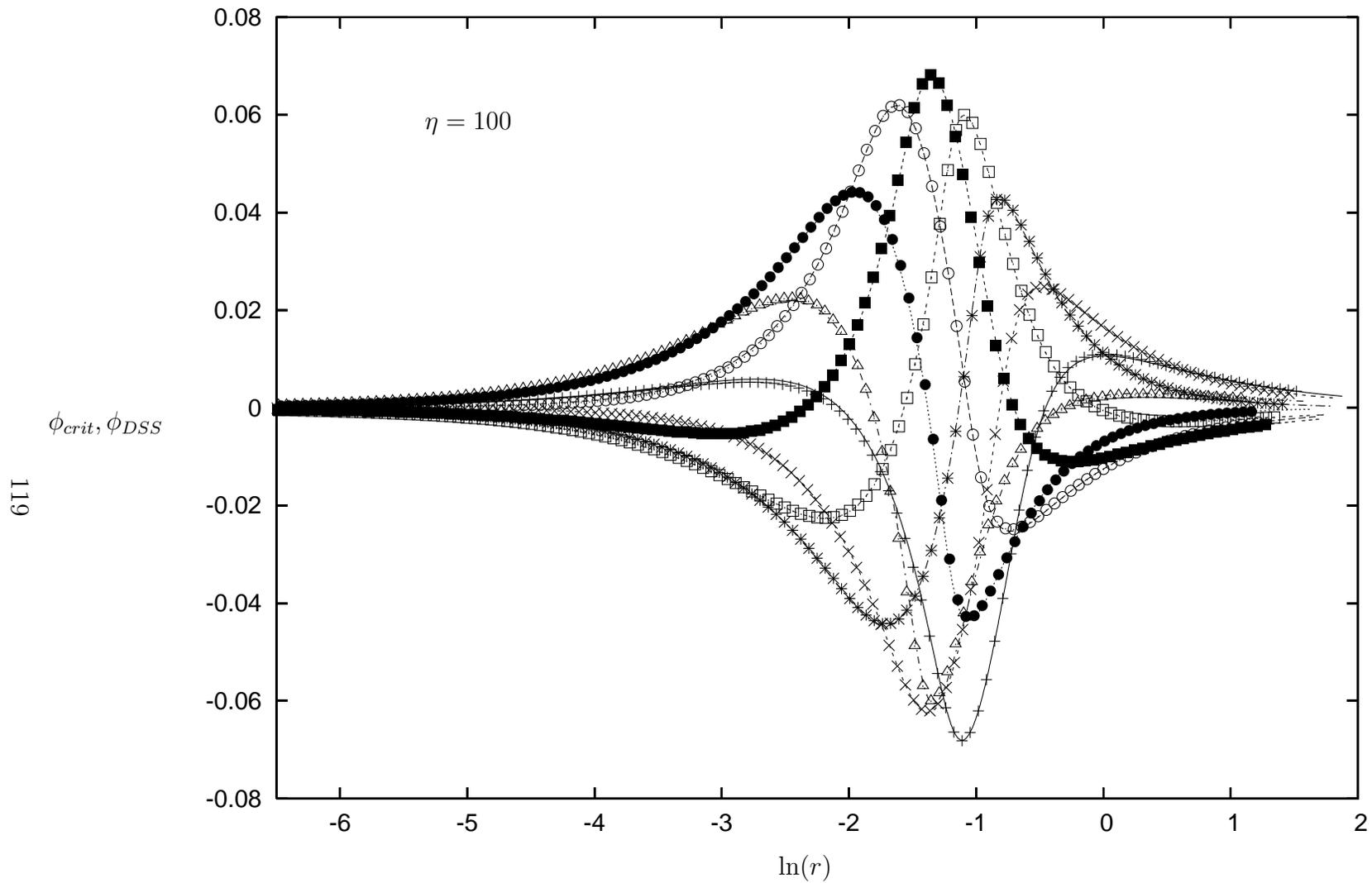}
\end{psfrags}
\end{center}
\caption{Snapshots of the critical solution $\phi_{crit}$ (solid lines)
at $\eta = 100$ compared to the DSS solution $\phi_{DSS}$ (dots).
The snapshots are taken at times $\tau_i = i \Delta/N$ for $N = 16$ and
$i=0,2,4,6,8,10,12,14$, i.e. spanning one period. 
}\label{fig::Crit_sol_DSS_eta.eq.100}
\end{sidewaysfigure}

For large couplings $0.2 \le \eta < 100$, we find that the critical solution
at the threshold of black hole formation is discretely self-similar
(See \cite{Husa-Lechner-Puerrer-Thornburg-Aichelburg-2000-DSS}).
Figs.~\ref{fig::Max_2mr_eta.eq.100} -- 
\ref{fig::Crit_sol_DSS_eta.eq.100} illustrate this for $\eta = 100$.
All the runs for these figures were done with 
the family (\ref{eq::initial_data_gaussian}) (width $\sigma = 1.0$), 
with $r_{outer} = 30.0$
as the initial spatial extension of the grid and with $N = 2000$ grid points
initially.

\begin{sidewaysfigure}[p]
\begin{center}
\begin{psfrags}
 \psfrag{lnr}[]{$\ln (r)$}
 \psfrag{eta.eq.0.2933}[]{$\eta = 0.2933$}
 \psfrag{Delta}[c][c][1][0]{$\Delta$}
 \psfrag{phi}[r][l][1][-90]{$\phi_{crit}, \phi_{DSS}$}
\includegraphics[width=8in]{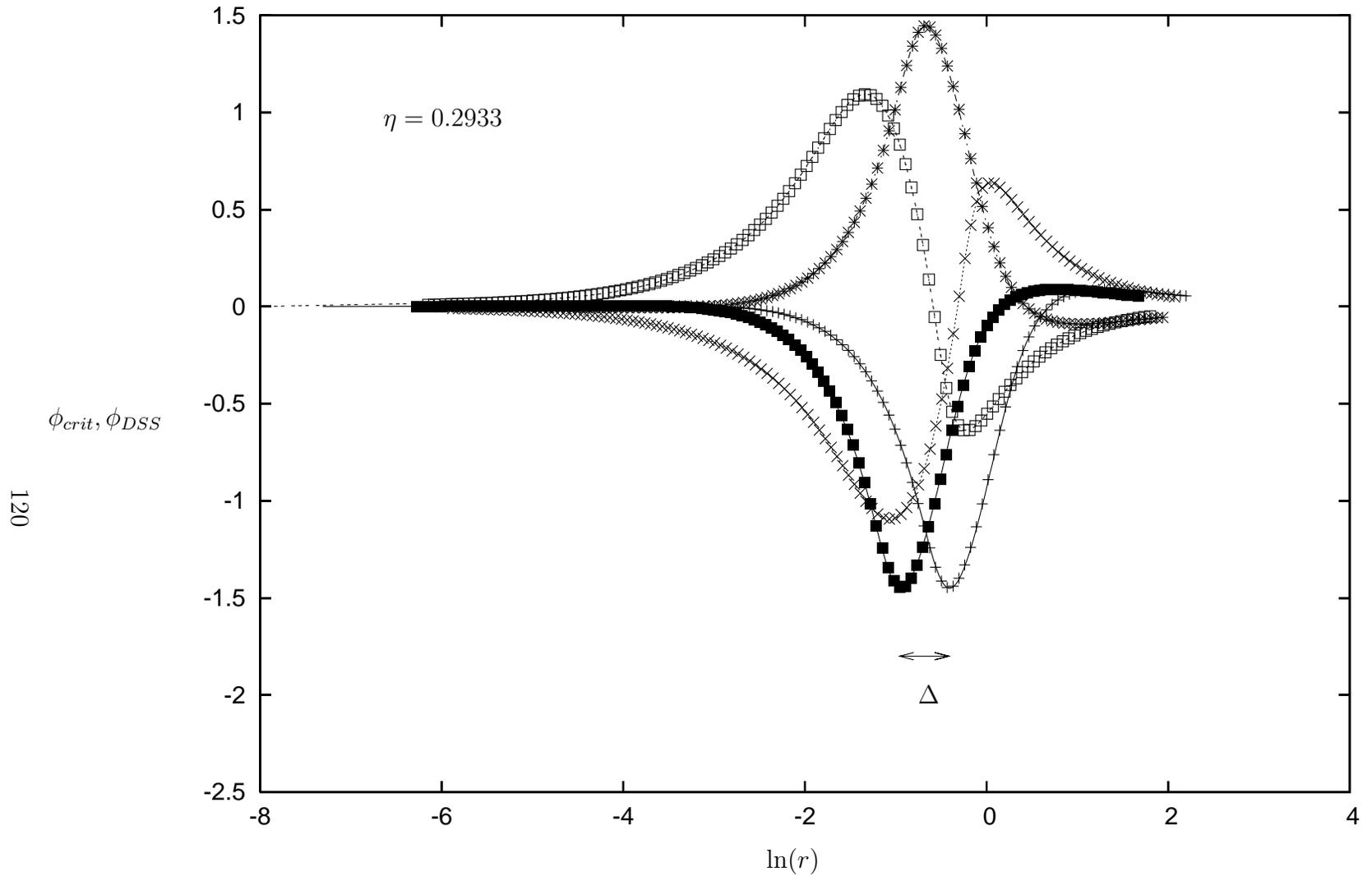}
\end{psfrags}
\end{center}
\caption{Snapshots of the critical solution $\phi_{crit}$ (solid lines)
at $\eta = 0.2933$ compared to the DSS solution $\phi_{DSS}$ (dots).
Snapshots are taken at times $\tau_i = i \Delta/N$ for $N=64$ and $i =
0,16,32,48,64$. Note that for $\tau = \Delta$ the critical solution
retains its shape at $\tau = 0$, but is shifted in $\ln (r)$ by $-\Delta =
-0.5399$.  
}\label{fig::Crit_sol_DSS_eta.eq.0.2933}
\end{sidewaysfigure}

Fig. \ref{fig::Max_2mr_eta.eq.100} shows
$\max\limits_{r}2m/r$ for $\eta = 100$, 
which if plotted vs. $ - \ln(u^* - u) = \tau$ is a periodic function 
with period  $\Delta/2$. 
(As the metric functions $\beta$ and $\Vr$ are periodic with
period $\Delta/2$, $\max\limits_r 2m/r$ shows the same periodicity.) 
Fig. \ref{fig::R_axis.eta.eq.100} shows the Ricci scalar at the origin 
$r=0$, $\mathcal R(u,r=0)$, which behaves 
as $e^{2 \tau} \tilde R(\tau)$, where $\tilde R$ denotes a periodic
function of its argument.
Figs.~\ref{fig::Mass_scaling_eta.eq.100} and 
\ref{fig::Ricci_scaling_eta.eq.100} show the scaling of the black hole mass
for super-critical initial data
and $\max\limits_u \mathcal R(u,r=0)$ for sub-critical data respectively.
Fig. \ref{fig::Mass_scaling_eta.eq.100} also shows the superimposed
wiggles in the mass scaling.
Finally Figs.~\ref{fig::Crit_sol_DSS_eta.eq.100} and 
\ref{fig::Crit_sol_DSS_eta.eq.0.2933} show that a near critical 
evolution comes close to the DSS solution
at intermediate times for $\eta = 100$ and $\eta = 0.2933$ respectively. 
Compared are the field $\phi$ 
as evolved from near critical initial data, and the DSS solution
itself, constructed as described in Sec.~\ref{sec::DSSsolutions}.

%\afterpage{\clearpage}

%Directory:Dice/new_runs/eta.eq.100/Critical_Solution

%Directory:Dice/new_runs/eta.eq.100/Critical_Solution

%Directory:Dice/new_runs/eta.eq.100/Scaling_Super/

%Directory:Dice/new_runs/eta.eq.100/Scaling_Sub/

%Directory:DSS/src/Compare_to_dice/eta.eq.100

%Directory:/DSS/src/Convergence_tests/64/

%Directory:/DSS/src/Figures
\begin{sidewaysfigure}[p]
\begin{center}
\begin{psfrags}
 \psfrag{eta}[]{$\eta$}
 \psfrag{Delta}[r][l][1][-90]{$\Delta_{crit}, \Delta_{DSS}$}
\includegraphics[width=8in]{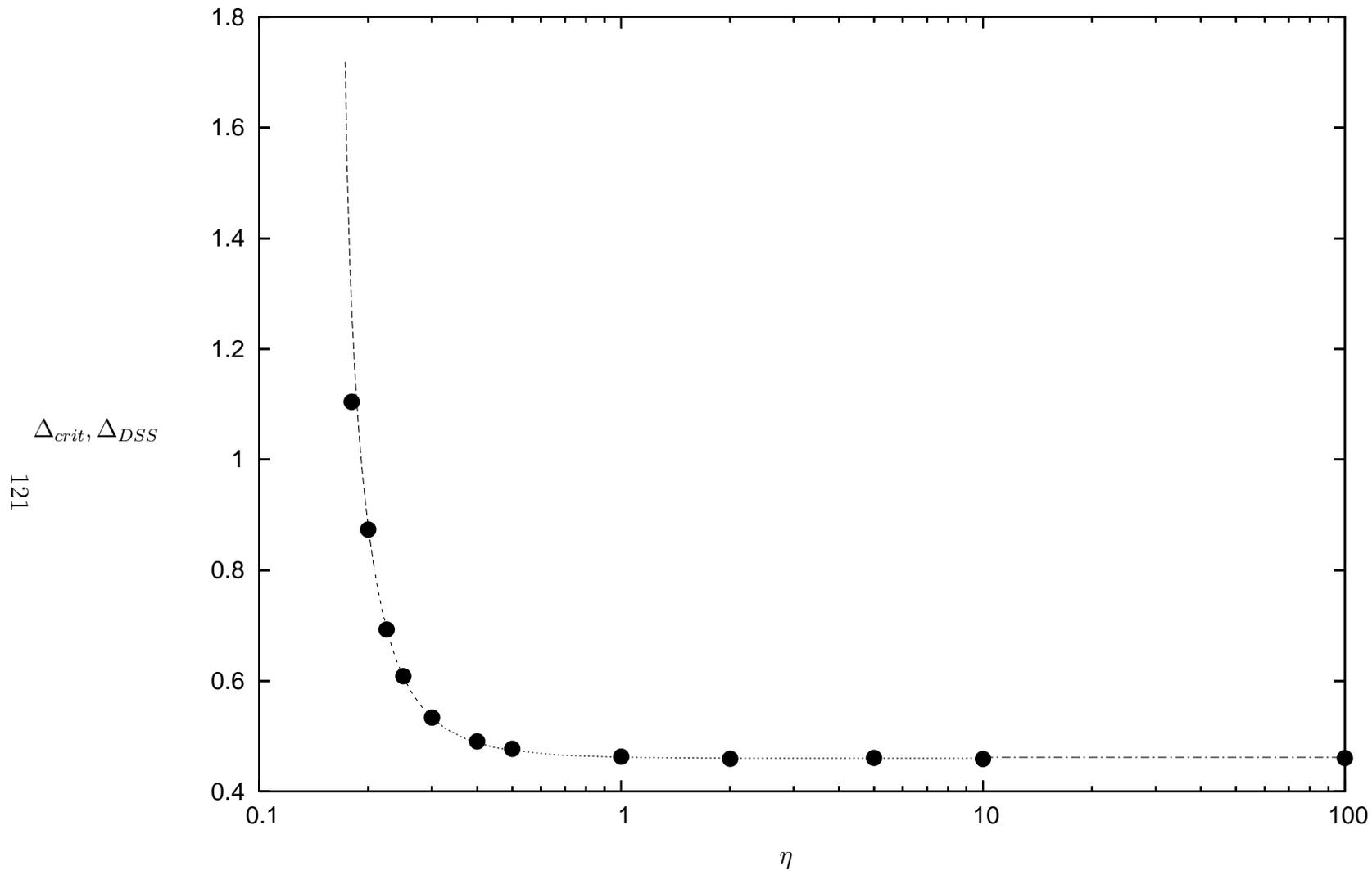}
\end{psfrags}
\end{center}
\caption{The echoing period $\Delta_{crit}$ (large solid dots)
as a function of the coupling constant $\eta$, compared to the period
$\Delta_{DSS}$ of the DSS solution (see Fig. \ref{fig::DSS_Period}).
(The values of $\Delta_{crit}$ are taken from
\cite{Husa-Lechner-Puerrer-Thornburg-Aichelburg-2000-DSS}.)
}\label{fig::Crit_sol_Delta_eta}
\end{sidewaysfigure}

\begin{table}[h]
\begin{center}
\begin{tabular}{cccccccc}
   & \multicolumn{3}{c}{Initial Data
      Family~\ref{eq::initial_data_gaussian}}
   & \multicolumn{4}{c}{Initial Data Family~\ref{eq::initial_data_4thpower}}
                                                                        \\
   & \multicolumn{3}{c}{Parameter is $A$}
   & \multicolumn{4}{c}{Parameter is $\sigma$}
                                                                        \\
\cline{2-4}\cline{5-8}
   $\ccbeta$
        & $A^*$                 & $\Delta/2$    & $\gamma$
        & $A$   & $\sigma^*$    & $\Delta/2$    & $\gamma$
                                                                        \\
\hline %----------------------------------------------------------------
\begin{tabular}{@{}r@{}l@{}}
  0&.18         \\
  0&.2          \\
  0&.225        \\
  0&.25         \\
  0&.3          \\
  0&.4          \\
  0&.5          \\
  1&            \\
  2&            \\
  5&            \\
 10&            \\
100&            %%%\\
\end{tabular}
%
% Gamssian initial data
% p* = A family
%
                % data from
                %    fgrep 16000 p_star.amp.dat
                % rounded to 8 significant figures to match table II
                & \begin{tabular}{@{}l@{}}
                  0.019\,523\,015   \\
                  0.018\,942\,512   \\
                  0.018\,241\,056   \\
                  0.017\,578\,042   \\
                  0.016\,392\,639   \\
                  0.014\,534\,866   \\
                  0.013\,167\,548   \\
                  0.009\,528\,975\,1  \\
                  0.006\,809\,778\,3  \\
                  0.004\,333\,205\,6  \\
                  0.003\,070\,144\,2  \\
                  0.000\,972\,589\,54 %%\\
                  \end{tabular}
                                % data from Delta.amp.16000.dat
                                % rounded to 4 significant figures
                                % based on the typical differences between
                                % values from different resolutions
                                & \begin{tabular}{@{}l@{}}
                                  0.5522        \\
                                  0.4367        \\
                                  0.3464        \\
                                  0.3043        \\
                                  0.2668        \\
                                  0.2452        \\
                                  0.2386        \\
                                  0.2314        \\
                                  0.2295        \\
                                  0.2304        \\
                                  0.2293        \\
                                  0.2302        %%%\\
                                  \end{tabular}
                                                % data from
                                                % gamma.amp.16000.dat
                                                % rounded to 4 significant
                                                % figures based on the
                                                % typical
                                                % differences between values
                                                % from different resolutions
                                                & \begin{tabular}{@{}l@{}}
                                                  0.1063      \\
                                                  0.1091     \\
                                                  0.1207     \\
                                                  0.1173     \\
                                                  0.1152     \\
                                                  0.1132     \\
                                                  0.1152     \\
                                                  0.1163     \\
                                                  0.1179     \\
                                                  0.1183      \\
                                                  0.1186     \\
                                                  0.1187     %%%\\
                                                  \end{tabular}
%
% derivative of 4th-power pseudo-Gamssian initial data
% p* = sigma at varying A family
%
        % data from
        %  fgrep -e -amp ../*=ccbeta/d4g-width=p/8000/doit.run \
        %     | cut -c4- | sort -n
        & \begin{tabular}{@{}l@{}}
          0.003         \\
          0.002         \\
          0.002         \\
          0.002         \\
          0.002         \\
          0.002         \\
          0.0015        \\
          0.0015        \\
          0.0010        \\
          0.0005        \\
          0.0005        \\
          0.0001        %%%\\
          \end{tabular}
                % data from
                %    fgrep 8000 p_star.d4g-width.dat
                % rounded to 8 significant figures
                & \begin{tabular}{@{}l@{}}
                  1.083\,153\,54        \\
                  0.615\,317\,49        \\
                  0.651\,519\,42        \\
                  0.688\,851\,73        \\
                  0.766\,003\,44        \\
                  0.929\,746\,89        \\
                  0.707\,335\,37        \\
                  1.210\,138\,07        \\
                  1.064\,744\,72        \\
                  0.734\,344\,76        \\
                  1.318\,800\,46        \\
                  0.631\,472\,58        %%%\\
                  \end{tabular}
                                % data from Delta.d4g-width.8000.dat
                                % rounded to 4 significant figures
                                & \begin{tabular}{@{}l@{}}
                                  0.5478        \\
                                  0.4327        \\
                                  0.3472        \\
                                  0.3046        \\
                                  0.2675        \\
                                  0.2445        \\
                                  0.2386        \\
                                  0.2313        \\
                                  0.2305        \\
                                  0.2308        \\
                                  0.2312        \\
                                  0.2311        %%%\\
                                  \end{tabular}
                                                % data from
                                                % gamma.d4g-width.8000.dat
                                                & \begin{tabular}{@{}l@{}}
                                                  0.1028        \\
                                                  0.1150        \\
                                                  0.1169        \\
                                                  0.1173        \\
                                                  0.1146        \\
                                                  0.1139        \\
                                                  0.1130        \\
                                                  0.1155        \\
                                                  0.1167        \\
                                                  0.1178        \\
                                                  0.1182        \\
                                                  0.1182        %%%\\
                                                  \end{tabular}
\end{tabular}
\end{center}
\caption{%%%
        This table shows two families of near-critical initial data
        parameters for various coupling constants $\ccbeta$.
        For the Gaussian-like initial data
        family~\protect\ref{eq::initial_data_gaussian}, we use the
        `amplitude''~$A$ as the parameter~$p$ (at a fixed
        `width''~$\sigma = 1$), with a numerical grid of
        \hbox{16\,000} grid points.
        For the family~\protect\ref{eq::initial_data_4thpower},
        we use the `width''~$\sigma$ as the parameter~$p$
        (with different `amplitudes''~$A$ for different
        coupling constants), with 8000 grid points.
        For each coupling constant and each family, the table also
        shows the $\max\,\, 2m/r$ echoing period $\Delta/2$ of the
        near-critical evolution, and the mass-scaling-law critical
        exponent $\gamma$ determined for the entire critical search.
        For $\eta = 0.18$ the DSS symmetry is only approximate
        (see Sec. \ref{sec::transitionregion} for details).
        This table is taken from 
        \cite{Husa-Lechner-Puerrer-Thornburg-Aichelburg-2000-DSS}. 
        All the runs for this
        table were done by J. Thornburg.
        }%%%
\label{tab::large_couplings}
\end{table}

Table \ref{tab::large_couplings} (taken from 
\cite{Husa-Lechner-Puerrer-Thornburg-Aichelburg-2000-DSS})
gives the scaling exponent $\gamma$, the
echoing period $\Delta/2$ and the value of the critical parameter
for families (\ref{eq::initial_data_gaussian}) and 
(\ref{eq::initial_data_4thpower}) for various values of the coupling constant
$\eta$. 
As can be seen from this table,
for very large couplings $\eta = 100$, the scaling exponent 
$\gamma \simeq 0.1185$
and the echoing period $\Delta \simeq 0.461$ 
are in very good agreement with the
results for $\eta = \infty$ reported in
\cite{Liebling-inside-global-monopoles} 
($\gamma = 0.119$, $\Delta = 0.46$).
When decreasing the coupling, the scaling exponent $\gamma$ hardly changes
(the variation is at most $5 \%$) whereas the echoing period $\Delta$
starts to increase at lower couplings, which is in good agreement 
with the results presented in 
Sec.~\ref{subsec::DSScode_results}. Indeed, as can be seen
from Fig. \ref{fig::Crit_sol_Delta_eta}, 
the echoing period $\Delta_{crit}$ of near critical
evolutions matches the period $\Delta_{DSS}$ 
of the ``exact'' DSS solution, as described 
in Sec.~\ref{subsec::DSScode_results} for $\eta \ge 0.2$.

For $\eta = 0.18$, although the DSS solution still exists, the 
observed DSS periodicity for near critical evolutions is only 
approximate.
This phenomenon will be described in more detail in
Sec.~\ref{sec::transitionregion}.

\section{Critical Phenomena for Small Couplings}\label{sec::smallcoupling}
%%%%%%%%%%%%%%%%%%%%%%%%%%%%%%%%%%%%%%%%%%%%%%%%%%%%%%%%%%%%%%%%%%%%%%%%%%

For small couplings we investigated critical phenomena between dispersion
and singularity formation ($0 \le \eta \lesssim 0.07$) and between 
dispersion and black hole formation ($0.07 \lesssim \eta \lesssim 0.14$).
We find that the critical phenomena in this range of couplings is governed
by the first CSS excitation described in Sec.~\ref{sec::CSSsolutions}.

For the couplings where black holes form ($0.07 \lesssim \eta$) the black
hole mass scales according to (\ref{eq::scaling_gen}) with a scaling
exponent $\gamma$ that corresponds to $1/\lambda_{CSS}$, the relative error
being at most $3 \ \%$.

For couplings where the end state is the CSS ground state ($\eta \lesssim
0.07$) the quantity $\max\limits_u \mathcal R(u,0)$ for sub-critical data
exhibits scaling according to (\ref{eq::Ricci_scaling}), although only for 
small values of $(p - p^*)$. Furthermore this time the scaling exponent
$\gamma$ differs from the theoretical prediction $1/\lambda_{CSS}$ by up to
$15 \ \%$. The reason for this inaccuracy is not clear to us at the moment.
One possible reason could be, that for these small couplings $\ln ((p -
p^*))/p^* < 10^{-14}$, which limits the determination of $p^*$ due to
floating point round off errors, does not allow to reduce the 
admixture of the unstable mode of the CSS solution in the initial data 
as much as is the case
for large couplings, where we observe a beautiful scaling law
(Sec.~\ref{sec::largecoupling}). Whether this is indeed the reason could be
checked by switching to higher numerical precision. 
Another possible reason might
be, that the code does not work as accurately for small couplings as it does
for large couplings. Convergence tests would be a first check of this.

%Directory:eta.eq.0.11/1_Width/Scaling_results/
\begin{figure}[p]
\begin{center}
\begin{psfrags}
 \psfrag{ln(p - p*)}[]{$\ln(p-p^*(N))$}
 \psfrag{eta.eq.0.11}[]{$\eta = 0.11$}
 \psfrag{ln (m_BH)}[r][l][1][-90]{$\ln (m_{BH})$}
\includegraphics[width=3.5in]{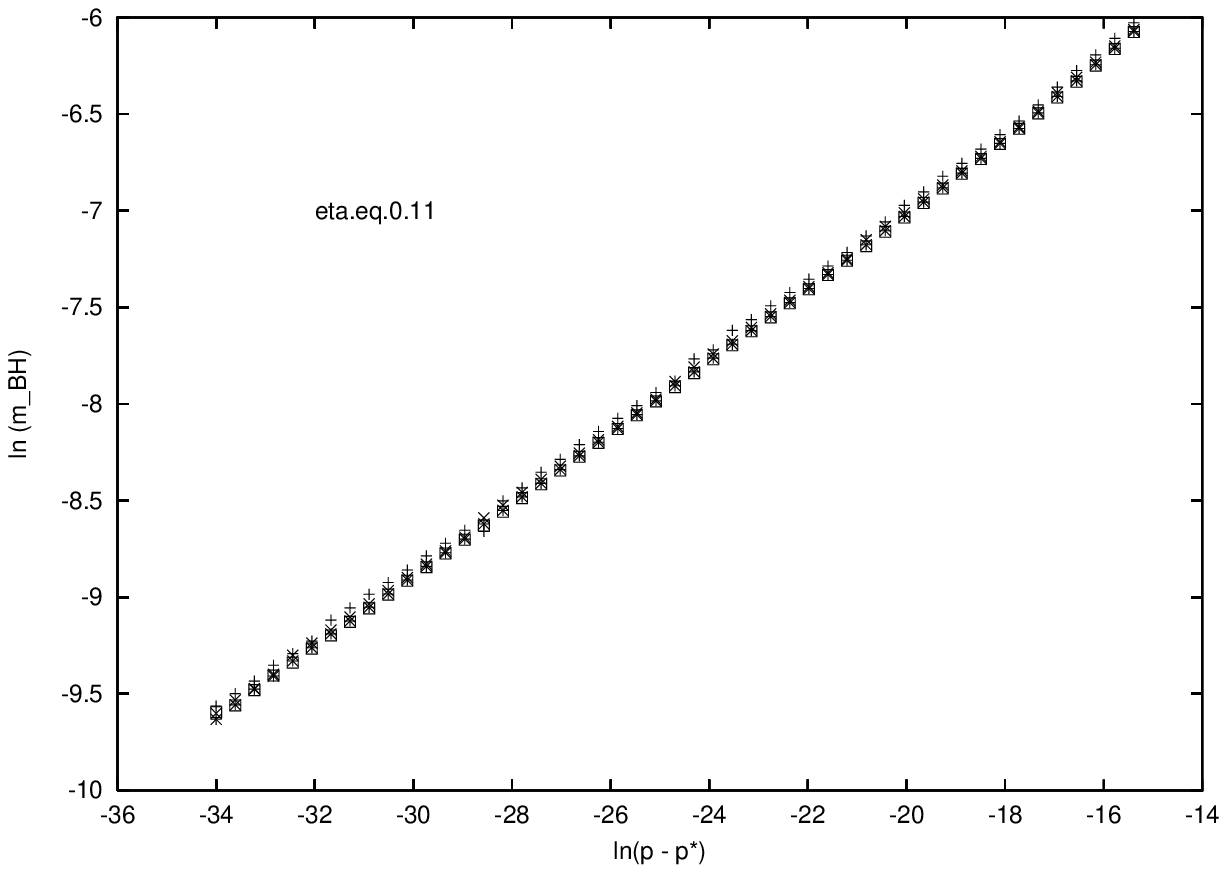}
\end{psfrags}
\end{center}
\caption{The mass scaling for $\eta = 0.11$ for the family
(\ref{eq::initial_data_gaussian}) (with fixed width $\sigma = 1$ and variable
amplitude). Different symbols denote different grid resolutions:
  ``+'' denote $N=500$ grid points initially, ``x'' $N=1000$, 
``$\divideontimes$''  $N=2000$ and ``{\footnotesize{$\square$}}'' 
 $N=4000$. The straight line has a slope of $\gamma \simeq 0.185$, which is
in good agreement with $1/\lambda_{1} = 0.181$. (The relative error is
$2 \%$.)
}\label{fig::mass_scaling_eta.eq.0.11}
\end{figure}

%Directory:eta.eq.0.11/1_Width/Scaling_results/
\begin{figure}[p]
\begin{center}
\begin{psfrags}
 \psfrag{ln (p - p*)}[]{$\ln(p-p^*(N))$}
 \psfrag{eta.eq.0.11}[]{$\eta = 0.11$}
 \psfrag{ln (m_BH)}[r][l][1][-90]
               { ${m_{BH}(1000) - m_{BH}(2000)\atop
                   4 \ \left(m_{BH}(2000) - m_{BH}(4000)\right)}$}
\includegraphics[width=3.5in]{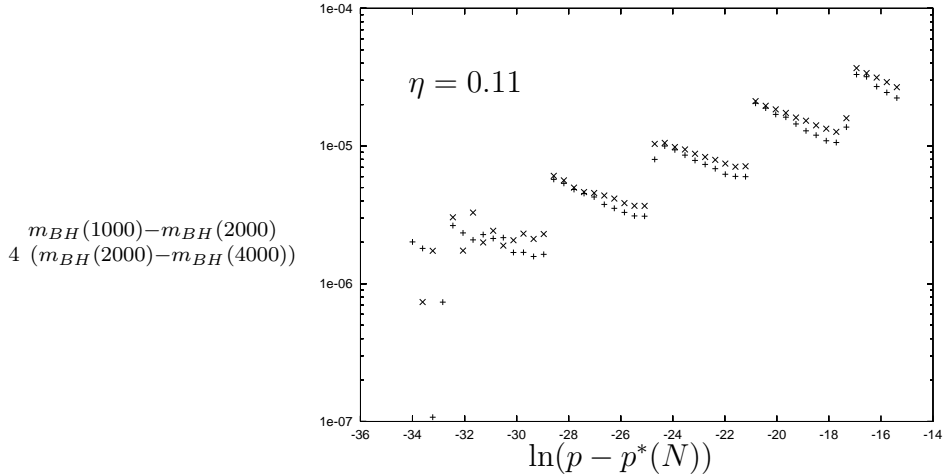}
\end{psfrags}
\end{center}
\caption{The differences of black hole masses of 
  Fig.~\ref{fig::mass_scaling_eta.eq.0.11} for different initial grid
  resolutions: ``+'' denotes the difference $(m_{BH})_{1000} -
  (m_{BH})_{2000}$,
  ``x'' denotes four times the difference $(m_{BH})_{2000} -
(m_{BH})_{4000}$. These quantities almost lie on top of each other, which
shows second order convergence of the black hole mass with the grid
resolution (see Sec.~\ref{app::convergence}).  
}\label{fig::mass_convergence_eta.eq.0.11}
\end{figure}

%Directory:eta.eq.0.11/1_Width/Scaling_results/
\begin{figure}[p]
\begin{center}
\begin{psfrags}
 \psfrag{ln \(p*-p\)}[]{$\ln(p^*[N]-p)$}
 \psfrag{eta.eq.0.11}[]{$\eta = 0.11$}
 \psfrag{ln \(max R\(u,0\)\)}[r][l][1][-90]{$\ln (\max\limits_u \mathcal
R(u,0))$}
\includegraphics[width=3.5in]{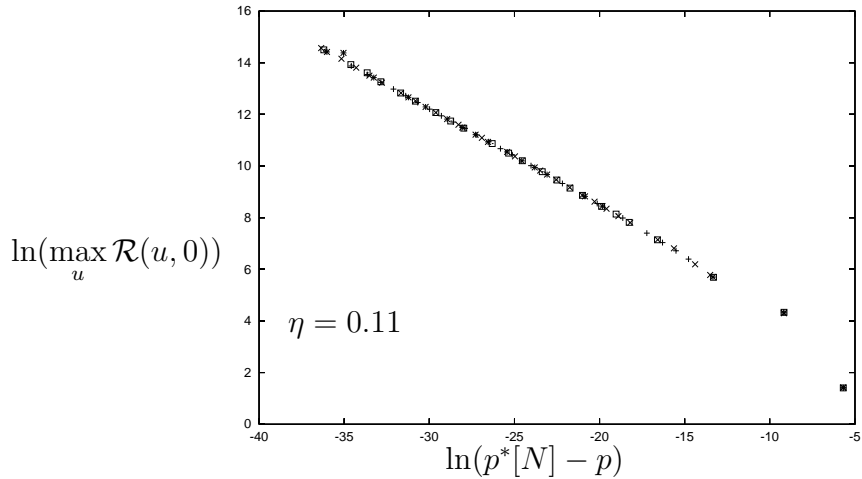}
\end{psfrags}
\end{center}
\caption{The quantity $\ln (\max\limits_u \mathcal R(u,0))$ for sub-critical
data of the same family as in Fig.~\ref{fig::mass_scaling_eta.eq.0.11}
and different grid resolutions: 
``+'' denote $N=500$, ``x'' $N=1000$, 
``$\divideontimes$'' $N=2000$ and ``{\footnotesize{$\square$}}'' 
$N=4000$. The straight line has a slope of $- 2 \gamma$ with $\gamma \simeq
0.185$. 
}\label{fig::ricci_scaling_sub_eta.eq.0.11}
\end{figure}

%Directory:eta.eq.0.11/1_Width/4000_Crit_Search/
\begin{figure}[p]
\begin{center}
\begin{psfrags}
 \psfrag{r}[]{$r$}
 \psfrag{eta.eq.0.1}[]{$\eta = 0.1$}
 \psfrag{phi}[r][l][1][-90]{$\phi$}
\includegraphics[width=3.5in]{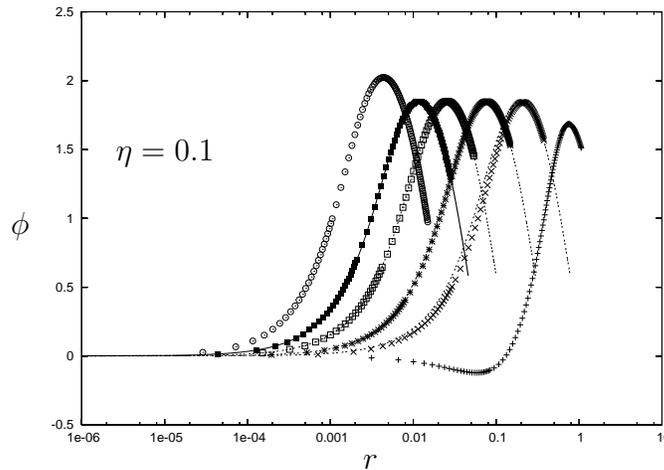}
\end{psfrags}
\end{center}
\caption{The field $\phi$ of a near critical evolution (dots) 
at intermediate times between $u = 12.723$ 
and $u=15.150$ (moving from right to left). The field approaches the CSS
solution (solid lines). The culmination time $u^*$ (determined from the last
but second  snapshot) is  $u^* = 15.167$. The past SSH of the CSS solution
is located where the CSS solution attains the value $\pi/2$ for the second
time. 
}\label{fig::crit_sol_eta.eq.0.1}
\end{figure}

%Directory:eta.eq.0.1/1_Width/New_dice/Critical_solution
\begin{figure}[t]
\begin{center}
\begin{psfrags}
 \psfrag{z}[]{$\ln(z)$}
 \psfrag{eta.eq.0.1}[]{$\eta = 0.1$}
 \psfrag{delta_phi}[r][l][1][-90]{$\delta \phi$}
\includegraphics[width=5in]{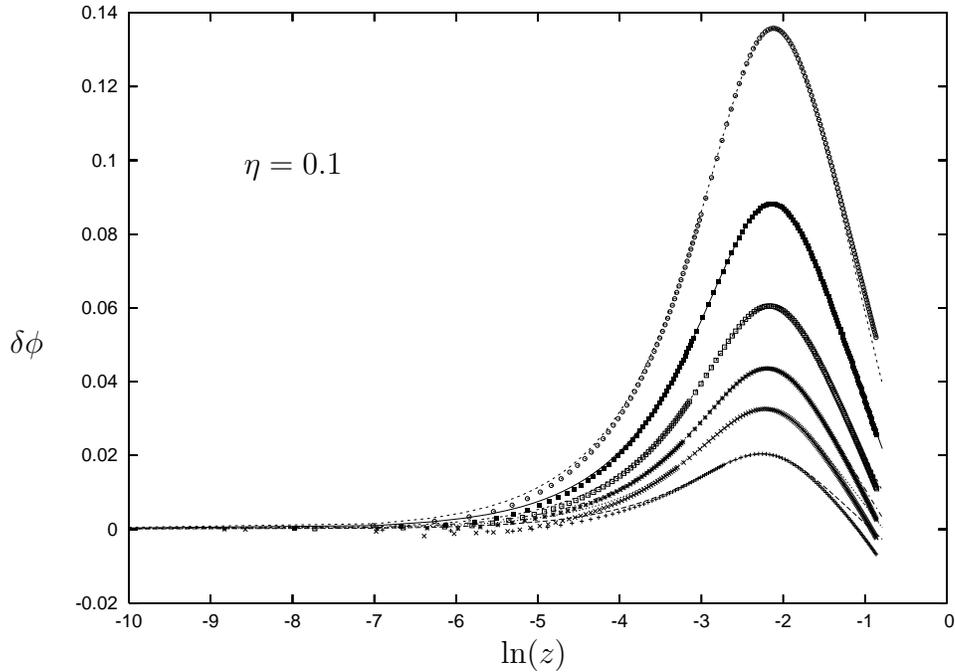}
\end{psfrags}
\end{center}
\caption{The deviation of the critical solution from the CSS solution for
several time steps, $\tau =  3.4169, \ 3.539, \
3.6069, \ 3.6791, 3.756, \ 3.838.$
Plotted is the difference between the evolved $\phi$ and the CSS solution
$\phi_{CSS}$, i.e. $\delta \phi(\tau, z) = \phi(\tau,z) - \phi_{CSS}(z)$
(dots). Overlaid is the
function $a(\tau) (y_{unst}(z) + b(\tau) y_{gauge}(z))$ (solid lines), 
where the parameters $a(\tau)$ and $b(\tau)$ were fitted with bare eye such
that the maxima of the two functions agreed. 
The gauge mode has to be taken into account, because of the error in
determining $u^*$. These parameters should depend on time $\tau$ as 
$a(\tau) = a_0 \ e^{\lambda\tau}$ and 
$b(\tau) = b_0 \ e^{(1 - \lambda) \tau}$. So from the fit the
eigenvalue $\lambda$ could be determined. 
From Fig. \ref{fig::Deviation_of_critsol_pars_eta.eq.0.1} one sees, that
the value of $\lambda$ computed from $a(\tau)$ is close to the 
``exact'' value of $5.5846$. Unfortunately $b(\tau)$ is not so well behaved
(see Fig.~\ref{fig::Deviation_of_critsol_pars_b_eta.eq.0.1}).
}\label{fig::Deviation_of_critsol_eta.eq.0.1}
\end{figure}

%Directory:eta.eq.0.1/1_Width/New_dice/Critical_solution
\begin{figure}[p]
\begin{center}
\begin{psfrags}
 \psfrag{tau}[]{$\tau$}
 \psfrag{eta.eq.0.1}[]{$\eta = 0.1$}
 \psfrag{lna}[r][l][1][-90]{$\ln (a(\tau))$}
\includegraphics[width=3.5in]{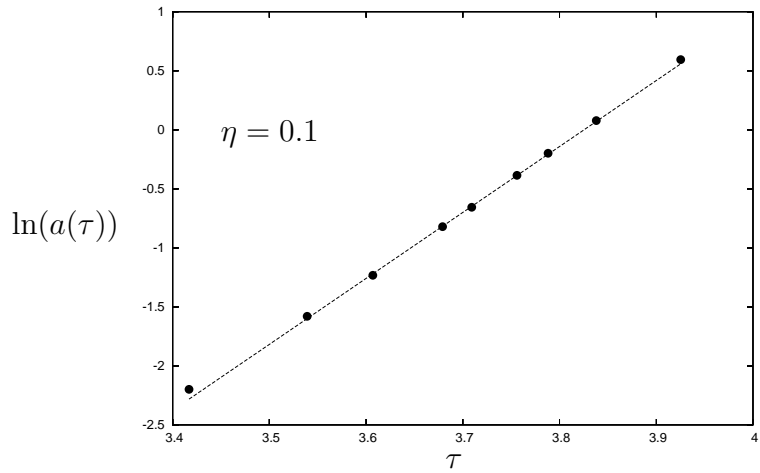}
\end{psfrags}
\end{center}
\caption{The parameter $a$ of Fig.  
\ref{fig::Deviation_of_critsol_pars_eta.eq.0.1}
in dependence of $\tau$ (dots). The solid line has a slope of $5.5846$,
corresponding to the eigenvalue $\lambda$ computed from the perturbation
analysis.
}\label{fig::Deviation_of_critsol_pars_eta.eq.0.1}
\end{figure}

%Directory:eta.eq.0.1/1_Width/New_dice/Critical_solution
\begin{figure}[p]
\begin{center}
\begin{psfrags}
 \psfrag{tau}[]{$\tau$}
 \psfrag{eta.eq.0.1}[]{$\eta = 0.1$}
 \psfrag{lnb}[r][l][1][-90]{$\ln (b(\tau))$}
\includegraphics[width=3.5in]{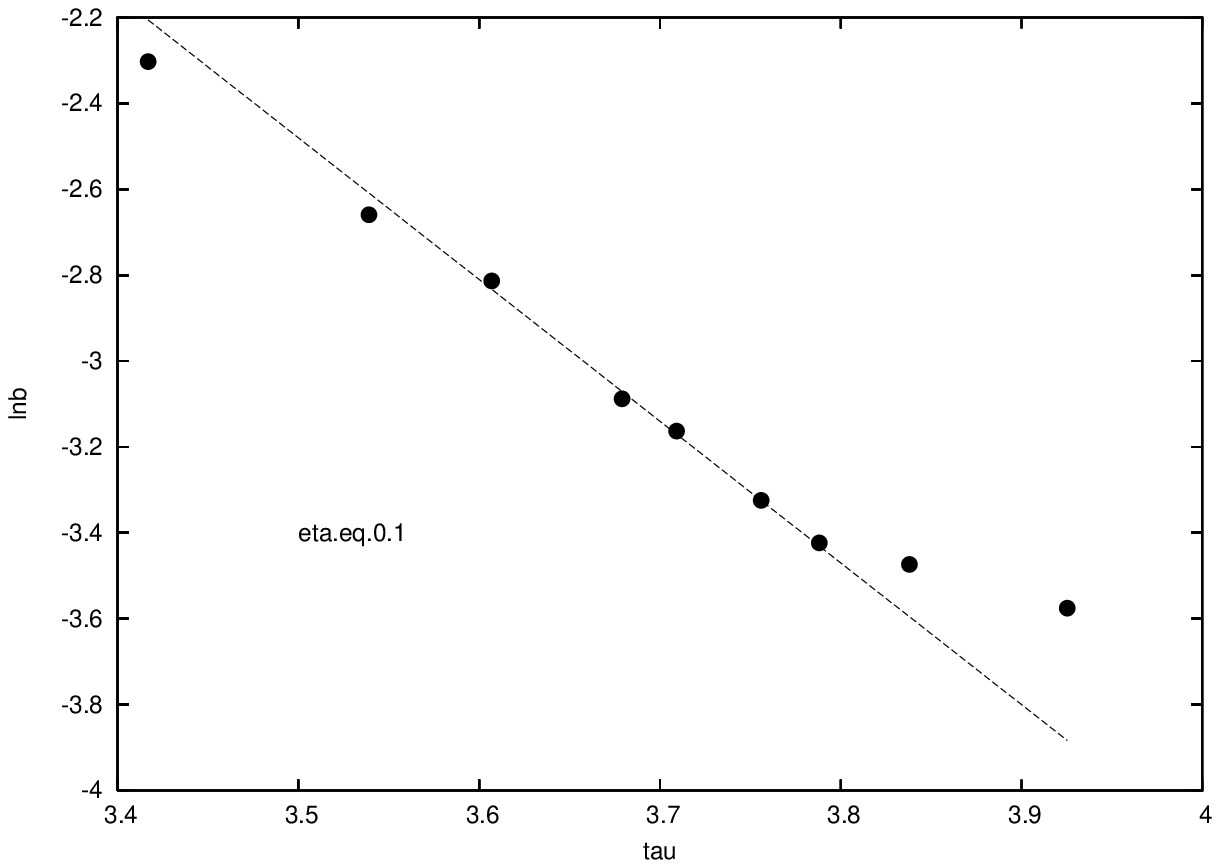}
\end{psfrags}
\end{center}
\caption{The parameter $b$ of Fig.  
\ref{fig::Deviation_of_critsol_eta.eq.0.1}
in dependence of $\tau$ (dots). The solid line has a slope of $-3.3$,
which would correspond to an eigenvalue $\lambda = 4.3$.
}\label{fig::Deviation_of_critsol_pars_b_eta.eq.0.1}
\end{figure}

\afterpage{\clearpage}

Figs.~\ref{fig::mass_scaling_eta.eq.0.11} -- 
\ref{fig::Deviation_of_critsol_pars_b_eta.eq.0.1} illustrate the critical
phenomena for the couplings $\eta = 0.11$ and $\eta = 0.1$.
Figs.~\ref{fig::mass_scaling_eta.eq.0.11} and 
\ref{fig::ricci_scaling_sub_eta.eq.0.11} show the scaling of the black hole
mass for super-critical data and of the quantity $\max\limits_u \mathcal
R(u,0)$ for sub-critical data. Both figures show the results for various
initial grid resolutions ($N=500, 1000, 2000, 4000$). 
As described in \cite{Husa-Lechner-Puerrer-Thornburg-Aichelburg-2000-DSS}
the critical value of the parameter $p^*$ depends on the grid resolution.
This was taken into account in Figs.~\ref{fig::mass_scaling_eta.eq.0.11} --
\ref{fig::ricci_scaling_sub_eta.eq.0.11}, i.e. for each resolution $N$ the
corresponding value $p^*(N)$ was used. In
\cite{Husa-Lechner-Puerrer-Thornburg-Aichelburg-2000-DSS} it was shown,
that for large values of the coupling the critical value $p^*(N)$ shows
second order convergence with the grid resolution. We note that this is also
the case for small (presumably $\eta \gtrsim 0.07$) couplings. For $\eta =
0.11$ and the initial data family (\ref{eq::initial_data_gaussian})
with fixed width $\sigma = 1$ and variable amplitude the respective values
are
\bea
p^*(500) & = & 0.0214965187393766, \quad p^*(1000) =
0.0214974987387296,\nonumber\\
p^*(2000) & = & 0.0214977388532529, \quad p^*(4000) =
0.0214977981419972.\nonumber
\eea
The differences are
\bea
\delta p_1 & = & p^*(1000) - p^*(5000) = 9.79999352997835 \ 10^{-07},
\nonumber\\
\delta p_2 & = & p^*(2000) - p^*(1000) = 2.40114523302609 \ 10^{-07}, 
\nonumber\\
\delta p_3 & = & p^*(4000) - p^*(2000) = 5.92887442994738 \ 10^{-08}, 
                                          \nonumber\\
\eea
and the ratios thereof
\be
\frac{\delta p_1}{\delta p_2} = 4.08138308136726, \quad 
\frac{\delta p_2}{\delta p_3} = 4.0499175035613.
\ee

In addition to the convergence of $p^*(N)$ we demonstrate convergence of the
black hole mass. Fig.~\ref{fig::mass_convergence_eta.eq.0.11}
shows the differences of the black hole masses $m_{BH}(N)$ 
of Fig.~\ref{fig::mass_scaling_eta.eq.0.11}. The difference between runs
with $N$ and $2N$ grid points, and the difference between 
runs with $2N$ and $4N$ grid points multiplied by a factor of four are close
throughout the shown interval of $p - p^*$.  

Finally Figs.~\ref{fig::crit_sol_eta.eq.0.1} --
\ref{fig::Deviation_of_critsol_pars_b_eta.eq.0.1}
deal with the intermediate asymptotics of near critical data for $\eta =
0.1$. Fig.~\ref{fig::crit_sol_eta.eq.0.1} shows, that near critical 
data evolve towards the first CSS excitation, stay there for some 
time and then deviate again. Fig.~\ref{fig::Deviation_of_critsol_eta.eq.0.1}
investigates the deviation of the evolved field $\phi$ from the CSS
solution. According to the theory this deviation should be dominated by the
unstable mode of the CSS solution. An additional complication arises
from switching to adapted coordinates, in that an error in determining the
culmination time $u^*$ brings the gauge mode (Sec.~\ref{subsec::gauge_modes})
into the game. In Fig.~\ref{fig::Deviation_of_critsol_eta.eq.0.1}
we have taken this fact into account in the following way: according to the
theory the evolved field should behave
as $\phi(\tau,z) = \phi_{CSS}(z) + e^{\lambda \tau} y(z)$, where we assume
that all the stable modes already have damped out.
In adapted coordinates with slightly different $u^*$ this would read
$\phi(\tau,z) = \phi_{CSS}(z) + e^{\lambda \tau} y(z) + e^{\tau}
y_{gauge}(z)$. Therefore we try to fit $\delta \phi(\tau,z) = \phi(\tau,z) -
\phi_{CSS}(z)$ to the function 
$f(\tau,z) = a(\tau) (y(z) + b(\tau) y_{gauge}(z))$ adjusting $a$ and $b$
such that the maxima agree. The fitted parameters then should behave as
$a(\tau) = a_0 e^{\lambda \tau}$ and $b(\tau) = b_0 e^{(1 - \lambda)\tau}$.
Figs.~\ref{fig::Deviation_of_critsol_pars_eta.eq.0.1} and 
\ref{fig::Deviation_of_critsol_pars_b_eta.eq.0.1} show that $a(\tau)$ is in
good agreement with the above formulae, whereas $b(\tau)$ is slightly
off.

\section{Critical Phenomena for Intermediate Couplings -- Transition from
        CSS to DSS}\label{sec::transitionregion}
%%%%%%%%%%%%%%%%%%%%%%%%%%%%%%%%%%%%%%%%%%%%%%%%%%%%%%%%%%%%%%%%%%%%%%%%%

Finally we describe the region of couplings $0.14 \lesssim \eta \lesssim 0.18$,
where the transition from CSS to DSS as critical solution takes place. We
know from the last sections (Sec.~\ref{sec::largecoupling} and 
Sec.~\ref{sec::smallcoupling}) that for $\eta < 0.14$ the critical solution
is the first CSS excitation whereas for $\eta \ge 0.2$ the critical solution
is DSS. Furthermore in Sec.~\ref{subsec::CSS-DSS} we proposed the hypothesis, that the DSS
solution merges with the CSS solution at $\eta \simeq 0.17$ in a homoclinic
loop bifurcation and does not exist for smaller $\eta$. 
While our numerical results for the stability of the DSS solution at $\eta =
0.1726$ are not conclusive (Sec.~\ref{subsec::StabilityDSS}) it is
reasonable to assume that the DSS solution does not change stability. 

We start by describing the intermediate asymptotics of near critical
evolutions. In the whole range of couplings we find a behavior, which we
call ``episodic CSS'', that is: the near critical solution approaches the
CSS solution $\phi_{CSS}$, goes away, approaches its negative $-\phi_{CSS}$,
goes away etc.~until after a small number of such episodes it finally parts
to form either a black hole or to disperse. The culmination times $u^*$
associated which each episode increase with the episodes. 

In the
following we will describe the critical phenomena we find
for the two coupling constants $\eta = 0.1726$, where we have
constructed the DSS solution, and $\eta = 0.16$, where we think, that the
DSS solution does not exist.

For $\eta = 0.1726$ the evolution could be compared to the DSS solution. We
find that the DSS solution is approached better and better during the 
time the solution stays in the neighborhood of the ``critical
hyper-surface'', although not as good as at higher couplings 
e.g.~at $\eta = 0.2$. In \cite{Lechner-Thornburg-Husa-Aichelburg-prl-2001} we will
quantitatively give the ``distance'' of the near critical solution to the DSS
solution in some norm\footnote{The discretized version of
$\int\limits_{0}^{r_{max}} r^2 \ dr |\phi(u_0,r) - \phi_{DSS}(u_0,r)|^2 /
\int\limits_{0}^{r_{max}} r^2 \ dr$, where the DSS solution is taken at a 
time, that minimizes this error and is appropriately shifted in $\ln r$ and
$r_{max}$ is $\min(r_{outer}(u_0),r_{SSH}(u))$.}
for various coupling constants. These investigations show clearly that for
$\eta = 0.2$ the near critical data approach the DSS solution quickly
(roughly within one cycle), stay in the vicinity (with a distance in the
above norm of $\sim 10^{-3}$) before they deviate. For $\eta = 0.1726$ on the
other hand the approach to the DSS solution absorbs all the time the
solution stays in the neighborhood and the  
field only comes as close as $\sim
10^{-2}$, before it deviates. 

A possible reason for this slow approach (as compared to larger couplings)
could be that the stable modes of the DSS solution damp out much more slowly
at $\eta = 0.1726$ than at $\eta = 0.2$. It would be very interesting to
test this behavior further, in using a higher numerical precision (quartic
precision). This way the admixture of the unstable mode in the initial data
could be reduced, which would prolong the time the solution stays close to
the DSS solution. It should then be possible to observe a closer approach
to the DSS solution at $\eta = 0.1726$.

Combining the approach to the DSS solution with the fact that CSS and DSS
lie ``close'', it is clear that we observe the above described CSS episodes.

\begin{figure}[p]
\begin{center}
\begin{psfrags}
 \psfrag{ln\(p*-p\)}[]{$\ln(p^* - p)$}
 \psfrag{eta.eq.0.16}[]{$\eta = 0.16$}
 \psfrag{ln (max R(u,0))}[r][l][1][-90]{$\ln (\max\limits_u \mathcal R(u,0))$}
\includegraphics[width=5in]{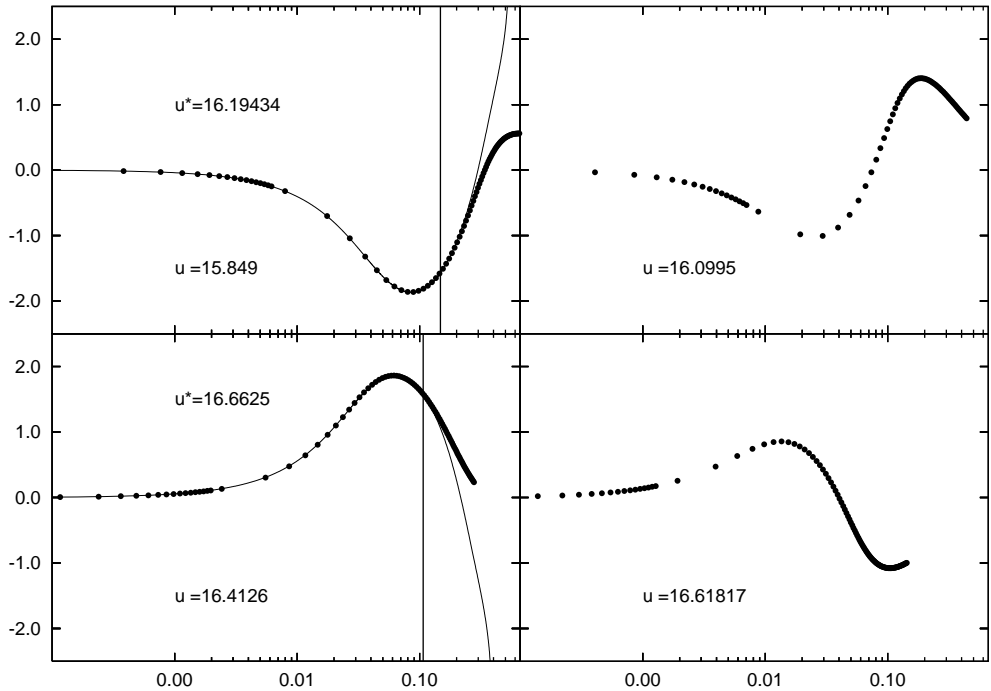}
\end{psfrags}
\end{center}
\caption{For $\eta = 0.16$ the intermediate asymptotics of near critical
data (family (\ref{eq::initial_data_gaussian}) with fixed width $\sigma =
1$; number of grid points initially $N = 8049$)
are shown (dots; not every grid point is plotted). 
The four plots are snapshots at various times $u$, where $u$
increases from left to right and from top to bottom.
On the first and third plot the CSS solution is
superimposed (lines), where it has been shifted (in $\ln r$) 
horizontally such that the first monotonic region agreed best with the
evolved data. (The best fit was determined automatically by a 
fitting script by J.~Thornburg). Given the horizontal shift in $\ln
r$, the corresponding culmination time $u^*$ is determined, as well as  the
location of the past SSH, which is denoted by a vertical line in these
plots. One clearly sees that the evolved data approach the CSS solution,
turn away and then approach its negative. The culmination times associated
which each CSS episode increase.
}\label{fig::episodic_css}
\end{figure}

\begin{figure}[p]
\begin{center}
\begin{psfrags}
 \psfrag{lnz}[]{$\ln(z)$}
 \psfrag{eta.eq.0.16}[]{$\eta = 0.16$}
 \psfrag{deltaphi}[r][l][1][-90]{$\delta \phi$}
\includegraphics[width=5in]{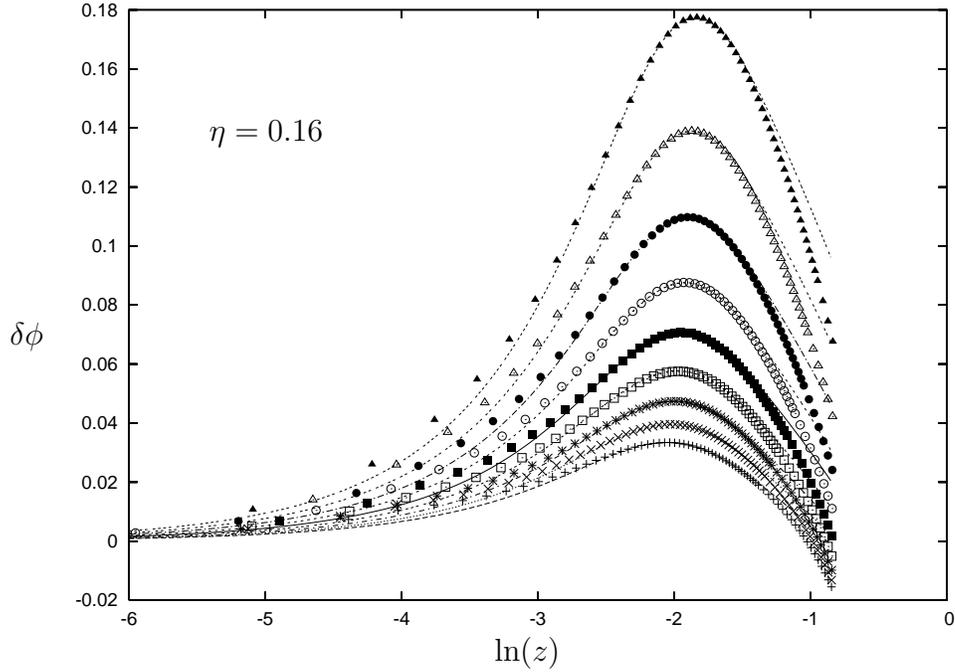}
\end{psfrags}
\end{center}
\caption{The deviation of the critical solution from the CSS solution
at the last but one episode. Plotted is $\delta \phi(\tau,z) = \phi(\tau,z)
- \phi_{CSS}(z)$ at several time steps between $\tau = 0.1576$ and 
$\tau = 1.991$ (dots). As in Fig. 
\ref{fig::Deviation_of_critsol_eta.eq.0.1} the functions $a(\tau) (y_{unst}
+ b(\tau) y_{gauge})$ are overlaid (solid lines) 
with fitted values of the parameters $a$
and $b$, such that the maxima agree.
Again the gauge mode is taken into account in order to correct for the
uncertainty in $u^*$.  
Although the agreement of the shapes is not very good,
on can infer
from Figs.~\ref{fig::Deviation_from_CSS_pars_a_eta.eq.0.16} and 
\ref{fig::Deviation_from_CSS_pars_b_eta.eq.0.16}, 
that (at least for a short time) 
the maximum of $\delta \phi$ grows exponentially with a rate, 
which is close to the eigenvalue of the unstable mode $\lambda = 5.202$. 
}\label{fig::Deviation_from_CSS_eta.eq.0.16}
\end{figure}

\begin{figure}[p]
\begin{center}
\begin{psfrags}
 \psfrag{tau}[]{$\tau$}
 \psfrag{eta.eq.0.16}[l][r][1][0]{$\eta = 0.16$}
 \psfrag{lna}[r][l][1][-90]{$\ln (a(\tau))$}
\includegraphics[width=3.5in]{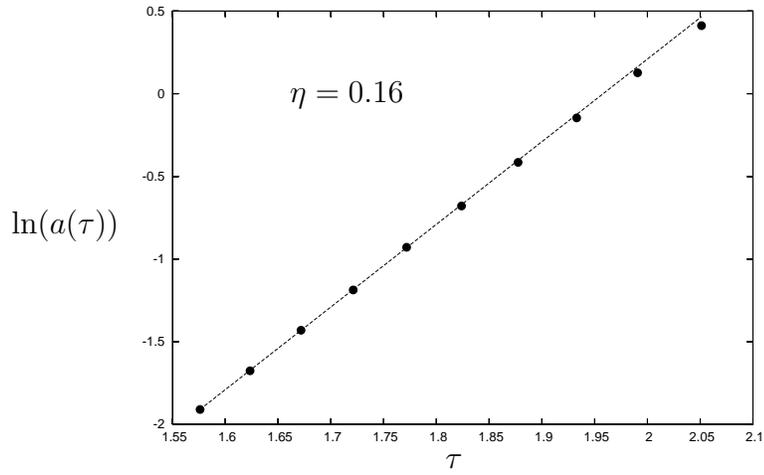}
\end{psfrags}
\end{center}
\caption{The parameter $a$ of Fig.  
\ref{fig::Deviation_from_CSS_eta.eq.0.16}
in dependence of $\tau$ (dots). The solid line has a slope of $5$, which is
close to the eigenvalue $\lambda = 5.202$ 
computed from the perturbation analysis. 
}\label{fig::Deviation_from_CSS_pars_a_eta.eq.0.16}
\end{figure}

\begin{figure}[b]
\begin{center}
\begin{psfrags}
 \psfrag{tau}[]{$\tau$}
 \psfrag{eta.eq.0.16}[l][r][1][0]{$\eta = 0.16$}
 \psfrag{lnb}[r][l][1][-90]{$\ln (b(\tau))$}
\includegraphics[width=3.5in]{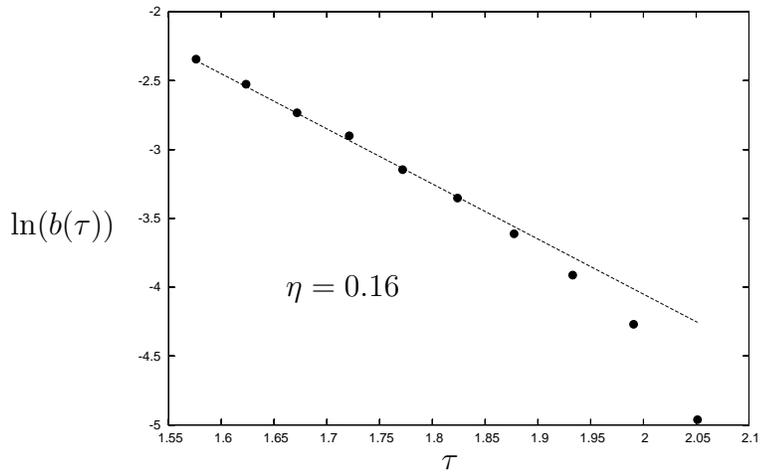}
\end{psfrags}
\end{center}
\caption{The parameter $b$ of Fig.  
\ref{fig::Deviation_from_CSS_eta.eq.0.16}
in dependence of $\tau$ (dots). The solid line has a slope of $1 - 5 = 4$,
according to $\ln b(\tau) = (1 - \lambda) \tau + const$.
}\label{fig::Deviation_from_CSS_pars_b_eta.eq.0.16}
\end{figure}

For $\eta = 0.16$ the CSS episodes are illustrated in 
Fig.~\ref{fig::episodic_css}. We also studied the way the evolved solution
deviates from the CSS solution at the last but one episode.
Fig.~\ref{fig::Deviation_from_CSS_eta.eq.0.16} shows the deviation $\delta
\phi$ together with the fitted functions $f(\tau,z) = a(\tau)
(y_{unstable}(z) + b(\tau) y_{gauge}(z))$ defined as in 
Sec.~\ref{sec::smallcoupling}. Clearly the fits are not as good as for
$\eta = 0.1$ (see Fig.~\ref{fig::Deviation_of_critsol_eta.eq.0.1}).
Nevertheless the fits 
(Fig.~\ref{fig::Deviation_from_CSS_pars_a_eta.eq.0.16} and 
\ref{fig::Deviation_from_CSS_pars_b_eta.eq.0.16}) show that the maximum
grows exponentially according to the unstable mode of the CSS solution.

The explanation for the episodes in this case is not as straightforward as
for $\eta = 0.1726$. A possible explanation would be, that although 
the DSS solution does not exist, there are still orbits in phase space,
which ``mimic'' a DSS critical solution, i.e. orbits, that
do not close exactly, but nevertheless act as intermediate attractors.

\begin{figure}[p]
\begin{center}
\begin{psfrags}
 \psfrag{ln\(p*-p\)}[]{$\ln(p^* - p)$}
 \psfrag{eta.eq.0.16}[]{$\eta = 0.16$}
 \psfrag{ln (max R(u,0))}[r][l][1][-90]{$\ln (\max\limits_u \mathcal R(u,0))$}
\includegraphics[width=3in]{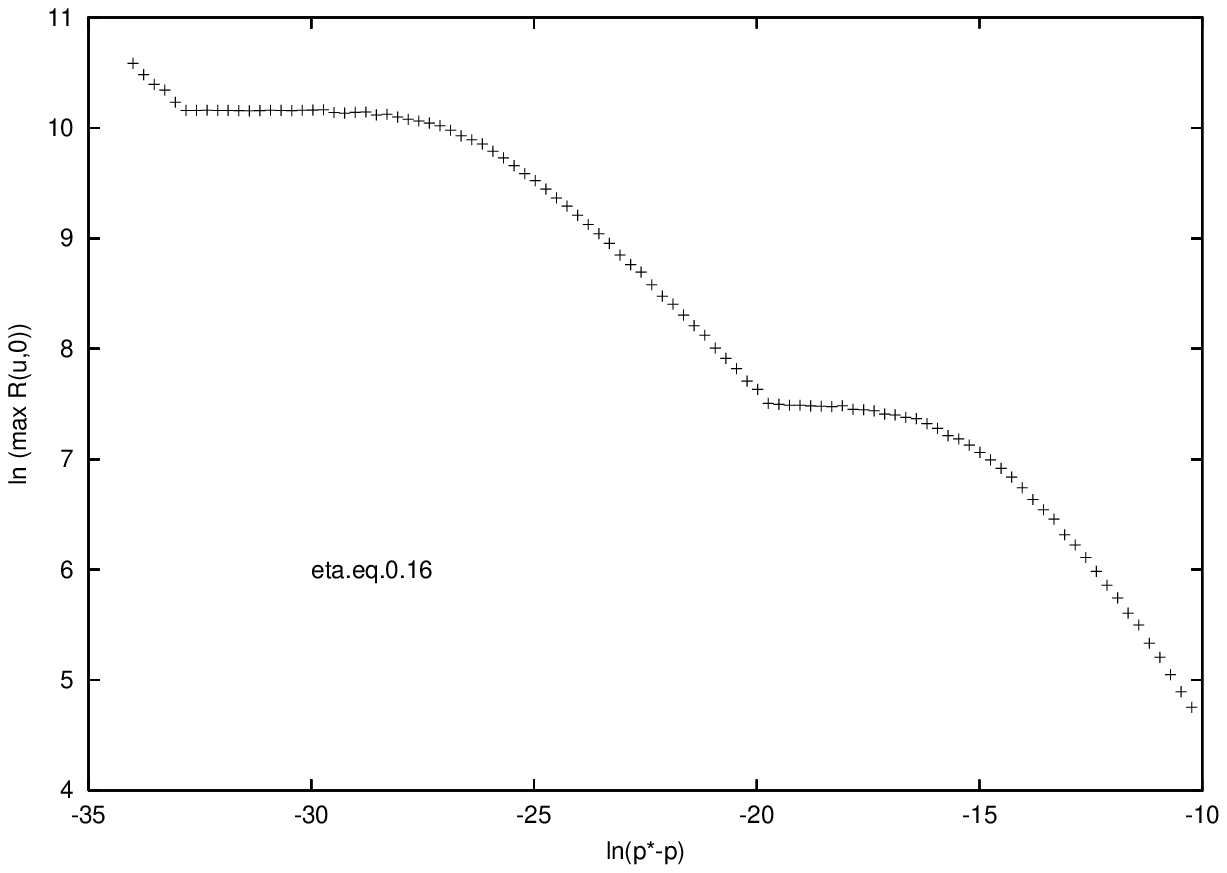}
\end{psfrags}
\end{center}
\caption{Scaling of $\max\limits_u \mathcal R (u,0)$ for sub-critical data
at $\eta = 0.16$. The family of initial data was
(\ref{eq::initial_data_gaussian}) with fixed width $\sigma = 1$, the
number of grid points was $N = 8049$.
}\label{fig::ricci_scaling_sub_eta.eq.0.16_1}
\end{figure}

\begin{figure}[p]
\begin{center}
\begin{psfrags}
 \psfrag{ln\(p*-p\)}[]{$\ln(p^*[N] - p)$}
 \psfrag{eta.eq.0.16}[l][c][1][0]{$\eta = 0.16$}
 \psfrag{ln (max R(u,0))}[r][l][1][-90]{$\ln (\max\limits_u \mathcal R(u,0))$}
\includegraphics[width=3in]{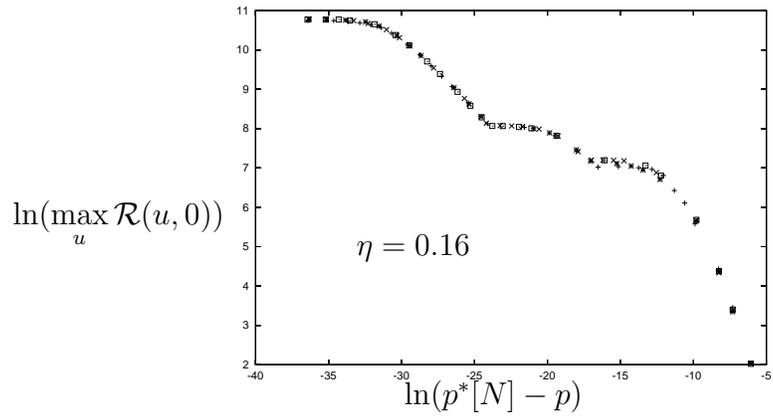}
\end{psfrags}
\end{center}
\caption{The family \ref{eq::double_gaussian} qualitatively yields the same 
scaling as in Fig.~\ref{fig::ricci_scaling_sub_eta.eq.0.16_1}.
The resolutions shown are $N = 500, 1000, 2000, 4000$.
}\label{fig::ricci_scaling_sub_eta.eq.0.16_2}
\end{figure}

Taking this, one would expect, that the black hole mass as well as the
scalar curvature exhibit scaling, which is similar to a typical ``DSS
scaling''. Figs.~\ref{fig::ricci_scaling_sub_eta.eq.0.16_1} and 
\ref{fig::ricci_scaling_sub_eta.eq.0.16_2} 
show the scaling of the scalar curvature
for the families (\ref{eq::initial_data_gaussian}) and 
(\ref{eq::double_gaussian}) at $\eta = 0.16$.
The logarithm of the scalar curvature as a function of 
$\ln(p^*-p)$ shows oscillations, but not enough of them in order to judge,
whether these wiggles are superimposed on a straight line, or whether the
wiggles are almost periodic. Using a higher numerical precision probably
would yield a clearer picture.

Unfortunately, we were not able to produce reliable results
concerning the scaling of the black hole mass at $\eta = 0.16$.
In all the evolutions, we have looked at, we find that there are two peaks
in the function $2m/r$, which come close to the threshold ($0.995$)
towards the end of the evolution. The inner peak is afflicted
with numerical errors, nevertheless it slows down the evolution, such that 
in some cases the outer peak cannot reach the threshold anymore. 
The result in some cases is a ``broken'' mass scaling.  
Clearly further work has to be invested here, before any conclusions can be
drawn.

\begin{figure}[p]
\begin{center}
\begin{psfrags}
 \psfrag{ln\(p*-p\)}[]{$\ln(p^* - p)$}
 \psfrag{eta.eq.0.1726}[]{$\eta = 0.1726$}
 \psfrag{lnR}[r][l][1][-90]{$\ln (\max\limits_u \mathcal R(u,0))$}
\includegraphics[width=3in]{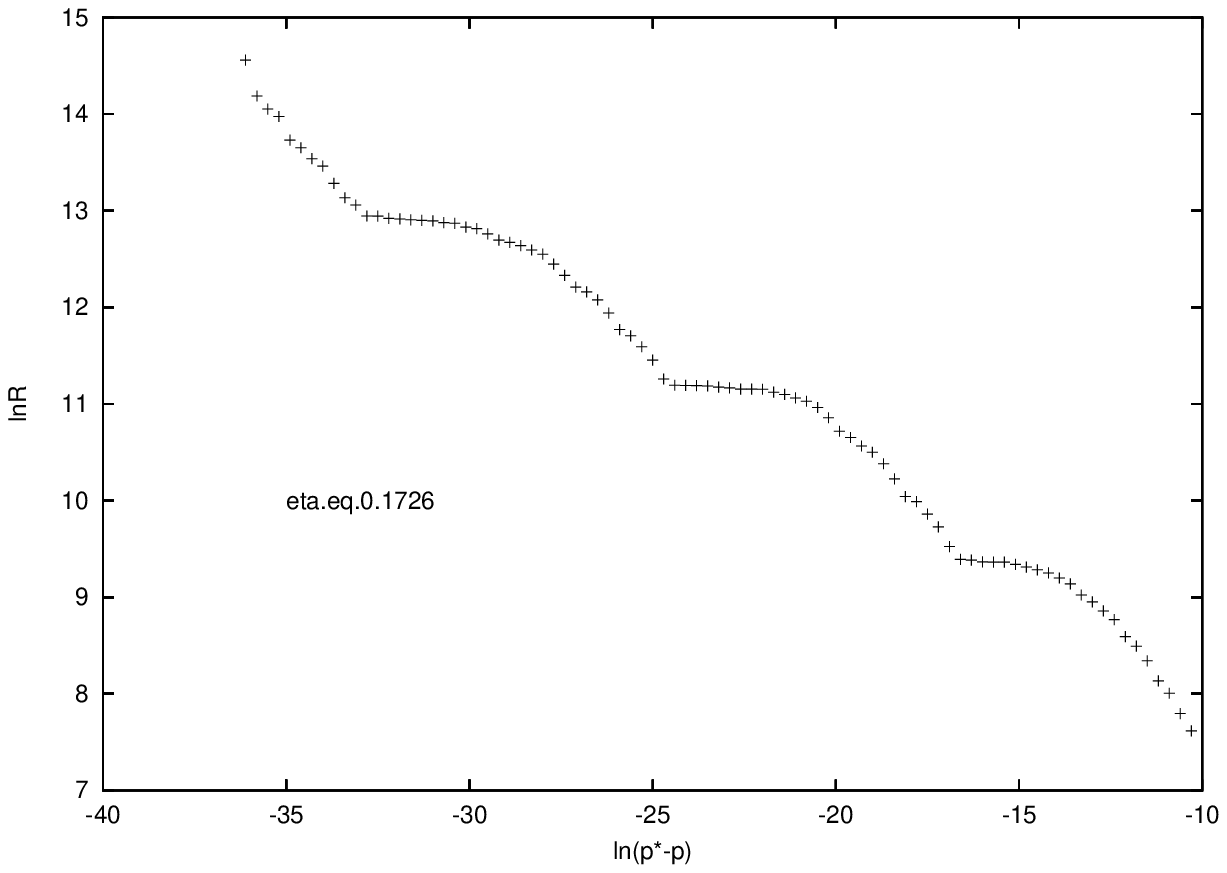}
\end{psfrags}
\end{center}
\caption{Scaling of $\max\limits_u \mathcal R (u,0)$ for sub-critical data
at $\eta = 0.1726$. The family of initial data was
(\ref{eq::initial_data_gaussian}) with fixed width $\sigma = 1$, the
number of grid points was $N = 2000$.
}\label{fig::ricci_scaling_sub_eta.eq.0.1726}
\end{figure}

\begin{figure}[p]
\begin{center}
\begin{psfrags}
 \psfrag{ln\(p*-p\)}[]{$\ln(p^* - p)$}
 \psfrag{eta.eq.0.1726}[]{$\eta = 0.1726$}
 \psfrag{lnR - f}[r][l][1][-90]{$\ln (\max\limits_u \mathcal R(u,0)) -
f(\ln(p^*-p))$}
\includegraphics[width=3in]{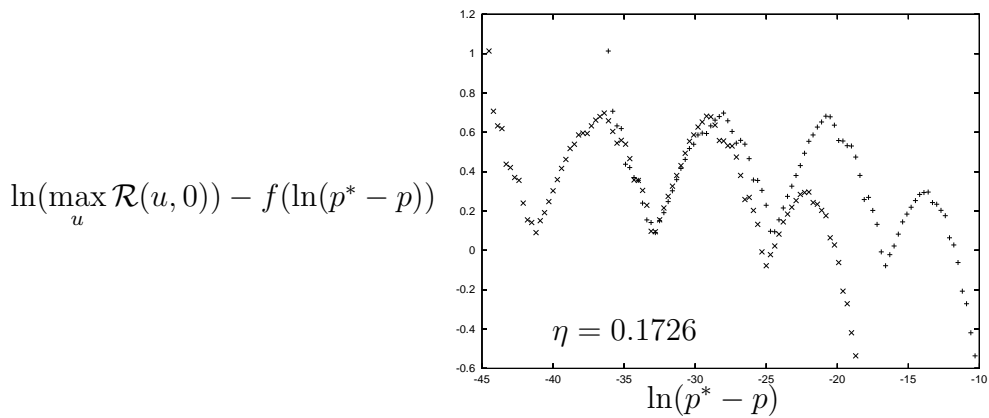}
\end{psfrags}
\end{center}
\caption{The straight line $f(\ln (p^* -p)) = - 2 \gamma \ln(p^*-p) + k$ was
fitted to $\ln \max_u \mathcal R(u,0)$ in
Fig.~\ref{fig::ricci_scaling_sub_eta.eq.0.1726} with naked eye. The fit gave
$\gamma \simeq 0.1045$. In order to look at the fine structure
this function was subtracted from the scalar curvature. The result is 
shown in this figure (``+''). In order to check for periodicity, the 
same data were re-plotted (``x''), 
with a shift in $\ln(p^*-p)$ of $\Delta/2\gamma$.
As can be seen, the periodicity is not exact, but close. 
}\label{fig::ricci_sub_plusf_eta.eq.0.1726}
\end{figure}

For $\eta = 0.1726$, where we assume the DSS solution to be the
critical solution, although not approached as closely as at
higher couplings, the scaling should be more conclusive.
Fig.~\ref{fig::ricci_scaling_sub_eta.eq.0.1726} shows the quantity
$\ln(\max\limits_u \mathcal R(u,0))$ for sub-critical data of the family
(\ref{eq::initial_data_gaussian}). A straight line with slope $-2 \gamma$
and $\gamma = 0.1045$ was fitted to these data with naked eye.
In order to investigate the fine structure more closely, this straight 
line was subtracted from the data. The result is shown 
in Fig.~\ref{fig::ricci_sub_plusf_eta.eq.0.1726}. The difference
$\ln (\max\limits_u \mathcal R(u,0)) - f$ is almost periodic,
the period being roughly $\Delta/2 \gamma$, 
where the echoing period $\Delta$ was
taken from the directly constructed DSS solution. 
The reason for this periodicity not being exact might lie in the fact, 
that the DSS solution is
not approached close enough. Again higher numerical precision could clarify
things.

\begin{figure}[p]
\begin{center}
\begin{psfrags}
 \psfrag{ln\(p-p*\)}[]{$\ln(p - p^*)$}
 \psfrag{eta.eq.0.1726}[]{$\eta = 0.1726$}
 \psfrag{lnMBH}[r][l][1][-90]{$\ln (m_{BH})$}
\includegraphics[width=3in]{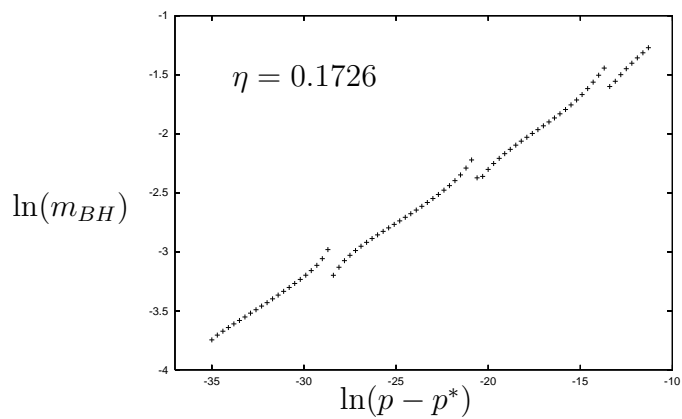}
\end{psfrags}
\end{center}
\caption{Scaling of the black hole mass for super-critical data
at $\eta = 0.1726$ (family (\ref{eq::initial_data_gaussian}) 
with fixed width $\sigma = 1$, number
of grid points was $N = 2000$). Almost all of the runs in this plot stopped
because of $du < 10^{-14}u$.
}\label{fig::mass_scaling_eta.eq.0.1726}
\end{figure}

\begin{figure}[p]
\begin{center}
\begin{psfrags}
 \psfrag{ln\(p-p*\)}[]{$\ln(p - p^*)$}
 \psfrag{eta.eq.0.1726}[]{$\eta = 0.1726$}
 \psfrag{lnMBH}[r][l][1][-90]{$\ln (m_{BH}) - f$}
\includegraphics[width=3in]{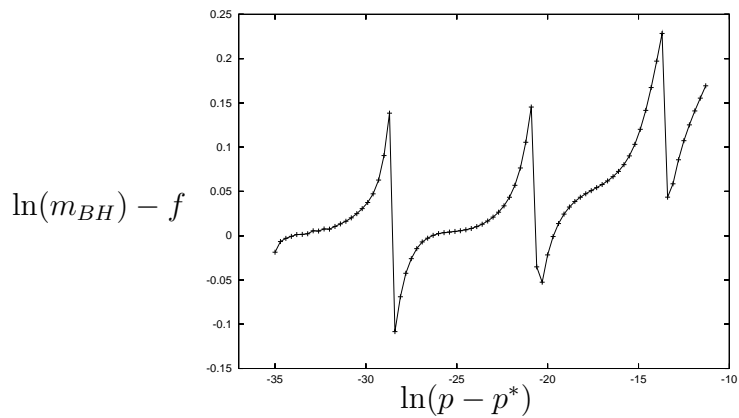}
\end{psfrags}
\end{center}
\caption{The straight line $f(x) = \gamma x + k$ with $\gamma = 0.0965$ was
subtracted from the data in Fig.~\ref{fig::mass_scaling_eta.eq.0.1726}.
The result is an almost periodic function of $\ln(p-p^*)$ with a period
roughly equal to $\Delta/2 \gamma$.
}\label{fig::mass_super_minusf_eta.eq.0.1726}
\end{figure}

The scaling of the black hole mass at $\eta = 0.1726$ is shown in
Fig.~\ref{fig::mass_scaling_eta.eq.0.1726}. The ``worms'' displayed
approximately align along a straight line with slope $\gamma = 0.0965$.
The wiggles again are only close to periodic, as shown in 
Fig.~\ref{fig::mass_super_minusf_eta.eq.0.1726}.
We don't know, whether the discontinuities in the mass scaling in
Fig.~\ref{fig::mass_scaling_eta.eq.0.1726} stem from a systematic error in
the measurement of the black hole mass.

%Directory:eta.eq.0.16/1_Width/Again_Scaling_super/

%Directory:eta.eq.0.16/Test_1_Width/CriticalSolution1000/

%Directory:eta.eq.0.16/Test_1_Width/CriticalSolution1000/

%Directory:eta.eq.0.16/Test_1_Width/CriticalSolution1000/

%Directory:eta.eq.0.16/1_Width/Scaling_SUB/

%Directory:eta.eq.0.16/Double_Guglhupf/Scaling_Results/

%Directory:eta.eq.0.1726/1_Width/Scaling_Sub/

%Directory:eta.eq.0.1726/1_Width/Scaling_Sub/

%Directory:eta.eq.0.1726/1_Width/Bigger_router/Scaling_Super_2000

%Directory:eta.eq.0.1726/1_Width/Bigger_router/Scaling_Super_2000

%%%%%%%%%%%%%%%%%%%%%%%%%%

\chapter{Discussion and Outlook}\label{chap::discussion}
%%%%%%%%%%%%%%%%%%%%%%%%%%%%%%%%%%%%%%%%%%%%%%%%%%%%%%%

In this thesis we have reported on our work on 
the self-gravitating SU(2) $\sigma$ model in spherical symmetry.
We have described our results concerning static solutions in the presence
of a positive cosmological constant $\Lambda$,
on self-similar solutions and on type II critical phenomena.  

We have shown numerically that the model
(with a positive cosmological constant) admits a discrete one-parameter
family of static spherically symmetric regular solutions.
These solitonic solutions are characterized by an integer excitation
number $n$.
A given excitation will only exist up to a critical value of the
coupling constant $\eta$; the higher $n$, the lower the corresponding
 critical value.
 Our calculations indicate that the infinite tower of solitons present on
 a de Sitter background persists at least up to a value of 
$\eta = 1/2$. Thus 
 there exists a $\eta \ge 1/2$ beyond which the number of excitations 
is finite and decreases with the strength of the coupling.
Qualitatively the $\sigma$ model under consideration 
shows striking
 similarities to the EYM system as studied in detail by 
Volkov et.al.~\cite{VSLHB-1996-cosmological-EYM-BMcK-solutions}.
 The main difference being that the static solutions to the EYM-system 
 depend
 on the value of the cosmological constant while in our case $\Lambda$ 
 scales
 out from the equations and $\eta$ plays the role of a 
``bifurcation'' parameter.
Another difference concerns the globally regular static solutions with
compact 
spatial slices. For the EYM system these appear for definite values of 
$\Lambda(n)$
while for the $\sigma$ model the corresponding solutions exist only in the 
(singular) limit as $\Lambda$ goes to zero and definite values of 
$\eta$. Thus in 
our case there are closed static universes with vanishing cosmological 
constant, the lowest excitation being the static Einstein cosmos. 
This is possible because in this case the stress-energy tensor of the
$\sigma$ field is of the form of a perfect fluid with the equation of state 
$p = -\mu/3$.
Another interesting aspect is the geometry of a given excitation as a
function of the coupling strength: the static region is always 
surrounded by a Killing
horizon separating the static from a dynamical region, which for small 
couplings becomes asymptotically de Sitter. As the coupling is increased the 
two-spheres of symmetry beyond the horizon are first past and then become 
future trapped and a cosmological singularity develops. 
Finally, for even 
stronger couplings, again the region beyond the horizon collapses, but
within the static region the in- and outgoing directions 
(as defined by the sign of 
the expansion for null geodesics) interchange.

An important question to be answered is whether these solitons are stable
under small radially symmetric time dependent perturbations.
Here we have presented preliminary results, stating 
that for $\Lambda > 0$ all excitations are unstable 
with their number 
of unstable modes increasing with $n$. This was to be expected at least for 
small coupling. The lowest excitation thus has a single unstable mode and it 
is known, from other models, that such a solution can play the role of a 
critical solution in a full dynamical treatment of spherically symmetric
collapse.

As a further step,
it would be interesting to study existence and stability of static
solutions to this model, that have no regular center of spherical symmetry,
but rather a static region, which is bounded by two horizons, 
in analogy to the
Schwarzschild-de-Sitter (Kottler) spacetime. 
This is currently investigated by N.
M\"ullner \cite{Nikolaus-Muellner-Thesis}.

In a further part of this thesis
we have numerically reproduced the results of Bizon and Wasserman 
\cite{Bizon-Wasserman-2000-CSS-exists-for-nonzero-beta}
concerning continuously self-similar solutions to the self-gravitating
$\sigma$-model in spherical symmetry. We also supplemented this work with
a stability analysis.
As in \cite{Bizon-Wasserman-2000-CSS-exists-for-nonzero-beta}
we find that a countably infinite spectrum of CSS solutions exists up to a
maximal coupling $\eta = 0.5$. 
For vanishing coupling $\eta = 0$, the ground state of this 
family was already given in closed form by Turok and Spergel
\cite{Turok-Spergel-1990}.
The homothetic Killing vector, generating
continuous self-similarity, is timelike inside the 
past self-similarity horizon (the backwards light cone of the culmination
point, where in the coupled case a space time singularity occurs).
For small couplings all members of the family are regular up to the future
self-similarity horizon. For larger couplings -- the critical coupling
depending on the excitation number --
these solutions contain marginally trapped surfaces.
A linear stability analysis, carried out with two different (numerical)
methods, revealed that the number of unstable modes corresponds to the
excitation number of the solution. In particular the ground state is stable
and the first excitation has one unstable mode.

The stability properties of the CSS ground state and the first CSS excitation
make these solutions relevant for the dynamics of the system. 
We have shown, that for very 
small couplings the CSS ground state is the global 
``end state'' for a set of initial data. The singularity at the culmination
point in general is not shielded by a horizon, such that the CSS ground
state gives rise to the formation of naked singularities.
Although the formation of naked singularities is a general end state
in this model for small couplings, this behavior cannot 
be viewed as a violation of the cosmic censorship
hypothesis. The blow up (of the energy density)
also occurs in flat space, the formation of naked singularities therefore 
is not due to gravity, but to the matter model itself.
On the contrary gravity regularizes these singularities in the sense, that
for bigger couplings the only possible end states for ``strong'' initial
data are black holes. 

The first excitation, which lies at the threshold of singularity formation
in flat space \cite{Bizon-Chmaj-Tabor-1999-sigma-3+1-evolution} 
also is the critical solution between black hole formation
and dispersion for small couplings (respectively between the formation of
naked singularities and dispersion for very small couplings).

For large couplings on the other hand, the solution at the threshold of
black hole formation is discretely self-similar.
Both our results of time evolution and critical searches on one hand 
and the ``direct construction'' using the (discrete) symmetry and 
pseudo-spectral methods on the other hand 
show, that the period $\Delta$ of the DSS solution
rises sharply below $\eta \simeq 0.3$. We were able to construct
the DSS solution down to a coupling of 
$\eta = 0.1726$\footnote{Proceeding further
down would need more Fourier coefficients and therefore would increase the 
computational costs considerably.}.
At this lowest coupling we compared the DSS solution to the first CSS
excitation, and found that there exists a phase of the DSS solutions, where
the shapes of both functions agree rather well.
At $\eta = 0.18$ the agreement is not so good. This suggests that the DSS
solution bifurcates from the CSS solution at some coupling $\eta_C$.
Due to the fact, that the CSS solution does not change its stability around
$\eta_C$ we suggest, that the bifurcation is a global bifurcation 
(in contrast to a local bifurcation), and furthermore that we deal with a
homoclinic loop bifurcation. The theory then predicts, that the period
$\Delta$ scales as the logarithm of $\eta_C - \eta$. A fit determined
$\eta_C \sim 0.17$.
The results of the stability analysis of the DSS solution are
in good agreement with the scaling of the black hole mass and the 
Ricci scalar: the DSS solution has one unstable mode with an eigenvalue of
$\lambda \sim 9.0$, which is almost independent of $\eta$.
At $\eta = 0.1726$ the numerical results of the stability analysis
are not conclusive.

In view of the above described bifurcation scenario, we can expect the
critical phenomena in the transition region, where the critical solution
changes from CSS to DSS, to be rather complicated.
Our results support the following view: at couplings $\eta \gtrsim 0.17$, 
where the DSS solution exists, it is the critical solution. 
Due to the ``closeness'' of the DSS and CSS solution, the CSS solution is
approached (and left) in several ``episodes''.
Presumably at
couplings close to $\eta_C$ the stable modes of the DSS solution 
damp out more slowly than at
larger couplings. With the given numerical precision near critical data
therefore cannot approach the DSS solution as close as at higher couplings.
This results in a periodicity of quantities like $2m/r$, which is not exact,
as well as in a scaling of the Ricci scalar, which has not 
an exactly periodic fine structure. Increasing the
numerical precision, we could reduce the admixture of the unstable mode 
in the initial data, thereby prolonging the ``life time'' of the critical
solution. We speculate, that then it would be possible to see an approach to
the DSS solution which is as close as at larger couplings.

At couplings below the bifurcation value of $\eta_C \sim 0.17$, we see
near critical evolutions approach a configuration which still has some
resemblance to discrete self-similarity and which shows CSS episodes.
Speaking in terms of dynamical systems it is possible, that the critical
solution now has not an exact symmetry, but is an invariant manifold of 
orbits, that ``almost'' close. In phase space
this invariant manifold lies in the vicinity of the CSS solution.
The smaller the coupling, the more pronounced are the CSS episodes, until
at a coupling of $\eta \sim 0.14$ the CSS solution is approached only once
and is the critical solution.

To our knowledge this kind of transition from CSS to DSS in the critical
solution and the phenomenon of a homoclinic loop bifurcation, has not been
observed up to now in the context of type II critical collapse.
Liebling and Choptuik for example \cite{Liebling-Choptuik-1996} 
report on a transition from CSS to DSS in the Brans-Dicke 
model\footnote{It might be 
interesting to note, that this Brans-Dicke model in spherical symmetry 
can be viewed as a $\sigma$ model
with a two-dimensional target manifold of constant negative curvature.}. 
But this transition is connected to a change in
the stability of the CSS solution. 

Further work has to concretize some of the above
results mainly with numerical methods. The code would need some improvement
for the treatment of the origin, and it has to be tested with respect to
convergence for (very) small couplings. 
Concerning the possible end states for strong initial data at very small
couplings it would be interesting to determine the transition (in $\eta$)
from naked singularities to black holes more precisely.
For small couplings, where the critical solution is CSS, it would be
necessary to trace down why the critical exponent deviates
by a few precent from the theoretically predicted value.
Convergence tests and using a higher numerical precision could give
first hints. In order to investigate the phenomena of episodic CSS in more
detail, the first step could consist again in using a higher numerical
precision. 

The ``DSS code'' could be used to study discretely self-simlar
or time periodic solutions in other models. In particular it would be
interesting to investigate, whether gravity is responsible
for the existence of discretely self-similar solutions.
As suggested by P.~Bizon \cite{Piotr-private-communication-2001}, 
a first step into this direction would be to
study a certain artificial matter model in flat space
with a self-interaction which mimics the interaction with gravity.

%%%%%%%%%%%%%%%%%%%%

\begin{appendix}

\chapter{The Shooting and Matching Method}\label{app::SM}
%%%%%%%%%%%%%%%%%%%%%%%%%%%%%%%%%%%%%%%%%%%%%%%%%%%%%%%%%%%%%%%%

As the shooting and matching method is used at several places in this work,
we give a short description of it here.
A good description of this method can be found e.g. in  
\cite{Numerical-Recipes}.
Consider the following ODE boundary value problem, given by 
a coupled system of $N$ ODEs,
\be\label{ODEgen}
\dot y{}^i (t) = F{}^i(y,t), \qquad i=1,\dots N.
\ee
and the following $N$ Dirichlet boundary conditions at the ends of the 
interval $[a,b]$,
\bea
g_{l}{}^j(y(a)) = 0,& \qquad & j = 1, \dots, M, \nonumber\\
g_{r}{}^k(y(b)) = 0,& \qquad & k = 1, \dots, N-M. 
\eea
So $M$ variables at the left boundary depend on the remaining
$N-M$ variables, which are free parameters $a_{k}, k = 1,\dots,N-M$, 
and on the right boundary there are $N-M$ variables, 
that depend on $M$ free parameters $b_{j}, j=1,\dots,M$.

After choosing the values of $N-M$ variables at the left boundary
and $M$ variables at the right boundary, which we subsume with
$\vec c = (a_{k}, b_{j})$, Eqs. (\ref{ODEgen}) can be
integrated from both ends of the interval to some matching point $t_{match}
\in (a,b)$. Of course the values of the solutions $y(t_{match})$ at the
matching point resulting from the integration from left and from right will
in general not agree, but there will be a ``miss distance''
\bea\label{eq::matching_condition}
f{}^i(\vec c)  = y_{left}{}^i(t_{match}; a_k) -
y_{right}{}^i(t_{match};b_j) & \quad & i = 1, \dots N, 
              k=1,\dots N-M, \nonumber\\
             & &  j=1,\dots,M.
\eea 
If $F$ is smooth, then $f$ will depend smoothly on the
parameters $\vec c$. The aim now is to find those values of the parameters
for which $f$ evaluates to zero\footnote{It might be convenient to replace
the solutions $y^i$ on the right hand side of 
(\ref{eq::matching_condition}) by some function thereof}. 
This can be achieved via a Newton
iteration. For values $\vec c$ close to the initial guess $\vec c_0$, 
the miss distance can be expanded as
\be
f{}^i(\vec c) = f{}^i(\vec c_0) + 
    \frac{\partial f{}^i(\vec c_0)}{\partial c^j} 
    (c^j - c_0^j)
 + O((\vec c - \vec c_0)^2).
\ee
If the initial guess $\vec c_0$ is close to a zero of
$f$ a first approximation to this zero is given by $\vec c$ with
\be
0 = f{}^i(\vec c_0) + 
    \frac{\partial f{}^i(\vec c_0)}{\partial c^j} 
    (c^j - c_0^j),
\ee
neglecting the higher order terms.
If furthermore the Jacobian $J^i{}_j(\vec c_0) = 
\partial f{}^i/\partial c^j |_{\vec c_0}$
is invertible we can solve for $\vec c$,
\be
c{}^i = c_0{}^i - (J^{-1})^i{}_j (\vec c_0) f{}^j(\vec c_0).  
\ee 
If $f$ depends linearly on the parameters, then one step is enough, for
nonlinear relations several steps have to be applied in order to approach
the zero.

Numerically, the Jacobian is computed by first integrating the ODEs
(\ref{ODEgen}) with the initial guess $\vec c_0$, and then with the
$N$ perturbed values $\vec c_0 + \delta \vec c_k, k= 1,\dots,N$ and
$(\delta c_k)^i = \epsilon \delta^i{}_k$. The Jacobian can then be obtained 
by e.g. forward differencing
\be
J^{i}{}_{k}(\vec c_0) = 
          \frac{f{}^{i}(\vec c_0 + \delta \vec c_k) - f{}^i(\vec c_0)}
                 {\epsilon}. 
\ee

%%%%%%%%%%%%

%\documentclass[12pt,a4paper]{report}
%\usepackage{german,a4,bbm,graphicx,psfrag}
%
%
%\include{diss_macros}

%\textwidth=15cm
%\textheight=22cm
%\topmargin=-2cm
%\oddsidemargin=1cm
%\pagestyle{plain}
%\parindent=0pt                    % no indentation for paragraphs,
%\parskip=5pt plus 2pt minus 1pt   % but do a little skip

\chapter{Discrete Fourier Transform}\label{app::DFT}
%%%%%%%%%%%%%%%%%%%%%%%%%%%%%%%%%%%%%%%%%%%%%%%%%%%%%%%%%%%%%%%%

In suitably chosen coordinates discrete self-similarity manifests itself 
in a periodic dependence on one of the coordinates. This 
suggests to work with Fourier expansions. 
The problem of constructing a DSS solution then reduces
to an ODE boundary value problem for the Fourier coefficients.
In Sec.~\ref{app::truncated_Fourier_series}
we review the basic properties of Fourier series and truncated Fourier
series. In practice we don't work with truncated Fourier series, but with 
the discrete Fourier transform, which establishes a relation 
between $N$ discrete Fourier coefficients and the values of the function
at $N$ ``grid points'' in real space. The discrete Fourier transform can be
viewed as the discrete approximation to the (continuous) Fourier transform.
(see Sec.~\ref{sec::DFT}). In Sec.~\ref{sec::Differentiation}
we define differentiation within this framework.
Sec.~\ref{sec::aliasing} explains how algebraic manipulations are executed
in ``real'' space, involving a pair of forward and backward 
transformations. There has to be taken special care in order to reduce
aliasing errors, that result from such a process.

This appendix follows closely 
\cite{Canuto-Hussaini-Quarteroni-Zang:pseudospectral}, which gives a good
description of discrete Fourier transform, pseudo-spectral
methods and aliasing errors. 
We only give the basic definitions and cite the main
results. For further details and proofs we refer to 
\cite{Canuto-Hussaini-Quarteroni-Zang:pseudospectral}.
Our discussion concerns functions defined on the interval
$[0,2 \pi]$. For functions, that are defined on the interval $[0,\Delta]$
any occurrence of the independent variable $x$ has to be replaced by
$2 \pi x/\Delta$.

\section{Truncated Fourier Series}\label{app::truncated_Fourier_series}
%%%%%%%%%%%%%%%%%%%%%%%%%%%%%%%%%%%%%%%%%%%%%%%%%%%%%%%%%%%%%%%%%%%%%%%

Consider the Hilbert space of (Lebesgue-) square integrable functions 
$L^2(0,2 \pi)$ with scalar product
\be\label{eq::scalar_product}
\left( u, v\right) = \int\limits_0^{2 \pi} u(x) \bar v(x) \ dx
\ee
and norm
\be\label{eq::norm}
||u|| = \int\limits_0^{2\pi} |u(x)|^2 \ dx.
\ee
The functions $\phi_k(x) = e^{i k x}, \ k \in \mathbbm Z$ form an
orthogonal system with respect to this scalar product
\ref{eq::scalar_product}
\be
\int\limits_0^{2 \pi} \phi_k(x) \bar \phi_l (x) \ dx = 2 \pi \delta_{kl}.
\ee
For $u \in L^2(0,2\pi)$ the Fourier coefficients of $u$ are given by
\be\label{eq::cont_Fourier_coeffs}
\hat u_k = \frac{1}{2\pi} \int\limits_0^{2\pi} u(x) e^{-i k x} \ dx   \qquad
k \in \mathbbm Z.
\ee
If $u$ is real then $\hat u_{-k} = \bar{\hat u}_k$.

The (formal) Fourier series of $u$ is given by 
\be
S u(x) = \sum\limits_{k=-\infty}^{\infty} \hat u_k \phi_k(x).
\ee 
The {\em truncated Fourier series of order N}\footnote{The convention to
discuss truncated Fourier series in terms of the trigonometric polynomial
of degree $N/2$ rather than $N$ is not common in the literature but is 
special to 
\cite{Canuto-Hussaini-Quarteroni-Zang:pseudospectral}} is the trigonometric
polynomial of degree $N/2$
\be
P_{N} u(x) = \sum\limits_{k = - N/2}^{N/2 -1} \hat u_k e^{i k x}
\ee
Defining the space of trigonometric polynomials of degree $N/2$ as 
\be\label{eq::span}
S_{N} = span\{ e^{ikx} | - N/2 \le k \le N/2 -1\}
\ee
$P_Nu$ is the orthogonal projection of $u$ upon the space $S_N$ with respect
to the scalar product \ref{eq::scalar_product},
\be
\left( P_N u, v \right) = \left (u, v\right) \qquad \textrm{for all } v \in
S_N.
\ee
Equivalently $P_N u$ is the closest approximation of $u$ within $S_N$ with
respect to the norm \ref{eq::norm}.

For $u \in L^2(0,2\pi)$  the Fourier series of $u$, $Su(x)$ converges to $u$
in the $L^2$ norm \ref{eq::norm}, that is
\be\label{eq::convergence_in_l2norm}
\int\limits_0^{2\pi} | u(x) - P_N(x) |^2 \ dx \to 0 \qquad \textrm{as } N
\to \infty.
\ee
The Parseval identity  states that
\be\label{eq::Parseval}
||u||^2 = 2 \pi \sum\limits_{k = - \infty}^{\infty} |\hat u_k|^2,
\ee
in particular the series on the right hand side converges.
If $u$ satisfies additional criteria, the convergence
\ref{eq::convergence_in_l2norm} can be improved. E.g.~ if $u$ is continuous,
periodic ($u(0^+) = u(2\pi^-)$), and of bounded variation on $[0,2\pi]$, then
$Su$ is uniformly convergent, i.e.
\be
\max\limits_{x \in [0, 2\pi]} | u(x) - P_N(x)| \to 0 \qquad \textrm{as } N
\to \infty.
\ee
Concerning the rate of convergence the Parseval identity \ref{eq::Parseval}
gives the following
\be\label{eq::error_l2}
|| u - P_N || = \biggl( 2 \pi \sum\limits_{{k < -N/2}\atop{k \ge N/2}} 
 |\hat u_k|^2  \biggr)^{1/2}.
\ee
On the other hand for $u$ sufficiently smooth and periodic we have
\be\label{eq::error_maxnorm}
\max\limits_{x \ \in \ [0, 2 \pi]} | u(x) - P_N(x) | \le 
   \sum\limits_{{k < -N/2}\atop{k \ge N/2}} 
 |\hat u_k|^2.
\ee
So the rate of convergence of the Fourier series is connected to how fast
the Fourier coefficients of $u$ decay.

We are interested in the following result: 
if $u$ is smooth ($C^{\infty}$) and periodic with all its derivatives on 
$[0, 2\pi]$ then the Fourier coefficients $\hat u_k$ decay faster than any
negative power of $k$. Of course this only applies for $k$ bigger than some
$k_0$, the minimal frequency which is needed to represent the 
essential structure of $u$. 

Combining this with the formulae for the error (\ref{eq::error_l2}),
(\ref{eq::error_maxnorm})
one finds that for $u$ satisfying the above conditions 
the error between $u$ and the truncated Fourier series 
decays faster than any negative power of $N$.
This is called {\em spectral accuracy}, {\em exponential convergence}
or {\em infinite order accuracy}. 

\section{Discrete Fourier Transform}\label{sec::DFT}
%%%%%%%%%%%%%%%%%%%%%%%%%%%%%%%%%%%%%%%%%%%%%%%%%%%%

For any integer $N > 0$ consider the ``grid points''
\be\label{eq::grid_points}
x_j = \frac{2 \pi j}{N} \qquad j = 0,\dots,N-1,
\ee
where for our purposes we assume $N$ to be even.
If $u$ is known at these grid points, then the 
{\em discrete Fourier transform} (DFT) of $u$ is given by
\be\label{eq::DFT}
\tilde u_k = \frac{1}{N} \sum\limits_{j=0}^{N-1} u(x_j) e^{-i k x_j} \qquad
-N/2 \le k \le N/2 - 1.
\ee
As the $e^{i k x_j}$ satisfy the orthogonality relation 
\be\label{eq::discrete_orthogonality}
\frac{1}{N} \sum\limits_{j=0}^{N-1} e^{i p x_j} = 
        \left\{ 
          \begin{array}{c c}
           1 & \textrm{if } p = N m, m \in \mathbbm Z \\
           0 & \textrm{otherwise},
          \end{array}
         \right.
\ee
the {\em inverse transform} is given by
\be\label{eq::inverse_DFT}
u(x_j) = \sum\limits_{k = - N/2}^{N/2 - 1}\tilde u_k e^{i k x_j} \qquad 
    j = 0, \dots, N-1.
\ee

Form a computational point of view, the discrete Fourier transform
(\ref{eq::DFT}) involves $N^2$ multiplications. It is therefore
an ``$O(N^2)$-process''. Fortunately there exists a less expensive
way to compute the $N$ coefficients $\tilde u_k$, which is called
{\em Fast Fourier Transform} (FFT). If $N$ is an integer power of $2$, 
then the computational costs for the discrete Fourier transform using FFT
are only of order $N \log_2 N$.
A good description of the FFT can be found 
e.g.~in \cite{Numerical-Recipes}. 

The polynomial 
\be\label{eq::trigonometric_interpolant}
I_N u(x) = \sum\limits_{k = -N/2}^{N/2-1} \tilde u_k e^{i k x}
\ee
is the $N/2$ degree {\em trigonometric interpolant} at the grid points
\ref{eq::grid_points}, i.e.~ $I_N u(x_j) = u(x_j)$. 
This polynomial is also called the {\em discrete Fourier series} of $u$.

The discrete Fourier coefficients $\tilde u_k$ can be regarded as a
discrete
approximation to the continuous Fourier coefficients $\hat u_k$, 
in that using the trapezoidal rule to evaluate the integral
\ref{eq::cont_Fourier_coeffs} gives $\tilde u_k$.  

The discrete approximation to the inner product \ref{eq::scalar_product}
on the space $S_N$ (\ref{eq::span}) is given by
\be\label{eq::scalar_product_discrete}
\left(u,v\right)_N = \frac{2 \pi}{N} \sum\limits_{j=0}^{N-1} u(x_j) \ 
     \bar v(x_j).
\ee
Due to (\ref{eq::discrete_orthogonality}) 
it coincides with the inner product \ref{eq::scalar_product}
if $u, v \in S_N$, i.e.~
\be
\left(u,v \right)_N = \left( u, v\right) \qquad \textrm{for all } u,v \in
S_N.
\ee 

The interpolation operator $I_N$ can be regarded as an orthogonal projection
operator upon the space $S_N$ with respect to the scalar product 
(\ref{eq::scalar_product_discrete}), as
\be
\left( I_N u, v \right)_N = (u, v)_N \qquad  \textrm{for all } v \in S_N  
\ee
trivially.
Therefore $I_N u$ is the best approximation to $u$ within the space $S_N$
with respect to the norm $||u||_N = \sqrt{ (u,u)_N}$.

The discrete Fourier transform of $u$ (\ref{eq::DFT}) can be expressed in
terms of the continuous Fourier coefficients of 
$u$ (\ref{eq::cont_Fourier_coeffs}): if $Su(x_j) = u(x_j)$ 
one obtains by using the relation
(\ref{eq::discrete_orthogonality})
\be\label{eq::relation_tilde_u_hat_u}
\tilde u_k = \hat u_k + \sum\limits_{{m = -\infty}\atop{m \ne 0}}^{\infty}
\hat u_{k + N m}. 
\ee
This means that the $k$-th mode of the trigonometric interpolant does not
only depend on the $k$-th mode of $u$ but also on all the $(k + N m)$-th
modes (which cannot be distinguished from the $k$-th frequency at the grid
points (\ref{eq::grid_points})). This effect is called {\em aliasing}.

We can write
\be
I_N u  = P_N u + R_N u,  
\ee
where
\be
R_N u (x) = \sum\limits_{k = -\infty}^{\infty} 
     \left( \sum\limits_{{m = -\infty}\atop{m = 0}}^{\infty}
            \hat u_{k + N m}\right) e^{ i k x}.
\ee
$R_N u$  is orthogonal 
to $u - P_N u$ with respect to the scalar product (\ref{eq::scalar_product})
and therefore we have
\be
|| u - I_N u||^2 = ||u - P_N u||^2 + || R_N u||^2.
\ee
So the error of the discrete Fourier series is always larger than the error
of the truncated Fourier series, due to $R_N u$, 
which is called {\em aliasing error}.
Nevertheless it can be shown, that asymptotically the truncation errors and
the interpolation errors decay at the same rate.

The sequence of interpolation polynomials shows similar convergence
properties as the sequence of truncated Fourier series.
E.g.~for $u$ continuous, periodic and of bounded variation on
$[0,2\pi]$, $I_N u$ converges uniformly to u on $[0,2 \pi]$.
Concerning the fall off of the discrete Fourier coefficients
we have e.g.~for $u$ being $C^{\infty}$ and periodic with all its 
derivatives: for any fixed $k \ne 0$ and any positive $N$ such that $N/2 >
|k|$, let $\tilde u_k = \tilde u_k^{(N)}$ be the $k$-th Fourier coefficient of
$I_N u$. Then Eq.~\ref{eq::relation_tilde_u_hat_u} shows, that
$|\tilde u_k^{(N)}|$ decays faster than algebraically in $k^{-1}$, uniformly
in $N$. Using analogous arguments as in
Sec.~\ref{app::truncated_Fourier_series}, we therefore get, that the error
between a periodic $C^{\infty}$ function and its discrete Fourier series
decays faster than any negative power of $N$.

In this work we use a slightly modified interpolating polynomial.
First we define 
\begin{eqnarray}
a_0 &=& \frac{1}{N} \sum\limits_{j=0}^{N-1} u(x_j) \\
a_l &=& \frac{2}{N} \sum\limits_{j=0}^{N-1} u(x_j) \cos(\frac{2 \pi l j}{N})
\quad l= 1,\dots N/2 -1 \\
b_l &=& \frac{2}{N} \sum\limits_{j=0}^{N-1} u(x_j) \sin(\frac{2 \pi l j}{N})
\quad l= 1,\dots N/2 -1 \\
a_{N/2} &=& \frac{1}{N} \sum\limits_{j=0}^{N-1} u(x_j) \cos(\pi k),
\end{eqnarray}
so 
\bea
\tilde u_k & = & \frac{1}{2}(a_k - i b_k), \qquad - N/2 +1 \le  k \le N/2 -1,
k\ne 0 \nonumber\\
\tilde u_0 & = & a_0, \qquad \tilde u_{-N/2} = a_{-N/2}.
\eea
As $a_{-k} = a_k$ and $b_{-k} = - b_{k}$ we can re-write the expression for
$u$ at the grid points (\ref{eq::inverse_DFT})
\be
u(x_j) = a_0 + \sum\limits_{k=1}^{N/2 -1}a_k \cos(k x_j)
+ \sum\limits_{k=1}^{N/2 -1}b_k \sin(k x_j) + 
a_{N/2} \cos(\frac{N}{2} x_j).
\ee
We now define the interpolating polynomial $\tilde I_N u$ to be 
\be
\tilde I_N u (x) = a_0 + \sum\limits_{k=1}^{N/2 -1}a_k \cos(k x)   
+ \sum\limits_{k=1}^{N/2 -1}b_k \sin(k x) +    
a_{N/2} \cos(\frac{N}{2} x)
\ee
By definition $\tilde I_N u(x)$ agrees with $I_N u(x)$ at the
grid points, but differs from the latter in between (the difference arising
solely in the $N/2$ frequency term).

\section{Differentiation}\label{sec::Differentiation}
%%%%%%%%%%%%%%%%%%%%%%%%%%%%%%%%%%%%%%%%%%%%%%%%%%%%%
The derivative of $\tilde I_N u(x)$ with respect to $x$ is given by
\be
(\tilde I_N u )' (x) = \sum\limits_{k=1}^{N/2 -1}a_k 
   (-k) \sin(k x)
+ \sum\limits_{k=1}^{N/2 -1}b_k  k
  \cos(k x) - 
a_{N/2} \frac{N}{2} \sin(\frac{N}{2} x)
\ee
As $\sin(\frac{N}{2}x)$ vanishes at the grid points $x_j$, we define 
the connection between the coefficients of the {\em collocation derivative}
of $u$ and the coefficients of $\tilde I_N u$ to be
\begin{eqnarray}
(ap)_0 & = & 0 \nonumber\\
(ap)_k & = & k \ b_k \qquad k = 1, \dots , N/2 - 1   \nonumber\\
(bp)_k & = & -k \ a_k \qquad k = 1, \dots , N/2 - 1  \nonumber\\
(ap)_{N/2} & = & 0,
\end{eqnarray}
where $(ap)_k$ and $(bp)_k$ denote the coefficients of 
the collocation derivative. Note that interpolation and 
differentiation do not commute, unless $u \in \tilde S_N$. One can show
that collocation differentiation is spectrally accurate.

\section{Pseudo-spectral methods and Aliasing}\label{sec::aliasing}
%%%%%%%%%%%%%%%%%%%%%%%%%%%%%%%%%%%%%%%%%%%%%%%%%%%%%%%%%%%%%%%%%%%

The way we construct the DSS solutions (see Sec.~\ref{sec::DSSsolutions})
our basic variables are the discrete Fourier coefficients (with respect to
$\tau$) of the field $\phi$, its derivative $\phi'$ 
and the metric functions $\beta$ and $\Vr$. In order to set up the
ODEs (in the spatial variable $z$) we have to compute the discrete Fourier
coefficients of e.g.~products of these functions or e.g.~the sine of $\phi$.
This can be done by applying the inverse Fourier transform
(\ref{eq::inverse_DFT}) to the coefficients, carrying out the
algebraic manipulations and taking the sine in ``$\tau$-space'' and then
transforming the result back to Fourier space.
The overall computational scheme therefore includes operations carried out
in Fourier space as well as operations carried out in ``$\tau$-space''.
Such a method is called {\em pseudo spectral method}.

This transforming back and forth has to be carried out with care if one
wants to keep the aliasing errors as small as possible. To see this
consider the smooth and periodic functions $u(x)$ and $v(x)$, with their 
Taylor series expansions $u(x) = \sum\limits_{k = -\infty}^{\infty} \hat u_k
e^{ikx}$ and $v(x) = \sum\limits_{k=-\infty}^{\infty} \hat v_k e^{ikx}$. We
denote their product by $w(x)$
\be
w(x) = u(x) v(x).
\ee
The Fourier coefficients of $w$, then are given by
\be\label{eq::product_exact}
\hat w_k = \sum\limits_{m+l = k} \hat u_m \hat v_n  \qquad 
             -\infty < m,l < \infty.
\ee

Given the discrete Fourier coefficients $\tilde u_k$ and $\tilde v_k$, 
and the corresponding inverse transforms $u_j = \sum\limits_{k = -
N/2}^{N/2-1} \tilde u_k e^{i k x_j}$ and the analogous expression for $v_j$,
we define $w_j$ to be their product in real space,
\be
w_j = u_j v_j.
\ee
The discrete Fourier coefficients of $w_j$ then are given by
$\tilde w_k = \frac{1}{N} \sum\limits_{j = 0}^{N - 1} w_j e^{-i k x_j}$ and
therefore
\be\label{eq::product_DFT}
\tilde w_k = \sum\limits_{m+l=k} \tilde u_m \tilde v_l + \sum\limits_{m+l =
k \pm N} \tilde u_m \tilde v_l \qquad -N/2 \le m,l,k \le N/2 - 1.
\ee
We assume for a moment, that the coefficients $\hat u_k, \hat v_k$ for $k
< - N/2, k \ge N/2$ are negligible. Then we have
$\tilde u_k \simeq \hat u_k$, $\tilde v_k \simeq \hat v_k$, i.e. the
aliasing errors due to interpolation are negligible, and the sum in 
(\ref{eq::product_exact}) equals the first sum in (\ref{eq::product_DFT})
for $-N/2 \le k \le N/2$. Then clearly the second sum in
(\ref{eq::product_DFT}) introduces an error into the product, which again is
called aliasing error.

In the following, we describe a method, which reduces this error.
The basic idea is to increase the number of Fourier coefficients to $M$ 
before transforming to real space, carry  out the manipulations in 
real space with this higher number of grid points, transform back to 
Fourier space and then throw away the additional modes. If $M$ is chosen
appropriately the second sum in (\ref{eq::product_DFT}) does not contribute
to the relevant frequencies of $w$.

Let $M > N$. We define the following $M$ discrete Fourier coefficients
\be
\tilde U_k = \left\{
               \begin{array}{cc}
               \tilde u_k  &   -N/2 < k < N/2 -1 \\
                    0      &    -M/2 \le k < -N/2; N/2 \le k \le M/2  
               \end{array} 
             \right.
\ee
and the analogous expressions for $\tilde V_k$.
Given these Fourier coefficients, we get the inverse transforms
$U_j = \sum\limits_{k = -M/2}^{M/2-1} \tilde U_k e^{i k x_j}$, where
$j$ now runs from $0$ to $M-1$ and the grid points $x_j$ are given by $x_j =
2 \pi j /M$ and the analogous expression for $V_j$. Let again $W_j = U_j
V_j$ and $\tilde W_k = \frac{1}{M}\sum\limits_{j = 0}^{M-1} 
W_j e^{- i k x_j}$, then
the coefficients $\tilde W_k$ are given according to (\ref{eq::product_DFT})
\be\label{eq::product_DFT_2}
\tilde W_k = \sum\limits_{m+l=k} \tilde U_m \tilde V_l + \sum\limits_{m+l =
k \pm M} \tilde U_m \tilde V_l \qquad -M/2 \le m,l,k \le M/2 - 1.
\ee
The strategy now is to consider only the coefficients $\tilde W_k$ for
$-N/2 \le k \le N/2 -1$ and throw away the higher frequencies.
As $\tilde U_m$ and $\tilde V_l$ are nonzero only for 
$-N/2 \le m,l \le N/2-1$, the first sum in (\ref{eq::product_DFT_2})
equals the first sum in (\ref{eq::product_DFT}) for 
$-N/2 \le k \le N/2 -1$. The aim is now to choose 
$M$ such, that the second sum in
(\ref{eq::product_DFT_2}) does not contribute to the frequencies of
interest. We have
$-N \le m+l \le N-2$ and $M-N/2 \le k + M \le M+ N/2 -1$ and $- M - N/2 \le
k - M \le -M + N/2 -1$. So if $M > 3N/2 -1$ the second sum in
(\ref{eq::product_DFT_2}) does not 
contribute to $\tilde W_k$ for $-N/2 \le k \le N/2 -1$.

For products of higher order $M$ has to be chosen larger.

%%%%%%%%%%%%%

%\documentclass[12pt,a4paper]{report}
%\usepackage{german,a4,bbm,graphicx,psfrag}
%
%
%\include{diss_macros}

%\textwidth=15cm
%\textheight=22cm
%\topmargin=-2cm
%\oddsidemargin=1cm
%\pagestyle{plain}
%\parindent=0pt                    % no indentation for paragraphs,
%\parskip=5pt plus 2pt minus 1pt   % but do a little skip

\chapter{The ``Diamond-Integral-Characteristic-Evolution'' Code -- DICE}
\label{app::dice}
%%%%%%%%%%%%%%%%%%%%%%%%%%%%%%%%%%%%%%%%%%%%%%%%%%%%%%%%%%%%%%%%%%%%%%%%

The DICE code evolves the self-gravitating SU(2) $\sigma$ model in 
spherical symmetry (optionally with cosmological constant)\footnote{It also
optionally evolves the self-gravitating massless Klein-Gordon field in
spherical symmetry}. Given initial data $\phi_0(r) = \phi(u_0,r)$ at the
initial null slice, Eqs.~(\ref{eq::phi}), (\ref{eq::betap}) and
(\ref{eq::Vp}) are solved numerically, while the ``additional'' Einstein
equations, Eqs.~(\ref{eq::Euur}) and (\ref{eq::Ethth}) 
merely serve for tests of consistency and accuracy (see Sec.~\ref{app::convergence}).
The diamond integral scheme by \Gomez{} and Winicour 
\cite{Gomez-Winicour-1992-in-dInverno,
Gomez-Winicour-1992-sssf-2+2-asymptotics,
Gomez-Winicour-1992-sssf-2+2-numerical-methods}
is used to integrate the wave equation (\ref{eq::phi}) (see
Sec.~\ref{NSWE}). Grid points freely fall along ingoing null geodesics. The
hypersurface equations (\ref{eq::betap}) and (\ref{eq::Vp}) as well as the
geodesic equation (\ref{eq::ingoing_null_geodesic}) in the latest version 
are integrated using a
second order iterated Runge-Kutta scheme 
(see Sec.~\ref{app::integrate_hypersurface_eqs}). In the vicinity of the
center of spherical symmetry the integration schemes (except for the
geodesic equation) are replaced by Taylor series expansions (this follows 
\cite{Garfinkle-1995-sssf-2+2-self-similarity}; see Sec.~\ref{app::Origin}).

The ``kernel'' of the DICE code was originally developed by Sascha Husa. 
Further improvement, development and testing of the code -- 
including improvement of the integration scheme for the ODEs,
measurement of the black hole mass, 
convergence tests, critical search modus etc. --
was done by Michael \Purrer{} and Jonathan Thornburg. M.~\Purrer{} 
also added the feature to evolve the massless Klein Gordon field with a
compactified radial coordinate (see \cite{Michaels-Diploma-Thesis}). 
I helped to develop the physical and analytical foundations of the code,
but did not actually participate in writing the code.
A description of the DICE code can be found
in \cite{Husa-Lechner-Puerrer-Thornburg-Aichelburg-2000-DSS} and 
in \cite{Michaels-Diploma-Thesis}.

\section{The NSWE Algorithm}\label{NSWE}
%%%%%%%%%%%%%%%%%%%%%%%%%%%%%%%%%%%%%%%%

The ``central'' evolution algorithm uses the NSWE scheme by 
Gomez and Winicour \cite{Gomez-Winicour-1992-in-dInverno,
Gomez-Winicour-1992-sssf-2+2-asymptotics,
Gomez-Winicour-1992-sssf-2+2-numerical-methods}. 
It is based on the fact, that the 
wave operator $\dal_{g}$ in spherical symmetry can be expressed in
terms of the wave operator in 2-dimensions $\dal_{h}$, which is defined
with respect to the two dimensional metric
\be
ds^2_h = - e^{2 \beta} du \left( \Vr du + 2 dr \right).
\ee

Setting $\psi := r \phi$, we can write
\be\label{eq::dal_g}
\dal_{g} \phi = \frac{1}{r} \dal_h \psi - \frac{e^{-2 \beta}}{r^2} \left(
  \partial_r \Vr \right) \psi.
\ee
Now we use the fact, that any two dimensional metric is conformally flat, 
i.e. by setting $d \tilde u = \Vr d u$ we have
\be
ds^2_h = \frac{e^{2 \beta}}{\Vr} d s^2_{\tilde h} \qquad \textrm{with}
   \quad d s^2_{\tilde h} = - d \tilde u \left ( d\tilde u + 2 dr \right).
\ee
The wave operator transforms under this conformal transformation as
\be\label{eq::dal_h}
\dal_h \psi = e^{-2\beta} \Vr \dal_{\tilde h} \psi.
\ee

\begin{figure}[h,t,b]
\begin{center}
\begin{psfrags}
 \psfrag{Origin}[]{Origin}
 \psfrag{N}[]{\bf N}
 \psfrag{S}[]{\bf S}
 \psfrag{W}[]{\bf W}
 \psfrag{E}[]{\bf E}
 \psfrag{ut=ut0}[l][c][1][0]{$\tilde u = \tilde u_0, \ u = u^k$}
 \psfrag{u=u0}[l][c][1][0]{}
 \psfrag{ut=ut1}[l][c][1][0]{$\tilde u = \tilde u_1, \ u = u^{k+1}$}
 \psfrag{u=u1}[l][c][1][0]{}
 \psfrag{vt = vt0}[l][c][1][0]{$\tilde v = \tilde v_0, \ v = v_{i-1}$}
 \psfrag{v=v0}[l][c][1][0]{}
 \psfrag{vt=vt1}[l][c][1][0]{$\tilde v = \tilde v_1, v = v_i$}
 \psfrag{v=v1}[l][c][1][0]{}
 \psfrag{Sig}[]{$\Sigma$}
\includegraphics[width=3.5in]{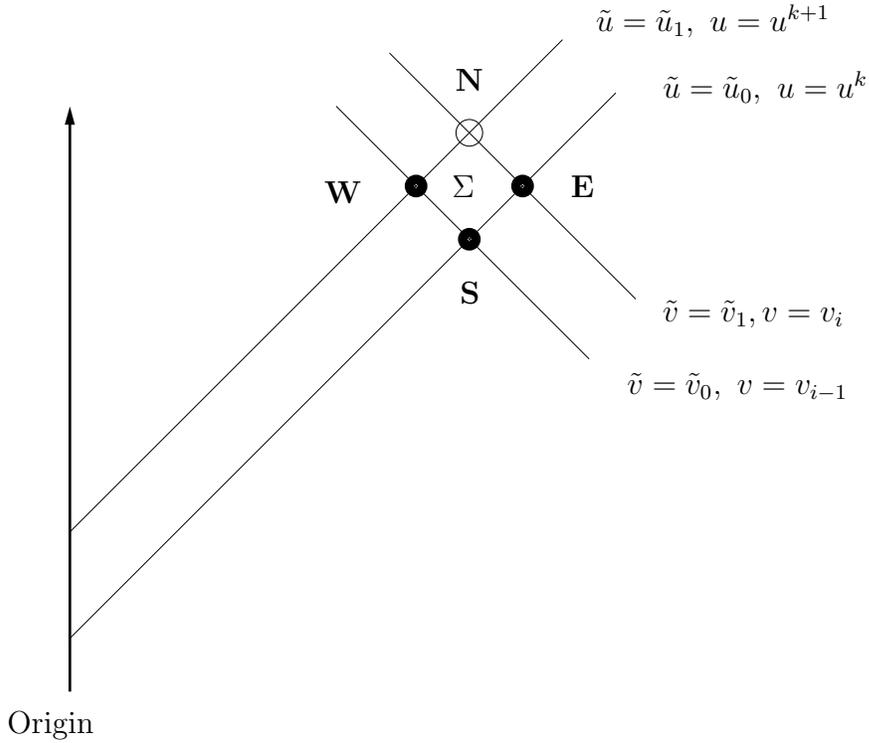}
\end{psfrags}
\end{center}
\caption{A schematic diagram illustrating the NSWE-algorithm. The two
dimensional wave operator in flat space $\dal_{\tilde h}$ acting on $\psi =
r \phi$ is integrated over the null parallelogram $\Sigma$.
$\psi_{N}$ therefore is given by $\psi$ at the points S, E, W
plus an integral over $\Sigma$.
}\label{fig::NSWE}
\end{figure}

Consider now the parallelogram $\Sigma$ spanned by the four null lines
$\tilde u = \tilde u_0, \tilde u= \tilde u_1, \tilde v= \tilde v_0, 
\tilde v= \tilde v1$ (see Fig. \ref{fig::NSWE})\footnote{$\Sigma$ is
chosen such that it is bounded by 
two null slices, separated by one numerical time step, and the two ingoing
null geodesics along which two neighbouring grid points move.}. 
If we integrate Eq. \ref{eq::dal_h} over $\Sigma$ we get
\be\label{eq::int_sigma}
\int\limits_{\Sigma} d^2 x \sqrt{- h}\dal_h \psi = 
    \int\limits_{\Sigma} d^2 x \sqrt{- \tilde h} \dal_{\tilde h} \psi.
\ee
In double null coordinates $(\tilde u, \tilde v = \tilde u + 2
r)$, $\dal_{\tilde h}$ reads
\be
\dal_{\tilde h} \psi = 
   - 4 \partial_{\tilde u} \partial_{\tilde v} \psi,
\ee
so (\ref{eq::int_sigma}) gives
\bea
\int\limits_{\Sigma} d^2 x \sqrt{- h}\dal_h \psi & = & - 2
  \int\limits_{\tilde u_0}^{\tilde u_1} 
  \int\limits_{\tilde v_0}^{\tilde v_1} 
    d {\tilde u} d {\tilde v} \partial_{\tilde u} \partial_{\tilde v}
    \psi = \nonumber\\
 & = & 2 (- \psi_{N} + \psi_W + \psi_E - \psi_S). 
\eea
From Eqs. (\ref{eq::phi}) and (\ref{eq::dal_g}) we have
\be
\dal_h \psi = \frac{1}{r} e^{-2 \beta} \left( \partial_r \Vr \right) \psi
   + \frac{\sin (2 \psi/r)}{r}
\ee
and therefore
\bea\label{eq::NSWE}
\psi_N & = & \psi_W + \psi_E - \psi_S - \frac{1}{2} 
   \int\limits_{\Sigma} d^2 x \sqrt{-h} \left( \frac{1}{r} 
     e^{-2 \beta} \left( \partial_r \Vr \right) \psi
   + \frac{\sin (2 \psi/r)}{r} \right) = \nonumber\\
  & = & 
  \psi_W + \psi_E - \psi_S - \frac{1}{2} 
   \int\!\!\!\int_{\Sigma} du \ dr \left( \frac{1}{r} 
     \left( \partial_r \Vr \right) \psi
   + \frac{e^{2 \beta} \sin (2 \psi/r)}{r} \right).
\eea
The integral on the right hand side can be approximated to second order by 
\be\label{eq::approx_integral}
\int\!\!\!\int_{\Sigma} du \ dr  f(u,r) \simeq 
   \frac{1}{2} \left(f_E + f_W \right) \Delta u \ \Delta r,  
\ee
where $\Delta u = u^{k+1} - u^{k}$ and 
    $\Delta r = \frac{1}{2} (r_E - r_S + r_N - r_W)$.

Assuming now, that the fields $\psi, \beta, \Vr$ and $\Vr'$ are known 
at the points $S, E, W$, and that furthermore the $r$ coordinate
of $N$ is known, then the field $\psi$ at $N$ can be computed 
via \ref{eq::NSWE} and \ref{eq::approx_integral}

\section{Treatment of the Origin}\label{app::Origin}
%%%%%%%%%%%%%%%%%%%%%%%%%%%%%%%%%%%%%%%%%%%%%%%%%%%%

The values of the field $\psi$ and the metric functions $\beta$ and 
$\Vr$ close to the origin are computed using a 
Taylor series expansion.
Using the fact that $\psi(u, 0) = \psi'(u,0) = 0, 
\partial_u \psi(u,0) = \partial_u \psi'(u,0) = 0$, 
etc.~and the field equation (\ref{eq::phi}), $\psi(u_0 + \Delta u, r)$ is
given by
\be\label{eq::Taylor_psi}
\psi(u_0 + \Delta u, r) = \frac{1}{2}\psi''(u_0,0) r^2 +  
      \frac{1}{6}\psi'''(u_0,0) r^3 + 
      \frac{1}{6}\psi'''(u_0,0) r^2 \Delta u 
+ O(\Delta^4). 
\ee
Using the hypersurface Eqs. (\ref{eq::betap}) and (\ref{eq::Vp})
together with the Taylor expansion (\ref{eq::Taylor_psi})
the expansions of $\beta$ and $\Vr$ at the time $u_0 + \Delta u$ read
\bea\label{eq::Taylor_metric}
\beta(u_0+\Delta u, r) & = & 
      \eta \Bigl( \    \frac{1}{16} \psi''(u_0,0)^2 r^2
                   + \frac{1}{24}\psi''(u_0,0) \psi'''(u_0,0 ) r^2 du +
                     \nonumber\\
         & + & \frac{1}{18} \psi''(u_0,0) \psi'''(u_0,0) r^3 \ \Bigr)
                   + O(\Delta^4)
                                                          \\
\Vr(u_0 + \Delta u, r) & = &   1 
                  -  \eta  \Bigl( \ \frac{1}{8} \psi''(u_0,0)^2 r^2 
                 + \frac{1}{12} \psi''(u_0,0) \psi'''(u_0,0) r^2 du +
                     \nonumber\\
          & + & \frac{1}{18} \psi''(u_0,0) \psi'''(u_0,0) r^3 \ \Bigr)
                   + O(\Delta^4).
\eea

At the time step $u^k$ the coefficients $\psi''(u^k,0)$ and 
$\psi'''(u^k,0)$ are extracted from $\psi$ by fitting the cubic 
polynomial $c_1 r^2 + c_2 r^3$ to $\psi$ at the 5 
innermost non origin grid points (a further version uses a fit
at the time steps $u^k$ and $u^{k-1}$ to compute $\psi''(u^k,0)$ and
$\psi'''(u^k,0)$).

After integrating the geodesic equation (\ref{eq::ingoing_null_geodesic}) 
the field $\psi$ at the
first three non origin grid points at the time step $u^{k+1}$ is
computed from (\ref{eq::Taylor_psi}). 
The metric functions $\beta$ and $\Vr$
at these grid points then are given by (\ref{eq::Taylor_metric}).

\section{Integrating the Hypersurface Equations and the 
%%%%%%%%%%%%%%%%%%%%%%%%%%%%%%%%%%%%%%%%%%%%%%%%%%%%%%%
       Geodesic Equation}\label{app::integrate_hypersurface_eqs}
       %%%%%%%%%%%%%%%%%%%%%%%%%%%%%%%%%%%%%%%%%%%%%%%%%%%%%%%%%

As the 
hypersurface equations (\ref{eq::betap}) and (\ref{eq::Vp})
at any fixed slice $u=u^k$ are ODEs, 
they can be integrated using a second order 
iterated Runge-Kutta scheme
\cite{Hyman-1989-MOL-in-Buchler}. 
For a general ODE ${\dot y} = F(y,t)$, defining
\be\label{eq::y_pred}
y_{pred}^{k+1} = y^k + \Delta t F(y^k, t^k),
\ee
$y^{k+1}$ is given as follows
\be\label{eq::y_k+1}
y^{k+1} = y^k + \frac{1}{2} \Delta t 
    \left(F(y^k, t^k) + F(y^{k+1}_{pred},t^{k+1}) \right).
\ee

At the initial hypersurface, given the initial configuration $\phi(u=0,r)$,
the metric functions can be computed by integrating the hypersurface
equations with the above scheme.
At any further slice $u^k$, the field $\psi$ has to be computed
at the $i$-th grid point, using the NSWE scheme, before the metric functions 
can be updated at this grid point, using (\ref{eq::y_pred}) and
(\ref{eq::y_k+1}).

The geodesic equation (\ref{eq::ingoing_null_geodesic}), 
again an ODE, also is integrated
with the above method. One difficulty arises due to the fact, that
at the time, we integrate the geodesic, the metric function $\Vr$ is not 
yet known at the next time step. This is solved by taking 
$\Vr$ and $\Vr'$ at the grid point $i-1$ at slice
$u^{k+1}$ and doing a linear extrapolation to the $i$-th grid point.

\section{Grid Refinements and Adaptive Time Steps}
%%%%%%%%%%%%%%%%%%%%%%%%%%%%%%%%%%%%%%%%%%%%%%%%%%
In order to be able to ``see'' type II critical collapse, where
the solution at the threshold of black hole formation is self-similar,
it is essential for the code to resolve widely varying scales both in
space and in time.
There are two features that enable the 
DICE code to manage this resolution: first, the grid points freely fall
along the ingoing null geodesics $v = const$. These geodesics 
tend to focus in regions of strong curvature, as they necessarily 
occur if the evolved solution stays close to a self-similar solution for
some time. This helps to increase the spatial resolution in such regions. 
Second, as the grid points eventually hit the origin and are dropped
from the grid, the number of grid points is doubled each time half of the 
grid points are lost (this follows
\cite{Garfinkle-1995-sssf-2+2-self-similarity})
(the values of the grid functions at these additional
grid points are obtained by interpolation).
This method is most effective, if in addition the outer boundary of the grid
is fine tuned to be (slightly outside but) close to the past SSH of the
self-similar solution (again this was used by Garfinkle 
in \cite{Garfinkle-1995-sssf-2+2-self-similarity}).
  
Concerning the time step, the scheme used is not restricted by a 
Courant-Friedrichs-Lewy (CFL) limit, as the numerical domain of dependence 
always equals the physical domain of dependence of the grid. 
Nevertheless, we need an upper bound for the time steps in order to 
get enough resolution in time. Following 
Refs.~\cite{Goldwirth-Piran-1987-sssf-2+2,
Goldwirth-Ori-Piran-1989-in-Frontiers} this is achieved by requiring
\be
\Delta u \le \frac{C \Delta r}{\Vr} ,
\ee
for all grid points. This restricts the time steps such that
no grid point is allowed to fall inwards more than
$C/2$ grid spacings within a single time step. 
(Usually $C$ is set to $1.5$).

Apart from providing enough resolution in time this restriction on the time
step also prevents the scheme to run into an apparent horizon
$\Theta_+ = 0$, where the code would crash due to $\beta \to \infty$.
Instead the time evolution slows down before the formation of an 
apparent horizon due to both $\Delta r$ going to zero at 
the apparent horizon (because $r$ fails to parameterize the outgoing null
rays $u=const$ there) and an increase in $\Vr$. 
This increase in $\Vr$ at a null slice close
to an apparent horizon can be explained as follows: first of all according
to Eq.~\ref{eq::betap} $\beta$ is non decreasing with $r$. So if it gets
large somewhere on the slice it stays large for all larger $r$.
Furthermore at large enough $r$ the field $\phi$ and its derivative are
small, and therefore the quantity $2m/r$ is decreasing. 
$\Vr$ can be written as $\Vr = e^{2 \beta}(1 - 2m/r)$, so $\Vr$ increases
monotonically and gets large outside the outermost peak of $2m/r$.

\section{Diagnostics, Accuracy and Convergence Tests}\label{app::convergence}
%%%%%%%%%%%%%%%%%%%%%%%%%%%%%%%%%%%%%%%%%%%%%%%%%%%%%%%%%%%%%%%%%%%%%%%%%%%%%

In order to test the accuracy of the code we use the following quantities.
First we compute the mass function \ref{eq::mass_ur}
in two different ways, namely
\be\label{eq::m_MS}
m_{MS} = \frac{r}{2} \left(1 - \Vr e^{-2 \beta} \right)
\ee
and 
\be\label{eq::m_rho}
m_{\rho} = \frac{\eta}{2}\int\limits_0^{r} dr r^2 \left(
        \Vr e^{-2 \beta} (\phi')^2 + 2 \frac{\sin^2(\phi)}{r^2} \right).
\ee
For a solution to the Einstein equations, both expressions are identical,
for a numerically computed solution nevertheless, they will differ by a
small amount due to finite differencing errors. Defining
\be
\delta m = \frac{m_{MS} - m_{\rho}}{m_{total, initial}},
\ee 
where $m_{total,initial}$ is the mass contained in the grid at 
the initial slice, we have a measure of accuracy, which should always be 
$\ll 1$.

The other quantities that serve as a check for accuracy are the Einstein
equations $E_{uur}$ (\ref{eq::Euur}) and 
$E_{\theta \theta} (\ref{eq::Ethth})$, which are not used
to compute the solution.
Here the question of normalization remains open in part. 
One possibility would be to define the sum of the absolute values
of each term contributing to the expressions $E_{uur}$ and
$E_{\theta\theta}$, calling them $E_{|uur|}, E_{|\theta\theta|}$
and use this as the normalization 
(this is e.g. done in 3+1 numerical relativity). 
The expressions $E_{uur}/E_{|uur|}, 
E_{\theta\theta}/E_{|\theta, \theta|}$ then should always be 
$\ll 1$. 
J.~Thornburg chose a different normalization, which does
not only consist of the sum of the absolute values but taking the maximum
over the whole slice of the absolute value of each term, and then taking the
sum. Calling this expression $E_{max|uur|}$  
he uses $E_{uur}/(1 + E_{max|uur|})$ and the analogous
expression for $E_{\theta\theta}$. 
The reason for taking the maxima was,
that as we are working on null slices and integrate outwards from the 
origin, errors occurring at some location in the slice are transported
outwards. The additional $1$ in the denominator reduces 
the error norm to the pure expression $E_{uur}$ near the origin, and was
introduced to make this norm better behaved close to the origin.
What speaks in favor of this normalization is the fact, that the 
error norms constructed like this in most of the experienced situations 
are small, when the solution is well behaved,
whereas they get large, when something is wrong.

A number of convergence tests was done by J. Thornburg and M. P\"urrer
to test, whether the accuracy really corresponds to second order.
Assuming that the fields are globally second order accurate, 
i.e. that the quantity $\Psi_N$, computed numerically with $N$ 
grid points differs from the true continuum solution $\Psi$ by
$\Psi_{N} = \Psi + c/N^2$, where $c$ is independent of the
resolution, and all higher order terms are neglected, 
then $\Psi_{2N}$, computed with
twice the resolution differs from $\Psi$ 
by $\Psi_{2N} = \Psi + c/(4 N^2))$.
Assume now that $\Psi = 0$ \footnote{If the value of $\Psi$ is not known in
advance, one has to compare
three different solutions, with $N$, $2N$ and $4N$ grid points. Second order
convergence then is given if
\be
\Psi_{4N} - \Psi_{2N} = \frac{c}{16 N^2} - \frac{c}{4 N^2} =
\frac{1}{4}(\Psi_{2N} - \Psi_{N}).
\ee}, 
as is the case e.g.~for the above defined quantities
$\delta m$, $E_{uur}$ and $E_{\theta \theta}$. 
Then
\be
\Psi_{2N} = \frac{c}{4 N^2} = \frac{1}{4} \Psi_{N}.
\ee
Three excellent examples of these convergence tests can be found
in the appendix of 
\cite{Husa-Lechner-Puerrer-Thornburg-Aichelburg-2000-DSS}, 
where second order convergence is shown for the quantity 
$\delta m$ for near critical solutions, and for the quantity $p^*$.
Further convergence tests of the DICE code can be found in
\cite{Michaels-Diploma-Thesis}.

An additional test on the code was done by M.~P\"urrer who 
re-investigated critical collapse of the massless Klein-Gordon field
with the DICE code, and found the critical solution to be DSS with the
reported \cite{Choptuik-1993-self-similarity} values of 
$\Delta$ and $\gamma$.
The fact that the Klein-Gordon DSS solution can be resolved with this code
is a good test for resolution, as the echoing period $\Delta \approx 3.43$ 
in this case is much bigger than for the $\sigma$ model for large couplings, 
and therefore is harder to resolve numerically.

We also mention the fact that the critical solution of the $\sigma$-model
for large couplings agrees rather well with the directly constructed
DSS solutions (see Figs.~\ref{fig::Crit_sol_DSS_eta.eq.100} and 
\ref{fig::Crit_sol_DSS_eta.eq.0.2933}).
This provides a good test for the DICE code as well as for the ``DSS code''.
 
\section{Measurement of the black hole mass}
%%%%%%%%%%%%%%%%%%%%%%%%%%%%%%%%%%%%%%%%%%%%

The formation of a black hole is signalled by the formation of an apparent
horizon $\Theta_+ = 0$ 
(or $\beta \to \infty$ or $\frac{2m}{r} \to 1$).
As explained in the last section,
the code slows down and stops before an apparent horizon forms.
The code  ``detects'' the black hole whenever $\frac{2m}{r}$ 
exceeds some threshold
close to $1$ anywhere in the grid. This threshold is usually set to be
$0.995$. At each time step the ``momentary'' black hole mass is defined
as the mass $m_{MS}$ of the outermost such grid point. 
At each further time step this ``momentary mass'' should increase,
corresponding to additional matter falling into the black hole.
If the numerical grid extends to large enough values of $r$ initially
and if the mass does not change substantially from one time step to 
the other, we have a good estimate for the ``final'' black hole mass.
After detecting a black hole the code either stops because 
the number of time steps (after black hole detection) exceeds some limit
or because the time steps shrink to $du/u < 10^{-15}$ (which corresponds to
the order of floating point roundoff errors). 

For large couplings the runs were made with an initial spatial extension
of the grid, which lay substantially outside the backwards light cone of
the critical solution. For smaller couplings on the other hand, the outer
boundary of the grid had to be fine tuned to be close to the backwards
light cone, in order to get enough resolution. In this case clearly
the final numerical estimate of the mass does not 
correspond to the final mass of the black hole, but rather measures the
apparent horizon close to the past SSH of the critical solution.

%%%%%%%%%%%%%%

%
%
%   Konventionen etc.
%   
%last changed: DO, 4.10.
%
%

\chapter{Conventions}\label{app::conventions}
%%%%%%%%%%%%%%%%%%%%%%%%%%%%%%%%%%%%%%%%%%%%%

Conventions concerning curvature quantities and the signature of the metric
are the same as in \cite{Wald}. In particular we have

\begin{description}

\item[Signature:]
          \be
          (-+++)
          \ee

\item[Riemann tensor:]
     \be
     {\mathcal R}^{\sigma}_{\rho\mu\nu}=\Gamma^{\sigma}_{\nu\rho,\mu}
   -\Gamma^{\sigma}_{\mu\rho,\nu} + \Gamma^{\alpha}_{\nu\rho}
   \Gamma^{\sigma}_{\alpha\mu} - \Gamma^{\alpha}_{\mu\rho}
   \Gamma^{\sigma}_{\alpha\nu}
   \ee

\item[Ricci tensor:]
\be
{\mathcal R}_{\rho\nu} = {\mathcal R}^{\sigma}_{\rho\sigma\nu}
\ee

\item[Scalar curvature:]
\be
\mathcal R = g^{\rho\nu}\mathcal R_{\rho\nu}
\ee

\item[Action:]
\be
S = \int d^{4}x \sqrt{-g} \{ \frac{1}{2\kappa} 
   (\mathcal R - 2 \Lambda) + \mathcal{L}_{\mathcal{M}}
    \}
\ee

\item[Stress energy tensor:]
\be
2\int d^{4}x \delta_{g}(\sqrt{-g}\mathcal{L}_{\mathcal{M}}) =
\int d^{4}x \sqrt{-g}(-T_{\mu\nu})\delta g^{\mu\nu},
\ee
where $\delta g^{\mu\nu} \equiv (\delta g^{-1})^{\mu\nu}$

\item[Einstein equations:]
\be
G_{\mu\nu} + \Lambda g_{\mu\nu} = \kappa T_{\mu\nu}
\ee

\end{description}

Throghout this work the speed of light is set to unity, $c=1$.

%%%%%%%%%%%%%%%%%%%%%

\end{appendix}

%\include{bibl}
%%%%%%%%%%%%%%%

%\bibliography{cl}
%%%%%%%%%%%%%%%%%

%\include{Curr.vit}
%%%%%%%%%%%%%%%%%%

\end{document}